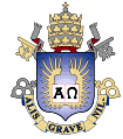

**Kleinner Silva Farias de Oliveira**

# Empirical Evaluation of Effort on Composing Design Models

## TESE DE DOUTORADO

Thesis presented to the Programa de Pós-Graduação em Informática of the Departamento de Informática, PUC-Rio as partial fulfillment of the requirements for the degree of Doutor em Informática

Advisor: Prof. Alessandro Garcia
Co-Advisor: Prof. Carlos José Pereira de Lucena

Rio de Janeiro
March 2012

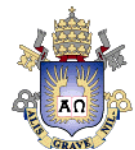

PONTIFÍCIA UNIVERSIDADE CATÓLICA
DO RIO DE JANEIRO

**Kleinner Silva Farias de Oliveira**

## Empirical Evaluation of Effort on
## Composing Design Models

**Thesis presented to the Programa de Pós-Graduação em Informática, of the Departamento de Informática do Centro Técnico Científico da PUC-Rio, as partial fulfillment of the requirements for the degree of Doutor.**

**Prof. Alessandro Garcia**
Advisor
Departamento de Informática – PUC-Rio

**Prof. Carlos José Pereira de Lucena**
Co-Advisor
Departamento de Informática – PUC-Rio

**Prof. Arndt von Staa**
Departamento de Informática – PUC-Rio

**Profª. Karin Koogan Breitman**
Departamento de Informática – PUC-Rio

**Prof. Toacy Cavalcante de Oliveira**
Universidade Federal do Rio de Janeiro – UFRJ

**Profª. Christina von Flach Garcia Chavez**
Universidade Federal da Bahia – UFBA

**Prof. José Eugenio Leal**
Coordinator of the Centro Técnico Científico da PUC-Rio

Rio de Janeiro, 21 March 2012

**Kleinner Silva Farias de Oliveira**

He received his BSc in Computer Science from the Federal University of Alagoas in 2006 and in Information Technology from the Federal Institute of Alagoas in 2006. He received his MSc in Computer Science from the Pontifical Catholic University of Rio Grande do Sul in 2008.

Bibliographic data

CDD: 004

To my family

# Acknowledgments

I am honored and grateful to have counted with excellent professionals along the way of this work. Especially, I would like to thank my supervisor, Prof. Dr. Alessandro Garcia, who provided constant encouragement, guidance, and freedom to develop this thesis. His competent and constructive criticism was essential to my success, my sincere admiration to him.

I am very grateful to have Prof. Dr. Carlos Lucena as one of my supervisors. I would like to thank him for all support on my research path by guiding me with his extensive knowledge, providing advice, and encouragement in the early stages.

I am indebted to Jon Whittle (Lancaster University, UK) for his in-depth reviews of many papers, and numerous interesting discussions and suggestions about the critical points throughout this work, lending a sense of reality to the research being conducted. I received so much from him to enlarge my vision on software engineering.

I could not forget to thank Dr. Toacy Oliveira, who trusted me fully from the beginning of my research career. Thanks to his mentoring, friendship, and all frequent discussions that help me to guide my professional life.

My gratefulness also goes to all my colleagues and professors from the Computer Science Department at PUC-Rio, especially to my friends from the Software Engineering Laboratory (LES) who gave me opportunities to grow as professional putting my ideas into action. It has been a privilege working on that stimulating environment.

During the development of this work, I was lucky to have collaborated with a number of research colleagues who contributed to this thesis in different ways and who allowed me to collaborate with their works as well. It was a pleasure to work with all of them on a number of papers.

I am also thankful to the members of my examination committee, who has generously contributed their time and expertise.

Thanks to my friends (whose names I did not quote to avoid being unfair to any of them right now) for giving invaluable tips, and giving lots of very useful advice. Especially, I am deeply grateful to my fiancée Carla Pedroso for putting up with me, and giving me so much that I could not even itemize them. Your love and understanding were instrumental in this journey. I love you so much.

Finally, I would like to thank my family for the constant support and belief in me in every conceivable way. This thesis is dedicated to my grandmother, Valdenira (in Memoriam), and my parents, Sandra and Carlos, my sisters, Kelyne and Klyvia, and my brother, Kleberson, with all my love.

I would like to express here my gratitude to CAPES/CNPq for the financial support of this doctoral study.

# Resumo

Composição de modelos desempenha um papel fundamental em muitas atividades de engenharia de software como, por exemplo, evolução e reconciliação de modelos conflitantes desenvolvido em paralelo por diferentes times de desenvolvimento. Porém, os desenvolvedores têm dificuldades de realizar análises de custos e benefícios, bem como entender o real esforço de composição. Sendo assim, eles são deixados sem qualquer conhecimento prático sobre quanto é investido; além das estimativas de evangelistas que frequentemente divergem. Se o esforço de composição é alto, então os potenciais benefícios tais como aumento de produtividade podem ser comprometidos. Esta incapacidade de avaliar esforço de composição é motivada por três problemas: (i) as abordagens de avaliação atuais são inadequadas para mensurar os conceitos encontrados em composição, por exemplo, esforço e conflito; (ii) pesquisadores não sabem quais fatores podem influenciar o esforço de composição na prática. Exemplos de tais fatores seriam linguagem de modelagem e técnicas de composição que são responsáveis para manipular os modelos; (iii) a falta de conhecimento sobre como tais fatores desconhecidos afetam o esforço de composição. Esta tese, portanto, apresenta uma abordagem de avaliação de esforço de composição de modelos derivada de um conjunto de estudos experimentais. As principais contribuições são: (i) um modelo de qualidade para auxiliar a avaliação de esforço em composição de modelos; (ii) conhecimento prático sobre o esforço de composição e o impacto de fatores que afetam tal esforço; e (iii) diretivas sobre como avaliar esforço de composição, minimizar a propensão a erros, e reduzir os efeitos negativos dos fatores na prática de composição de modelos.

## Palavras-chave

Composição de modelos, esforço de desenvolvimento, estudos empíricos.

# Abstract

Model composition plays a central role in many software engineering activities such as evolving models to add new features and reconciling conflicting design models developed in parallel by different development teams. As model composition is usually an error-prone and effort-consuming task, its potential benefits, such as gains in productivity can be compromised. However, there is no empirical knowledge nowadays about the effort required to compose design models. Only feedbacks of model composition evangelists are available, and they often diverge. Consequently, developers are unable to conduct any cost-effectiveness analysis as well as identify, predict, or reduce composition effort. The inability of evaluating composition effort is due to three key problems. First, the current evaluation frameworks do not consider fundamental concepts in model composition such as conflicts and inconsistencies. Second, researchers and developers do not know what factors can influence the composition effort in practice. Third, practical knowledge about how such influential factors may affect the developers' effort is severely lacking. In this context, the contributions of this thesis are threefold: (i) a quality model for supporting the evaluation of model composition effort, (ii) practical knowledge, derived from a family of quantitative and qualitative empirical studies, about model composition effort and its influential factors, and (iii) insight about how to evaluate model composition efforts and tame the side effects of such influential factors.

## Keywords

Model composition, development effort, empirical studies.

# Table of Contents

# List of Figures

# List of Tables

# List of Acronyms and Abbreviations

IBM – International Business Machine

UML – Unified Modeling Language

AO – Aspect-Oriented

AOM – Aspect-Oriented Modeling

CBO – Coupling Between Object Classes

EMF – Eclipse Modeling Framework

GQM – Goal Question Metric

IBM – International Business Machine

IDE – Integrated Development Environment

LCOM – Lack of Cohesion in Methods

MDD – Model Driven Development

MVC – Model View Controller

OCL – Object Constraint Language

OMG – Object Management Group

OO – Object-Oriented

RQ – Research Question

SPL – Software Product Line

UML – Unified Modeling Language

ECL – Epsilon Comparison Language

EML – Epsilon Merge Language

*Believe in your dreams*

Kleinner Farias

# 1
# Introduction

Model composition plays a central role in many software engineering activities, e.g., evolving design models to add new features (Thaker et al., 2007; Jayaraman et al., 2007) and reconciling models developed in parallel by different development teams (Wagner et al., 2003; Perry et al., 1998; Berzins, 1994). In fact, developers use model composition throughout the software development process, from the initial stage by integrating abstract design models (e.g., conceptual models) to the final stage by composing more detailed ones (e.g., UML class and sequence diagrams). In collaborative software development, for example, separate development teams may concurrently work on specific parts of an overall design model that are more relevant to them. However, it is necessary at some point to bring these models together in order to create a "big picture view" of the overall design model. For this reason, to date, there has been a significant body of research about model composition in the areas of model management (IBM, 2012), integration of software product lines (Jayaraman et al., 2007), and software merge (Mens, 2002).

The term *model composition* can be briefly defined as a set of tasks that should be performed to combine two (or more) input models, $M_A$ and $M_B$, in order to produce an output intended model, $M_{AB}$ (Brunet et al., 2006; Mens, 2002; Clarker, 2001). However, an output composed model, $M_{CM}$, is usually produced instead of $M_{AB}$. While the $M_{CM}$ would be the model produced by a model composition technique, the $M_{AB}$ is, in fact, the model intended by developers. The $M_{CM}$ often needs to be reviewed and changed to become compliant with $M_{AB}$. These models seldom match ($M_{CM} \neq M_{AB}$) as some properties of the $M_A$ and $M_B$ conflict with each other. If not properly handled, these conflicts may cause syntax and semantic inconsistencies in $M_{CM}$. Therefore, in order to transform $M_{CM}$ into $M_{AB}$, developers must also invest effort to identify and resolve these inconsistencies.

In practice, developers use model composition if they understand the effort to obtain $M_{AB}$. However, developers are unable to grasp the composition effort and realize any cost-effectiveness analysis. Hence, they are left without any practical knowledge about the effort to be invested in order to compose the design models apart from evangelists' anecdotal feedback, which often diverge from each other. If model composition is an error-prone and effort-consuming activity, then the potential benefits, e.g., gains in productivity, can be compromised. This inability of evaluating composition effort is due to three problems. First, the current measurement approaches are inadequate to assess the concepts found in model composition, such as specific effort dimensions, conflicts, and inconsistencies. Second, researchers and developers do not know the factors that can influence the composition effort in practice. Examples of key factors would be: (i) the design decomposition (e.g., object-oriented design or aspect-oriented design) represented by a certain modeling language, and (ii) the selected composition technique (e.g., IBM Rational Software Architecture) that is responsible for supporting the composition of design models. Third, practical knowledge about how the influential factors may affect the developers' effort is severely lacking. To date, there exists a clear need for addressing these problems as software modeling is increasing collaborative (France & Rumpe, 2007). If the effort on model composition is high, then the potential benefits (e.g., effectiveness in producing $M_{AB}$) of using model composition can be hindered in real projects.

It is important to address these problems due to several other reasons. First, before adopting, for example, a model composition technique in practice, developers need appropriate evaluation frameworks to reveal the actual effort to obtain $M_{AB}$ in practical settings. This decision should be supported by practical knowledge rather than evangelists' estimation. Second, by knowing the influential factors on model composition effort, they can make decisions more effectively. For example, at the early stages of software projects, developers need to choose which design decomposition will be used (e.g., object-orientation or aspect-orientation), which design characteristics will be applied to the design models (e.g., stability), and which composition technique will be adopted (e.g., IBM RSA or Epsilon). In addition, developers can reduce side effects of such decisions if they can rely on such knowledge up front. For example, developers can use a particular type of composition technique in software evolution scenarios where they are known to be more cost-effective than others can. Third, by empowering researchers with lessons learned from empirical studies, they can precisely improve existing modeling languages and composition techniques, thereby reducing the error likelihood and effort of composing design models.

With these issues in mind, it is particularly important, albeit challenging, to measure effort and understand the factors that can jeopardize the composition of design models. The definition of software metrics and the execution of empirical studies have been pointed out as a powerful way to gather empirical evidence in software engineering fields (Fenton & Pfleeger, 1997) as well as to derive lessons learned (Kitchenham et al., 2008; Wohlin et al., 2000). The remainder of this Chapter is organized as follows. Section 1.1 presents the problem statement. Section 1.2 describes the limitations of the related work. Section 1.3 describes the study methodology. Section 1.4 elaborates the key contributions of this thesis. Finally, Section 1.5 describes how the next chapters are organized.

## 1.1.
## Problem Statement

The problem of empirical evaluation of model composition effort is rooted in the inadequate support for measuring this effort and the lack of practical knowledge to design empirical studies in this context. In fact, current studies on model composition neither explicitly take into account effort as a measurement unit nor even provide indicators about how developers invest effort in practice. The current measurement methods for software design aim at simply quantifying specific properties of object-oriented (OO) decompositions (such as, degree of inheritance) and general properties of design models (e.g., coupling and cohesion), thereby failing to provide effective indicators for model composition effort. For example, from a sequence of output composed models, developers should be able to identify those models that are likely to have a high concentration of inconsistencies, which require a higher effort to produce the intended model. Indicators can help developers to identify those critical models.

Unfortunately, researchers are unable to properly evaluate model composition efforts nowadays. Hence, developers often make misinformed decisions without empirical knowledge about factors affecting model composition effort. For instance, the effort of applying a particular composition technique to compose UML models might be higher depending on the type of software change being realized. In addition, it might be that the composition effort of more modularized models might be substantially reduced. If so, this means that developers should invest more effort on improving the modularity of input design models before they are composed. If empirical knowledge of these factors is not available, designers are likely to invest much higher effort than what is needed when carrying out model composition. They are also likely to spend undesirable effort to detect and resolve inconsistencies because of misinformed decisions.

In addition, before adopting model composition in practice, it is necessary to have actual evidence of the effort that developers should invest to compose design models. The lack of appropriate measurement approaches jeopardizes the execution of empirical studies. In other words, without experimental investigations, model composition cannot be widely accepted in practice. This means that researchers are unable to properly test hypotheses, analyze correlations between variables, and perform comparative analysis of two or more empirical studies. Then, it is not possible to create a credible body of knowledge on composition effort supported by empirical evidence.

These shortcomings become more apparent in an age that model composition is starting to play a central role in many software engineering activities. In fact, model composition techniques are essential to support the evolution of design models in order to add new features (Thaker et al., 2007; Jayaraman et al., 2007) and reconcile models developed in parallel by different development teams (IBM, 2011; Wagner et al., 2003; Perry et al., 1998; Berzins, 1994). Unfortunately, model composition may become an effort-consuming task as the lack of knowledge about the influential factors (such as type of composition technique, design modeling language, and design characteristic) can bring harmful effects to the composition effort. The absence of a cost-effectiveness analysis, supported by effort indicators and experimental investigations, makes challenging the activity of composing design models. Therefore, researchers and developers need guidance for assessing model composition effort quantitatively and qualitatively.

## 1.2.
## Limitations of Related Work

To the best of our knowledge, this thesis is the first work aimed at: (i) carrying out a series of empirical studies on model composition effort so that a body of empirical knowledge in this field can be created and refined in the future; and (ii) defining support for the evaluation of model composition effort. In fact, it is well known that empirical studies in model composition are severely lacking. A previous roadmap study of model-driven software development (France & Rumpe, 2007) highlights that the state of the practice in assessing model composition provides evidence that the composition of design models is still in the "craftsmanship era." In (Mens, 2002), the author also points out the need to empirically evaluate the effort that developers invest to compose software artifacts, in particular, when using the most commonly used design models, such as component diagrams and class diagrams.

This thesis identified two critical limitations in the current related work. First, the traditional measurement approaches are unable to support the analysis of model composition effort. Second, the current literature in model composition fails to provide empirical knowledge about how developers spend effort to produce an output intended model. These limitations are described as follows.

**Limitation of Traditional Measurement Approaches**

Researchers and developers are increasingly concerned with defining software metrics for different software engineering fields (Basili, 2007). This need is attested by the high number of many measurement approaches proposed over the last decade, e.g., (Chidamber & Kemerer, 1994; Fenton & Pfleeger, 1997; Chidamber et al., 1998). These measurement approaches focus on quantifying particular properties of software products. As far as evaluation of model composition effort is concerned, the conventional measurement approaches suffer from two types of major criticisms.

First, most of the existing product metrics is focused on supporting the assessment of particular forms of design decomposition, such as object-oriented (OO) software design. Typically, such metrics suites aim at quantifying attributes of OO systems, such as data abstraction, encapsulation, polymorphism, and inheritance usage. Such attributes often require more than one metric to be entirely characterized. Each metric quantifies properties of an object-oriented decomposition, such as classes and their relationships. The operational definition of these metrics relies on the constructs of the OO programming languages (e.g., Java and C++) and OO design modeling languages (e.g., UML). Examples of these constructs are UML packages, components, classes, and relationships that are specified in the UML metamodel.

For instance, Chidamber and Kemerer proposed a metrics suite to quantify some of these attributes in OO designs or programs (Chidamber & Kemerer, 1994). Examples of such metrics are coupling between objects, cohesion in methods, depth of inheritance, and so forth. In 1998, Chidamber and colleagues evaluated those metrics in order to assess their usefulness for practicing managers (Chidamber et al., 1998). In 1997, Fenton and Pfleeger formally analyzed the same metrics by applying basic criteria from measurement theory; their goal was also to offer an accessible and comprehensive introduction to software metrics with an emphasis on real-world applications (Fenton & Pfleeger, 1997). However, the aforementioned measurement approaches do not take into account the particularities of model composition activities. They only quantify static attributes of object-oriented software artefacts. Therefore, they cannot be directly used to improve our empirical understanding about model composition effort. These quantification methods are in stark contrast with the needs required by the effort measurement addressed in this thesis.

A second limitation of the existing measurement approaches is their inability to evaluate specific activities of model composition. During the composition process, developers execute a set of tasks to combine two input models ($M_A$ and $M_B$) and produce an output intended model ($M_{AB}$). Examples of these tasks would be the application of the composition techniques and the resolution of inconsistencies in the composed model. The execution of each task consumes effort. By knowing the effort invested in each model composition task, developers may identify forms of alleviating the overall composition effort. Unfortunately, the traditional measurement approaches are unable to capture effort spent on specific model composition activities. Researchers do not know which and how model composition artefacts, produced in each task, should be quantified. This lack of effective measurement approaches for model composition effort also hinders the design and execution of empirical studies.

**The Lack of Practical Knowledge on Model Composition Effort**

Researchers and developers acknowledge the importance of practical knowledge about the model composition effort. In general, the current works propose new model composition techniques and superficially assess the proposed solutions. Reviewing the current literature, existing works make use of and evaluate software composition techniques in the realm of configuration management (Aiello, 2010a; Perry et al., 2001; Grinter, 1997; Rochkind, 1975). These studies focus on the composition of code and assess the technical feasibility of the techniques. Perry and colleagues investigated the composition of code in the context of collaborative software development (Perry et al., 2001). The authors realize an observational case study to understand how concurrent changes in large-scale software systems happen. The main results indicate that the degree of parallelism is very high, i.e., higher than considered by tools; and there is a significant correlation between the degree of parallel work on a given component and the number of quality problems it has.

However, little has been done to understand how developers invest effort in real-world settings. Today, it is well known that empirical studies on model composition are severely lacking. This scenario is still aggravated when considering composition effort. In fact, experts in the literature recently highlighted the scarcity of empirical studies (France & Rumpe, 2007). Additionally, the authors not only recognize but also recommend the execution of empirical studies to evaluate the impact of parallel changes on the development effort (Mens, 2002; Perry et al., 2001). In addition, they reinforce that empirical studies would allow researchers to evaluate the scalability of current composition techniques, to weigh the trade-offs in effort, and understand why and in what situations one approach might be better than another might.

In a broader context, we have also observed that many techniques have been proposed and incorporated into tools over the last decades. Examples of these techniques are SVN (SVN, 2011) and GIT (GIT, 2011). Using these tools, developers can control the evolution of software artefacts. In practice, these techniques help developers to check out artefacts for editing and then checking them back (Grinter, 1997; Rochkind, 1975). By controlling and registering these two activities, such techniques manage the evolution of the artefacts. In the seminar paper (Altmanninger et al., 2009), Altmanninger and colleagues apply the state-of-the-art versioning systems and analyze the challenges coming along with merging different versions of one model.

Other authors investigate the identification of conflicting changes by providing workspace awareness tools (Sarma et al., 2012; Burn et al., 2011a; Sarma et al., 2008). These tools are able to proactively identify overlapping changes between software artefacts such as code. The authors advocate that earlier contradicting changes are detected, the easier they are to resolve (Sarma et al., 2012). Sarma and colleagues propose a tool, named Palantír, which provides users with information about relevant ongoing parallel changes occurring in private workspaces, thereby enabling the early detection and resolution of potential conflicts.

Although these techniques are robust and broadly used in industry, nothing has been done to investigate about the effort to compose software artefacts. In (Uhl, 2008), Uhl points out that the model composition is more challenging than code composition. One of the reasons is because model composition involves the comparison and composition of graphical views, forms, dialogs, and property sheets as well as text. In fact, they are much more difficult to compare, mostly because visualizing the differences in a usable way is difficult. Moreover, Mens (Mens et al., 2002) also reinforces that the need for more empirical and experimental research regarding the amount of effort required resolving the composition inconsistencies.

To sum up, we observe that: (1) researchers do not even know which factors can, in fact, affect the composition effort; (2) nothing has been done to define how to evaluate the composition effort; and (3) there exists no cost effectiveness analysis about the model composition effort in order to support (or not) its well-informed use in practice.

## 1.3.
## Study Methodology

The main goal of this thesis is to define an evaluation approach for model composition effort, thereby gathering empirical knowledge about the effort of composing design models. Based on this empirical knowledge, we aim at generating insight about how to reduce the composition effort model. This aimed will be achieved by understanding the side effects of influential factors on model composition effort. With this in mind, the goal of this study is formulated based on the GQM template (Basili et al., 1994) as follows:

**General Goal:** *Analyze* the influential factors *for the purpose of* investigating their effects *with respect to* model composition effort *from the perspective of* developers *in the context of* the evolution of design models.

To address that general goal, we formulate an overall research question (RQ), which is presented below:

- **RQ_overall:** How can the composition of design models be evaluated, in particular, with respect to developers' effort?

This general research question is elaborated into more detailed research questions, which require proper measurement means and empirical studies on model composition effort. The first research question (RQ1) addresses the need for providing an approach to support model composition evaluation. RQ1 is designed as follows:

- **RQ1:** How can the evaluation of model composition be organized in terms of a comprehensive framework?

The composition effort may be affected by a wide range of influential factors. In this thesis, we decided to study three factors that are fundamental to produce an expected output composed model: (i) the composition technique being employed, (ii) the design decomposition techniques, and (iii) the structural characteristics of the design models involved in the composition. The first factor is the type of model composition technique, which can be categorized into heuristic-based composition techniques (IBM RSA, 2011) and specification-based composition techniques (Epsilon, 2011). This factor, discussed in Section 2.4, may affect the effort that developers invest to combine the input models in order to produce an output intended model.

The second research question (RQ2) aims at evaluating the relative effort of composing the input models by applying heuristic-based and specification-based composition techniques. Each of these alternative techniques might require less effort in specific or all scenarios involving software evolution – the context of our studies of model composition. Then, we investigate the effects of using different composition techniques to produce the output intended model. RQ2 is stated as follows:

- **RQ2:** What is the effort of composing design models with specification-based composition techniques and heuristic-based composition techniques?

The third research question (RQ3) analyzes the effort of detecting inconsistencies. Detection of inconsistencies requires that developers inspect the elements of the composed model, which are structured according to the selected design decomposition. Therefore, we analyze the effects of significantly different forms of design decomposition (i.e., object-orientation and aspect-orientation) on the quality of the output composition. In particular, our goal is to understand how different design decompositions affect the inconsistency rate, the inconsistency detection effort, and the degree of misinterpretations of the output composed models. RQ3 is presented below:

- **RQ3:** What is the effect of design decomposition techniques in particular with respect to misinterpretation, inconsistency rate, inconsistency detection effort, and inconsistency resolution effort?

The fourth research question (RQ4) analyzes the effort of resolving inconsistencies. That is, we investigate the effort that developers invest to transform an output composed model into an intended model. Additionally, we analyze if well known design characteristics (Martin, 2003; Meyer, 1997), such as model stability (Section 2.6.1), may be used as an indicator of the presence of inconsistencies and of the effort to resolve inconsistencies. RQ4 is stated as follows.

- **RQ4:** What is the impact of design characteristics on the inconsistency rate and inconsistency resolution effort?

Our studies to answer these research questions are viewed as the key original contribution of this work. No previous work has studied these different dimensions of model composition effort until now. It is important to highlight that we aim at investigating these research questions in the context of composing well-known design models, including UML class diagrams and architectural models, which are the most used design models in practice (Dobing & Parsons, 2006). While we mostly focus on structural design models in our studies, behavioral models were also involved in one of the studies. The next section discusses the thesis contributions more carefully.

## 1.4.
## Thesis Contributions

The previous sections discussed the limitations of related work, stated the research problem being addressed, and then presented the study methodology. This section describes the thesis contributions, which consist of an evaluation approach and the production of empirical knowledge about model composition effort. All contributions are derived from a series of empirical studies, including controlled experiments, quasi-experiments, case studies, interviews, and observational studies. These qualitative and quantitative studies evaluate the composition effort from different perspectives in realistic and controlled contexts by collecting multiple sources of evidence. More specifically, the contributions of this thesis are the following:

1. *A quality model for model composition effort (RQ1).* Some quality models for design modeling have been previously proposed. Some examples are described in (Lange, 2007a; Krogstie, 1995; Lindland et al., 1994). However, these quality models aim at software modeling in general rather than model composition effort. The contribution of this thesis is, therefore, the extension of the existing quality models for model composition effort. The extension is based on practical knowledge derived from our experience in conducting a range of empirical studies, including two controlled experiments, five industrial case studies, three quasi-experiments, interviews, and seven observational studies. Therefore, our evidence-based quality model provides guidance to developers and researchers about how to plan empirical studies in model composition. The guidance is characterized by: (i) a unifying terminology for activities and artefacts involved in model composition tasks, and (ii) the systematic relation between quality notions and metrics for the qualitative and quantitative assessment in the realm of model composition.

These elements of the quality model can also help to identify and empirically evaluate possible factors or indicators of model composition effort. For instance, the quality model helped us to select metrics and procedures to evaluate how the three influential factors (i.e., design decompositions, the design characteristics, and the composition techniques) affect model composition. The quality model can also serve as a reference frame to structure empirical studies performed by other researchers in the future. Without a reference frame, the replication and comparison of empirical studies as well as the generalization of their results are jeopardized. Chapter 3 elaborates the quality model.

2. *Insight and practical knowledge on model composition effort (RQ2-4).* The quality model guides the investigation about the effects of factors on the model composition effort. As previously mentioned, three factors are considered in this thesis: (1) the composition techniques (Section 2.4), (2) the design modeling technique used to decompose the design models (Section 2.5), and (3) the model stability (Section 2.6). The evaluation is performed by a series of experimental studies including: two controlled experiments, five industrial case studies, three quasi-experiments, more than fifty interviews, and seven observational studies. The empirical findings enhance the knowledge about the impact of the influential factors on: (i) the effort to apply model composition techniques; (ii) effort to detect inconsistencies; and (iii) the effort to resolve inconsistencies. Additionally, we gather insight about how to evaluate the developers' effort, reduce error proneness in model composition, and tame side effects of the influential factors in practice. The current body of knowledge on model composition is improved as our studies allowed to: (i) test out recurring claims, which were formulated by the experts in the literature, but that were never evaluated; (ii) identify correlations between key dependent and independent variables involved in model composition; for instance, identify which types of changes make model composition an error-prone and effort-consuming task; (iii) build a clear understanding to further support the formulation of theories on model composition; (iv) provide a solid background to inspire the creation of the next-generation model composition techniques and tools; and (v) pinpoint when the model composition techniques work and when they do not work.

These contributions are presented and discussed throughout the next chapters, and refined in Chapter 7. They have been reported in a number of papers, where part of them were already published in international conferences and workshops or submitted to journals. Table 1 shows the list of publications that are related to the thesis directly and indirectly.

## 1.5.
## Thesis Outline

This section outlines how the contributions are reported in each chapter, and makes explicit the relation between the chapters and the research questions.

*Chapter 2: Background and Related Work.* It defines the main concepts used throughout this thesis. These definitions are essential to understand the contributions and the results achieved. In addition, this chapter discusses related work, contrasting the commonalities and differences with respect to our research.

*Chapter 3: A Quality Model for Model Composition (RQ1).* This chapter sets up the context for proposing a quality model for model composition effort by discussing the limitations of existing quality models. After that, the chapter introduces the quality model, which provides the basis for all empirical studies realized throughout this research. This quality model takes into account the elements relevant to the three influential factors investigated in our empirical studies: the model composition techniques (Section 2.4), the design modeling languages (Section 2.5), and the design characteristics (e.g., model stability) (Section 2.6). More specifically, the quality model relates composition metrics and a series of quality notions, such as semantic, syntactic, and social quality notions. The quality model also serves as a practical guideline to select metrics and procedures to evaluate how the influential factors affect the model composition. This chapter elaborates on initial ideas reported in (Farias et al., 2008a).

*Chapter 4: Effort on the Application of Composition Techniques (RQ2).* This chapter reports upon the effects of composition techniques — both specification-based techniques and heuristic ones — on the developers' effort and its relation to the correctness of the output composed models. This cost-effectiveness analysis of the techniques is realized based on a range of quantitative and qualitative empirical studies including one controlled experiment, five industrial case studies, observational studies, and interviews. These combined studies allow building a body of knowledge about the effort that developers invest to compose design models. It is expected that the specification-based techniques reduce the developers' effort and assure the correctness of the compositions when compared to the heuristic-based techniques. However, the results, supported by a comprehensive set of statistical analyses, reveal the opposite, the specification-based techniques increase the developers' effort and do not assure the correctness of the compositions when compared to the heuristic-based techniques. The results presented in this chapter are presented in three papers (Farias, 2011a; Farias et al., 2012a; Farias et al., 2012c).

*Chapter 5: Effort on the Detection of Inconsistencies (RQ3).* This chapter investigates the effects of significantly different forms of design decomposition (i.e., object-oriented modeling and aspect-oriented modeling) on the effort to detect inconsistencies in the output composed model. The results provide insight about the impacts of using different modeling languages on the effort of detecting inconsistencies. As in the previous studies, this insight is generated from a family of experimental investigations including one controlled experiment, five industrial case studies, observational studies, and interviews. These studies allowed investigating RQ3 from different perspectives, i.e., varying the artifacts analyzed, the context (in vivo and in vitro), and the cultural biases in applying the evaluation (companies and university in different locations). Elements of this chapter were reported in three papers (Farias et al., 2012b; Farias, 2011a; Medeiros et al., 2010).

*Chapter 6: Effort on the Resolution of Inconsistencies (RQ4).* This chapter investigates the effort that developers spend to resolve inconsistencies. In particular, we study the influence of modeling languages and model stability on the production of inconsistencies and on the effort to resolve these inconsistencies. As in the previous chapter, the findings and lessons learned are gathered from a multiple studies, including two quasi-experiments in the context of evolving design models. All results are supported by statistical tests. Elements of this chapter are reported in papers as well (Farias et al., 2012d; Farias et al., 2010a; Farias et al., 2010b; Farias et al., 2011).

*Chapter 7: Conclusions.* This chapter presents a summary of our research, a refinement of the contributions, and the final remarks.

| **Direct Publications** | **RQ** |
|---|---|
| 1. Kleinner Farias, Alessandro Garcia, and Carlos Lucena, *Evaluating the Impact of Aspects on Inconsistency Detection Effort: A Controlled Experiment*. In: 15th International Conference on Model-Driven Engineering Languages and Systems (MODELS), Foundations Track, Austria, 2012. | RQ3 |
| 2. Kleinner Farias, Alessandro Garcia, Jon Whittle, Christina Chavez, and Carlos Lucena, *Evaluating the Effort of Composing Design Models: A Controlled Experiment*, In: 15th International Conference on Model-Driven Engineering Languages and Systems (MODELS), Applications Track, Austria, 2012. | RQ2 |
| 3. Kleinner Farias, Alessandro Garcia and Jon Whittle, *Assessing the Impact of Aspects on Model Composition Effort*, In: 9th International Conference on Aspect-Oriented Software Development (AOSD'10), Saint-Malo, France, 2010 (Indicated to Best Paper Award - Accept. Rate < 30%). | RQ3, RQ4 |
| 4. Kleinner Farias, Alessandro Garcia, Carlos Lucena, *Evaluating the Effects of Stability on Model Composition Effort: an Exploratory Study*, Journal of Software and Systems Modeling, 2012. | RQ4 |
| 5. Kleinner Farias, Alessandro Garcia, Jon Whittle, and Carlos Lucena, *Analyzing the Effort on Composing Design Models of Large-Scale Software*, IEEE Transactions on Software Engineering, 2012. (Submitted) | RQ2 |
| 6. Kleinner Farias, *Empirical Evaluation of Effort on Composing Design Model*, In: Doctoral Symposium at the International Conference on Software Engineering (ICSE'10), pages 405-408, South Africa, 2010. | All |
| 7. Kleinner Farias, Alessandro Garcia and Jon Whittle, *On the Quantitative Assessment of Class Model Compositions: An Exploratory Study*, In: Empirical Studies of Model-Driven Engineering (ESMDE'08) at MODELS'08, v. 1, pages 1-10, 2008. | all |
| 8. Kleinner Farias, Alessandro Garcia, Carlos Lucena, *Evaluating the Effects of Stability on Model Composition Effort: an Exploratory Study*, In: VIII Experimental Software Engineering Latin American Workshop at XIV Iberoamerican Conference on Software Engineering, April, Rio de Janeiro, pages 81-91, 2011. | RQ4 |
| 9. Kleinner Farias, *Analyzing the Effort on Composing Design Models in Industrial Case Studies*, In: 10th International Conference on Aspect-Oriented Software Development Companion, pages 79-80, Porto de Galinhas, Brazil, 2011. | all |
| 10. Ana Luisa Medeiros, Kleinner Farias, Alessandro Garcia, and Thais Batista, *Evaluating Composition Techniques for Architectural Specifications: A Comparative Study*, In: Empirical Evaluation of Software Composition Techniques (ESCOT 2010) at AOSD'10, Rennes, France, 2010. | RQ2, RQ3 |
| 11. Everton Guimarães, Alessandro Garcia, and Kleinner Farias, *Analyzing the Effects of Aspect Properties on Model Composition Effort: A Replicated Study*, In: 6th Workshop on Aspect-Oriented Modeling at MODELS'10, Oslo 2010. | RQ2, RQ3 |
| 12. Kleinner Farias, Alessandro Garcia and Carlos Lucena, *On the Comparative Evaluation of Aspect-Oriented Model Composition Techniques*, In: III Latin-American Workshop on Aspect-Oriented Software Development (LA-WASP´09) at XXIII Brazilian Symposium on Software Engineering, pages 45-49, Ceará, Brazil, 2009. | all |

## Indirect Publications

1. Kleinner Oliveira, Karin Breitman, Toacy Oliveira, *A Flexible Strategy-Based Model Comparison Approach: Bridging the Syntactic and Semantic Gap*, Journal of Universal Computer Science, v. 15, p. 2225-2253, 2009.

2. Kleinner Farias, Ingrid Nunes, Viviane Silva, Carlos Lucena, *MAS-ML Tool: Um Ambiente de Modelagem de Sistemas Multi-Agentes*, In: Workshop on Software Engineering for Agent-oriented Systems at XXIII Brazilian Symposium on Software Engineering, Ceará, Brazil, 2009

3. Enyo Gonçalves, Kleinner Farias, Mariela Cortes, Viviane Silva, Ricardo Feitosa, *Modelagem de Organizações de Agentes Inteligentes: uma Extensão da MAS-ML Tool*, In: 1st Workshop on Autonomous Software Systems, CBSoft 10, 2010, Salvador, Bahia, 2010.

4. Enyo Goncalves, Kleinner Farias, Mariela Cortes, Alexandre Feijo, Fabiano Oliveira, Viviane Silva, *MAS-ML Tool: A Modeling Environment for Multi-Agent Systems*, In: 13th International Conference on Enterprise Information Systems (ICEIS), 2011, Beijing, China 2011.

5. Kleinner Oliveira, Karin Breitman, Toacy Oliveira, *Ontology Aided Model Comparison*, In: Fourteenth IEEE International Conference on Engineering of Complex Computer Systems (ICECCS`09), p. 78-83, Potsdam, Germany, 2009.

Table 1: List of direct and indirect publications

# 2
# Background and Related Work

Empirical studies are essential to evaluate the composition effort of design models in practice. These studies allow building a body of knowledge supported by empirical evidence, testing out hypotheses, identifying important context variables, and understanding how influential factors may affect developers' effort when composing models. Without these studies, it is not possible to realize effective improvements for the current state of the art of model composition.

The goal of this Chapter is to provide an overview of the main concepts and definitions required understanding the empirical studies of model composition presented in this thesis. This chapter also describes the relevant elements underpinning the three model composition factors investigated in this thesis. Finally, it also provides an overview of the limitations of related work considering the topics addressed in our research questions (Section 1.3).

The remainder of this chapter is organized as follows. To begin with, Section 2.1 presents the purpose of using model composition in practice. After that, the main characteristics of the design modeling languages are presented (Section 2.2) and the purpose of using design models is also discussed (Section 2.3). Then, the elements of the three influential factors are explained in the next sections. Section 2.4 describes the types of composition techniques. Section 2.5 presents the modeling languages used to represent design decompositions. Section 2.6 elaborates on the design characteristics studied, more specifically those related to model stability. In all previous three sections, the related works are discussed and contrasted.

## 2.1.
## Purpose of Using Model Composition

Model composition is a fundamental activity that addresses the limitations of humans for simultaneously dealing with a plurality of artefacts and tasks (Mistrík et al., 2010; Whitehead, 2007). Dijkstra advocates to master complexity someone should deal with one important issue at a time (Dijkstra, 1976). With this in mind, software developer tends to work on simple tasks rather than on complex tasks; but each task manipulating small artefacts rather than big, complex ones. For example, developers work on small parts of an overall design model in order to focus on part of the model relevant to them. Unfortunately, they are unable to create a "big picture" view from the small parts created in parallel by different software development teams. The composition of the parts can be performed by using a model composition technique. Many academic and industrial composition techniques (Section 2.4) have been proposed to help developers to use model composition for different purposes.

In this thesis, we investigate the composition effort in the context of the evolution of design models. We identify three particular purposes of using model composition, which are presented based on the degree of relevance for the study. They are described below:

1. *Change of design models.* Developers use model composition to systematically change design models in collaborative development environment. Typically, they add, modify, remove, or even refine model elements of some existing design model in parallel. By using a more systematic way of bringing together changes, developers aim at implementing the changes rather than concerning on integrating the parts of even grasping the impact of the changes. Consequently, this absence of concerns on composing the models helps developers to effectively change the models.

2. *Reconciliation of design models.* Usually developers create design models in parallel and parts of these models conflict with each other. Thus, the model composition techniques can identify these contradicting parts and help developers to reconcile them. In (Clarke, 2001), Clarke defines a mechanism for identifying and reconciling these conflicts. This mechanism provides guidance to developers explaining how reconciling contradicting models.

3. *Analysis of overlapping parts.* Design models are realized in multiple ways, and hence at some point developers must converge on a single one. As humans, developers are unable to recall all myriad of changes performed during the composition time (Whitehead, 2007). Hence, they cannot foresee when the changes are going to overlap. Therefore, the composition technique helps developers to identify the overlapping parts. This identification is critical because developers must decide which part will remain into the output composed model.

Regardless of the usage scenario, developers are always concerned with making the use of the composition technique to correctly produce the output composed model. The composition techniques studied in this thesis are explained in Section 2.4.

## 2.2.
## Properties of the Design Modeling Languages

Popular modeling languages, such as the UML (OMG, 2011), have particular properties and different diagrams that can play a role on model composition effort. Some relevant properties are described as follows.

*Lack of a rigorous definition.* The design modeling languages are defined by a metamodel, which specifies the syntax and semantics of the language' constructs. This specification is aided by a set of well-formedness rules that enable a more precise definition of the constructs. These rules can be expressed by using OCL (OMG, 2011), for example. Unfortunately, these rules are seldom represented in a formal way (Larman, 2004; OMG, 2011). Rather, they are usually expressed using natural language. If well-formedness rules are not formally specified, then they can jeopardize the benefits of using of model composition (Section 2.1). For example, if a composition incorrectly reports a high number of conflicts, then developers will invest some unnecessary effort to deal with them. A high amount of conflicts makes the composition unmanageable (Mens, 2002), increasing the likelihood of inconsistencies in the output composed model. Incorrect composed models jeopardize the communication between the developers, as misinterpretation may become inherent (Broy & Cengarle, 2011; Maoz et al., 2011a; Maoz et al., 2011b; Lange & Chaudron, 2004). If the syntax and semantics are formally specified, the conflicts and inconsistencies are reduced or even localized more quickly. Therefore, given the state of practice on software modeling, this thesis attempts to investigate model composition effort when rigorous definition is not available. We study the identification of conflicts and inconsistencies in scenarios where developers need to deal with the lack of formal information. All the studies follow this strategy (Chapter 4, Chapter 5, and Chapter 6).

*Multi-view design modeling languages.* The design modeling languages also define a range of structural and behavioral diagrams to represent static and dynamic aspects of software systems. The elements of complementary diagrams (e.g., UML class and sequence diagrams) should have a precise consistency with each other; otherwise, conflicting information in different views of the same system may lead to misinterpretations. For example, an abstract class in a class diagram cannot be used in a sequence diagram, as abstract classes cannot be instantiated. Otherwise, developers may not observe the inconsistency and make different interpretations about this class. Some of them may infer that the class is concrete, while others will infer that the same class is abstract. The rate of conflicting information typically increases when developers evolve design models in parallel or even when the synchronization of design models is not fully realized. Different developers tend to assign values to the model's properties that are conflicting. This thesis attempts to investigate how this lack of agreement between the models leads to problems during the composition. Essentially, we are concerned on understanding how these multi-view inconsistencies influence the effort of composing design models and how developers deal with such inconsistencies in practice.

*Complexity of the design modeling languages.* The size and complexity of the design models have grown in recent years (Lange, 2007b) as developers are increasingly creating systems that are more complex. To deal with these problems, the design modeling languages have also grown and delivered new constructs. For example, the UML and its extensions provide 13 diagram types, totaling more than 150 constructs (Dori, 2002). However, the high number of diagrams and constructs has led the language to become more complex than it was originally planned. If design models are complicated, then their compositions can also tend to be more complicated. Consequently, developers tend to modularize the design models in such a way that the size and complexity of the design models can be minimized. For example, developers may use object-oriented or aspect-oriented modeling in order to better modularize design models. This thesis attempts to understand how the use of different modeling languages can minimize the complexity of the design models; hence, reducing the composition effort (Chapter 4, Chapter 5, and Chapter 6). For example, we are concerned with knowing how different forms of decomposing designs can influence the composition of such models.

Therefore, this thesis studies model composition effort in the presence of imprecise model semantics as well as non-trivial, multi-view design models.

## 2.3.
## Purpose of Using Design Models

Many modeling languages have been proposed in recent years, such as the UML (OMG, 2011) and its extensions (Clarke & Banaissad, 2005; Baniassad & Clarke, 2004). These languages provide a set of modeling resources to developers so that they can represent concepts and their relationships. According to (Rumbaugh et al., 1999), the representations created by using these resources are abstractions in essence from a reality observed and reported at a specific level of detail. Developers can use these modeling resources in a range of situations such as specifying software architectures, communicating design decisions, and documenting software systems. In this thesis, we use UML class diagrams and UML component diagrams, and their respective extensions in aspect-oriented modeling. These two modeling languages (and diagrams) were chosen because some reasons.

First, UML is de fact  the standard design modeling language adopted by researchers and professionals in practice. The UML class and sequence diagrams are the most used diagrams (Dobing & Jeffrey, 2006). Second, most modeling tools are dedicated to create and manage UML models and its extensions such as IBM Rational Software Architect (IBM, 2011). Third, the AO modeling is the state-of-the-art modeling language for the modularization of software systems (Clarke & Walker, 2005; Clarke & Banaissad, 2005). Fourth, the UML is a general-purpose modeling language for systems engineering applications. It supports the specification, analysis, and design of a broad range of systems (OMG, 2011). Fifth, as the UML is the basis of most modeling languages today, the results can be possibly transferable to other modeling languages based on it. Sixth, both languages define notations to allow developers to graphically represent static and dynamic views of a software system. These notations are available in thirteen diagram types described in (OMG, 2011; Clarke & Walker, 2005). The UML and AO models were used for three proposes during the empirical studies:

1. *Communication.* Developers use design models to communicate design decisions between teamwork members.

2. *Comprehension.* Developers use design models to comprehend the modules of a software system before implementing them.

3. *Documentation for maintenance.* The UML's diagrams are used during maintenance to locate system elements that are affected by a maintenance request.

Additionally, design models can be also used for other purposes such as code generation (Schmidt, 2006), effort estimation (Mohagheghi et al., 2005; Uemura et al., 1999), quality prediction (Genero et al., 2003; Cortellessa et al., 2002), and testing (Briand & Labiche, 2002). However, we do not use models for these specific purposes during the empirical studies. In the next section, we present the model composition techniques investigated in this thesis.

## 2.4.
## Model Composition Techniques

Academia and industry have proposed many model composition techniques in recent years. These techniques differ in their manner of expressing the compositions. While some of them require the explicit specification of how the compositions should be carried out, others rely on composition heuristics to "guess" how the elements of the input models will be composed. Therefore, the techniques can be grouped into two broad categories as follows:

- *Specification-based technique.* This category brings together the techniques with which developers express the compositions by explicitly determining the manner how the input model elements will be matched and composed. Two state-of-the-art examples of this category are the MATA (Whittle et al., 2009) and Epsilon (Epsilon, 2011) techniques.

- *Heuristic-based techniques.* Techniques in this category are characterized by a set of predefined composition heuristics, which are responsible for "guessing" the correspondence between the input model elements. Based on such guessed similarities, the techniques can then combine the input model elements. Two examples of the heuristic-based techniques are the IBM RSA (IBM, 2011) and conventional composition algorithms of model elements, including merge, union, and override (Clarke & Walker, 2005).

The specification-based technique used in our study was the Epsilon technique (Kolovos et al., 2011), and the heuristic-based techniques were the one supported by the IBM RSA tool (IBM, 2011) and traditional composition algorithms (Clarke, 2001; Clarke & Walker, 2001). They are explained in the next sections. Figure 1 shows an illustrative example that will be used to support the discussion of the studied composition techniques.

## 2.4.1.
## Traditional Composition Algorithms

We have studied three manual, heuristic-based composition algorithms: *override*, *merge*, and *union*. These algorithms were proposed and analyzed in (Clarke & Walker, 2005). There are some reasons that motivated the use of these algorithms in this study. First, evolution scenarios can be decomposed into one (or more) canonical operation supported by these algorithms. Typically, these operations are *additions*, *modifications*, and *removals* (Section 3.3).

Second, these algorithms can be also seen as basic "rules of the thumb" for developers to compose models; they do not need to be strictly realized for each instance of model composition in a software project. They provide general descriptions of how the compositions should be performed and guide developers to combine model elements. For example, these general composition guidelines may be useful to accommodate the specificities of particular model compositions and lead to fewer inconsistencies in the output composed model.

Third, they have been applied in a wide range of model composition scenarios, such as evolution and integration of software product lines (Jayaraman et al., 2007), and composition of design models (Clarke & Baniassad, 2005), and aspect-oriented modeling (Clarke & Baniassad, 2005). They have been recognized as candidate algorithms to compose well-modularized design models, such as aspect-oriented design models, e.g., Theme/UML (Clarke & Baniassad, 2005).

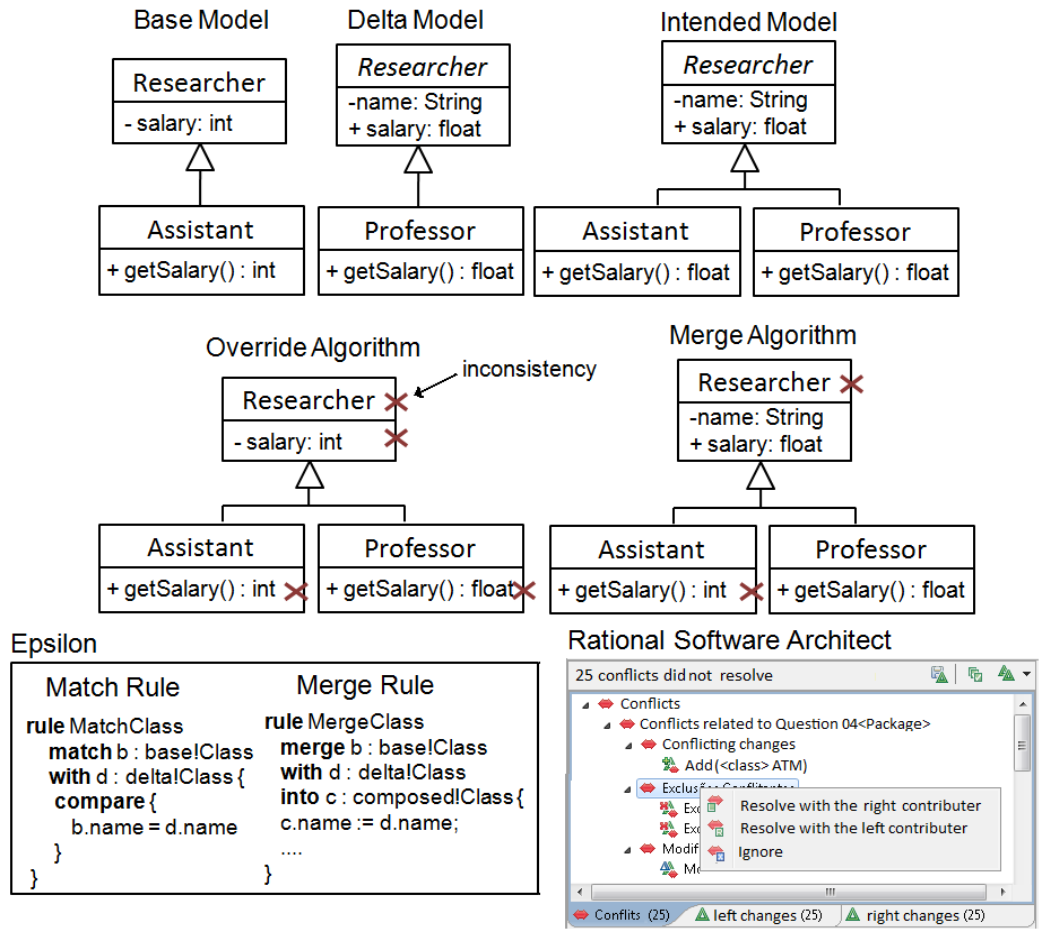

Figure 1: Illustrative example

In the following, we briefly define override, merge, and union algorithms, using a simple example to illustrate them. We assume the presence of two input model, $M_A$ and $M_B$. We consider that two elements from $M_A$ and $M_B$ are corresponding if they have been identified as equivalent in the matching process.

*Override (direction: $M_A$ to $M_B$).* For all pairs of corresponding elements in the base model ($M_A$) should override its similar element in the delta model ($M_B$). Elements not involved in the correspondence remain unchanged. They are then inserted into the output model. Figure 1 shows the application of this algorithm. The concrete class *Researcher* (*isAbstract* = false) overrides the abstract class *Researcher* (*isAbstract* = true), and the concrete classes *Assistant* and *Professor* were just inserted into the output composed model. However, the intended model was not produced. Rather, the output composed model has three inconsistencies. This implies that the algorithm was not able to properly accommodate the changes from the delta into the base model, as would be expected. Note that the algorithm was applied in the direction from the base model to the delta model.

*Merge.* For all corresponding elements in $M_A$ and $M_B$, such elements should be composed instead of overridden as in the override algorithm. The composition depends on the element type. Elements in $M_A$ and $M_B$ that are not involved in a correspondence match remain unchanged and, consequently, are inserted into the output model directly. In Figure 1, the merge algorithm is applied from the base model to the delta model; hence, a composed model is produced with two inconsistencies. Again, the intended model is not produced. Although the attribute *Researcher.name* has been correctly inserted into the class *Researcher*, it is a concrete class (*isAbstract* = false) instead of abstract (*isAbstract* = true), as it would be expected (according to the intended model). This problem affects the method *Assistant.getSalary():int* as a ripple effect. To produce the intended model, the merge algorithm should be applied from the delta model to the base model. Given this inverse order on the application of the algorithm, the changes in the delta model will predominate over the model elements in the base model.

*Union.* For all elements in the base and delta model that are corresponding elements, they should be manipulated in order to preserve their distinguished identification. It means that they should coexist in the output models with different identifiers; elements in the $M_A$ and $M_B$ that are not involved in a correspondence match remain unchanged, and they are inserted into the output model, $M_{AB}$. For example, we will have two classes *Researcher* in the composed model. However, both classes will carry identifiers that somehow preserve their original identities e.g., *BaseModel.Reseacher* and *DeltaModel.Researcher*.

## 2.4.2.
## IBM Rational Software Architect

IBM RSA is a comprehensive modeling and development environment that relies on the UML language artefacts (Norris & Letkeman, 2011). We choose IBM RSA due to some reasons.

First, it is the most robust composition techniques adopted in industry (Norris & Letkeman, 2011). In (Altmanninger et al., 2009), this superior quality is supported by empirical studies. Second, IBM RSA's model validation mechanism allows us to the automated identification of syntactic inconsistencies. This means that developers are expected to localize inconsistencies more quickly than manually, minimizing the detection effort. Third, it provides an adequate composition environment to report the conflicting between the input model elements.

Fourth, it allows creating all thirteen UML diagrams and executing some important operations such as model transformation and reverse engineering. In particular, it supports model-to-code (e.g., UML to Java) and code-to-model (e.g., Java to UML) transformations. In addition, it supports reverse transformations go from Java to UML, C++ to UML, and .NET to UML. IBM RSA is designed on top of the open-source Eclipse development platform. Therefore, it gives the developers a complete IDE for model-driven software development. In addition, it provides a disciplined control of shared design models in evolving software projects. Finally, empirical studies (Altmanninger et al., 2009) indicate that IBM RSA's composition technique has a considerable level of precision compared with other related technologies such as Subversion (SVN, 2012), EMF compare (EMF, 2012), and UNICASE (Unicase, 2012). More importantly, it enables model management in collaborative software development e.g., splitting, comparing and composing models created in cooperation.

Although IBM RSA implements a robust and precise model composition technique, it does not ensure that the intended model will be always produced. This means that developers should necessarily interact with models via the tool facilities to produce an output composed model. Figure 1 depicts an example of conflict report produced by RSA. For example, when conflicting changes emerge, developers should decide which changes — from the base model (*Researcher.isAbstract* = false) or from the delta model (*Researcher.isAbstract* = true) — will be inserted into the output composed model.

### 2.4.3.
### Epsilon

Epsilon is a *flexible* platform for model management (Kolovos et al., 2011) defined as an Eclipse Plug-in. This flexibility is achieved by providing a set of consistent task-specific languages for developers so that they can perform some tasks such as model comparison and model composition. To date, seven interoperable, but with different purposes, languages have been proposed to help developers to manage design models. Although there is a wide diversity of modeling languages, we put our attention on two specific languages: the epsilon comparison language (ECL, 2012) and the epsilon merge language (EML, 2012). They are two hybrid, rule-based languages used to compare and merge design models, respectively (EML, 2012). These two languages were chosen because three reasons.

First, they are responsible for executing the two most common tasks in model composition: comparison and composition of models. Second, these languages define a set of constructs expressive enough to seamlessly specify how the input model elements are going to be compared and integrated. Third, by using these languages, developers can master the complexity of dealing with inherent composition problems, i.e., the imprecise specifications of commonalities and differences between the input model elements. Lastly, they are intuitive and expressive enough so that we empirically investigate the effort of developers invest to compose two design models (Kolovos et al., 2011).

Additionally, the Epsilon platform also presents some interesting characteristics to support the use of those two languages. To begin with, the feature of syntax highlighting differs in colors and fonts the language constructs improving the readability of the composition specifications. Next, the code completion steeps the learning curve, i.e., the learning related to composition specification may be achieved more quickly. This resource can improve the quality of the composition specification by decreasing the initial difficulty of creating and editing the composition specifications. Developers can become more familiar with the languages; hence, improving the definition of the correspondence and composition relations. Thirdly, the syntax highlighting and code completion are two crucial characteristics, for example, to foster the use of model composition by novices. To sum up, the Epsilon is an Eclipse-based IDE provides important resources to developers, so that the comparison and composition rules can be carefully created and edited. Figure 1 shows an example of these rules. The *MatchRule* determines that there can be correspondence relations between the input classes if their names are similar. The *MergeRule* specifies that the name of the output composed classes should be equal to the name of the input class of the delta model, i.e., *c.name := d.name*.

To sum up, these three techniques (i.e., Epsilon, IBM RSA and Traditional Algorithms) are good candidates for comparisons because: (1) they are robust and usable tools, which are two prerequisites for an experiment like this; (2) IBM RSA is an industry leading model composition tool; and (3) traditional algorithms such as merge/override are well mentioned in the academic literature as a technique and have been used to build tools.

### 2.4.4.
### Limitations of Related Work on Model Composition Techniques

Model composition is a very active research field in many research areas, such as merging of state charts (Whittle & Jayaraman, 2010), composition of software product lines (Clarke, 2001), aspect-oriented modeling (Clarke & Walker, 2005), and mainly composition of UML design models (Farias et al., 2011a). In doing so, there has been more research on proposing model composition techniques or even creating innovative model composition techniques, such as traditional composition algorithms (Clarke, 2001; Clarke et al., 2005), IBM RSA (IBM RSA, 2011), Epsilon (Kolovos et al., 2011), MATA (Whittle & Jayaraman, 2011), Kompose (Kompose, 2011) rather than evaluating them.

Clarke and colleagues (Clarke, 2001; Clarke et al., 2005) propose three conventional algorithms, namely override, merge, and union, to compose UML design models such as UML class diagrams. These algorithms are the basis for other composition techniques such as Epsilon (Kolovos et al., 2011), Araxis Merge (Araxis, 2011), KDiff3 (KDiff3, 2011), and MergePlant (MergePlant, 2011). Araxis Merge is a 2/3-way file comparison, merging and folder synchronization for Windows and Mac OS X. The focus of the techniques is on synthesizing text-like files rather than design models (Araxis Merge, 2011). KDiff3 (KDiff3, 2011), MergePlant (MergePlant, 2011). They are useful for determining what has changed between versions, and then merging changes between versions.

Kolovos and colleagues (Kolovos et al., 2011) propose the Epsilon Platform in order to compose homogenous and heterogeneous design models. That is, the tool is able to combine input design models that are instanced from a particular metamodel or from different metamodels. Epsilon offers an innovative, flexible platform to promote compositions of design models.

However, none of these approaches has investigated the effort that developers should invest to compose design models. As a matter of fact, the current literature in composition techniques points out the absence of empirical studies and does highlight the importance of empirical evidence (Dingel et al., 2008; Apel et al., 2011; Uhl, 2006; Mens, 2006; France & Rumpe, 2007). This absence of knowledge may cause serious consequences. First, it is not possible to grasp if the effort invested by developers is cost-effective (or not). Cost-benefits analysis in terms of effort is crucial before applying any technique in practice. If the effort of applying a particular technique is high, then developers will not use in practice. Second, the composition techniques are improperly used due to the influential factors that directly (or indirectly) affect the use of the techniques are unknown.

The current works have notably aimed at evaluating the use of design models rather than the consequences of the application of composition techniques on them. In fact, there existing studies concentrate on investigating UML models in terms of quality attributes such as comprehensibility (Ricca et al., 2010) and completeness (Langes & Chaudron, 2004). These works are very important, as the current standard modeling language is the UML.

In addition, we have also observed that most of the research on the interplay of effort and composition techniques rests on subjective assessment criteria (France & Rumpe, 2007). Even worse, they depend on the expert judgments, who have built up an arsenal of mentally held indicators to analyze the growing complexity of models and then evaluate the effort on composing them. Therefore, to date, developers rely on feedback from experts to determine "how good" the input models and their compositions are.

According to (France & Rumpe, 2007), the state of the practice in assessing model quality provides evidence that modeling is still in the craftsmanship era and when we assess model composition the problem be aggravated. More specifically, to the best of our knowledge, our results are the first to empirically investigate the research questions in a controlled way by using specification-based and heuristic-based techniques.

To sum up, there are two critical gaps in the literature. First, practical knowledge about the relative effort of composing design models is lacking. That is, developers do not know very little about what they invest in terms of effort to apply the composition techniques as well as detecting and resolving inconsistencies. Second, insight about the potential influential factors is also lacking. Hence, developers are unable to improve the composition process (i.e., the execution of the composition activities) once they do not know which, in fact, jeopardize the execution of the activities. Second, the lack of empirical evidence about the correctness of the output models produced using these techniques in practice.

## 2.5.
## Design Modeling Languages

In this research, we focus our investigations on the Unified Modeling Language (UML) (OMG, 2011) and one of its extensions to Aspect-Oriented Modeling (AOM) (Clarke & Walker, 2005).

### 2.5.1.
### Unified Modeling Language

The Unified Modeling Language (UML) is a general-purpose modeling language adopted as the standard modeling language in practice (OMG, 2011). The UML models are by far the most widely used in object-oriented software engineering (OMG, 2011; Dobing & Parsons, 2006). In fact, most of its diagrams are primarily tailored to support object-oriented software development. It is used to specify, communicate, and document the artifacts of software-intensive systems under development.

UML is defined using a metamodeling approach, i.e., a metamodel is used to specify the models that comprise UML. The UML metamodel is defined based on a 4-layer metamodel pattern. While this approach lacks some of the rigor of formal specification techniques, it offers the advantages of being more pragmatic for most researchers and developers (OMG, 2011). The UML metamodel defines thirteen diagrams, such as the component diagram, the class diagram, the sequence diagram, and the use case diagram (OMG, 2011). Together the UML diagrams represent two different views of a system model: (1) *structural view*: it emphasizes the static structure of the system using objects, attributes, operations, and relationships. Examples of these diagrams are the class diagram and component diagram, and (2) *behavioral view:* it emphasizes the behavior of the system by showing collaborations among objects and changes to the internal states of objects. Examples of these diagrams are the sequence diagram, the activity diagram, and the state machine diagram.

In this research, we use three UML diagrams: class, sequence, and component diagrams. This choice is not an arbitrary choice, but based on observations drawn on empirical studies reported by Dobing and Parsons in (Dobing & Parsons, 2006). These researchers conducted an OMG-supported survey to investigate which UML diagrams are used in real-world projects more frequently. The survey identified the frequency of use of UML diagrams. The main result of the study was that class diagram is the most-used UML diagram used followed by use case diagram and sequence diagram. Consequently, these diagrams tend to be the diagrams that developers compose.

Additionally, developers usually compose these diagrams in practice (Norris & Letkeman, 2011). The key reason for using these diagram types is their usefulness and adequacy of information as perceived by the models' users. Their selection for this research is also motivated for the fact that there are aspect-oriented counterparts for these diagrams. The aspect-oriented versions of these diagrams are also used in some of our studies. Aspect-oriented modeling is discussed in the following subsection.

## 2.5.2.
## Aspect-Oriented Modeling

Separation of concerns is a fundamental principle that addresses the limitations of human cognition for dealing with complexity. Dijkstra advocates to master complexity, one should deal with one relevant concern at a time (Dijkstra, 1976). Parnas reinforces that complexity of software systems should be tamed by decomposing their modules into smaller, clearly separated modular units, each dealing with a single concern (Parnas, 1972). The principle of separation of concerns is employed through the decomposition and modularization of software systems. The expected benefits are an improved understandability and reuse in complex software systems. In software modeling, the achievement of separation of concerns depends largely on the suitability of abstractions and notations of modeling languages to represent these concerns. Typically, components, classes, and methods are examples of modular units in object-oriented modeling languages, such as UML and its profiles.

Unfortunately, object-orientation has some limitations in dealing with concerns that address global constraints and widely scoped functionalities, such as persistence, error handling, logging, among many others (Sant'Anna, 2008). These concerns have been commonly called *crosscutting concerns* since they naturally crosscut the boundaries of modular units that implement other concerns. Without proper means for separation and modularization in the UML, crosscutting concerns tend to be scattered over a number of modular units (e.g., components and classes) and tangled up with other concerns. Consequently, the cohesion in the modular units tends to decrease, while the coupling between them tends to increase. This can jeopardize the comprehensibility and evolvability of design models. Aspect-orientation (Kiczales et al., 1997) is an approach that supports a new flavor of separation of concerns. It introduces new modularization abstractions and composition mechanisms to improve separation of crosscutting concerns at different levels of abstraction. Aspect-orientation defines a new modular unit, called *aspect*, for separating crosscutting concerns, and provides new mechanisms for composing aspects with other modular units at well-defined points. In the following, we briefly describe the main aspect-oriented abstractions and mechanisms. After that, we illustrate the use of aspect-oriented modeling in the light of an example.

*Aspects*

*Aspect* is the term used to denote the abstraction that aims at supporting improved isolation of crosscutting concerns (Kiczales et al., 1997). Aspects are modular units of crosscutting concerns that crosscut a set of modular units — i.e., components, classes, interface, and so on (Sant'Anna, 2008). An aspect can affect, or crosscut, one or more modular units in different ways. Thus, aspect-oriented design models can be decomposed into components, classes, interfaces, and aspects. While aspects modularize crosscutting concerns and the other modular unit modularize non-crosscutting concerns. In addition to conventional attributes and methods, an aspect includes pointcuts and pieces of advice as described as follows.

*Join Points and Pointcuts*

Essential to the process of composing aspects and classes is the concept of *join points*, the elements that specify where aspects and other modular units are related. Join points are well-defined points in the dynamic execution of a system (Kiczales et al., 1997). Examples of join points are method calls, method executions, attributes sets and reads, and object initialization. Each aspect defines one or more first-order logic expressions, called *pointcut expressions* (or just pointcuts), to select the join points that will be affected by the aspect's crosscutting behavior (Kiczales et al., 1997).

*Advice*

When execution of the system reaches a join point, selected by some pointcut expression, an *advice*, can be executed before, after or around it (Filman et al., 2005). Advice is a special method-like construct attached to pointcuts (Kiczales et al., 1997). There are three basic forms of advice supported by most aspect-oriented languages (Kiczales et al., 1997): (i) a before advice runs whenever a join point is reached and before the actual computation proceeds, (ii) an after advice runs after the computation under the join point finishes, i.e., after the method body has run, and just before control is returned to the caller, and (iii) an around advice runs whenever a join point is reached, and has explicit control whether and when the computation under the join point is allowed to run at all.

Therefore, aspect-oriented (AO) modeling languages aim at improving the modularity of design models by providing a range of notations to represent these concepts. It is important to highlight that there are many approaches proposed for AO modeling. Most of them are aimed at representing basic AO concepts also supported by most aspect-oriented programming models. Approaches that are more conservative propose UML profiles (Losavio et al., 2009; Clarke & Banaissad, 2005; Chavez & Lucena, 2002) for supporting AO modeling (Losavio et al., 2009; Clarke & Banaissad, 2005). These techniques are more aligned to classic AO programming models, such as the one realized by AspectJ (Laddad &

Johnson, 2009) and dialects. In these profiles, the modularization of crosscutting concerns, for instance, is achieved by the definition of a new model element, called *aspect*. In general, the notation enables to explicitly distinguish between *aspects* and *classes*. An aspect can crosscut several classes in a system. These relations between aspects and other modules are then called *crosscutting relationships*. Typically, these relationships are motivated by crosscutting concerns.

Having the goal of this work in mind (Chapter 1), we opted for carrying out our investigation regarding UML profiles. Another reason for using AO UML profiles is that the real developers will participate in the empirical studies and these subjects tend to have previous experience with AspectJ (Laddad & Johnson, 2009) rather than with any other AO modeling approach. Thus, the UML profile for aspect-orientated tends to be the best choice for this typical characteristic of aspect-oriented software developers.

These profiles have the advantage of supporting classical AOP concepts at a higher abstraction level. This means that AO key concepts are usually represented via conventional extension mechanisms of the UML such as UML stereotypes. This alternative followed in our studies prevented, for example, classical side effects related to the learning curve in empirical studies. Otherwise, it would not be possible to investigate any causal relationships between design model languages and composition effort without any high overhead to the subjects involved.

It is also important to highlight that UML is the standard for designing software systems. The use of stereotypes reduces the gap between subjects with low and high skilled (or experienced) subjects (Ricca et al., 2010). The other consequence of using UML profiles for AO modeling is that the model reading technique used by the subjects would not be much influenced by new notation issues. Therefore, the use and interpretation of the models are exclusively influenced by the use of the concepts in object-oriented and aspect-oriented modeling. As UML profiles are supported by academic and commercial modeling tools, such as IBM Rational Software Modeling (IBM RSA, 2011), developers are familiar with stereotype notations. Additionally, learning the current state-of-the-art of AO modeling is not a trivial task for developers in early adoption of aspect-oriented programming. Finally, UML profiles for aspect-oriented design is the approach more common for structural and behavioral diagrams. Based on these reasons, the AOM language used in our study is a UML profile described in (Losavio et al., 2009; Clarke et al., 2005; Chavez & Lucena, 2002).

Figure 2 presents illustrative examples of some aspect-oriented models used in our study: class and sequence diagrams. The notation supports the visual representation of aspects, crosscutting relationships and other aspect-oriented modeling concepts. The stereotype <<aspect>> represents an aspect, while the dashed arrow decorated with the stereotype <<crosscut>> represents a crosscutting relationship. Inner elements of an aspect are also represented, such as pointcut (<<pointcut>>) and advice. An advice adds behavior before, after, or around the selected join points (Losavio et al., 2009; Clarke & Walker, 2005). The stereotype associated with an advice indicates when (<<before>>, <<after>> or <<around>>) a join point is affected by the aspect. The join point is a point in the base element where the advice specified in a specific pointcut is applied.

With this in mind, we discuss the limitations of the related work regarding the effort of detecting inconsistencies and empirical studies on software modeling.

### 2.5.3.
### Limitations of Related Work on Design Modeling Languages

Many design modeling languages have been proposed in recent years, such as UML and its extensions (OMG, 2011). Some empirical studies have also been performed with these languages in order to understand their usefulness in different contexts. For instance, AOM languages will be considered useful compared to traditional modeling techniques if the claimed improved modularity of aspectual design decompositions actually leads to practical benefits, such as reduction of inconsistency detection effort and misinterpretations. Unfortunately, it is well known that empirical studies of AO modeling are rare in the current literature, which confirms that it is still in the craftsmanship era (France & Rumpe, 2007).

Research has been mainly carried out in two areas: (1) defining new AOM techniques, and (2) proposing new weaving mechanisms for design models. First, several authors have proposed new modeling languages, focusing on the definition of constructs, such as <<aspect>> and <<crosscut>>. These constructs represent concepts of aspect-orientation as UML based extensions (Losavio et al., 2009;

Chavez & Lucena, 2002). In addition, Clarke and Baniassad (Clarke & Banaissad, 2005) make use of UML templates to specify aspect models.

On the other hand, the chief motivation of some works is to provide a systematic method for weaving aspect and base models (e.g., (Whittle & Jayaraman, 2010; Jézéquel, 2008; Klein et al., 2006). For example, Klein and colleagues in (Klein et al., 2006) present a semantic-based aspect-weaving algorithm for hierarchical message sequence charts (HMSC). They use a set of transformations to weave an initial HMSC and a behavioral aspect expressed with scenarios. Moreover, the algorithm takes into account the compositional semantics of HMSCs.

Unfortunately, most of empirical studies on aspect-orientation are focused on assessing implementation techniques. For example, Hanenberg and colleagues (Hanenberg et al., 2009) compare the time invested by developers to implement crosscutting concerns using object-oriented and aspect oriented programming techniques. Other studies focus on the assessment of aspect-oriented programming under different perspectives, such as software stability (Ferrari et al., 2010; (Greenwood et al., 2007) and fault-proneness (Burrows et al., 2010). However, empirical studies about AO modeling have not been conducted in particular in the context of modeling inconsistencies (or defects). Only the literature on OO modeling does highlight that empirical studies have been done on identifying defects in design models (Langes & Chaudron, 2004). Lange (Langes & Chaudron, 2006a) investigates the effects of defects in UML models. The two central contributions were: (1) the description of the effects of undetected defects in the interpretation of UML models, and (2) the finding that developers usually detect more certain kinds of defects than others do.

In particular, in this thesis, we aim at studying certain effects on model composition from one of the most prominent and recently proposed approaches to achieve separation of concerns at design level: aspect-oriented modeling language (Clark & Walker, 2005; Losavio et al., 2009). In addition, our other focus is on analyzing the empirical studies on UML and AO modeling. We reinforce that aspect-oriented modeling supports early separation of otherwise crosscutting concerns in software design. An improved modularization may ameliorate one of the main purposes of using of design models: communication. If developers communicate properly, so the interpretation of the models is also proper. Thus, we analyze empirical studies investigating the side effects of inconsistencies on the interpretation of the design models and the effort invested by developers to detect them. In conclusion, there are two critical gaps in the current understanding about AOM that are addressed in this thesis: (1) the lack of practical knowledge about the developers' effort to localize inconsistencies, and (2) the lack of empirical evidence about the detection rate and misinterpretations when understanding AO and OO models.

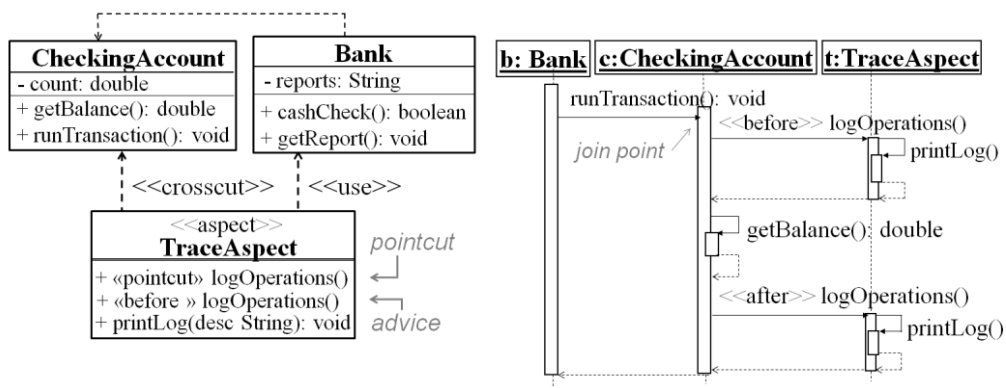

Figure 2: An illustrative example of AO models used in our study.

## 2.6.
## Design Characteristics

Researcher investigates how design characteristics, such as design stability, can influence the evolution of software artifacts (Kelly, 2006; Martin, 2003). In this thesis, we study whether the model stability can affect the composition effort. In the next section, we discuss how model stability is addressed in this thesis.

## 2.6.1.
## Model Stability

Developers need an indicator to identify the most severe composition cases in which the output composed models produced have a high number of inconsistencies and require a great deal of the developers' effort to be transformed into an output intended model. Without this indicator, it is particularly challenging for developers to exam hundreds of output composed models produced in a collaborative software development environment. In this thesis, we investigate if the model stability can be this indicator.

In practice, the stability of the output composed model can be computed based on the internal design characteristics of (evolving) models. According to (Kelly, 2006), a design characteristic (e.g., coupling and cohesion) is stable if, when observed over two or more versions of the software, the differences in the metric associated with that characteristic are considered small. With this in mind, we can consider the output composed model as stable if its design characteristics have a low variation regarding the characteristics of the output intended model.

In our study, we define low variation as being equal to (or less than) 20 percent. This choice is based on previous empirical studies (Kelly, 2006) on software stability that has demonstrated the usefulness of this threshold. For example, if the measure of a particular characteristic (e.g., coupling and cohesion) of the output composed model is equal to nine, and the measure of the output intended model is equal to 11. So the output composed model is considered stable in relation to the output intended model (because nine is 18% lower than 11) with respect to the measure under analysis. Following this stability threshold, we can systematically identify whether (or not) the output composed model remains stable in a particular evolution scenario or not. This threshold has been used more as a reference value rather than a final decision maker. Although its effectiveness has been demonstrated in (Kelly, 2006), we will also analyze in our empirical studies if this threshold can be, in fact, used to indicate the most severe composition cases in which an elevated number of inconsistencies and require a great deal of the developers' effort to resolve these inconsistencies. This investigation is realized in Chapter 6.

We will carry out this new analysis because this threshold plays a crucial role in the identification of the output composed models that will be reviewed by the developers. The identification of stable and unstable output composed models is based on the study of the differences between the measures of the design characteristics of the output composed model and the output intended model. These differences are calculated comparing the measures of each characteristic of the design models. We use a suite of design metrics to quantify such characteristics of the models used in our study. The metrics can be seen in the next Chapter 3 (Table 5, Table 6, and Table 7), and Chapter 6.

These metrics were used because they are conventional metrics and they have been used previous works e.g., (Martin, 2003; Kelly, 2006; Fenton & Pfleeger, 1997), which have tested the effectiveness of these indicators for the quantification of design characteristics. We are also interested in identifying evolution scenarios where composition techniques are able to effectively accommodate changes from the delta model in the base model. The quantification method of model stability is presented later in Section 6.1.2.4. With this in mind, the next step is to discuss the limitations of related works considering the subject.

## 2.6.2.
## Limitations of Related Work on Design Characteristics

The current literature in software design has defined a set of characteristics that can be used to measure the quality of a design in terms of the interdependence between the modules of that design (Martin, 2003). A pivotal example of such characteristics is the software stability as previously mentioned in Section 2.6.1. According to (Martin, 2003), when we design software, we strive to make it stable in the presence of change. In fact, stability is at the very heart of all software design discipline.

Some works about design stability have been conducted in recent years such as (Kelly, 2006; Martin, 2003). Kelly has demonstrated the usefulness of stability to software maintenance. For this, she presents a method for examining software systems that have been actively maintained and used over the long term. The method relies on a criterion of stability and a definition of distance to flag design characteristics that have potentially contributed to the software maintenance (Kelly, 2006). The main contribution is the demonstration that the method is useful to provide insight into the relative importance of individual elements of a set of design characteristics for the long-term evolution of software. On the other hand, Martin (Martin, 2003) provides a definition of software stability and shows how the characteristic can be applied.

Unfortunately, we have observed that the existing literature in model composition and software design has failed to provide metrics or studies for empirically revealing the effects of stability on model composition effort. Thus, we see our work as the first step to investigate empirically the interplay between stability and model composition effort. In other words, nothing has been done to investigate the use of stability as an indicator of severe cases of composition effort.

The absence of studies exploring this relationship prevents developers from understanding the influence of stability on the developers' effort. Without this knowledge, developers end up relying on the evangelist feedback, rather than empirical data, to comprehend how well the composition effort can be. In conclusion, these works differ in their aims to the work presented in this thesis. This thesis does not propose how to come up with a good guidance to design software, neither proposes any particular method to quantify stability. Rather, we empirically evaluate how stability influences the developers' effort when composing models (Section 6.1). We defer further consideration about this topic to Section 6.2.4.

## 2.7.
## Concluding Remarks

In this chapter, we have presented the main concepts discussed throughout this thesis. To begin with, we describe the three purposes of using model composition. After that, we analyzed the characteristics of  design modeling languages that can affect the use of model composition. Three characteristics are discussed: the lack of a rigorous definition, the multi-view design modeling languages, and the complexity of the design modeling languages.

We also revisit the purpose of using design models. The empirical studies use design models for different particular purposes. This happens because we need to investigate the effort of composing design models from alternative perspectives. More specifically, we study the use of design models for three purposes: communication, comprehension, and documentation for maintenance.

Moreover, following the description of the basic terminology used in this thesis, we present the concepts associated with three key factors potentially influencing mode composition effort: composition techniques, design modeling languages, and design characteristics. After mentioning these three factors, we try to discuss how each factor can affect the effort of composing design models in practice.

Observing the related works, the major conclusion is that nothing has been done to evaluate the impact of such three influential factors on model composition effort. In fact, some works such as (France & Rumpe, 2007) emphasize the need for further researches in order to generate a clear understanding about the effects of these factors on model composition effort. For example, several composition techniques have been proposed and used in practice. However, little has been done to quantify the effort invested by developers to compose design models. Without studies that evaluate whether the effort invested is worthwhile or not, it is not possible to recognize the benefits of using composition techniques. This lack of knowledge about the effects of the composition on the developers' effort is also extended as to the other two factors: design modeling languages and design characteristics. To date, the literature fails to provide insight on the influence of these two factors on the composition effort. For example, researchers and developers do not know if by using a particular design modeling language, they will minimize the composition effort on the parts of the design model created in parallel by different software development teams.

# 3
# A Quality Model for Model Composition Effort

Software quality is defined as "conformance to requirements" (Boehm, 1978). Therefore, the quality of a software system can be seen as the characteristics that lead its comprising artifacts or its development activities to satisfy a set of requirements. A *software quality model* defines and organizes the concepts required to characterize or evaluate the quality of a software system (Lange & Chaudron, 2005b; Boehm et al., 1978). Certain quality models are intended to be general — i.e., they can be used to evaluate certain quality attributes in any software engineering context. However, in order to be useful in practice, each quality model should support the evaluation of a particular category of software artifacts and/or software development activities relevant to a certain software engineering context, such as model composition.

In this context, a quality model for model composition effort should: (i) define the conceptual elements required to characterize and evaluate model composition effort, and (ii) define and structure the quality notions (Lange, 2007; Boehm et al., 1978) that are relevant to model composition artifacts and activities. A quality model with these components is proposed in this thesis. The goal of this quality model is to fill the gap in the current literature that fails to provide adequate quality frameworks for model composition.

Therefore, the goal of this chapter is to define a quality model for model composition effort. This quality framework serves as a guideline for researchers and developers to carry out qualitative investigations considering model composition effort and to assess any quality achievements. The proposed quality model (Section 3.5) is a practical quality framework built from evidence-based knowledge acquired throughout the execution of a series of empirical studies (Table 1). The empirical studies range from controlled experiments, case studies, quasi-experiment, and observational study. These studies will be described in Chapters 4, 5, and 6. Additionally, this quality model is also based on (1) experience obtained from previous works performed over the past six years (Table 1), and (2) previous quality models such as (Marín et al., 2010; Lange, 2007a; Lindland et al., 1994; Boehm et al., 1979; McCall et al., 1977). Although the proposed quality model overcomes the limitations of related work (Section 3.2) and it can be applied to any design models, it does not aim to be a final and complete one. With this in mind, it has been designed to be extensible so that other researchers can tailor it for different purposes.

The creation of this quality model requires answering some open questions. First, what are the artifacts and activities involved in model composition? What do we expect from model composition? Developers do not know which tasks should be performed and what models participate in a model composition process (Section 3.3). Second, how can we evaluate the model composition effort? Researchers do not know which evaluation criteria should be used (Section 3.5), and how they can contribute to achieve the required quality (Fitzpatrick, 1999). Therefore, the proposed quality model addresses the first research question of this thesis (RQ1): *How can the evaluation of model composition be organized in terms of a comprehensive framework?*

The remainder of this chapter is organized as follows. First, Section 3.1 provides some additional motivation for our quality model. Then, Section 3.2 discusses the limitations of the related work. Section 3.3 defines how model composition effort can be evaluated. Section 3.4 defines composition conflicts and inconsistencies. Finally, Section 3.5 brings forward the quality model, which serves as the reference frame for the empirical studies conducted throughout this research.

## 3.1.
## Motivation

Although researchers and developers recognize the importance of evaluating model composition (France & Rumpe, 2009; Farias et al., 2010), the practice of this evaluation is not a trivial task (Basili & Lanubile, 1999; Basili et al., 1999). This can be explained by some reasons. First, the current quality models fail to define the concepts (and their relations) required to characterize and evaluate model composition. Examples of these concepts are conflicts, inconsistencies, types of modeling languages, and model composition techniques. These concepts are not even mentioned in the current quality models. Hence, it is not possible to study the interplay of these concepts and model composition effort.

Second, because of the aforementioned problem, the use of prevailing quality models, discussed in Section 3.2, does not enable developers to distinguish between: (i) general quality notions that are typically associated with the design models in general, and (ii) quality notions that are specifically relevant to the evaluation of model composition effort. Rather, they only take into account well-known general concepts in software modeling. The imprecise specification of specific quality notions for composition effort causes misunderstanding about what should be evaluated in this context. Even worse, researchers cannot properly formulate and test hypotheses as well as replicate studies. If researchers cannot replicate studies, then the generalization of the results is hindered.

Third, the lack of a quality model jeopardizes the understanding about how conclusions can be drawn and related. According to (Basili and Lanubile, 1999; Wohlin et al., 2000), the degree of validity of any finding of empirical studies depends on how conclusions are drawn — i.e., the degree of confidence in a cause-effect relationship between the study variables and to what the extent the conclusions can be extrapolated to other contexts. A quality model guides researchers to investigate cause-effect relationships and promote the alignment between the results of empirical studies. Without a quality model, the conclusions across multiple studies are weakly connected, and a body of knowledge about model composition cannot be built.

Finally, the understanding of model composition is based on common wisdom, intuition, evangelist feedback, or even proofs of concepts. All these sources of information are not reliable sources of knowledge (France & Rumpe, 2007). Therefore, the lack of a quality model for model composition is a key factor for the empirical evaluation of effort on composing design models. In fact, without an adequate quality model the problem stated in Section 1.1 cannot be addressed. In the following section, we discuss the limitations of the related work.

## 3.2.
## Limitations of Related Work

Researchers recommend the use of quality models in empirical investigations (Runeson & Höst, 2009; Wohlin et al., 2000). In (Runeson & Höst, 2009), Runeson and Höst highlight the need for a reference frame (e.g., quality model or theory) to plan and execute case studies. The authors emphasize, for example, that quality models make the context of the empirical study clearer, and help researcher to conduct as well as review the results obtained. In (Wohlin et al., 2000), Wohlin and colleagues also confirm the importance of a quality model for empirical investigations.

To date, most approaches involving model composition rest on subjective assessment criteria. They depend on experts who build up an arsenal of mentally held indicators to evaluate the growing complexity of the produced design models (France & Rumpe, 2007). Consequently, developers ultimately rely on feedback from experts to determine "how good" the input models and their compositions can be. According to (France & Rumpe, 2007; Uhl, 2008), the state of the practice in assessing model quality provides evidence that modeling is still in the craftsmanship era and when we assess model composition this problem is accentuated. Finally, to the best of our knowledge, the need for methods for qualitative evaluation during a model composition process neither have been pointed out nor even proposed by current model composition techniques (Brun et al., 2011a; Maoz et al., 2011; Apel et al., 2011; Sarma et al., 2011; Dingel et al., 2008; Zito, 2006).

Some quality models in the area of modeling have been proposed through the last decades, such as (Marín et al., 2010; Lange, 2007; Lindland et al., 1994; Boehm et al., 1979; McCall et al., 1977). In (Boehm et al., 1979) and (McCall et al., 1977), the authors present quality models for conceptual modeling. However, both of them do not convey any concept related to model composition, such as conflicts and inconsistencies. In (Lange, 2007), Lange aims at proposing an extension of (Boehm et al., 1979) and (McCall et al., 1977) in the context of software modeling; they provide guidelines for selecting metrics and rules to quantify the quality of UML models. The purpose of this quality model is to support a broad quality evaluation of UML models. Although the Lange's quality model has been created based on a literature review and on experiences from industrial case studies, it is not suitable to evaluate model composition effort due to the reasons described in the previous section.

Moreover, we have also observed that previous works have been structuring and specifying the quality model in different ways. Although Boehm (Boehm et al., 1978), McCabe (McCabe, 1976), and Lange (Lange, 2007a) structure their proposed quality models following a hierarchical approach, they differ as to the manners of the hierarchical levels are defined. Each level defines a different set of concepts of the quality model. For example, McCall defines the quality framework in three hierarchical levels containing *Uses*, *Factors*, and *Criteria*, respectively. Boehm uses a different vocabulary but similar meaning for these levels. On the other hand, Lange proposes his quality model with four hierarchical levels containing *Use*, *Purpose*, *Characteristics*, and *Indicators*. Our proposed quality model adopts these four levels as the relation between quality notions and the indicators can be better specified and understood.

As mentioned in Section 3.1, the current quality models fail to specify the relations between the concepts found in software modeling and the ones defined in model composition. Hence, it is not trivial to grasp how developers' effort can be quantified only considering the concepts defined by Lange (Lange, 2007a). They are *User*, *Modeling Language*, *Domain*, and *Design Model*. It is not possible to answer whether, in fact, there are (or not) relations between those concepts and those found in the realm of model composition. For example, the related works do not discuss how the above concepts would relate to concepts such as *Conflict*, *Inconsistency*, and *Model Composition Techniques*. Understanding if these relations are possible, or even how it would occur, is important when studying model composition effort.

In 2010, Marín proposes a quality model based on the metamodeling standard (Marín et al., 2010). This type of specification offers some advantages concerning the previous ones. First, the elements of a quality model are defined by a description, syntax abstract, and semantics constraints. Second, the UML metamodel is also defined following a metamodeling approach. This means that the use of metamodeling can favor the comprehension of the quality model as developers are often familiarized with the UML specification. More specifically, the purpose of the quality model is to formalize the elements involved in the identification of the different types of defects relevant to Model-Driven Development (MDD). This not only encapsulates common defect types usually found in MDD, but also takes advantage of current standards in order to automate defect detection in MDD environments (Marín et al., 2010).

According to Boehm (Boehm et al., 1978), McCall (McCabe, 1976), and Lange (Lange, 2007a), researchers can evaluate software systems by relating metrics to quality attributes. Today, there are many works defining metrics in order to measure source code and design models such as (Fenton et al., 1996; Chidamber & Kemerer, 1994; McCabe, 1976; Martin, 2003). However, none of them explores the relation of metrics and quality notions in the context of model composition assessment. For example, in (Chidamber & Kemerer, 1994), the authors define a set of canonical metrics for OO designs, such as coupling between object classes (CBO) and the lack of cohesion in methods (LCOM).

Martin in (Martin, 2003) proposes another metrics and discusses design characteristics, such as stability. Although these works are effective to assess quality attributes of both source code and design models, they are inadequate to assess the model composition effort. For example, these quality models do not consider important elements in model composition, such as conflicts, inconsistencies, and composition techniques. That is, the current quality models are unable to guide researchers during the planning of empirical studies about model composition effort. This thesis, therefore, extends the previous quality models so that researchers and developers are able to characterize and evaluate model composition tasks. We structure the proposed quality model by using a four-level framework following a metamodeling standard, as in Marín's work (Marín et al., 2010). The proposed extensions are described in the next sections. Nevertheless, the main differences are (1) an abstract syntax is defined to represent the concepts that are the basis of the quality model, (2) new concepts are included in the model (such as conflict, inconsistency, composition technique, and design characteristic), and (3) four quality notions are added (such as effort, application, detection, and resolution notions).

## 3.3.
## A Quality Model for Model Composition Effort

After motivating the quality model (Section 3.1) and contrasting the related works (Section 3.2), this section describes the quality model for model composition effort, which is based on previous works (Lange, 2007; Krogstie, 1995; Lindland et al., 1994; Marín et al., 2010).

### 3.3.1.
### Model Composition Effort and Change Categories

In this section, we define model composition effort and the types of changes that are applied to the design models during the empirical studies. Moreover, this section answers some questions that have motivated the creation of the quality model (Section 3.1).

To begin with, we identify the different types of effort that developers can invest to produce an output intended model. Model composition effort can refer to the time invested (or the number of activities required) to produce the output intended model. In Figure 3, an effort equation summarizes three complementary facets of model composition effort. The equation makes explicit that developers invest effort to realize three activities to compose the base model, $M_A$, i.e. the model to-be changed, and the delta model, $M_B$, so that the intended model, $M_{AB}$, can be produced. However, some additional effort may be invested to solve inconsistencies in the composed model, $M_{CM}$:

1. $f(M_A, M_B)$: effort to apply composition technique to produce $M_{CM}$ from $M_A$ and $M_B$.

2. $diff(M_{CM}, M_{AB})$: effort to detect inconsistencies in $M_{CM}$.

3. $g(M_{CM})$: the effort to resolve inconsistencies i.e., the effort to transform the composed model ($M_{CM}$) into the intended model ($M_{AB}$). Note that if $M_{CM}$ is equal to $M_{AB}$, then $diff(M_{CM}, M_{AB}) = 0$ and $g(M_{CM}) = 0$. Otherwise, $diff(M_{CM}, M_{AB}) > 0$ and $g(M_{CM}) > 0$.

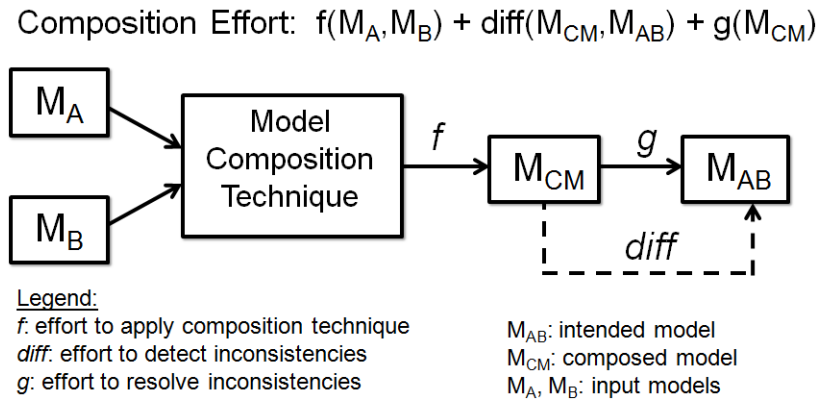

Figure 3: Overview of model composition effort: an equation

Developers spend effort to accommodate changes from the $M_B$ to the $M_A$. We have identified four types of changes that usually happen during this composition, which are widely accepted by researchers (Mens, 2002). Note that the quality model is not limited to be used to these changes. The changes are described as follows:

- *Addition*: new model elements from some delta model are inserted into the base model; for instance, the new attribute – *name: String* is inserted into the class Researcher (Figure 4).

- *Removal*: a model element in the base model is removed; for example, the attribute, *+salary: int* is removed from the class *Researcher*.

- *Modification*: a model element has some properties modified; for instance, the class Researcher in the base model has its property *isAbstract = false* modified to true in the delta model (name in italic style).

- *Derivation*: model elements are refined and/or moved to accommodate the changes (Mens, 2002); for example, the class *Researcher* in the intended model (Figure 4) has the attributes *name* and *salary* moved to the classes *Assistant* and *Professor*.

When developers accommodate these different types of changes into the base model ($M_A$) some conflicts between the properties of the design models can arise. We present the concept of conflicts and inconsistencies in the next section.

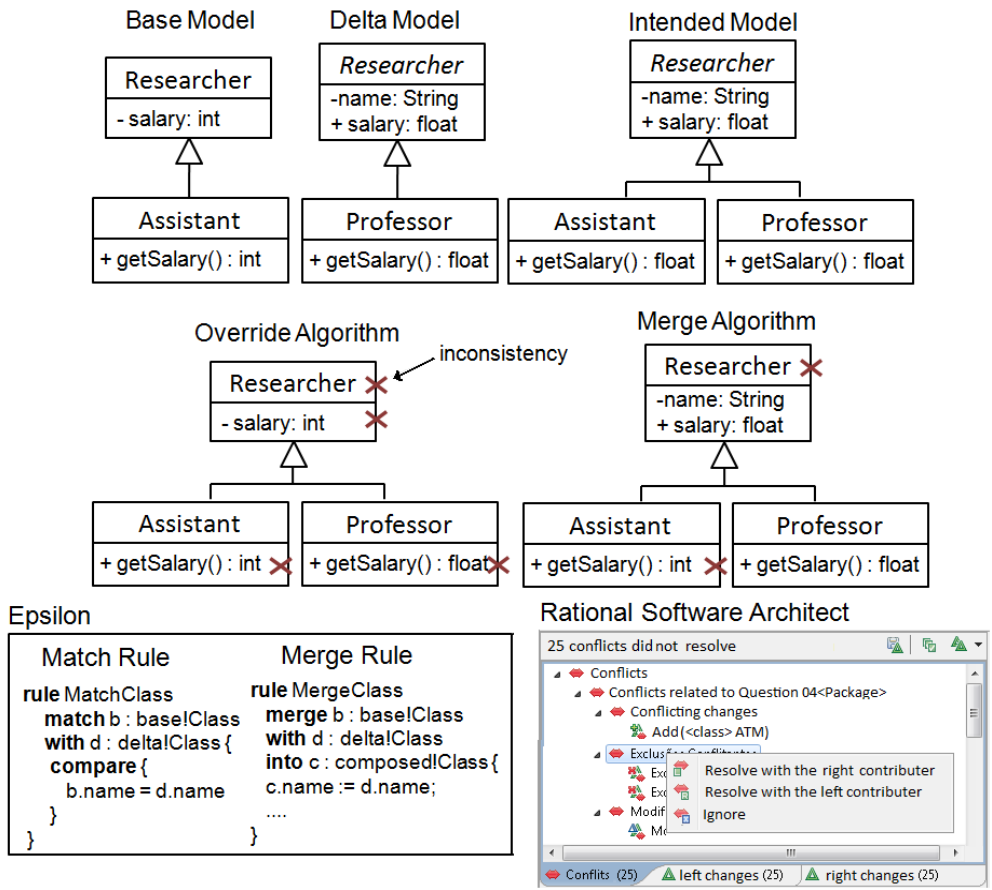

Figure 4: Illustrative example

### 3.3.2.
## Composition Conflicts and Inconsistencies

Composition conflicts consist of contradictions between the values assigned to the properties of the design models (Mens, 2002). They emerge when the input models $M_A$ and $M_B$ need to be composed and their overlapping parts have contradicting values. Figure 4 shows a practical example of conflicting changes when we try to compose the classes *Researcher* of the base and delta model.

In the base model, the UML class *Researcher* is defined as a concrete class (i.e., *Researcher.isAbstract* = false) whereas in the delta model class *Researcher* is set as an abstract class (i.e., *Researcher.isAbstract* = true). That is, we have contradicting values assigned to the same class. Then, the developers need to properly answer the question: should class *Researcher* be abstract or not? In this particular case, the correct answer is that the *Researcher* is abstract — i.e., *Researcher.isAbstract* = true. This can be observed in the intended model in Figure 4.

However, if this question is not properly answered, inconsistencies are inserted into the output composed model. Inconsistencies are unexpected values assigned to the properties (or characteristics) of the design models. For example, *Researcher.isAbstract* = false represents an inconsistency as the expected value is true. Note that when the conflicts are incorrectly resolved they are converted into inconsistencies in the output composed model. Figure 4 shows the class *Researcher* produced by the override and merge algorithms (Section 2.4.1) as a concrete class (*isAbstract* = false) instead of abstract (*isAbstract* = true) as would be expected. Note that these inconsistencies lead the model to-be considered not compliant with the intended model. Two categories of inconsistencies can emerge as follows:

- o *Syntactic inconsistency* emerges when any output composed model elements do not conform to the rules defined in the modeling language's metamodel. For example, a class must have attributes with different names.

- o *Semantic inconsistency* arises when the meaning of the elements of a composed model does not match with the elements of the intended model. For instance, a class in $M_{CM}$ has an unexpected method or it requires functionality from another class that no longer exists.

We consider both categories of inconsistencies throughout this thesis. The composition techniques, such as IBM RSA (Section 2.4.2), are able to automatically detect syntactic inconsistencies while the semantic inconsistencies can be only detected manually. The composition techniques are unable to detect semantic inconsistencies because semantic information about the model elements is rarely represented in a formal way.

| Metric | Description |
|---|---|
| NFCon | The number of inconsistent functionalities |
| NCCon | The number of model elements that are not compliant with the intended model |
| NDRCOn | The number of dangling reference inconsistencies |
| NASCon | The number of abstract syntax inconsistencies |
| NUMECon | The number of meaningless model elements |
| NBFCon | The number of behavioral feature inconsistencies |

Table 2: Metrics of semantic inconsistencies (Farias et al., 2008)

Hence, the composition techniques cannot proactively localize such inconsistencies. With this in mind, six metrics are proposed. Table 2 briefly presents these metrics. These inconsistencies were chosen because we have observed from empirical studies that they are the most common types of inconsistencies faced by developers in practice (Farias et al., 2008; Mens, 2002).

### 3.3.3.
### Abstract Syntax of the Quality Model

The goal of the abstract syntax is to define the quality model more precisely, thereby identifying the main concepts and their relationships. As this quality model is based on previous works (Lindland et al., 1994; Krogstie, 1995; Lange, 2007), the extensions are based on the creation of four new model elements, and six relationships, which are discussed as follows.

Figure 5 shows the abstract syntax of the proposed quality model, which relies on the metamodeling pattern used in the UML metamodel (OMG, 2011). Note that the numbers in Figure 5 correspond to the numbers in brackets of the quality notions to be discussed in Section 3.5.2. We adopted the UML metamodel as a reference because the UML is in fact the standard modeling language in both academia and industry (Dobing & Parsons, 2006). It is important to highlight that each association represents some effort that developers should invest. With this in mind, the elements of the abstract syntax (Figure 5) are presented as follows.

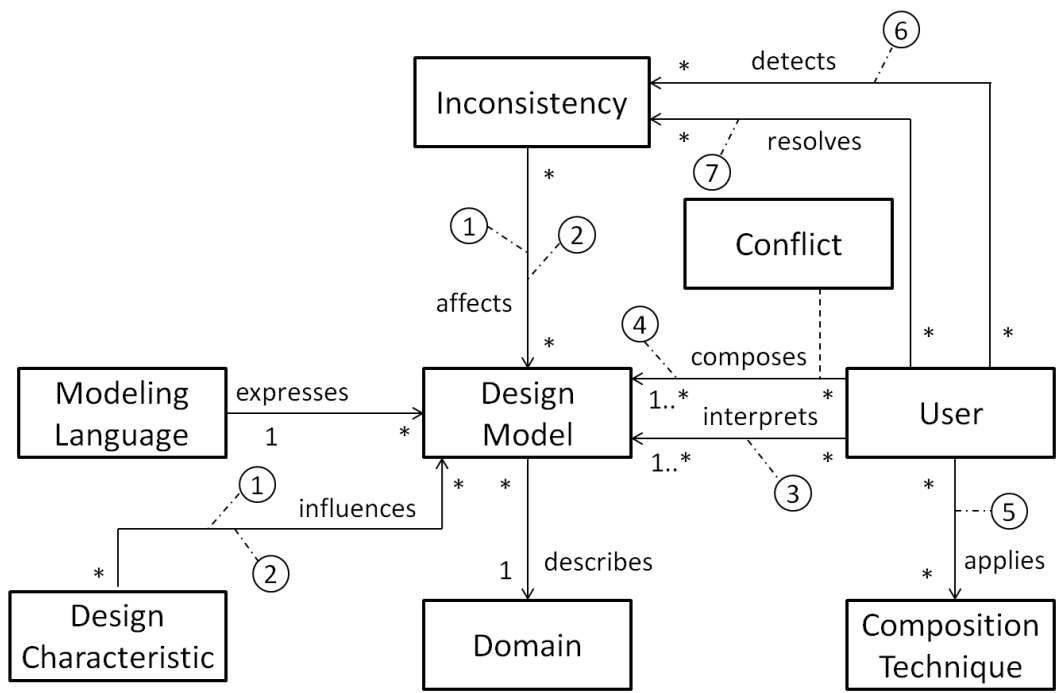

Figure 5: Abstract syntax of the quality model for model composition (based on (Lange, 2007))

### a. Domain

The first element to be discussed is the concept of *domain*. This concept represents an area of expertise or application that needs to be examined to solve a problem. The solution of the problem is represented in a design model. In other words, a domain consists of a reality to be represented by using a modeling language. Supply chain, finance, and telecommunications are three examples of domains. Typically, it can be stated as a conceptual model where a set of concepts and relations are represented.

*Association*

- Without a directed relationship

### b. Modeling Language

Modeling language is the concept that represents the language used to design a software system. Object-oriented modeling languages and aspect-oriented modeling languages are two examples of typical categories of languages used to represent significantly different forms of design decompositions. Modeling languages are commonly used in practice to improve the communication between development teams and provide alternative means for achieving design modularity. Different modeling languages – such as object-oriented and aspect-oriented ones – may influence the structure of a design. Software engineers use these languages to communicate design decisions and check the feasibility of implementing the envisaged design. Example of a premier software modeling tool is the IBM Rational Software Architect (IBM RSA, 2011). The modeling languages define a set of constructs that are used to create instances of the design models.

*Association*

- expresses: Design Model[*]

  Each *expresses* represents the statement of design models. An *expresses* means that the constructs of the design modeling language are instantiated to create a *Design Model* concerning some *Domain*.

UML and its profiles are examples of design modeling language used in practice. This is an ordered association from *Modeling Language* to *Design Model*.

### c. Design Model

Design model refers to the diagram used to represent static and dynamic aspects of a software system. UML class and sequence diagrams are examples of these design models. Developers commonly use these two diagrams, for example, to design structural and dynamic aspects of an application. Moreover, a design model represents the concepts (and their relations) from a domain. This representation helps to describe this domain.

*Association*

- describes: Domain[1]

  Each *describes* represents a particular domain. This representation defines that every design model should describe a particular domain. This is an ordered association from *Design Model* to *Domain*. *Design Models* can describe just a domain.

### d. User

User is a person who interprets design models to get an understanding of the domain (Lange, 2007a). A user can interpret one (or more) design model and compose design models for any particular purpose. Additionally, the user detects and resolves inconsistencies that arise from the compositions. Typical categories of users are software developers and researchers.

*Association.*

- composes: Design Model[2..*]

  Each *composes* represents the instance of a composition that is realized by *User*. A *composes* declares that there may be composition between instances of two (or more) design models. A composition is a tuple with two (or more) design models for each end of the association, where each design model is an instance of the type of the end (i.e., *Design Model*). This is an ordered association from *User* to *Design Model*. Users can compose tow (or more) design models.

- detects: Inconsistency[*]

  Each *detects* represents the detection of inconsistencies by the *User*. A *detects* specifies that there can be detection of inconsistencies when a *User* realizes composition of design models. This is an ordered association from *User* to *Inconsistency*. *User* can detect anything to many inconsistencies.

- resolves: Inconsistency[*]

  Each *resolves* represents the resolution of inconsistencies by *User*. A *resolves* specifies that there can be resolution of inconsistencies when a *User* realizes composition of design models. This is an ordered association from *User* to *Inconsistency*. *User* can resolve from none to many inconsistencies.

- interprets: Design Model[1..*]

  Each *interprets* represents the interpretation of design models by *User*. A *resolves* specifies that there can be resolution of inconsistencies when a *User* realizes composition of design models. This is an ordered association from *User* to *Inconsistency*. *User* can interpret no or many inconsistencies.

- applies: Composition Technique[*]

  Each *applies* represents the application of model composition technique to compose design models by *User*. A *applies* specifies that there can be the use of composition technique when a *User* realizes composition of design models. This is an ordered association from *User* to *Composition Technique*. *User* can apply no or many composition techniques.

*e. Conflict*

Conflict is the concept that represents the contradictions between different design models to be composed. Since *User* tends to assign contradicting values to the properties of the *Design Models* (Section 3.4). Conflicts arise why the design models receive conflicting changes. These contradictions happen when the ordered association *composes*: Design Model [2..*] from *User* to *Design Model* is instantiated. Thus, conflict is a derived concept from the association *composes*.

For example, a developer defines that a class is abstract (i.e., *isAbstract* = true) while another developer specifies that the same class is concrete (i.e., *isAbstract* = false). *User* should grasp and tame these conflicts in order to able to produce an intended design model.

> *Association*

- Without a directed relationship

### f. Inconsistency

Inconsistency is the concept that represents the defects found in the output composed model (Section 3.4). It usually arises because *User* tends to incorrectly resolve the *Conflicts*. For example, developers can incorrectly tame the conflict whether a class should be abstract or not.

> *Association*

- affects: Design Model[*]

  Each *affects* consists of problems jeopardizing quality notions of the *Design Model*. When the *affects* takes place implies to say that an output composed model and the output intended model do not match ($M_{CM} \neq M_{AB}$). This is an ordered association from *Inconsistency* to *Design Model*.

### g. Design Characteristic

A design characteristic is the concept that illustrates the strategies used by developers to structure design models such as coupling and cohesion. Design characteristics are used to improve, for example, the capability of design models to be (more straightforwardly) composed. The design characteristics are also used as indicators (Martin, 2003) of prone to problems. An example of this design characteristic is model stability (Section 2.6).

> *Association*

- influences: Design Model[*]

  Each *influences* represents that the design characteristics modify the manner of the design model is created or can act as an indicator such as stability. This is an ordered association from *Design Characteristic* to *Design Model*.

### h. Composition Technique

Composition technique is the concept that represents the technique used by developers to compose the design models. Examples of these techniques are Epsilon and IBM Rational Software Architect. A model composition technique defines a set of operators that are used to manipulate the input model elements. More detail about this concept can be found in Section 2.4.

*Association*

- Without a directed relationship.

## 3.3.4.
## Quality Notions

After presenting the basic elements of the quality model, we discuss the quality notions associated somehow with each one of them. In our study, quality notions can be seen as non-functional requirements used to evaluate the effort of a composition. Our quality model focuses on seven quality notions, namely syntactic, semantic, social, effort, application, detection, and resolution notions. We propose four quality notions effort, application, detection, and resolution notions. Each of them captures a fundamental dimension of quality related to model composition activities. The other quality notions are tailored from previous works (Lindland et al., 1994; Krogstie, 1995; Lange, 2007a). Lindland (Lindland et al., 1994) proposed three quality notions — i.e., syntactic, semantic, and pragmatic ones. Krogstie (Krogstie, 1995) and Lange (Lange, 2007) add the social and communicative quality notion to the Lindland's quality notions, respectively. All these notions were tailored to the context of evaluation on model composition effort. These extensions are discussed as follows:

- **Syntactic Quality (1).** Krogstie originally proposed this quality notion (Krogstie, 1995) to represent the correctness of design models produced by a design modeling language (Lange, 2007a). If a design modeling language is not properly used, then some syntactic inconsistencies may emerge. This quality notion is relevant to our quality model as syntactic inconsistencies can also arise during model compositions (Mens, 2002). Developers need to be concerned with checking the syntactic consistency of the output composed model. The degree of correctness should be evaluated in terms of the presence or absence of inconsistencies of the composed model. In other words, syntactic quality is computed by measuring the inconsistencies resulting from conflicts between the input models. For this, inconsistency metrics (Farias et al., 2008a) are used. This notion helps developers to identify the number of deviations in the output composed model with respect to the language specification. This quality notion is studied in empirical studies presented in Chapters 5, 6, and 7.

- **Semantic Quality (2).** This notion deals with the degree of correspondence between the design model and the problem domain (Lange, 2007a). If the semantics of the model elements are affected, the main purpose of use of the design models — i.e., communication between the team members can be damaged. Thus, developers and designers need to be concerned with checking the meaning of the model elements in the output composed model. In a similar way to the syntactic notion, the degree of correctness should be evaluated in terms of the presence or absence of inconsistencies. That is, semantic quality is calculated by measuring the conflicting correspondence between the design model and the problem domain (Chapter 2). This inadequate representation may occur by two reasons (but not limited to): (i) the inability of the developers to represent the concepts and the relationship of the domain, and (ii) the inaccuracy of the composition techniques that inadequately manipulate the semantics of the model elements (Mens, 2002). To quantify these semantic inconsistencies, some metrics defined in (Farias et al., 2008a) are used. This quality notion is studied in Chapters 5, 6, and 7.

- **Social Quality (3).** Design models are essentially used to communicate design decisions between the software development teams (Larman, 2004; Dobing et al., 2006). If there is a disagreement between the interpretations of the design models, the communication between the developers is severely harmed. With this in mind, researchers should elaborate studies in order to understand the effects of the misinterpretations on the implementation. For example, if the degree of misinterpretations is high, the diverging understanding may be converted into defects in code. These two reasons can in fact damage the interpretation of the output composed models. The social quality notion, therefore, matches the interpretations of the developers and checks the degree of disagreement between them. Therefore, the focus of such social notion is to evaluate the threats to the agreement of interpretations of the design models by the developers. The evaluation aims at comprehending how the misinterpretation may be motivated by (but not limited to): (1) the inadequate layout of the model elements caused by the incorrect positioning of the model elements, and (2) the representations of the constructs of the current modeling languages are not friendly. The method described in (Lange, 2007a) to measure the degree of the misinterpretations is used. This quality notion is studied in Chapter 6.

- **Effort Quality (4).** This quality notion addresses the effort of producing an output intended model. It is expected that the practices of applying a composition technique, detecting, and resolving inconsistencies are not effort-consuming tasks. However, they will inevitably require extensive effort to produce an indented model in several cases. Therefore, this quality notion deals with the cost of obtaining an expected output model. This quality notion is studied in Chapters 5, 6, and 7. The next three quality notions refine this quality notion by addressing the easiness (or difficulty) in the tasks of applying composition techniques, detecting, and resolving composition inconsistencies.

- **Application Quality (5).** This notion represents the applicability of a particular model composition technique. In other words, it addresses the ease of producing an output composed model by applying a model composition technique. Ideally, developers expect to be able to effortlessly compose design models by using either heuristic-based or specification-based composition techniques. However, two difficulties make the practice of applying composition techniques not trivial. The first difficulty arises from the inherent challenge of making use of different categories of model composition techniques. Each of them imposes different burdens on software designers. For instance, developers need to manually specify rules in order to define the equivalence and composition relations between the input model elements. On the other hand, they may also compose the models using heuristic-based composition techniques. The second difficulty consists of the accidental problems that emerge from the practice of bringing design models together. Usually developers need to resolve conflicting changes performed in parallel. This quality notion is studied in Chapter 5.

- **Detection Quality (6).** After producing an output composed model, developers should review it to assure its correctness. That is, developers should check if some inconsistency was produced as the result of the composition. When inconsistencies arise, developers should be able to quickly localize them. If the detection of inconsistencies is hard, then the assurance of the correctness of the models may also be hard. Unfortunately, the localization of inconsistencies is not always a trivial task. This can be explained by at least two reasons (but not limited to): (i) the composition techniques cannot often help developers to automatically detect all kinds of inconsistencies. Since, the meanings of the model elements are rarely represented in a formal way; and (ii) developers cannot understand specific inconsistencies, mainly semantic inconsistencies, given the problem at hand and their knowledge about the meaning of the model elements. With this in mind, researchers should study the degree of difficulty that developers face to localize inconsistency so that the consistency of the output composed model can be assured. In particular, it is expected that researchers provide a clear understanding about the effort to detect inconsistencies in practice. Therefore, the focus of this quality notion is on evaluating the cost to localize inconsistencies in the output composed model. This evaluation is important because it allows researchers to understand, for example, if design modeling languages such as UML and aspect-oriented modeling can significantly affect the detection effort, or if alternative composition techniques such specification-based or heuristic-based ones can influence the detection. This quality notion is studied in Chapters 5 and 6.

- **Resolution Quality (7).** After detecting inconsistencies, developers should resolve them in order to transform the output composed model into the output intended model. That is, developers should invest some additional effort (apart from producing the output composed model) trying to find some solution to the inconsistencies already localized. Otherwise, the practice of composing design model can become prone to inconsistencies or even require more effort than it would be expected. This additional effort can make the practice of assuring the consistency of the composed models difficult and costly. Unfortunately, the resolution of inconsistencies is not always an easy task. This can be explained by the lack of accuracy of the composition techniques to understand the meaning of the model elements and the incapability of the developers to find an adequate solution to the inconsistencies (Mens, 2002). This notion, therefore, addresses the degree of difficulty to resolve inconsistencies. This difficulty of resolving inconsistency can be calculated considering the time invested to resolve them or even the number of activities that developers should perform. Moreover, it copes with the inherent and accidental difficulties of solving composition anomalies e.g., syntactic and semantic inconsistencies. The first complexity arises from the need to reason and then make decision about how to tame inconsistencies. The accidental difficulty is caused by the modeling technique such as OO or AO modeling used to represent the design models and by the manner as they are structured i.e., more modularized or not. This quality notion helps understanding the difference between how the developers think about inconsistency resolution and how in fact they resolve inconsistencies. This quality notion is studied in Chapters 5 and 7.

Table 3 describes how the quality notions that are addressed through the empirical studies presented in the next chapters.

| Chapter | Quality Notion | Description |
|---------|----------------|-------------|
| 3 | all quality notions | Definition of the quality model for model composition effort |
| 4 | effort, application, detection, resolution, syntactic, semantic | Empirical studies address the quality notions in practice |
| 5 | effort, detection, social, syntactic, semantic | A controlled experiment is performed to investigate the five quality notions |
| 6 | effort, resolution, syntactic, semantic | Quasi-experiments were realized to study the four quality notions |
| 7 | all quality notions | All quality notions are discussed based on the series of empirical studies performed |

Table 3: Definition of chapters where quality notions are investigated

### 3.3.5.
### Levels of the Quality Model

The quality model is organized following a 4-level specification pattern. To define the quality model with levels, we need to consider: (1) when model composition is used i.e., in which phase of the development process it is used; (2) why model composition is applied i.e., the purpose of using the model composition; (3) what can be used to characterize model composition i.e., the characteristics that are directly related to model composition; and (4) how such characteristics can be quantified i.e., the definitions of rules and metrics used to measure the characteristics. These four levels are hierarchically organized and this fine-grained partitioning allows separating concerns across layers of abstractions, and providing flexibility to future studies so that they may extend the quantity model.

This section, therefore, brings forward the levels of the quality model and the concepts that belong to the levels. Recall that this thesis attempts to investigate the effort that developers invest to use model composition in the context of design model evolution; however, that does not mean that the model cannot be tailored to other contexts. The model has four levels (based on (Lange, 2007a)), which are described as follows:

#### a. Level 1: Use of Composition

The top level of our quality model describes the high-level use of model composition in practice. These uses are:

- **Development:** developers use model composition to incrementally create the design models before the implementation phase. This use combines quality characteristics that concern the composition before the design model of a system has been completely finished.

- **Evolution:** developers make use of composition techniques to evolve design models. This use combines quality characteristics that concern the product when it is changed.

### b. Level 2: Purposes of Composition

The second level defines the purposes of using that model composition is applied. These purposes are directly related to the purposes discussed in Section 2.1. In practical terms, it specifies why developers use composition. Thus, we identify three purposes of using that are described as follows:

- **Analysis:** Users identify overlapping parts between the model to-be composed. This allows them to analyze possible conflicting changes that are strong candidate to become inconsistencies.
- **Change:** Users essentially use composition techniques to add, modify, remove, or even refine model elements of some existing design model.
- **Reconciliation:** Users use the resource of model composition techniques to reconcile contradicting changes (Clarke, 2001).

### c. Level 3: Characteristics of Composition

The third level of our quality model contains the inherent characteristics of the design model and model composition technique. The characteristics are described in Table 4. According to the distinction between the characteristics of the design model and the characteristics of the model composition technique, we indicate for each characteristic whether it is a characteristic of the design model (column M) or a characteristic of the model composition technique (column T). Some characteristics are defined for both design model and composition technique.

| Characteristic | M | T | Description |
|---|---|---|---|
| Effort | | X | The effort to execute f, diff, and g. |
| Complexity | X | | The degree of difficulty to understand a model (Lange, 2007; Feton et al., 1994). |
| Modularity | X | | The manner by which a software system can be systematically structured and separated such that it can be understood in isolation (Parnas, 1972). |
| Stability | X | | The degree of changes that a module suffers given a need of change i.e. a module is stable if its design characteristics have a low variation (Kelly, 2006). |
| Size | X | | The number of model elements in a design model |
| Correctness | X | | The extent to which a design model is complaint with a reference design model. |
| Consistency | X | | The extent to which no inconsistency is contained (Easterbrook et al., 1996) |
| Communicativeness | X | | The degree of facility to communicate and assimilate content (Boehm et al., 1978; Lange, 2007). |

Table 4: Characteristics of design models

The composition effort that is applied to exclusively to the model composition is characterized by the effort to apply the composition techniques $(f(M_A,M_B))$, to detect $(diff(M_{CM}, M_{AB}))$ and resolve inconsistencies $(g(M_{CM}))$. With this in mind, the characteristics (in Table 4) describe the design models and the composition technique.

### d. Level 4: Metrics and Rules

The fourth level defines how the aforementioned characteristics are quantified. To allow the quantification of these characteristics, a suite of metrics and rules were used. Rules are special cases of metrics; being usually mappings of some observations from the empirical domain to a binary value: true or false (Wust, 2011; Lange, 2007a). These rules evaluate and measure design models, mainly checking well-formed rules and design rules. Two practical examples of well-formed rules would be "Abstract class must not be instantiated" and "Abstract class must not have a concrete class as superclass." Note that the consistency of the design model is affected if these two rules are not assured.

In our empirical studies, several elements appear in the models, depending on the types of diagrams used. Class, interface, and component and examples of elements in component diagrams, which were used in several studies of this thesis. Metrics can be defined to quantify these elements. In order to illustrate these specific metrics: (i) Table 5 describes the metrics for classes, (ii) Table 6 shows the metrics for interfaces, and Table 7 describes the metrics for components. These tables also describe the relations between the characteristics (level 3) and the metrics and rules (level 4) are specified.

The metrics and rules are defined in previous work (Chidamber & Kemerer, 1994; Lorenz & Kidd, 1994; Lee et al., 1995; Martin, 2003; Lorenz, 1994; Chidamber et al., 1998; McCabe; 1976). Although these metrics are often used in previous research, we do not claim that this list of metrics and rules is complete. These metrics were chosen because they are well-known indicators to quantify model characteristics, and are often supported by robust measurement tools, such as SDMetrics (Wust, 2011).

After presenting the concepts and describing the three levels, Figure 6 describes the three top levels of the quality model: *Use, Purpose*, and *Characteristic*. The fourth level Metrics and Rules and the relations to level three are depicted in Table 5, Table 6, and Table 7. Note that a checkmark indicates which characteristic of level three is related to the metric or rule in level four. In Figure 6, the arrows indicate relations between two concepts of different levels. The arrows can be interpreted as follows: a lower level concept is part of all higher-level concepts to which it is related by an arrow, and a higher-level concept contains the related lower level concepts. The interpretation of the relations is that a concept in a lower level in the quality model contributes to the related concepts of the higher level.

| Metric | Characteristic | Description |
|--------|----------------|-------------|
| NAttr | SI | The number of attributes in the class. |
| NOps | SI | The number of operations in a class. |
| IFImpl | CO, MO | The number of interfaces the class implements. |
| NOC | CO, CM | The number of children of the class. |
| NDesc | CO | The number of descendents of the class. |
| NAnc | CO | The number of ancestors of the class. |
| DIT | CO, CM | The depth of the class in the inheritance hierarchy. |
| OpsInh | CO | The number of inherited operations. |
| AttrInh | CO | The number of inherited attributes. |
| DepOut | CO, MO, CM | The number of elements on which this class depends. |
| DepIn | CO, MO, CM | The number of elements that depend on this class. |
| ECAttr | MO | The number of times the class is externally used as attribute type. |
| ICAttr | MO | The number of attributes in the class having another class or interface as their type. |

*SI: size, CO: complexity, MO: modularity, and CM: communicativeness*

Table 5: Metrics for class

| Metric | Characteristic | Description |
|--------|----------------|-------------|
| NOps | SI | The number of operations in the interface. |
| Assoc | CO | The number of elements the interface has an association with. |
| NAnc | CO | The number of ancestors of the interface. |
| NDesc | CO | The number of descendents of the interface. |
| NOps | SI | The number of operations in the interface. |
| ECAttr | CO | The number of times the interface is used as attribute type. |
| ECPar | CO | The number of times the interface is used as parameter type. |
| Assoc | CO | The number of elements the interface has an association with. |
| NDirClients | CO | The number of elements directly implementing the interface. |
| NIndClients | CO | The number of elements implementing a descendent of the interface. |
| NAnc | CO, MO | The number of ancestors of the interface. |
| NDesc | CO, MO | The number of descendents of the interface. |

*SI: size, CO: complexity, MO: modularity, CM: communicativeness*

Table 6: Metrics for interface

| Metric | Characteristic | Description |
|--------|----------------|-------------|
| NOps | SI | The number of operations of the component. |
| NComp | SI | The number of subcomponents of the component. |
| NPack | SI | The number of packages of the component. |
| NCCmp | SI | The number of classes of the component. |
| NIntCmp | SI | The number of interfaces of the component. |
| Connectors | CO | The number of connectors owned by the component. |
| ProvIF | CO, MO | The number of interfaces the component provides. |
| ReqIF | CO, MO | The number of interfaces the component requires. |
| DepOut | CO, MO, CM | The number of outgoing dependencies. |
| DepIn | CO, MO, CM | The number of incoming dependencies. |
| AssocOut | CO, CM | The number of associated elements via outgoing associations. |
| AssocIn | CO, CM | The number of associated elements via incoming associations. |

*SI: size, CO: complexity, MO: modularity, CM: communicativeness*

Table 7: Metrics for components

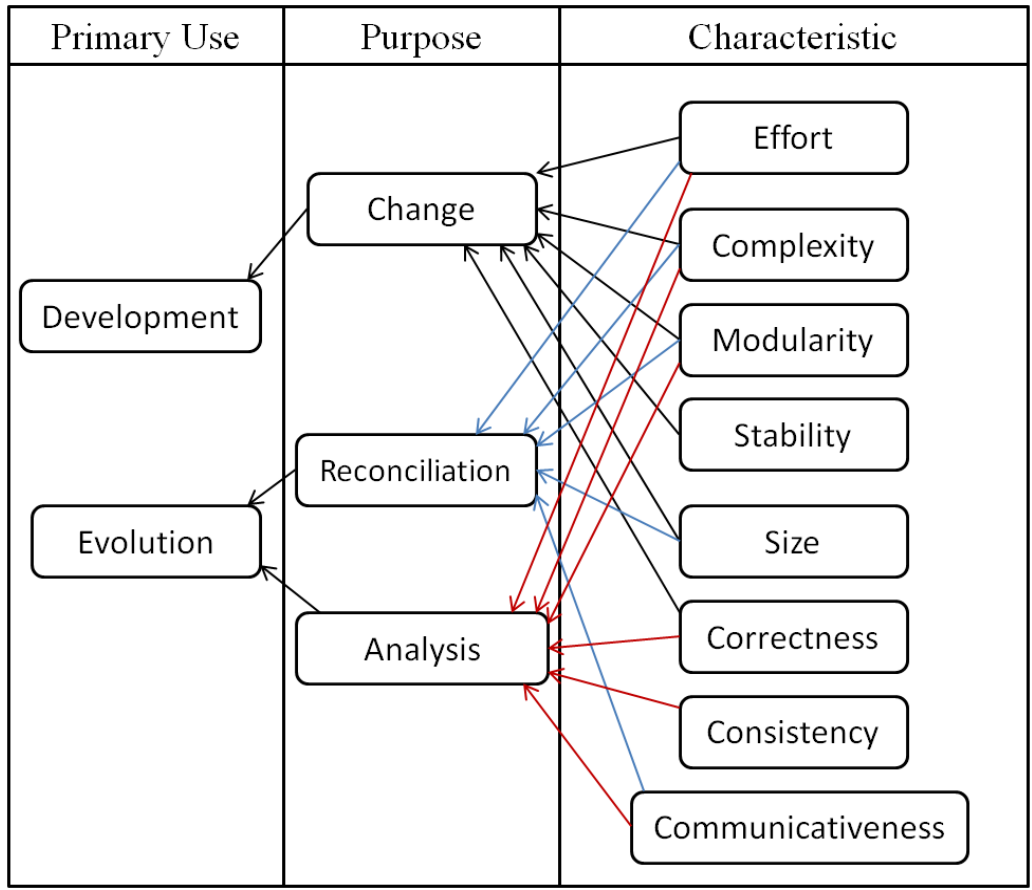

Figure 6: The purposed quality model (based on (Lange, 2007a))

## 3.4.
## Concluding Remarks

Developers need to evaluate model composition effort. However, the evaluation without any quality model is not a trivial task (Basili & Lanubile, 1999) as usually developers have no previous knowledge or experience about empirical evaluations of model composition. This chapter, therefore, presents a quality model for model composition effort. It is intended to help researchers and developers to carry out empirical studies of model composition.

The proposed model extends three previous quality frameworks for conceptual models proposed by Lindland (Lindland et al., 1994), Krogstie (Krogstie, 1995), and Lange (Lange, 2007a). The model is organized in a four-level structure. The first level defines the context where model composition is used in practice, being development and evolution the two usage scenarios proposed and investigated. The second level refers the purposes of using model composition. We identify and evaluate model composition for three purposes of using: change, analysis, and reconciliation. The third level refers to the characterization of the elements involved in model composition: the models and model composition techniques. That is, it considers the artefacts and the techniques responsible for manipulating them. The fourth level aims at quantifying the elements identified in the third level. To this end, metrics and rules are used.

By defining this quality model, we can solve the problems presented in Section 4.1 First, researchers and developers can make use of a unifying framework for the evaluation of model composition. As a result, the findings resulting from multiple studies can be compared, or even checked whether they are valid in a specific context or not. Finally, the use of the quality model serves as a reference frame for structuring empirical studies of model composition. In this context, the quality model guides all empirical studies performed throughout the thesis.

# 4
# Effort on the Application of Composition Techniques

The goal of this Chapter is to evaluate the effects of model composition techniques on the developers' effort. To this end, two studies are performed. The first study investigates the effort that developers invest to compose design models based on a controlled experiment. The second study evaluates the effort to compose design models from industrial case studies.

## 4.1.
## Effects of Composition Techniques on the Composition Effort

Model composition techniques can be classified in two categories (Chapter 2): (i) specification-based techniques, such as Epsilon (Epsilon, 2011) and MATA (Whittle & Jayaraman, 2010), and (ii) heuristic-based techniques, such as merge and override (Clarke, 2001; Clarke & Walker, 2001) and the three-way merge algorithm (Mens, 2002). The manual model composition is expected to be error-prone and time consuming. Then, developers apply model composition techniques with the aims of reducing the composition effort and improving the correctness of the composed model. The techniques in the first category primarily aim at producing correctly composed models, but it is questionable if they necessarily reduce composition effort. On the other hand, the second category aims at alleviating the developers' effort. However, its positive impact on the correctness of the composed models is expected to be worse than the first category.

By using the specification-based techniques, developers explicitly specify the correspondence and composition relations between the input model elements $M_A$ and $M_B$ to produce $M_{AB}$ (Section 2.4). On the other hand, by using the heuristic-based techniques developers apply a set of predefined heuristics, which "guess" the relations between model elements before composing $M_A$ and $M_B$. Specification-based techniques provide a systematic way to specify the relations between the input model elements, instead of trying to "guess" them. It is expected that these techniques  not only alleviate the composition effort, but also assure a higher rate of correctly composed models when compared to the heuristic techniques (Epsilon, 2011; Whittle & Jayaraman, 2010).

To date, however, there is little empirical evidence to confirm (or not) if these expectations hold; mainly, when developers try to: (1) select and apply the model composition techniques; (2) detect syntactic and semantic inconsistencies; and (3) resolve the identified inconsistencies in realistic settings. As described in Chapter 3, these three composition activities are required to obtain the intended model $M_{AB}$. Empirical studies in model composition are lacking, mainly ones considering the impact of the composition techniques on the following quality notions described in our quality model: effort, application, detection, resolution, syntactic, and semantic notions. In fact, the literature fails to provide such empirical evidence to software developers. As a result, developers are left without any practical knowledge to answer questions such as "*what are the effects of specification-based and heuristic-based techniques on the developers' effort and the correctness of the composed models?*" It is important to answer this question because, before adopting any composition technique in realistic settings, it is necessary to have practical knowledge about the effects of model composition techniques.

In fact, to date, both specification-based and heuristic-based techniques have been used without any empirical evidence. Currently developers rely on diverging feedbacks (Norris & Letkeman, 2011) from evangelists to evaluate how good techniques can be, rather than on practical, evidence-based knowledge derived from experimental studies. The practical knowledge about these effects (or even a trade-off analysis) can be viewed as the main impairment to the wide application of composition techniques in practice where resources and time are tight. Note that if a composition technique reduces effort but does not favor model correctness (or vice-versa), it is quite questionable whether it can be applied in industry. On the other hand, if the composition effort is high, the potential benefits of using composition techniques (e.g., gains in productivity) can be compromised.

The literature in model composition fails to provide assessments of model composition techniques (Apel et al., 2011; Sarma et al., 2011; Shao et al., 2011; Brun et al., 2011; Whittle et al., 2009; Klein et al., 2006). Apel (Apel et al., 2011). Mens (Mens, 2002) also reinforces the need for more empirical and experimental research. Burn and colleagues (Brun et al., 2011b) evaluate the composition of code in the context of a retrospective, quantitative study of the evolution of nine open-source systems. They concluded that inconsistencies in code are the norm rather than the exception, and that 16% of all merges required human effort to resolve them. However, even this kind of primary empirical analysis is lacking in the context of model composition.

With this in mind, this Chapter reports a controlled experiment performed with 24 subjects, which used Epsilon, IBM RSA and traditional composition algorithms to evolve design models. The techniques are investigated in 144 evolution scenarios and by about 2304 compositions of model elements (such as classes and relationships). The main results, supported by a complete statistical and qualitative analysis, are: (1) the IBM RSA and traditional composition algorithms require less effort to produce the intended model than Epsilon, and (2) there is no significant difference in the correctness of the output composed models generated by these techniques. Additionally, in some cases, the number of inconsistencies produced by Epsilon was significantly higher than one generated by IBM RSA and traditional composition algorithms. The techniques investigated are robust and representative and there are reasons to believe the results will generalize to broader scenarios. However, we do not claim generalization beyond these techniques and their use on other types of design models, in particular class and sequence diagrams.

The remainder of the chapter is organized as follows. Section 4.1.1 presents the experiment planning. Section 4.1.2 analyzes the results. Section 4.1.3 contrasts our work with related work. Section 4.1.4 presents the threats to validity. Finally, Section 4.1.5 describes some concluding remarks.

### 4.1.1.
### Experiment Planning

This section presents the experiment planning followed to carry out a controlled experiment. This planning is based on practical and conventional guidelines of empirical studies such as (Wohlin et al., 2000; Kitchenham et al., 2008; Shadish et al., 2002; Sjober et al., 2002). We have opted to conduct a controlled experiment to investigate the hypotheses formulated in Section 4.1.1.2 due to a number of reasons (Basili et al., 2007). First, it allows us to conduct well-defined, focused studies, with the potential for gathering statistically significant results, which is not possible with non-controlled case studies. Moreover, it helps to formulate hypotheses by forcing us to clearly state the question being studied and allow us to maximize the number of questions being asked.

Second, as controlled experiments require well-formulated dependent and independent variables as well as null and alternative hypotheses, it also allows us to understand the relations of specific variables and measures.

Third, by running a controlled experiment, we are forced to state clearly what questions the investigation is intended to address and how we will address them, even if the study is exploratory (Basili, 2007). Moreover, we can create a study design in such a way that maximizes the chance for replication of the study in order to test the hypotheses in different contexts and by independent researchers.

Fourth, controlled empirical studies can better investigate the cause-effect relationships between variables, allowing us to understand, for example, the effects of the independent variables on the dependent variables. Additionally, a controlled study provides insight into why relationships and results do and do not occur. It also forces us to analyze the threats to validity, leading to the identification of where replications or alternate studies are needed and where variations might show different effects. It also allows us to build a body of knowledge in model composition that helps researchers to build theories supported by clear empirical evidence.

### 4.1.1.1.
### Experiment Definition

This study aims at evaluating the effects of model composition techniques on six quality notions, namely syntactic, semantic, effort, application, detection, and resolution ones. For this, we control two variables: the *effort* to compose design models and the *correctness* of the output models. Correctness is also controlled, as the evaluation of effort needs to be put in the perspective of the quality of the produced models. Otherwise, the cost-effective analysis cannot be fully drawn. These effects are investigated through a controlled experiment in which developers use specification-based and heuristic-based techniques to evolve design models. With this in mind, the objective of this study is stated based on the GQM template (Basili et al., 1994) as follows:

*Analyze* composition techniques

*for the purpose of* investigating their effects

*with respect to* effort and correctness

*from the perspective of* developers

*in the context of* the evolution of design models.

Therefore, this controlled experiment addresses the research question RQ2, as stated in Section 1.3.

- **RQ2:** What is the relative effort of composing design models with specification-based composition techniques and heuristic-based composition techniques?

Based on this, we further decompose the RQ2 into two research subquestions described below:

- **RQ2.1:** What is the relative effort of composing two input models by using specification-based composition techniques with respect to heuristic-based composition techniques?

- **RQ2.2:** Is the number of correctly composed models higher with specification-based techniques than with heuristic techniques?

### 4.1.1.2.
### Hypothesis Formulation

Table 8 describes the hypotheses for testing the effects of composition techniques on effort and correctness. These hypotheses are elaborated throughout this section.

*Hypothesis 1.* The first hypothesis of this section is that, although the specification-based composition technique provides a more systematic way to compose the input models, it does not reduce the composition effort. Our expectation is that developers invest more effort to write down the specifications rather than using the heuristic-based composition techniques. This can be explained based on the expectation that they are not intuitive or flexible enough to express the change requests. Moreover, the presence of inconsistencies in the output composed model may have a detrimental effect on the composition effort.

As developers should examine all points in the input models (affected by the specifications) or even "guess" which input model elements are incorrectly combined. Consequently, this additional effort may increase the composition effort rather than minimize it. However, it is by no means obvious that this hypothesis holds. It may be, for example, that they help developers to match and then compose the input models more quickly. With this in mind, the null hypothesis states that the specification-based technique requires less (or equal) effort to compose the input models than the heuristic-base technique. On the other hand, the alternative hypothesis states that the effort is significantly higher. These hypotheses are summarized as follows. Note that our expectation has a specific direction, which leads, in turn, to the definition of one-tailed hypotheses.

**Null Hypothesis 1, $H_{1-0}$:** The specification-based composition techniques require less (or equal) effort than the heuristic-based composition techniques to produce $M_{AB}$ from $M_A$ and $M_B$.

$H_{1-0}$: Effort$(M_A, M_B)_{Specification} \leq$ Effort$(M_A, M_B)_{Heuristic}$

**Alternative Hypothesis 1, $H_{1-1}$:** The specification-based composition techniques require more effort than the heuristic-based composition techniques to produce $M_{AB}$ from $M_A$ and $M_B$.

$H_{1-1}$: Effort$(M_A, M_B)_{Specification} >$ Effort$(M_A, M_B)_{Heuristic}$

For a more detailed investigation, we break this hypothesis in three subhypotheses ($H1_2$, $H1_3$, and $H1_4$). The goal is to evaluate the relative efforts (f, diff, and g) defined in the composition effort equation (see Figure 3). A complete formulation of these hypotheses can be seen in Table 8.

*Hypothesis 2.* The second hypothesis is that the use of specification-based composition techniques increases the number of correctly composed models. This is because developers can explicitly specify the composition relations between the input models. However, it is not clear whether this manner of realizing model composition promotes higher correctness of the output model. The need to explicitly take into consideration each of the models' properties (such as *isAbstract*), when specifying the relations, may cause difficulties to properly write down the specifications. If this is confirmed, then inconsistencies are inserted into the output composed model, compromising its correctness (i.e., $M_{CM} \neq M_{AB}$). With this in mind, the null hypothesis assumes that the specification-based composition technique produces a lower (or equal) number of correctly composed models than the heuristic-based composition technique. On the other hand, the alternative hypothesis states that the specification-based technique produces a higher number of correctly composed models than the heuristic-based technique. In other words, the correctness (Cor) of the output composed models is usually assured when they are produced by the specification-based techniques. These hypotheses are presented as follows:

> **Null Hypothesis 2, $H_{2-0}$:** Specification-based techniques produce a lower (or equal) number of correctly composed models than the heuristic-based techniques.
>
> **$H_{2-0}$:** $Cor(M_{CM})_{Specification} \leq Cor(M_{CM})_{Heuristic}$
>
> **Alternative Hypothesis 2, $H_{2-1}$:** Specification-based techniques produce a higher number of correctly composed models than heuristic-based techniques.
>
> **$H_{2-1}$:** $Cor(M_{CM})_{Specification} > Cor(M_{CM})_{Heuristics}$

| Null Hypothesis | Alternative Hypothesis |
|---|---|
| $H1_{1-0}$: $Effort(M_A,M_B)_S \leq Effort(M_A,M_B)_H$ | $H1_{1-1}$: $f(M_A,M_B)_S > f(M_A,M_B)_H$ |
| $H1_{2-0}$: $f(M_A,M_B)_S \leq f(M_A,M_B)_H$ | $H1_{2-1}$: $f(M_A,M_B)_S > f(M_A,M_B)_H$ |
| $H1_{3-0}$: $diff(M_{CM},M_{AB})_S \leq diff(M_{CM},M_{AB})_H$ | $H1_{3-1}$: $diff(M_{CM},M_{AB})_S > diff(M_{CM},M_{AB})_H$ |
| $H1_{4-0}$: $g(M_{CM})_S \leq g(M_{CM})_H$ | $H1_{4-1}$: $g(M_{CM})_S > g(M_{CM})_H$ |
| $H2_{1-0}$: $Cor(M_{CM})_S \leq Cor(M_{CM})_H$ | $H2_{1-1}$: $Cor(M_{CM})_S > Cor(M_{CM})_H$ |
| $H2_{2-0}$: $Rate(M_{CM})_S \geq Rate(M_{CM})_H$ | $H2_{2-1}$: $Rate(M_{CM})_S < Rate(M_{CM})_H$ |
| **Dependent Variables** | |
| *Effort*: Effort to compose the input models (RQ3.1) | |
| *f*: Effort to apply the composition techniques (RQ3.1) | |
| *diff*: Effort to detect inconsistencies (RQ3.1) | |
| *g*: Effort to resolve the inconsistencies (RQ3.1) | |
| *Cor*: Correctness of the composition (RQ3.2) | |
| *Rate*: Inconsistency rate of the composed model (RQ3.2) | |

Table 8: Tested hypotheses

The correctness of the model compositions is influenced by the presence (or not) of inconsistencies in the output composed model. Thus, we attempt to investigate if the specification-based technique also produces a lower inconsistency rate than the heuristic-based techniques. The new elaborated hypotheses are stated in Table 8.

### 4.1.1.3.
### Context and Subject Selection

The subjects used the the traditional algorithms (Section 2.4.1), the IBM RSA (Section 2.4.2), and Epsilon (Section 2.4.3) to realize six evolution scenarios (Table 9). They had no previous knowledge about the design models or the changes. Thus, the evolution scenarios were typical tasks where developers were not the initial designers of the models. The design models used were fragments of industrial models captured from different application domains, such as financial applications and simulation of petrol extraction.

The experiment was conducted with 24 subjects (being 8 students) from Brazilian companies. All professionals held a Master's degree, Bachelor's degree, or equivalent, and had the required knowledge on software modeling and programming to participate in the experiment. Students were also invited to participate in the experiment because of the recognized importance of students in empirical studies (Host et al., 2000); they are important to enable us to have subjects with different levels of experience in the study. They are from two Master and Doctoral programs in Computer Science at two Brazilian universities: Pontifical Catholic University of Rio de Janeiro (PUC-Rio) and Federal University of Bahia (UFBA). These students attended to two courses: "empirical studies in software engineering" (PUC-Rio) and "software evolution" (UFBA). The experiments were part of the courses and were performed as practical laboratory exercises. In all cases, we had to ensure that every participant would undergo the same learning experiences and had previous experience with software evolution.

| Task | Models | Required Changes to the Base Model |
|------|--------|-------------------------------------|
| 1 | Oil Extraction | ***Add*** one class, one method, and one relationship. ***Modify*** one class from concrete to abstract. |
| 2 | Car System | ***Remove*** two methods and ***modify*** the direction of a relationship. |
| 3 | ATM | ***Add*** two classes and ***refine*** two classes from one. ***Remove*** this last class. |
| 4 | Supply Chain | ***Add*** two classes and one relationship. |
| 5 | Finance | ***Remove*** one class and ***add*** two methods to a particular class. ***Refine*** two classes from one and remove the last one. ***Remove*** one relationship. |
| 6 | Simulation of extraction | ***Modify*** the direction of five relationships. ***Modify*** the name of two methods. |

Table 9: The tasks of the evolution scenarios

### 4.1.1.4.
### Experimental Design

The experimental design of this study is characterized as a randomized complete block design with three treatments (Wohlin et al., 2000). The study had a set of activities that are organized in three phases (Figure 7). The subjects were randomly assigned and equally distributed to the treatments. The distribution follows a within-subjects design in which all subjects serve in the three treatments. This allowed us to compare the data collected. In each treatment, the subjects used a composition technique to carry out two experimental tasks. As three composition techniques were used, then six tasks were performed. Therefore, the experimental design was, by definition, a balanced design.

### 4.1.1.5.
### Operation and Material

**Operation**. Figure 7 shows through an experimental process how the three phases were organized. The subjects individually performed all activities to avoid any threat to the experimental process. The activities are further described as follows.

*Training*. All subjects received training to be sure of their familiarity with both software modeling and model composition techniques. It is important to highlight that the subjects were not aware about the research questions (and hypotheses) of the study in order to avoid biased results.

*Apply the techniques*. The participants were encouraged to compose $M_A$ and $M_B$ based on a document with the evolution descriptions, which define how the model elements should be changed. This document describes (in a more elaborated way) the experimental tasks shown in Table 9. The measure of application effort, video and audio records, and a composed model represent the results of this activity. Each subject performed it six times (for each task presented in Table 9). The video and audio records were later used during the qualitative studies. It is important to point out that a participant (subject x) produced $M_{CM}$ and in the second phase other (subject n-x) detected and resolved the inconsistencies in $M_{CM}$ in order to produce $M_{AB}$.

*Detect inconsistencies*. Subjects reviewed $M_{CM}$ in order to detect inconsistencies. For this, they checked if $M_{CM}$ had the changes described in the evolution descriptions and if the contradicting changes between $M_A$ and $M_B$ were correctly solved. As a result, we have the measure of the detection effort, video and audio records, and a list of inconsistencies identified.

*Resolve inconsistencies*. The subjects resolved the inconsistencies previously localized to produce $M_{AB}$. The resolution effort was also measured and the video and audios were registered.

*Make interview.* Subjects reflected on their experiences on model composition using an in-depth semi-structured interviews. These interviews enriched the qualitative data collected. For example, it was possible to observe, for example, some non-verbal communication issues that help us to infer the study's findings.

*Answer questionnaire*. The subjects filled out a questionnaire. This allows us to collect their background (i.e., their academic background and work experience) and apply some inquisitive questions.

**Material**. The subjects used UML class diagrams in the experiment because they are the most used design models in practice. Each model had approximately eight classes and seven relationships. We have avoided using large models due to some reasons. First and more importantly, proper modeling practices determine that each model should not have much more than seven modular units. Second, experimental guidelines recommend that artifacts used in experiments should be simple; otherwise, the size and complexity may affect the results in undesirable ways (Wohlin et al., 2000).

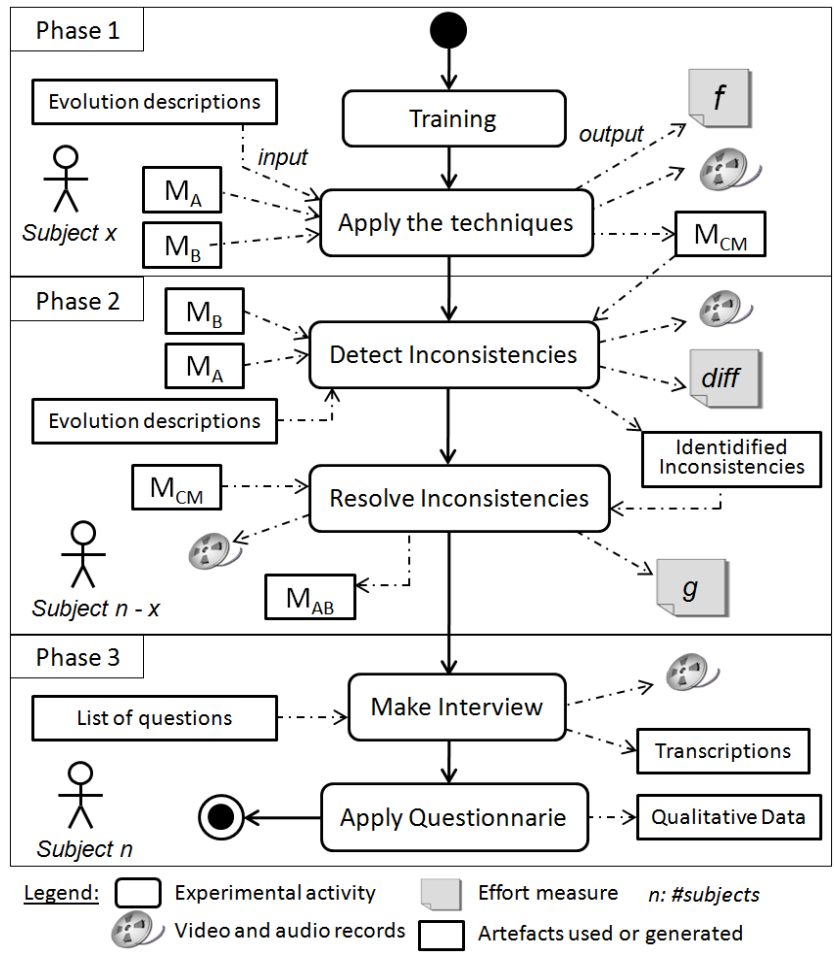

Figure 7: The Experimental process

Third, the delta model should be as small as possible; otherwise, the subjects would have conflict management problems (Mens, 2002). In (Asklund, 1994), Asklund recommends that software changes should be relatively small so that the number of conflicts is not very high. In (Perry et al., 1998), Perry confirms this idea from a statistical basis in a large-scale industrial case. As previously mentioned, the subjects used another material named evolution description. This file describes the changes that should be performed in $M_A$ to transform it into $M_{AB}$. Table 9 illustrates the changes.

### 4.1.1.6.
### Instrumentation and Measurement

The independent variable of this study is the choice of composition techniques. This variable is nominal and assumes two values: specification-based technique and heuristic technique. We investigate the impact of these independent variables on the following dependent variables.

- *Effort*. This variable measures the overall time (in minutes) invested by subjects to compose the input models ($H_{1-1}$). This measure is required by three other variables: effort to apply model compositions ($H_{1-2}$), effort to detect inconsistencies ($H_{1-3}$), and effort to resolve inconsistency ($H_{1-4}$).

- *Correctness*. The correctness of a composition ($H_{2-1}$) is asserted when the output composed model produced is correct with respect to intended model that fully satisfies the evolution description (i.e., $M_{CM} = M_{AB}$). The composed model produced may be rated as either correct or incorrect. Note that each composition performed by a subject produces a dichotomous data (correct or incorrect) defined from the comparison between $M_{CM}$ and $M_{AB}$. Therefore, this variable is a categorical one. Note that a composed model with one of the previously described inconsistencies (Section 3.3.4) would be deemed as incorrect. To promote a deeper understanding, we also investigate the inconsistency rate of the incorrectly composed model. It represents the ratio of the number of inconsistencies of a composed model divided by its number of model elements ($H_{2-2}$). The inconsistencies considered were previously described in Chapter 3.

### 4.1.1.7.
### Analysis Procedures

*Quantitative Analysis.* We performed descriptive statistics to analyze its normal distribution (Kitchenham et al., 2008) and statistical inference to test the hypotheses. The level of significance of the hypothesis tests was $\alpha = 0.05$. The analyses were carried out to test the hypotheses both individually for each experiment task and across all experiment tasks. To test $H_{1-1}$ (and its subhypotheses) we applied the non-parametric Wilcoxon signed-rank test (Wohlin et al., 2010) for the six tasks. This test is similar to the t-test, but does not require two separate sets of independent and identically distributed samples. Note that we have a same subject design. As a result, our samples are dependent. Moreover, the non-parametric Friedman ANOVA test (Conover, 1999) was also applied to reduce some potential threats to the validity of statistical conclusions. To test $H_{2-1}$ we applied the McNemar's test for marginal homogeneity (Wohlin et al., 2010; Devore etal., 1999). To test $H_{2-2}$ we consider the inconsistency rate produced during the evolution scenarios. As in $H_1$, we also applied the Wilcoxon signed-rank test and Friedman test.

*Qualitative Analysis.* Qualitative data were collected from some sources: questionnaire, audio/video records, and transcriptions, think aloud comments and interviews. This helped us to potentially obtain some complementary evidence to explain the quantitative results and then derive the conclusions from a chain of evidences (Jorgensen, 2005), which are formed from the systematic alignment of the quantitative and qualitative data.

## 4.1.2.
## Experimental Results

In this section, we present and interpret the experimental results about the RQ2.1 and RQ2.2. For this, a complete statistical analysis is presented, including descriptive statistics and statistical inference.

## 4.1.2.1.
## RQ2.1: Effort and Composition Techniques

*Descriptive Statistics.* The collected data indicate that the developers tend to spend less effort by using heuristic-based techniques rather than the specification-based techniques. In fact, they required less effort to-be applied (*f*), detect inconsistencies (*diff*), and resolve inconsistencies (*g*). Consequently, the general composition effort was also smaller. The traditional algorithms required less effort than the IBM RSA, which in turn required less than the Epsilon. This is a very interesting finding because the common sense would be otherwise i.e., developers would invest less effort by using the Epsilon and IBM RSA. Table 10 shows pieces of evidence through descriptive statistics of the collected data.

| | Effort | | | f | | | diff | | | g | | |
|---|---|---|---|---|---|---|---|---|---|---|---|---|
| | TRA | RSA | EPS | TRA | RSA | EPS | TRA | RSA | EPS | TRA | RSA | EPS |
| N | 46 | 46 | 46 | 46 | 46 | 46 | 46 | 46 | 46 | 46 | 46 | 46 |
| Min | *5* | 5 | 9 | 2 | 3 | 4 | 1 | 1 | 1 | 0 | 0 | 0 |
| 25th | 7 | 11 | 14 | 4 | 6 | 8.7 | 2 | 2 | 3 | 0 | *0* | *0.5* |
| Med | 11 | 14 | 21 | 6 | 8 | 12 | 3 | 4 | 4.5 | 0.5 | 2 | 3 |
| 75th | 18 | 24 | 34 | 9 | 11 | 17 | 5.2 | 8 | 8.7 | 4 | 7 | 9 |
| Max | 31 | 66 | 114 | 25 | 22 | 39 | 11 | 22 | 38 | 9 | 22 | 38 |
| Mean | 13.3 | 18.2 | 29.1 | 7.2 | 9.0 | 14.8 | 3.9 | 5.3 | 7.7 | 2.1 | 3.8 | 6.6 |
| St D. | 6.9 | 11.0 | 23.3 | 4.4 | 4.2 | 8.8 | 2.4 | 4.4 | 8.2 | 2.9 | 5.1 | 9.1 |

N: #compositions, Min: minimum, Med: median, Max: maximum,
StD: Standard Deviation, TRA: traditional, RSA: Rational Software Architect, EPS: Epsilon

Table 10: Descriptive statistic for the composition effort

Regarding the median of the general effort, it grew significantly from 11 to 14 and 21 by using RSA and Epsilon, respectively. This superior effort represents an increase by about 27.27 and 90.90 percent. This upward trend was not only observed in the measure of the general effort, but also in the *f, diff*, and *g*. Considering the mean of effort computed, this evidence was still clearer. The general effort increased from just over 13 minutes in the Traditional algorithms to 18.26 minutes in the IBM RSA, reaching almost 30 minutes in the Epsilon. This represents a rise of 36.88 and 118.66 percent, respectively. This evidence, therefore, demonstrate that the developers in fact tend to invest less effort with heuristic-based techniques than specification-based one. The next step it is to scrutinize whether this evidence are statistically significant to reject the null hypotheses ($H_{1-1}$, $H_{1-2}$, $H_{1-3}$ and $H_{1-4}$) stated in Section 4.1.1.2.

*Hypothesis Testing.* Since the Shapiro-Wilk test (Sheskin, 2007) indicated deviations from normality, the Wilcoxon signed-rank test and Friedman test were applied. While the Wilcoxon test allowed us to realize a pairwise comparison of the distributions, Friedman test allowed checking if there exist significant differences among the three techniques under investigation. We test $H_1$ (and its subhypotheses) to evaluate the RQ2.1 in the six experimental tasks (Table 11).

Table 11 and Table 12 show the p-values for the pairwise comparison. Bold p-values highlight statistically significant results (i.e., p-value < 0.05). They indicate the rejection of the respective null hypothesis. The main feature is that the general composition effort (*f*, *diff* and g) using heuristic-based techniques were significantly lower than using automated techniques in all cases. Still, by using the traditional algorithms this significance is higher. Thus, we can reject the H1 null hypotheses (and its $H1_{1-0}$, $H1_{2-0}$, $H1_{3-0}$ e $H1_{4-0}$). For example, in row 2 of Table 12, for measure Effort, between RSA and EPS, the W is negative (-544) and p-value is less than 0.05 (p = 0.0015) our selected significance level). This means that the composition effort by using the IBM RSA is significantly lower than one using Epsilon. Still in row 2 just a null hypothesis was not rejected in just one case: the effort to detect inconsistencies considering the IBM RSA and Epsilon (p-value = 0.0891). This means that the subjects did not spend substantially different effort to detect inconsistencies in IBM RSA and Epsilon. Therefore, our initial intuition that the specification-based technique would not reduce the composition effort is confirmed.

Given this surprising result, we were encouraged to apply the Friedman's test to eliminate threats to statistical conclusion validity. This test also confirmed the above conclusions. The results are shown in Table 13. Again bold p-value (<0.05) means that there is a significant difference between the mean ranks in repeated measures of effort. Hence, there is sufficient evidence to reject the null hypothesis, and conclude that there is a difference between the composition efforts at the 0.05 level of significance. For example, in row 1, a chi-Square ($\chi^2$)

| task | statistics | f($M_A$,$M_B$) | | | diff($M_{CM}$,$M_{AB}$) | | |
|------|-----------|----------------|----------------|----------------|----------------|----------------|----------------|
|      |           | TRA *vs* RSA | TRA *vs* EPS | RSA *vs* EPS | TRA *vs* RSA | TRA *vs* EPS | RSA *vs* EPS |
| All  | p-value   | **0.0269**   | **0.0001**   | **0.0003**   | **0.0337**   | **0.0003**   | 0.0891       |
|      | W         | -77          | -834         | -588         | -233         | -533         | -186         |
| 1    | p-value   | 0.4294       | 0.4062       | 0.3628       | 0.1438       | 0.5          | 0.3981       |
|      | W         | -4           | 5            | 6            | 16           | -1           | 4            |
| 2    | p-value   | 0.2305       | **0.0078**   | **0.0342**   | **0.0178**   | 0.2284       | 0.2303       |
|      | W         | -12          | -34          | -27          | -21          | -8           | 8            |
| 3    | p-value   | 0.3762       | **0.0171**   | 0.1548       | 0.2731       | 0.0526       | 0.1250       |
|      | W         | -4           | -26          | -16          | -8           | -20          | 8            |
| 4    | p-value   | 0.2931       | **0.0111**   | **0.0171**   | 0.2931       | 0.0634       | **0.0369**   |
|      | W         | -3           | -28          | -26          | 3            | -19          | -22          |
| 5    | p-value   | 0.0747       | **0.0039**   | **0.0177**   | **0.0207**   | .848         | 0.1982       |
|      | W         | -18          | -36          | -31          | -11          | -25          | -11          |
| 6    | p-value   | 0.2188       | 0.0750       | 0.1094       | 0.0672       | **0.0111**   | 0.1163       |
|      | W         | -9           | -18          | -13          | -12          | -28          | 15           |

W: sum of signed ranks, f: effort to apply the composition technique,
Diff: inconsistency detection effort, RSA: IBM rational software architect, EPS: Epsilon, TRA: traditional algorithm value of 26.21 and p = 0.001 (with p<0.05) indicates a statistically significant difference in the effort measures associated with the three techniques.

Table 11: Wilcoxon test results for application and detection effort

| task | statistics | General Effort | | | g($M_{CM}$) | | |
|------|-----------|----------------|---|---|-------------|---|---|
| | | TRA *vs* RSA | TRA *vs* EPS | RSA *vs* EPS | TRA *vs* RSA | TRA *vs* EPS | RSA *vs* EPS |
| All | p-value | **0.0056** | **0.0001** | **0.0015** | **0.0164** | **0.0003** | **0.0422** |
| | W | -420 | -900 | -544 | -261 | -423 | -248 |
| 1 | p-value | 0.3349 | 0.5 | 0.5 | 0.4661 | 0.3989 | 0.3054 |
| | W | 6 | 0 | 0 | -2 | -4 | -7 |
| 2 | p-value | **0.0149** | **0.0039** | 0.1462 | 0.0828 | 0.0528 | 0.2226 |
| | W | -32 | -36 | -16 | -14 | -24 | -10 |
| 3 | p-value | 0.2891 | **0.0156** | 0.1355 | 0.2303 | 0.0625 | 0.1238 |
| | W | -8 | -21 | -14 | -8 | -10 | 12 |
| 4 | p-value | 0.5 | **0.0111** | **0.0156** | 0.5 | **0.0178** | **0.0445** |
| | W | -1 | -28 | -26 | 0 | -21 | -17 |
| 5 | p-value | **0.0167** | **0.0071** | 0.977 | 0.2763 | 0.4326 | 0.5 |
| | W | -26 | -36 | -20 | -8 | -3 | -1 |
| 6 | p-value | 0.0452 | 0.0313 | 0.4228 | **0.0463** | 0.1250 | 0.4219 |
| | W | -21 | -23 | 3 | -17 | -28 | 28 |

W: sum of signed ranks, g: resolution effort, RSA: IBM rational software architect, EPS: Epsilon, TRA: traditional algorithm

Table 12: Wilcoxon test results for the resolution and general effort

| Task | Statistics | Effort | f($M_A,M_B$) | diff($M_{CM},M_{AB}$) | g($M_{CM}$) |
|------|-----------|--------|--------------|----------------------|-------------|
| all | p-value | **0.0001** | **0.0001** | **0.0048** | **0.0017** |
| | $\chi^2$ | 26.21 | 26.64 | 10.66 | 12.76 |
| 1 | p-value | 0.7682 | 0.8135 | 0.5690 | 0.3977 |
| | $\chi^2$ | 0.8571 | 0.4 | 1.1515 | 1.931 |
| 2 | p-value | **0.0048** | 0.0789 | 0.0789 | 0.1495 |
| | $\chi^2$ | 9.75 | 5.25 | 5.12 | 3.931 |
| 3 | p-value | 0.1916 | 0.1916 | 0.4861 | 0.3046 |
| | $\chi^2$ | 3.630 | 3.630 | 1.68 | 2.5454 |
| 4 | p-value | **0.0084** | **0.0036** | **0.0272** | **0.0207** |
| | $\chi^2$ | 8.615 | 9.333 | 6.333 | 7.5238 |
| 5 | p-value | **0.0099** | 0.0024 | 0.0024 | 1 |
| | $\chi^2$ | 8.968 | 10.516 | 10.51 | 0 |
| 6 | p-value | 0.0854 | **0.0272** | **0.0207** | **0.0003** |
| | $\chi^2$ | 5.429 | 6.231 | 7.6923 | 12.074 |

$\chi^2$: Friedman's Chi-Square, $\alpha = 0.05$

Table 13: Statistical test for the Friedman Test

## 4.1.2.2.
## RQ2.2: Correctness and Composition Techniques

*Descriptive Statistics.* Figure 8 shows the correctness of the compositions generated by using the three techniques: traditional algorithms, Epsilon, and IBM RSA during the six experimental tasks. The axis-y represents the proportions of numbers of $M_{AB}$ (the intended model) achieved by the number of compositions realized in each task using each composition technique, while the axis-x consists of the experimental tasks. Recall that the composition of $M_A$ and $M_B$ often $M_{CM}$ instead of $M_{AB}$. In this case, we calculate the rate of $M_{AB}$ produced in 46 compositions. Thus, the histogram shows how the correctly composed model happened throughout the experimental tasks.

The main outstanding feature is the lack of a distribution pattern of the proportions of correctly composed model in the tasks. For example, in task 1, TRA produced a lower proportion of correctly composed model than RSA and EPS. That is, the intended model was generated in 42.86 percent of the cases in TRA, whereas 57.14 percent of the cases in RSA and EPS. On the other hand, in task 2, TRA outnumbers RSA and EPS. It produced the intended model in 71.43 percent of the cases, while EPS and RSA produced 28.57 and 57.14 percent of the cases, respectively. Although TRA has obtained low measures in task 3 in comparison to task 2 (a decrease from 71.43 to 42.86 percent), it still got a superior value compared to EPS and RSA — i.e., value by about three times higher than the measure of EPS and RSA, comparing 42.86 and 14.29 percent.

Moreover, TRA and EPS had an equal proportion of correctly composed model in task 4, presenting an increase of around 20 percent considering RSA. On the other hand, in task 6, this superiority was reversed. RSA got double the value than TRA and EPS, comparing 28.57 and 57.14 percent. In task 5, the superiority of TRA and RSA considering EPS was evident. Still, subjects obtained the intended model by using TRA and RSA in all composition cases, while less than half of the cases in EPS. We have observed that TRA got a higher number of intended models than RSA and EPS. The subjects produced the intended model in 61.90 percent of the compositions using TRA against 59.52 and 42.86 percent using the RSA and Epsilon technique, respectively. Two interesting insight were that (1) the composition techniques require different effort in front of the categories of evolution changes, and (2) the specification-based technique does not guarantee superiority in terms of correctness in comparison with the heuristic-based techniques.

Table 14 shows the descriptive statistics of the inconsistency rate of the composed models. Our initial expectation was that the specification-based technique would minimize the inconsistence rate whereas also get lower measures than the heuristic-based techniques. However, this expectation was not confirmed. We have observed that, in most cases, the inconsistency rate was similar using specification-based and heuristic-based techniques. This means that developers will not produce correctly composed model by using a technique based on composition specifications. Rather, the output models will have equal (or even more) inconsistency rate. For example, on average, EPS produced a higher inconsistency rate than TRA and RSA. Table 14 shows evidence of the superiority of EPS compared to the TRA. In general, the mean of the inconsistency rate in Epsilon is two times higher than one TRA and RSA, increasing by about 123 and 176 percent, respectively. Still note that the inconsistence rate in RSA is also higher than in TRA. In short, the inconsistency rate in EPS is higher than RSA, which outnumber TRA. This suggests that the inconsistency rate have favored TRA in comparison with RSA and EPS in most cases. This implies that to some extent the number of inconsistencies is decreased whenever the composed model is produced by TRA and RSA. In the next section, we test H5 and H6 to check if whether the differences observed are substantially significant.

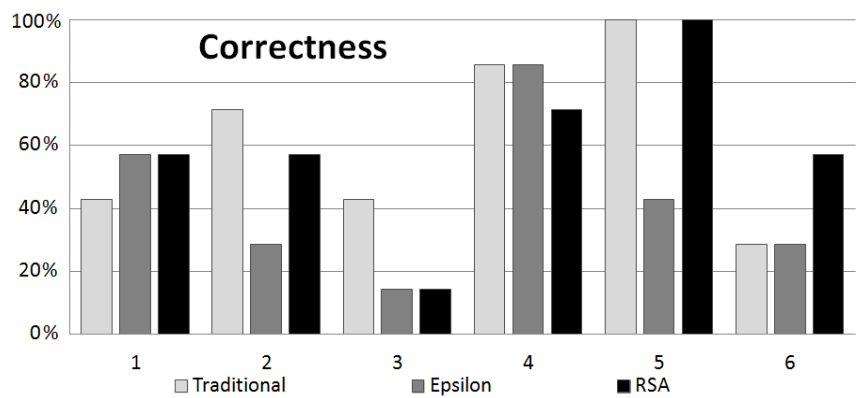

Figure 8: The correctness of the output composed model

|  | N | Min | 25th | Med | 75th | Max | Mean | St D. |
|---|---|---|---|---|---|---|---|---|
| TRA | 46 | 0 | 0 | 0 | 0.31 | 1.63 | 0.26 | 0.45 |
| RSA | 46 | 0 | 0 | 0 | 0.425 | 1.22 | 0.21 | 0.29 |
| EPS | 46 | 0 | 0 | 0.47 | 0.78 | 5.22 | 0.58 | 0.88 |

N: #compositions, Min: minimum, Med: median, Max: maximum,
StD: Standard Deviation,

Table 14: Descriptive statistic for the inconsistency rate

*Hypothesis Testing.* RQ2.2 evaluates if the specification-based techniques assure a higher number of correctly composed model than the heuristic-based techniques. We test $H2_1$ (and its sub hypothesis $H2_2$) to investigate RQ2.1. For this, we apply the McNemar test. Table 15 shows the chi-square statistic ($\chi^2$) and p-values for the pairwise comparisons. In all cases, the p-value is large (p > 0.05), so the null hypothesis of $H2_{1-0}$ cannot be rejected. Although the p-value to the six tasks is not shown in the table, the p-value took values greater than 0.05 in the six tasks. This implies that there is no significant difference between the proportions of correctly composed model of the composition techniques.

| Task | Comparison | $\chi^2$ | p-value |
|---|---|---|---|
| all | TRA vs RSA | 0.27 | 0.606 |
|  | TRA vs EPS | 0.75 | 0.387 |
|  | RSA vs EPS | 0 | 1 |

$\chi^2$: Friedman's Chi-Square, $\alpha = 0.05$

Table 15: McNemar test results for correctness

We test $H2_2$ by applying the Wilcoxon test. Table 16 depicts the pairwise p-values for each measure. Bold p-values point out statistically significant results. They also indicate the rejection of the null hypothesis. Note that the sum of signed ranks (W) shows the direction in which the result is significant. For example, in row 2, W is negative (-250) and p-value is lower than 0.05 (p = 0.0301) for the measure between TRA *vs* EPS. This means that the inconsistency rate for TRA is significantly lower than in EPS. RSA also obtained an inconsistence rate significantly lower (p = 0.001) than EPS. For instance, in row 1, the W is negative (-5) and p-value is higher than 0.05 for the inconsistency rate between TRA vs. RSA. This means that the inconsistency rate for TRA is lower, but no significantly lower than RSA.

| Task | Statistics | Rate |
|---|---|---|
| all | p-value | **0.0258** |
| | $\chi^2$ | 7.314 |
| 1 | p-value | 0.7682 |
| | $\chi^2$ | 0.4210 |
| 2 | p-value | 0.0854 |
| | $\chi^2$ | 4.666 |
| 3 | p-value | 0.4861 |
| | $\chi^2$ | 1.407 |
| 4 | p-value | 0.7682 |
| | $\chi^2$ | 0.666 |
| 5 | p-value | 0.4861 |
| | $\chi^2$ | 2 |
| 6 | p-value | 0.2366 |
| | $\chi^2$ | 3.3076 |

$\chi^2$: Friedman's Chi-Square, $\alpha = 0.05$

Table 16: Friedman test result for inconsistency rate

These results also encouraged us to apply the Friedman test (Table 17). We obtained a chi-square value ($\chi 2$) of 7.314 with p-value = 0.0258, which is lower than 0.05 hence is significant. This means that there exists a significant difference between the inconsistency rate by using TRA, RSA, and EPS. However, considering each experimental task, the results did not take significance (i.e., p > 0.05). This means that a technique did not significantly outperform the other two ones. For example, in task 1, the chi-square value ($\chi 2$) of 0.4210 with a p-value = 0.7682 indicates that there exist no significant difference between the three techniques in terms of inconsistency rate.

| Task | statistic | Inconsistency Rate | | |
|---|---|---|---|---|
| | | TRA *vs* RSA | TRA *vs* EPS | EPS *vs* RSA |
| All | p-value | 0.4851 | **0.0301** | **0.0011** |
| | W | -5 | -250 | 344 |
| 1 | p-value | 0.2188 | 0.2188 | 0.5000 |
| | W | 7 | 7 | -1 |
| 2 | p-value | 0.3750 | 0.2188 | 0.0781 |
| | W | 2 | -9 | 15 |
| 3 | p-value | 0.2002 | 0.1094 | 0.1355 |
| | W | -9 | -16 | 14 |
| 4 | p-value | 0.5000 | 0.5000 | 0.2071 |
| | W | -1 | 1 | -4 |
| 5 | p-value | 0.5000 | 0.1875 | 0.1250 |
| | W | 1 | -6 | 8 |
| 6 | p-value | 0.1982 | 0.1094 | **0.0469** |
| | W | 9 | -16 | 17 |

W: sum of signed ranks, g: resolution effort, RSA: IBM rational software architect, EPS: Epsilon, TRA: traditional algorithm

Table 17: Wilcoxon test results for the corretness

This finding can be explained based on two reasons captured during the interviews and analysis of the qualitative data (i.e., video records and audio). First, the specification-based technique adds a difficulty undesired to match and compose the input model elements, as it was not particularly challenging for the subjects write down the compositions. In particular, this was more often observed in compositions dominated by relations of the type one-to-many (1:N) or many-to-many (N:N) between the input model elements. The specification-based technique proved to be a highly intensive manual task and more prone to errors. Second, the IBM RSA shows the commonalities and differences between the input models in multiple views. This jeopardizes the subjects create a "big picture view" of the output intended model. Finally, we summarized three lesson learned as follows: (1) the model composition techniques should be more intuitive and flexible to express different categories of changes; (2) the techniques should represent the conflicts between the input models in more innovative views and report them soon after they arise; (3) new composition technique should be a mixture of specification-based and heuristic-based techniques; and (4) the heuristic-based techniques consumed less effort and were more effective than the specification-based technique. This suggests that the tools for specification-based techniques may be very rigid and need more flexibility so that, for example, developers can adjust the composition specification considering their experience.

### 4.1.3.
### Limitations of Related Work

Model composition techniques have been studied in many research areas such as merging of state charts (Whittle et al., 2010), composition of software product lines (Thaker et al., 2007; Jayaraman et al., 2007), composition of aspect-oriented models (Klein et al., 2006), and mainly composition of UML design models (Clarke, 2001; Dingel et al., 2008). Such research initiatives focus on proposing model composition techniques or even creating innovative modeling languages. However, the evaluation of the developers' effort on composing design models using the proposed techniques is still incipient. The lack of quantitative and qualitative indicators on composition effort hinders mainly the understanding of side effects peculiar to certain composition techniques.

Current works have notably aimed at evaluating modeling languages, such as UML in terms of some quality attributes such as comprehensibility (Ricca et al., 2010) and completeness (Lange et al., 2004). Although UML has been adopted as the industry standard modeling language, it is just a point of investigation in empirical studies considering model composition. In general, most of the research on the interplay of effort and composition techniques rest on subjective assessment criteria (France & Rumpe, 2007; Mens, 2002; Uhl, 2008; Farias, 2010a). Mens points out the need for studies aimed at investigating the effort to integrate software artifacts such as the source code. Uhl also highlights the superior difficulty of composing models compared to code and reinforce the need for studies reporting the effort required to compose design models.

Even worse, this has led to depend on feedback's experts, who have built up an arsenal of mentally held indicators to analyze the growing complexity of models and then evaluate the effort on composing them (Farias et al., 2010). Consequently, developers ultimately rely on feedback from experts to determine "how well" the compositions were performed. There are many examples of model composition techniques in the literature such as MATA (Whittle & Jayaraman, 2010), Kompose (Kompose, 2011), Epsilon (Epsilon, 2011), and IBM RSA (IBM RSA, 2011). Nevertheless, they will only be useful if the quality of the output composed models (e.g., correctness) is assured and the composition effort required is low. Unfortunately, these approaches do not offer any insight or empirical evidence about the effort required to compose design models. As a matter of fact, the current literature about composition techniques points out the absence of empirical studies and do highlight the importance of studies reporting empirical evidence (Farias et al., 2010; France et al, 2007; Whittle et al., 2010; Apel et al., 2011, Sarma et al., 2011; Mens, 2002; Nejati et al., 2007). To the best of our knowledge, our results are the first to empirically investigate the topics of the research questions in a systematic and controlled way.

### 4.1.4.
### Threats to Validity

This section discusses how the internal, statistical conclusion, construct, and external threats were mitigated.

*Internal Validity.* The inferences between the independent and dependent variables are internally valid if a causal relation is demonstrated (Wohlin et al., 2000; Kitchenham et al., 2008). Our study met the internal validity because: (1) the temporal precedence criterion was met; (2) the covariation was observed, i.e., the dependent variables varied accordingly, when the independent changed; and (3) there is no clear extra cause for the detected covariation.

*Statistical Conclusion Validity.* We checked if the independent and dependent variables were submitted to suitable statistical methods. For this, two points were analyzed. First, whether the presumed cause and effect covaries. The study of the normal distribution of the data collected reduced this threat; as it was possible to verify if parametric or non-parametric statistical methods might be used (or not). In doing so, the Shapiro-Wilk test (Sheskin, 2007) was used to determine how likely the collected sample was normally distributed. As the dataset did not assume a normal distribution, non-parametric statistics were used. Hence, the assumptions of the test statistics were not violated. Second, how strongly the inferences covary. The hypotheses were tested at significance level of 0.05 level (p-value $\leq 0.05$). In addition, some guidelines (Wohlin et al., 2000; Shadish et al., 2002; Sjoberg et al., 2002) were followed so that the assumptions of the statistical test were not violated and the homogeneity of the subjects' background was assured.

*Construct Validity.* It concerns the degree to which inferences are warranted from the observed cause and effect operations included in our study to the constructs that these instances might represent. That is, it answers the question: "Are we actually measuring what we think we are measuring?" All variables of this study were quantified based on previous studies (Farias et al., 2010). Thus, they were defined and independently validated. Moreover, the concept of effort used in our study is well known in the literature (Jorsengen, 2005). Therefore, we are sure that the quantification method used is correct, and the quantification was accurately done.

*External Validity.* We analyzed whether the causal relationships investigated in this study could be held over variations in people, treatments, composition techniques, and the design models. There are reasons to believe the results generalize beyond the three techniques used, but leave it to further work to fully test this.

### 4.1.5.
### Concluding Remarks of the First Study

The previous section represents a first controlled experiment to assess and compare the specification-based and heuristic-based techniques in terms of effort and correctness. By controlling these variables, we investigated the effects of model composition techniques on six quality notions, namely syntactic, semantic, effort, application, detection, and resolution ones. From the quantitative and qualitative analyses, we observed some findings.

First, developers tend to have an additional difficulty to match and compose the input model elements by using specification-based composition techniques, such as Epsilon. The main reason was that the creation of composition specifications has often been an effort-consuming task. Developers invested so much effort to define how the properties of the model elements should be related. This additional difficulty was converted into a superior effort to compose the design models. On the other hand, developers invested less effort to compose the design model by using the heuristic-based composition techniques, such as IBM RSA. The techniques did not require an extra effort to define the similarity between the model elements and realize the compositions.

Second, the composition techniques required different amount of effort in specific composition scenarios. That is, the type of change found in the delta model affected the composition effort. The compositions whose goal were to only accommodate new model elements from the delta model into the base model required similar effort between the heuristic-based and specification-based composition techniques. On the other hand, composition scenarios in which were not dominated by additions, the effort invested to compose the models were different. In particular, this was more often observed in compositions dominated by relations of the type one-to-many (1:N) or many-to-many (N:N) between the input model elements. The specification-based technique proved to be a highly intensive manual task and more prone to errors.

Moreover, we summarized three lessons learned as follows: (i) all the model composition techniques should be more flexible to express different categories of changes (Section 4.1.2.1); (ii) the techniques should report conflicts as soon as they arise (Section 4.1.2.1); such conflicts between the input models should be represented in more intuitive views; (iii) new composition technique should be a mixture of specification-based and heuristic-based techniques as if a set of adequate composition rules are defined and reused, the specification-based techniques can present better results compared to the heuristic-based techniques; and (iv) the heuristic-based techniques consumed less effort and were more effective than the specification-based technique. The latter finding suggests that the tools for specification-based techniques are hard to perform model composition, mainly due to the additional difficulty of manually specifying how the input models should be composed, given the problem at hand.

In addition, we found that the specification-based techniques neither reduce the developers' effort nor guarantee the correctness of the compositions. Even worse, the traditional composition algorithms outperformed the specification-based technique to some extent. Given that little is known about the real effort that developers invest to compose design models, this study might be seen as a first exploratory study that investigates the effects of the composition techniques on the effort in a systematic and controlled manner. However, further empirical studies are still required to better understand if these findings are confirmed or not in other contexts, considering other design models, having different evolution scenarios, and evaluating new composition techniques.

## 4.2.
## Analyzing the Effort of Composing Design Models of Large-Scale Software

As previously mentioned, there has been a significant body of research into defining model composition techniques in the area of governance and management of enterprise design models (Norris & Letkeman, 2011), software configuration management (Perry et al., 2001), composition of software product lines (Jayaraman et al., 2007; Thaker et al., 2007), aspect-oriented modeling (Whittle et al., 2009; Klein et al., 2006), and integration of state charts (Whittle & Jayaraman, 2010).

Unfortunately, both commercial and academic model composition techniques suffer from the *composition conflict problem*. That is, models to-be composed conflict with each other and developers are usually unable to deal with the conflicting changes. Hence, these conflicts are transformed into inconsistencies in the output composed model (Diskin et al., 2010). For example, two developers concurrently work on a same class diagram, which has two abstract classes *A* and *B*. The first developer creates an inheritance relationship between the abstract class *A* and *B* (i.e., *B.superclass = A*), while the second developer modifies the class *A* from abstract to concrete (i.e., *A.isAbstarct = false*). Although these are simple changes, usually the developers are not aware of these conflicting changes performed in parallel. Hence, the composition of the partial models produces an inconsistent class diagram i.e., an inheritance relationship between an abstract class *B* and a concrete class A. The current composition techniques cannot automatically resolve these inconsistencies (Egyed, 2010; Egyed, 2007); because inconsistency resolution relies on an understanding of what the models actually mean. This semantic information is typically not included in any formal way in the design models. Consequently, developers must invest some effort to manually detect and resolve these inconsistencies. The problem is that high effort compromises the potential benefits of using model composition techniques, such as gains in productivity.

To date, however, nothing has been done to (1) *quantify* the composition effort in key software development activities, including software evolution, and (2) *characterize* the influential factors that can affect the developers' effort in practice. Hence, developers cannot adopt or assess model composition based on practical, evidence-based knowledge from experimental studies. Rather, they rely on diverging feedbacks from evangelists; these feedbacks often diverge.

The goal of this second study, therefore, is to report on five industrial exploratory case studies that aimed at (1) providing empirical evidence about model composition effort, and (2) describing the influential factors that affected the developers' effort. These studies were performed in the context of the evolution of design models of five large-scale software systems. During 56 weeks, 297 evolution scenarios were performed, leading to 2.288.393 compositions between modules, classes, interfaces, and relationships. We draw the conclusions from quantitative and qualitative investigations including the use of metrics, interviews, and observational studies. We investigate the composition phenomena in their context, stressing the use of multiple sources of evidence, and making clear the boundary between the identified phenomenon and its context. While we believe this study is representative of the broader issues, we make no claims about the generality of our results beyond the composition of UML class and sequence diagrams of large-scale software.

The following subsections are organized as follows. Section 4.2.1 introduces the main concepts and knowledge that are going to be used and discussed throughout the thesis. Section 4.2.2 elaborates the composition scenario that will be used as a frame of reference. Section 4.2.3 describes the research methodology followed. Section 4.2.4 presents the analysis of composition effort. Section 4.2.5 contrasts our work with related work. Finally, Section 4.2.6 discusses some concluding remarks and future work.

## 4.2.1.
## Background

Three-way merge algorithm (Mens, 2002) is a well-known method to merge software artifacts. This method has increasingly been incorporated into the most popular and robust industrial modeling tools, such as IBM RSA (IBM RSA, 2011). This algorithm refines the specification of model composition cited previously. Instead of taking into consideration only two input models $M_A$, the local design model version, and $M_B$, the last design model release located in the enterprise repository, it also considers $M_P$, the parent of $M_B$. This means that it takes into account not only the differences between the two input design models $M_A$ and $M_B$ to conduct the composition, but also the contrast between them and $M_P$. For example, in Figure 10(A), the developer, Steve, produces a composed model, V3, merging the local version, S3, with its parent, V1, and with the last version of the repository, V2. Note that the more precise the match processes between the $M_P$, $M_A$, and $M_B$, the better the "best-guess" analysis to generate the resulting compositions.

Model composition following this algorithm can be represented as Merge($M_P$, $M_A$, $M_B$), where $M_P$ is the model version from which $M_A$ is descent, MA is the base model, and $M_B$ is the delta model. $M_P$ is used to better track the changes between $M_A$ and $M_B$. For example, revisiting the example in Section 4.2, the decision if the class $A$ should be (i.e., $A.isAbstract$ = false) or abstract (i.e., $A.isAbstract$ = true) may be supported by considering a previous version, $M_P$. This ancestral version will provide some addition information about how the class was previously. Based on this, developers can make decisions more effectively.

The merging session between $M_P$, $M_A$, and $M_B$ is typically executed as soon as an automated difference analysis between them is done. After identifying the commonalities and differences between the input models, they are merged so that a new release can be produced, $M_{AB}$. This type of composition is applied to collaborative working environment in order to enable more effective team collaboration. It is expected that this effectiveness can be transformed into gain of productivity, and sometimes this is possible because a couple of reasons (Mens, 2002). For example, it requires less user intervention, and in many cases, requires no intervention at all (depending upon the complexity of the composition). Hence, the expectation is that developers' effort invested in parallel increase their productivity proportionally. On the other hand, even though it has reached a high level of precision to compose UML design models, the three-way merge still remains one of the more taxing tasks of any collaborative software development team . This is due to the prior knowledge that developers should accumulate about the initial design model, $M_P$, the current version, $M_A$, and the intended changes, $M_B$.

## 4.2.2.
## Composition Scenario

After describing the main concepts used in our study, we describe the context where our study was carried out. In the absence of a theory about model composition (Sjøberg et al., 2008), this description is used as a frame of reference (Runeson & Höst, 2009) for our study. The goal is to illustrate the real-world settings in which the case studied happened. To this end, a motivating composition scenario is presented to carefully highlight the problems faced.

## 4.2.2.1.
## Collaborative Model Evolution

Figure 9 represents an ever-present collaborative software modeling scenario in our study. We explain three points about this scenario. First, developers work in parallel to increase productivity. They take part of the system functionalities represented in use cases, and then create UML classes, and sequence diagrams from them. The system functionalities described in these use cases overlap with each other; hence, the design models become to have some *critical overlapping points*. That is, diagrams that share model elements. This is a critical because if a model element is inconsistent, then all diagrams are affected. These points are a source of inconsistency propagation and developers are unable to trace the side effects of all propagations. For example, Peter, Steve and Bill produce UML class diagrams, named P1, S2 and B3, related to the first, second and third use case specification, respectively. However, it is by no means obvious (if not impossible) for the developers to foresee these overlapping points, detect the possible conflicts, and measure their consequences *at modeling time*. Steve cannot predict that changes performed in his model, S2, may give rise to conflicting changes into the Peter's model, P1, and Bill's model, B3. Similarly, it is an effort-consuming task for Peter to identify and grasp that conflicting changes between his model and the Steve's model may propagate into the Bill's model, B3, given the problem at hand. Consequently, the developers inevitably end up creating inconsistent models, since they are unable to effectively deal with a set of conflicting changes.

Second, to overcome this problem, the developers need to invest effort to localize and resolve the inconsistencies. For this, developers must understand the system functionalities and the reasons why the changes happened. For example, Steve would need to understand the semantics of the system functionalities described in the first and third use case specifications. This understanding is required to properly identify and resolve all composition inconsistencies present in his design models (S2). Finally, given the inherent complexity of composing design models it is particularly challenging for developers to: (1) objectively localize these critical overlapping points, (2) quantify the effort variables (*f*, *diff*, and *g*), (3) overcome the emerging inconsistencies, and (4) grasp which influential factors affects the effort variables.

## 4.2.2.2.
## Motivating Example

Given the need to evolve enterprise design models (e.g., UML class diagrams) and the time constraint (only three days), three developers (Peter, Steve, and Bill) work concurrently to increase the productivity. Firstly, developers check out the last version of the design model (V1) from the repository (Figure 10(A)). V1 is the base model represented in Figure 11(A). After that, they perform a set of modifications over their local versions (i.e., P1, S1, and V1) to evolve them. Figure 10(B) shows a timeline of the modifications and Figure 11(B) represents the delta model that brings together the changes. The developers perform four types of modifications:

(1) Add the stereotype <<MainClass>> to indicate that a class starts up a use case.

(2) Modify the color of a class from white to gray (and vice-versa) to indicate that is part of a framework (or not).

(3) Add the stereotypes <<use>> and <<instance>> to relationships to indicate that a class use and instantiate the other one, respectively.

(4) Add methods to represent that a class implements a new (part of) functionality.

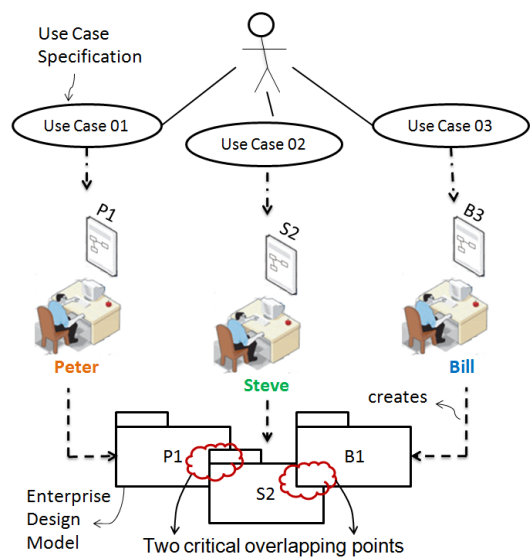

Figure 9: A real-world collaborative model composition leading to two critical overlapping points

(5) Delete some model element.

However, some composition conflicts between the V2 and S3 emerge when Steve submits its last local version, S3, to the repository. This composition session can be briefly represented by Merge(V1,V2,S3). These conflicting changes between the Peter and Steve versions are described as follows:

1) Peter sets correctly the color of the class *ApplicationType* to *gray* (step 1), while Steve sets the color to *white* (step 2).

2) Peter sets incorrectly the color of the class *Application* to *white* (step 2), while Steve updates the color of it to *gray* (step 3).

3) Peter adds the stereotype <<use>> to the relationship between the class *MarlimCore* and *EditPSDiagOptionsAction*, while Steve removes this relationship.

4) Peter removes the class *PSElementGroup*, while Steve creates an inheritance relationship between the class *PSElementGroup* and Production.

5) Peter creates a relationship of association between *PSDiagramOptionsDialog* and *MarlimInputData*, while Peter removes the attribute status: *StatusPanel* from the class *PSDiagramOptionsDialog* and transform it into a new class, and creates a relationship of aggregation between the new class *StatusPanel* and *MarlimInputData*.

6) Peter modifies the method *execute():void* to *runEditionPanel*, while Steve modifies the method's name to *executeEdition*().

To submit his changes, Steve should know to deal with these contradicting modifications so that the new model version, V3, can be produced. The problem is that, in general, the developers are not always able to understand the emerging conflicts or properly solve them. As a consequence, they realize (or let pass) some incoherent modifications over the input models.

To illustrate these incoherent actions, let us regard the conflicting change number one. If Steve does not accept Peter's changes, then the output composed model is going to have an unexpected change. That is, the class *AppliactionType* of the enterprise framework will have erroneously the color *white* instead of *gray*.

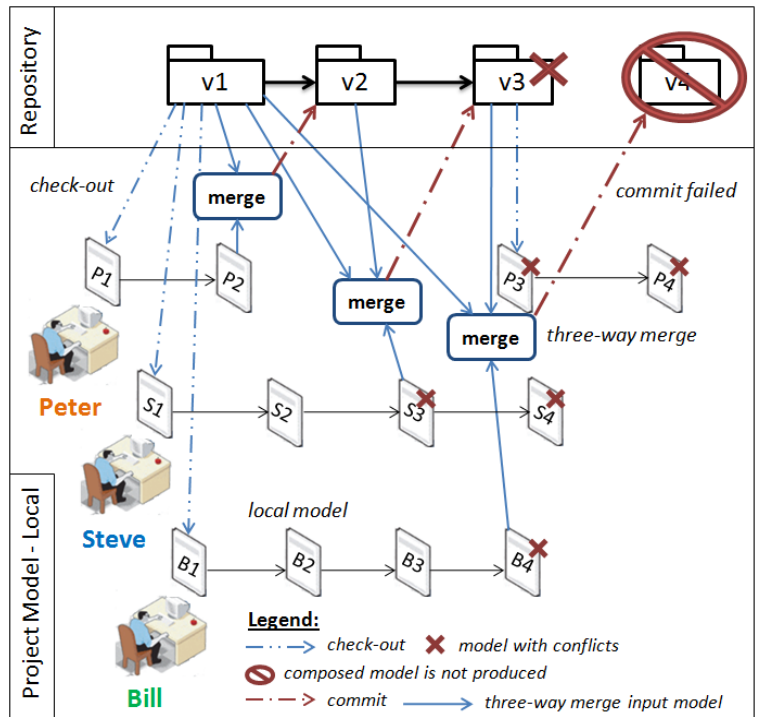

(A)

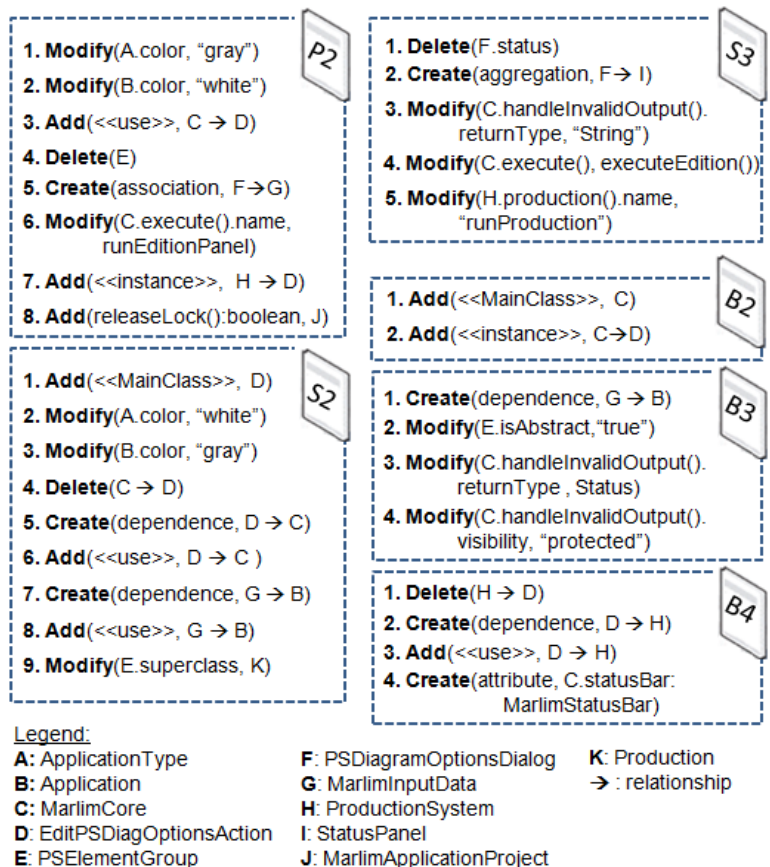

(B)

Figure 10: A real-world use scenario of model composition (A). The change descriptions performed by the developers (B).

Another example would be the conflicting change five. Peter and Steve propose two ambiguous modifications to allow the class *PSDiagramOptionsDialog* to access objects of the *MarlimInputData*. However, usually these ambiguities are neither properly localized nor understood. This leads the output composed model to have both changes. The result is, therefore, an output composed model with inconsistencies, which is produced from the local project to the enterprise repository V3. Even though, these inconsistencies are usually propagated downward to the developers' local projects. Peter's P3 and P4 local version in Figure 10(A), and the Steve's S4 local version represent this propagation. Bill follows the same submission procedures performed by Peter and Steve; then, he produces the composition session (see Figure 10(A)) represented briefly by Merge(V1,V3, B4) (see Figure 12). The problem is that, in this case, the output composed model, V4, could not be generated. The chief reasons were: the size of the delta model, once Peter's and Steve's changes are also considered during the composition session; and the amount and complexity of the conflicting changes that should be analyzed, since to produce V4 correctly, many semantic and syntactical issues need to be considered. That is, Bill inevitably needs to grasp the meanings of each modification accomplished previously by Peter and Steve. Even worse, this understanding cannot be always acquired. This problematic evolution scenario is described as follows:

1)   Bill assigns correctly the stereotype <<MainClass>> to the class *MarlimCore* (B2.step 1), while Peter attaches this stereotype to the class *EditPSDiagOptionsAction* (step 1).

2)   Bill attaches the stereotype <<instance>> to the dependence relationship (B2.step 2), while Peter attaches the stereotype <<use>> to this relationship (step 3) and Steve deletes this relationship (S2.step 4).

3)   Bill just creates the dependence relationship between the class *MarlimCore* and *EditPSDiagOptionsAction* (B3.step 1), while Steve correctly creates this relationship and attaches it to the stereotype <<use>> (S2.steps 7 and 8).

4)   Bill correctly transforms the concrete class *PSElemenGroup* to an abstract class (B3.step 3), while Peter removes this class (P2.step 4) and Steve creates an inheritance relationship between the classes *PSElemenGroup* and *Production*. This implies that if the change of Bill is accepted, then the change of Steve should be rethought, otherwise we will have a syntactically incorrect inheritance relationship between the now abstract class *PSElemenGroup* and the concrete class *Production.*

5) Bill modifies correctly the return type of the method *MarlimCore.handleInvalidOutput()* from *void* to *Status* (B3.step 4), while Steve modifies it wrongly to *String*.

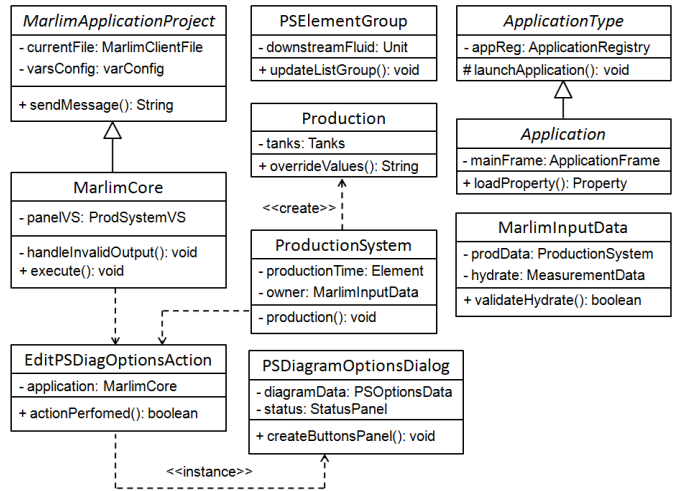

(A)

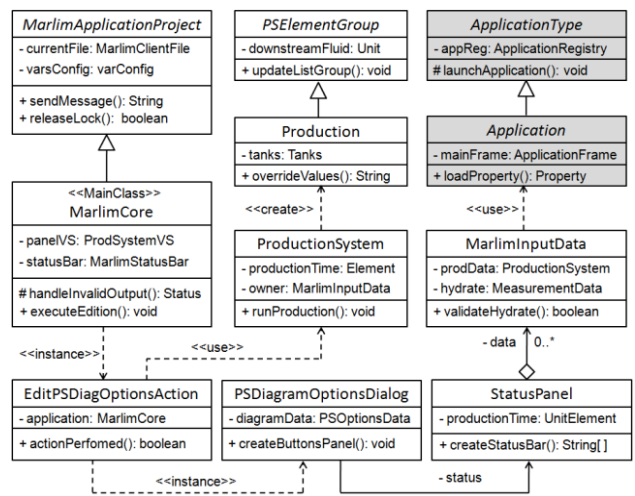

(B)

Figure 11: The Base Model (A) and the Intended model (B)

6) Peter attaches the stereotype <<instance>> to the dependence relationship between the classes *ProductionSystem* and *EditPSDiagOptionsAction* (P2.step 7), while Bill removes this relationship improperly (B4.step 1) (see Figure 13).

To resolve properly such conflicts, sometimes the developers must engage to seek solutions for conflicts that come from different sources. For example, the resolution of the second conflicting changes requires handling systematically the contradicting modifications created by not just one developer (Peter's changes), but by two developers (Peters' and Steve's changes). Moreover, this manipulation must necessarily involve the three developers so that semantic and syntactical issues can be carefully understood.

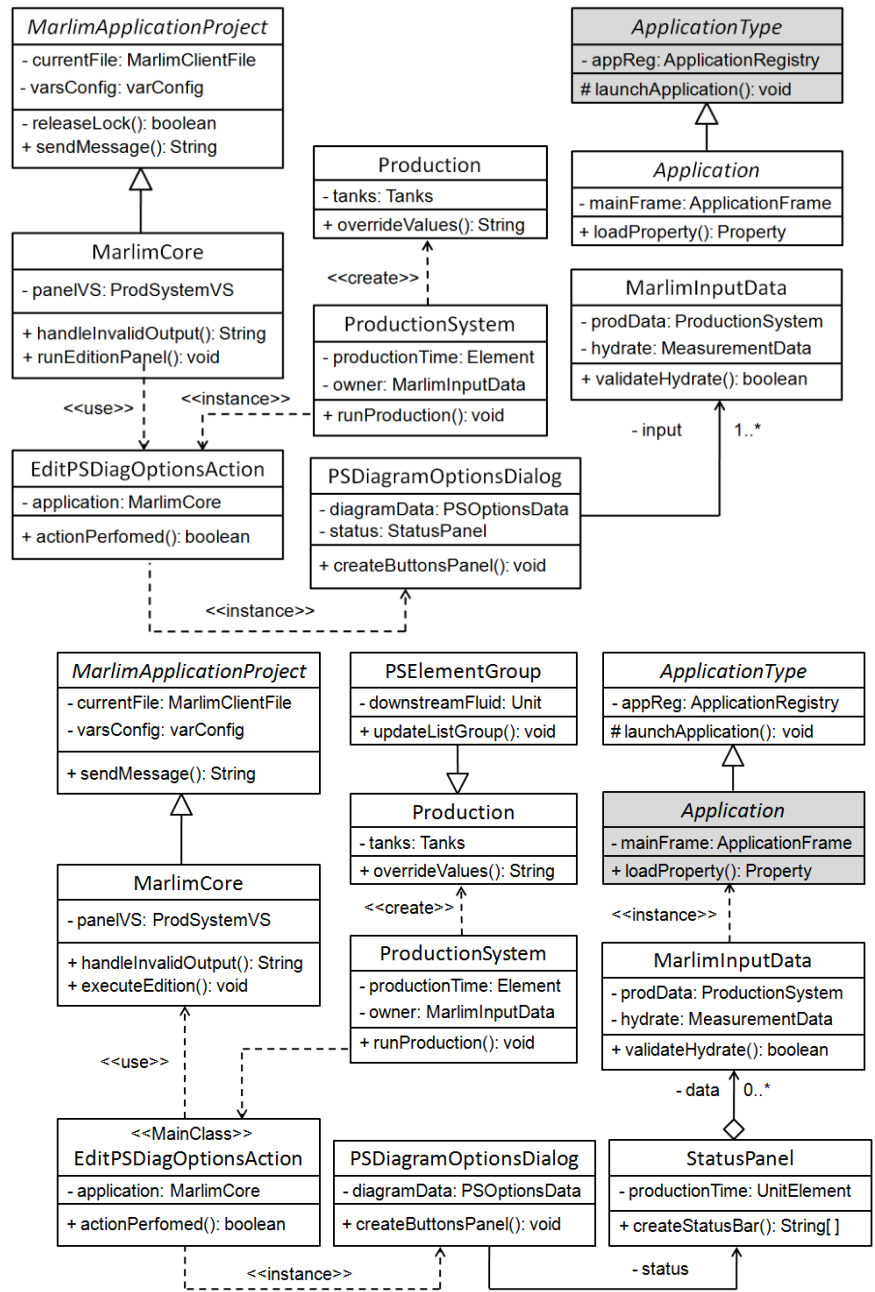

Figure 12: The model versions created by Peter (P2) (above) and Steve (S3) (below).

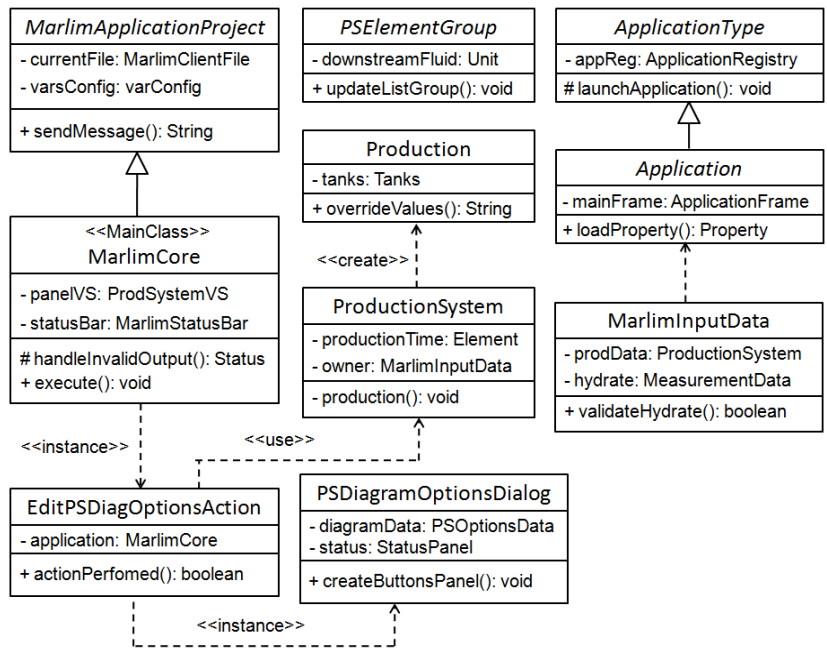

Figure 13: The model versions created by Bill (B4).

### 4.2.3.
### Study Methodology

This section presents the study methodology based on practical guidelines of empirical studies (Runeson & M. Höst, 2009; Wohlin et la., 2000; Kitchenham et al., 2008).

### 4.2.3.1.
### Objective and Research Questions

This study aims at evaluating the effects of model composition techniques on six quality notions, namely syntactic, semantic, effort, application, detection, and resolution ones. In particular, this Chapter focuses on generating practical knowledge about the values that the composition effort's variables assume in real-world settings. To this end, the research question (RQ2) defined in Section 1.3 is evaluated in this second study. As these variables may be affected by some influential factors, this work also attempts to understand and characterize these factors. With this in mind, we formulate two research questions:

- **RQ2.3:** What is the *effort* to compose design models?
- **RQ2.4:** What are the factors that affect composition effort?

### 4.2.3.2.
### Context and Case Studies

We performed five case studies to investigate RQ2.3 and RQ2.4 The context of the studies was collaborative modeling in industrial projects. Developers used model composition to evolve and reconcile design models. Table 18 presents a suite of metrics to characterize the models involved in the studies. Table 19 shows the collected measures for these metrics. As previously mentioned, during 56 weeks, 297 evolution scenarios were performed leading to 2.288.393 compositions between modules, classes, interfaces, and relationships.

All five cases differ in terms of their size, number of participants, and application domain. These cases are characterized as *holistic case studies* (Runeson & M. Höst, 2009; Wohlin et la., 2000; Kitchenham et al., 2008), where contemporary phenomena of model composition are studied as a whole in their real-life context. We present a brief description of the systems used as follows:

- *Alope***:** a system that controls and manages the import and export of Petroleum (and its derived products).
- *Bandeira*: a logistics system is responsible for the complement management of the flow of goods.
- *GeoRisco*: a system that supports forecast and controls of environmental catastrophes.
- *Marlim*: a system that simulates the design and extraction of Petroleum from deep ocean areas.
- *PlanRef*: a system that provides decision making support for logistics and planning processes in Petroleum refineries.

| Type | Metric | Description |
|------|--------|-------------|
| Size | NumClass | #classes |
| | NumAttr | #attributes |
| | NumOps | #operations |
| | NumInter | #interfaces |
| Inheritance | DIT | the sum of depth of the class in the inheritance hierarchy. |
| | OpsInh | #inherited operations. |
| | AttrInh | #inherited attributes. |
| Coupling | Dep_Out | #dependencies where the package is the client. |
| | Dep_In | #dependencies where the package is the supplier. |
| | NumPack | #packages |
| | R | #relationships between classes and interfaces. |
| | H | relational cohesion |
| | Ca | #afferent coupling of the packages |
| | Ce | #efferent coupling of the packages |
| | A | # abstractness (or generality) of the packages. |
| Project | NumWeeks | # weeks |
| | NumDev | # developers |
| | NumEvol | # evolutions scenarios |

#: the number or degree of all

Table 18: Metrics used

| Metrics | Alope | Bandeira | GeoRisco | Marlim | PlanRef |
|---------|-------|----------|----------|--------|---------|
| NumClass | 316 | 892 | 1394 | 2828 | 1173 |
| NumAttr | 1732 | 3349 | 8424 | 9689 | 3808 |
| NumOps | 3479 | 7590 | 10608 | 23722 | 9111 |
| NumInter | 18 | 83 | 143 | 223 | 93 |
| DIT | 140 | 216 | 1109 | 2528 | 871 |
| OpsInh | 3414 | 6620 | 12482 | 38181 | 16369 |
| AttrInh | 1507 | 1766 | 9003 | 9147 | 4406 |
| Dep_Out | 72 | 464 | 61 | 453 | 330 |
| Dep_In | 65 | 423 | 58 | 418 | 322 |
| NumPack | 34 | 166 | 175 | 345 | 187 |
| R | 1285 | 1360 | 3008 | 4493 | 2251 |
| H | 47.5 | 216.8 | 261.9 | 448.6 | 282.5 |
| Ca | 278 | 1147 | 1632 | 4044 | 2329 |
| Ce | 235 | 996 | 1278 | 2723 | 1451 |
| A | 9.58 | 50.45 | 36.9 | 66.5 | 51.9 |
| NumWeeks | 6 | 15 | 8 | 17 | 10 |
| NumDev | 3 | 7 | 2 | 7 | 4 |
| NumEvol | 6 | 95 | 55 | 64 | 77 |

Table 19: The collected measures of the case studies

These systems are featured as scientific software (Kelly, 2006) because they require knowledge from multiple application domains, and encompass a broad class of concepts of physical phenomena, including oil pressure, fluid density, logistic, temperature scale, dilatation of fluids, temperature, fluid pressure, geologic risk, and supply chain. They were chosen based on some reasons presented in the following. First, the cases used robust modeling tool (IBM Rational Software Architect) allowing developers to merge design models, work in parallel, and validate the design models. The IBM RSA was used due to: (1) the implementation robustness of its composition algorithms; (2) the tight integration with the Eclipse IDE; and (3) the tool had been already adopted in previous successful projects. In addition, we also required the UML CASE tools to have an XMI export facility, which will allow us to analyze the design models using metrics tool. Additionally, all cases used a bug tracking system, i.e., JIRA, with which it was possible to coordinate the developers' tasks, specifically during the creation of the design models and review of the models.

Finally, on average, four professional developers have participated in each case study, totaling more than 10 developers in all case studies. The advantage of using experienced professional developers is to avoid one of the main criticisms of most case studies in software engineering, in especial software modeling, regarding the degree of realism of the studies. Thus, we believe that the collected data are representative of developers with industrial skills.

### 4.2.3.3.
### Subjects

The background of the subjects was an ever-present concern in the experimental design. As the case studies were performed in vivo in a Brazilian company, the subject selection was based on *convenience* (Wohlin et al., 2000). In total, 12 subjects were recruited. Table 20 describes the subjects' background. We analyzed the level of theoretical knowledge and practical experience of these subjects.

Regarding the *theoretical knowledge* issues, we checked the quality of the education system that the subjects come from. We observed that this system, where the subjects were students, is a system that places a high value on

| Variables | Mean | SD | Min | 25th | Med | 75th | Max |
|---|---|---|---|---|---|---|---|
| Age | 25.3 | 4.47 | 21 | 22 | 24.5 | 27 | 38 |
| Degree | 2.16 | 1.06 | 1 | 1 | 2 | 3 | 4 |
| Graduation year | 2006.4 | 4.8 | 1992 | 2005.25 | 2006.5 | 2010 | 2010 |
| Years of study at university | 5.75 | 2.8 | 3 | 3 | 5 | 7.5 | 12 |
| YOEW UML | 1 | 1.4 | 1 | 1.25 | 3 | 4.75 | 5 |
| YOEW Java | 4.5 | 1.84 | 2 | 2.5 | 4 | 6.75 | 7 |
| Used IBM RSA (1 or 0) | 1 | 1 | 1 | 1 | 1 | 1 | 1 |
| YOEW soft. development | 5 | 3.6 | 2 | 2.25 | 4.5 | 5.75 | 16 |
| Hours of software modeling | 98.33 | 40.38 | 60 | 60 | 90 | 120 | 180 |
| Hours of OO programming | 156.66 | 89 | 80 | 80 | 130 | 225 | 360 |
| Hours of software design | 130 | 53.85 | 80 | 80 | 120 | 190 | 220 |

*Degree: 1 = Student, 2 = Bachelors, 3 = Masters, 4 = Ph.D.*
*YOEW = Year of Experience with, Med: Median*
*SD = Standard Deviation, $25^{th}$ = lower quartile, $75^{th}$ = upper quartile*

Table 20: Descriptive Statistics: Subjects' Background

theoretical issues about the foundational principles of software engineering and software modeling. Moreover, this educational system provides an academic formation with much more than 120 hours of courses (lecture and laboratory) exclusively dedicated to software engineering, object-oriented programming, and software modeling. This can be seen, in part, as an intensive UML-specific training. Furthermore, other important courses present in their formation are operating systems, databases, computer architecture, requirement engineering, and so on. Therefore, the subjects fulfilled the level of theoretical knowledge required.

Taking into consideration the *practical experience* of the subjects, we also observed that there are some even more compelling evidences about the level of practical experience of them. This knowledge was acquired from previous software development projects. This was confirmed by the analysis in which provides background data on the subjects that participated in the case studies. The data show that the subjects fulfill the requirements in terms of age, education, and experience. A benefit of the presence of a considerable theoretical and practical knowledge is that the members of the teamwork can learn from each other in terms of theoretical and practical issues. The main consequence of this knowledge sharing between team members is that the emerging problems can be solved more quickly and properly. If, for example, well-formedness rules of the design models are challenged, the subjects can work together to get it solved. Another point that is essential to emphasize is that, in all cases, the subjects were familiar with the software modeling tool they had to use, IBM RSA, and all subjects received training about merging design models. Lastly, based on this information (summarized in Table 20), we deemed that the subjects had the required training, theoretical knowledge and practical experience to perform the software modeling and merging tasks properly.

### 4.2.3.4.
### Study Design and Evaluation Procedures

Having presented the context of our studies and subjects, the next step is to describe precisely how the case studies were conducted.

### 4.2.3.4.1.
### Operation

The procedures of the study can be grouped into two phases: *creation* and *review*. In the first phase, the developers collaboratively created the design models. In the latter, they detected and resolved inconsistencies in the output composed models. Note that the intended model was produced after executing these two phases. Moreover, it is also important to emphasize that the effort variables (*f*, *diff*, and *g*) are incrementally measured as the phases are performed.

Figure 14 summarizes the procedures associated with both the production of the intended models and the measurement of the effort variables. Activities are represented using rounded rectangles, and the arrows indicate transitions between the activities. The diamonds are decisions (conditional branch), and the arrows connected to them are marked with the conditions. The initial state in an activity diagram is indicated by the black circle, while the final state is the encircled black circle. Following the simplest path of the procedure, issues are first submitted and examined (issue refers to general activities registered during the modeling project). Each issue is assigned to a developer. After opening the issue, the developer may execute three possible activities: creation of the design model, detection of inconsistencies, and resolution of inconsistencies. As these activities were carried out, the effort variables were quantified. Developers closed the issue after it has been validated.

*Creation of the Design Models.* First, the developers created a UML class diagram for each use case specification. In addition, sequence diagrams were created for the most important use cases, which represent around 30 percent of the full system specification. This percentage and the choice of the use cases were not made in an arbitrary manner, but based on the policies of the company. After that, the developers made use of the model composition technique to submit the created model to the repository. It is important to emphasize that developers created sequence diagrams only after its corresponding class diagram had been created and validated. To calculate the developer's effort to compose the local model with the repository version, the members of the team were stimulated to make a record of all composition sessions by using the software Camtasia Studio Pro (Camtasia, 2011). The generated videos were essentials to further analyses.

*Detection of Inconsistencies.* The developers reviewed the composed models in order to detect syntactic and semantic inconsistencies. For this, they performed a double checked model reviews by using the IBM RSA's model validation mechanism and by manually inspecting the models. During each review, the developers could read the use case specifications to check whether (or not) the generated models fulfill the requirements described in the specification. It is important to point out that a developer reviewed the models created by other developers, never the model created by him. Since the IBM RSA's validation mechanism can report false positive and false negative inconsistencies (Altmanninger et al., 2009), the teamwork members were encouraged to check if the reported inconsistencies were posing, in fact, a problem.

*Inconsistency Resolution.* Having identified the inconsistencies, the developers invested some effort to revolve them. In practical terms, they added, removed, or modified some existing model elements to solve them. After addressing the model inconsistencies, the developers submitted the intended model to the repository. Thus, the compositions were executed in two moments: after the original creation of the models and after the inconsistency resolutions. All model versions were registered in a version controlling control system, thereby allowing a systematic analysis of the history of the generated model versions.

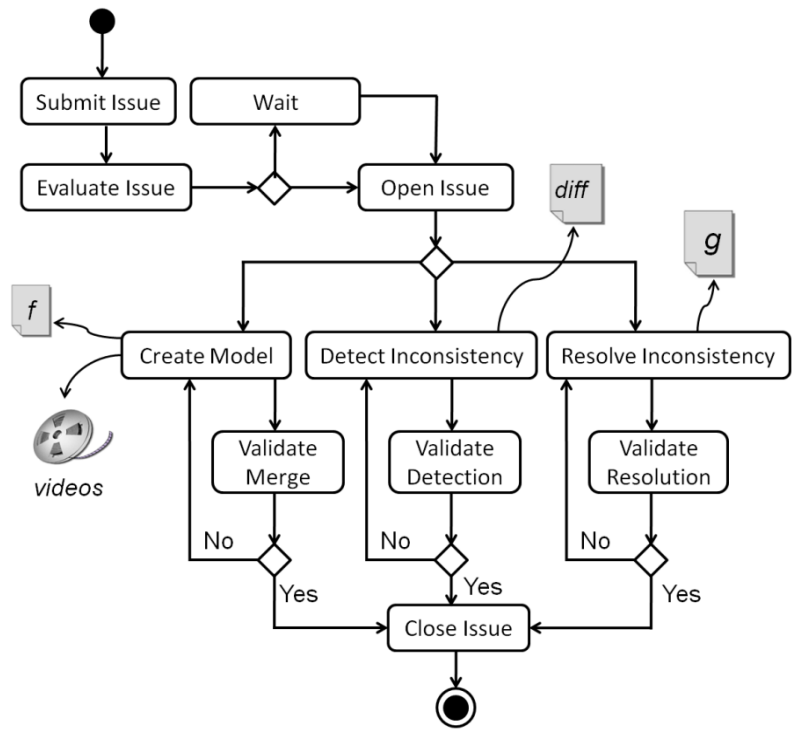

Figure 14: The flow of activities during the studies

## 4.2.3.4.2.
## Design Model Versions and Releases

The design models are semantically rich, have been evolving over the long term, can be checked for consistency. These features were carefully analyzed and elected as pre-requirements to perform the case studies. We feel, therefore, confident that the model releases are going to promote (1) more reliability and accuracy of our results, and (2) chiefly suitable conditions for yielding lessons about driving composition effort variables. Consequently, this enables us to grasp as the composition effort variables (*f*, *diff* and *g*) turn up in real-world settings, and identify and understand the factors that affect the production of the desired releases during the composition session.

*Deriving the Design Model Releases.* Given the collaborative environment work, the subjects incrementally created the releases using the IBM RSA's composition technique throughout the evolution scenarios. The creation steps are presented as follows. First, from a reverse engineering process, the team leader generates a set of elementary model elements, which will be used by other developers to create the design models. Note that this derivation of the model elements is indispensable in real-world settings; since the size of systems is considerably large (see Table 20).

Next, the developers make use of these elements to manually generate the design models. For example, the developers define which model elements should be inserted into the UML class diagram and what their relationships are. This decision is made from the information collected from the use case specification and the code. This creation process of the models is not only marked by intensive discussion among the members of the development team, but also by the constant submission of new model release increments to the repository so that the changes can be broadcasted to the other developers. To control the changes of the models and to facilitate collaboration, the version control system was intensively used during all case studies.

*Model Releases and Composition Specification.* For each evolution scenario, a new release was created. For each new release, the previous release was modified in order to incrementally accommodate the changes. To implement a new evolution scenario, a model composition specification can remove, add, derive, or modify the entities present in the previous release. During the design of all releases, a main concern was to follow the best practices of modeling and carefully realize the requirements described in the use case specifications.

### 4.2.3.4.3.
### Variables and Quantification Method

This section defines as the three effort variables (*f*, *diff*, and *g*) were quantified and their unit of measurement (time in minutes). Our analysis and quantification, therefore, rely on three effort measures described as follows.

*Application Effort Measure (f).* This measure represents the required time (in minutes) to match the input model element, resolve the conflicting changes, and submit the evolving changes to the repository. That is, the effort invested by developers to apply the model composition technique. This measurement only quantifies the effort to produce the composed model ($f(M_A, M_B)$) rather than the effort to detect ($diff(M_{CM}, M_{AB})$) and resolve inconsistencies ($g(M_{CM})$). This effort was calculated from recorded movies created by own developers, which were stimulated to record these videos throughout the case studies.

*Detection Effort Measure (diff).* The detection effort consists of the time needed to localize inconsistencies in the composed model for a given output composed model. Subjects were responsible for registering the time. This detection can be characterized as a semi-automated process; as developers make use of the IBM RSA's model validation mechanisms and manually go through the model to identify semantic problems. We consider all syntactic inconsistencies can be automatically detected. On the other hand, given that it is impossible to count all semantics inconsistencies automatically, we count only semantic inconsistencies that can be manually spotted. For example, relationships (e.g., association and inheritance) between model elements that no longer exist or a stereotype attached improperly. Usually these inconsistencies are not detected by tools upfront, but are visually by developers.

*Resolution Effort Measure (g).* It represents the time required to perform a set of activities (creations, removals, and modifications) needed to transform $M_{CM}$ into $M_{AB}$. Again, subjects were the responsible for registering the time.

## 4.2.3.4.4.
## Analysis Procedures

The analysis of the collected data was conducted with quantitative and qualitative methods. While the quantitative data concerns the measurements involving the study variables, objects, and units of the analysis, the qualitative data deals with the diagrams (pictures), descriptions, transcripts from interviews, and annotations. The goal of using a combination of qualitative and quantitative data is to exclusively provide a better understanding of the studied phenomena in their context.

a. *Quantitative Analysis*

The descriptive statistic is used so that the outstanding trends might be pinpointed. Box-plot graphically illustrates these trends. The presence of patterns in the data distribution, and lack thereof acted as a driver for further investigation allowing a deeper understanding. Note that we are not concerned with any correlation analysis or probabilistic formulation. Rather, our focus is only to describe and graphically present interesting aspects of the data. Further, these statistics were important to analyze and possibly remove outliers from the data. Outliers are extreme values of the measured variables that may influence the study's conclusions. To analyze the outliers we made use of box-plot. According to Wohlin (Wohlin et al, 2000), we should verify whether "the outliers are caused by extraordinary exceptions (unlikely to happen again), or whether their cause can be expected to happen again. For the first case, we should remove the outliers, and for the latter we should not remove the outliers." In our study, some outliers were identified. However, they did not represent any extraordinary exceptions, since they were expected to happen again. Consequently, they were not removed, as they did not compromise the results.

b. *Qualitative Analysis*

The qualitative analyses were concentrated on interviews, observational study, and archival data. Hence, the RQs were investigated from different viewpoints, subjects, artifacts, and projects.

*Interviews.* A *semi-structured* interview approach was performed following a *funnel model* (Runeson & Host, 2009), in which one initial open question is told and then directed towards to more specific one. It was organized in topics with *open* and *closed* questions (Runeson & Host, 2009). They were organized in such a way that research questions (*f*, *diff*, and *g*) could be exploited. An interview guide was created based on the authors' experience in model composition and on previous studies, together with the research questions of the study. The author of this thesis conducted the interviews. The interviews were recorded and transcribed into text; this was done by one else than the authors. Experienced subjects were selected for the interviews from the involved company and other Brazilian companies. That is, the interviewees (8) were not only developers that participated in the case studies, but also with other developers with different experiences of other companies. The selection was based on the interviewees' different experience in terms of model composition rather than their similarities. It was also assured that only anonymous data would be presented externally. Each interview lasted from 30 to 55 minutes, depending on how talkative the subjects were.

*Observational Study.* In order to investigate how model composition was performed in practice extensive observations were conducted through three different approaches. First, one of the authors worked in the modeling projects during the case studies taking part in everyday activities. This allowed a more effective observation. Secondly, the model composition tasks were recorded, and after analyzed. This allowed monitoring the task of the subjects. Thirdly, to obtain a feedback of the subjects about the task performed, they encouraged to "think aloud" by asking questions like "What is the key difficult to resolve the inconsistencies?", "What is your strategy to deal with conflicting changes?", and "What do you do to reduce composition effort?". In summary, data collected consisted of field notes, audio recordings of interviews and their transcriptions, videos, screenshots, and copies of artifacts.

*Archival Data.* The company's repository was an important source of data, since it enables us to access the different versions (specifically the evolution track) of the design models. The developers were encouraged to describe the evolution changes performed before executing the compositions. This description helped us to understand how the compositions were performed and reasons why the inconsistencies arose. For example, in the motivating example (Section 4.2.2.2), the developers, Peter, Steve, and Bill, should necessarily describe the changes performed by them. In total, more than 240 descriptions were created and the information stored in the repository. The comments were expressed in a free-text field, in which the subjects could report anything they thought might be relevant in explaining the changes that were being done. In addition, the developers were well aware the importance of these descriptions to understand the evolutions and the results obtained on each evolution scenario. For example, the comments helped us to identify when the composition had success (i.e., $M_{CM} = M_{AB}$) or failed (i.e., $M_{CM} \neq M_{AB}$), and grasp the rational what the developers thought at the time of composition session.

## 4.2.4.
## Study Results

In this section, we interpret the results about the RQ2.3 and RQ2.4. For this, we present and analyze quantitatively and qualitatively the collected data about the composition effort variables (Section 4.2.4.1) and explains the factors that influence these variables in practice (Section 4.2.4.2).

## 4.2.4.1.
## RQ2.3: Composition Effort Analysis

The composition effort analysis involves the examination across cases of a single variable, focusing on three characteristics: the distribution, the central tendency, and the dispersion.

**Application Effort  (f)**

This section investigates the variable concerning the effort to apply the composition technique. Table 21 shows a descriptive statistic about the application effort. These statistics will help us to pinpoint the central tendency and spread of values around it. A tally of 40 and 69 (N) compositions was registered in the Marlim and Bandeira project, respectively. The central tendency was calculated using the two most-used statistics: the mean and the median. The most interesting feature was that the composition of the large-scale industrial models used in our study required by about 4 minutes.

More specifically, the results indicate that effort to compose models was, on average, 3.17 minutes and 4.43 minutes in Bandeira and Marlim projects, respectively. Given the complexity and the size of the design models in question (Table 19), these central tendency measures are in fact low values. For example, a developer spent just around 4 minutes to submit the most complex evolving changes to the repository in the Marlim project. In addition, the median measures accompany these measures: 3 minutes and 3.12 minutes in the Bandeira and Marlim project, respectively. Thus, this implies that the required effort to apply the semi-automated model composition technique is low. Consequently, it is possible to advocate it as appropriate to collaborative software modeling in which resources and time are usually tight.

| Cases | N | Mean | SD | Min | 25th | Med | 75th | Max |
|---|---|---|---|---|---|---|---|---|
| Marlim | 40 | 4.73 | 4.52 | 0.25 | 2 | 3.2 | 6.79 | 22 |
| Bandeira | 69 | 3.29 | 1.93 | 0.83 | 2 | 3 | 4 | 14.25 |

*N = number of compositions, SD = standard deviation, Min = minimum, 25th = first quartile; Med = median, 75th: third quartile, Max: maximum.*

Table 21: Descriptive statistics for application effort

To understand the dispersion of the data around this tendency, not only the standard deviation, 25th and 75th percentiles were computed, but also the minimal and maximum values. Developers' effort tends to concentrate by around the central tendency rather than spreading out over a large range of values. Indeed, with 1.55 and 1.58 minutes, the standard deviation measures indicate that in the majority of the composition sessions the developers spend an effort near 3.17 minutes or 4.43 minutes. This information can help modeling mangers to: (1) systematically propose the effort estimation rather than essentially based on their judgment; and (2) check if the effort spent by developers is an expected value (or not), since it falls inside (or outside), these ranges of statistics that is expected to occur. Consequently, it is possible to improve the effort estimation, and hence a typical UML-based development, for example. Finally, this measure can be seen as the first step to overcome the lack of empirical evidence about the impact of model composition techniques on developers' effort in real-world settings.

To deepen our understanding about the application effort, Figure 15 distributes the collected sample in six effort ranges. These ranges in the histogram systematically group the application effort cases. The y-axis of the histogram represents the counts of merging, while the x-axis consists of the ranges of effort. The main outstanding feature is that: the presence of a distribution pattern of the application effort through the ranges of effort. The low-effort categories (i.e., $t < 2$, $2 \leq t < 4$, and $4 \leq t < 6$) represents the most likely range of effort that developers invest to compose the input models. The number of cases is equal to 29 (in Marlim) and 64 (in Bandeira), representing 72.5 percent and 92.75 percent of the composition cases, respectively. On the other hand, the number of cases in the high-effort categories (i.e., $6 \leq t < 8$, $8 \leq t < 10$ and $10 \leq t$) is equal to 12 (in Marlin) and 5 (in Bandeira), comprising 17.39 percent and 12.5 percent of the cases respectively. Thus, the number of composition cases in the low-effort categories outnumbers the amount of cases in the high-effort categories, comprising more than 70 percent and 90 percent of the cases in the Marlim and Bandeira project, respectively. On the other hand, the number of cases in the high-effort categories was by around 30 percent (in Marlin) and 7.25 percent (in Bandeira). In practice, this means that developers spent less than 6 minutes in 85.32 percent of the whole composition cases, and just 14.68 percent of the cases required more than 6 minutes.

Another even more compelling feature is that: there is a changing pattern among the effort categories. Although the changing pattern of the measures from a category to another one happens in different forms, it comes about with the same type of change in the most of the cases.

There are five changes in the number of counts of merging from one category to another being three of them similar as follows. From the first to the second category, the count of compositions had a gradual rise from nine to 13 (in Marlim) and from 10 to 33 (Bandeira). This means a growth of 44 percent and 230 percent, respectively. On the other hand, observing the third category, the count had a significant drop compared to the previous category.

The distribution of merging fell back from 13 to 6 and from 33 to 21 in the Marlim and Bandeira project, respectively. This implies into a significant drop of 53.84 and 36.3 percent. Following this same drop pattern, in the fifth category, the number of cases decreased abruptly from 7 to 1 (Marlim) and 3 to 1 (Bandeira), comprising a fall of 85.71 percent and 66.67 percent, respectively. However, the transitions from the third category to the fourth one as well as from the fifth category to the sixth one had different changing pattern. In the fourth category, the count kept stable (seven cases) in Marlim project and a decrease of 85.71 percent in Bandeira project was observed, from 21 to 3. In the sixth category, the count did not change, stagnating in 1 (Marlim), and, however, quadruplicated its value from 1 to 4 in the Bandeira project. This implies, therefore, that there is to some extent a particular behavior of change between the ranges of effort.

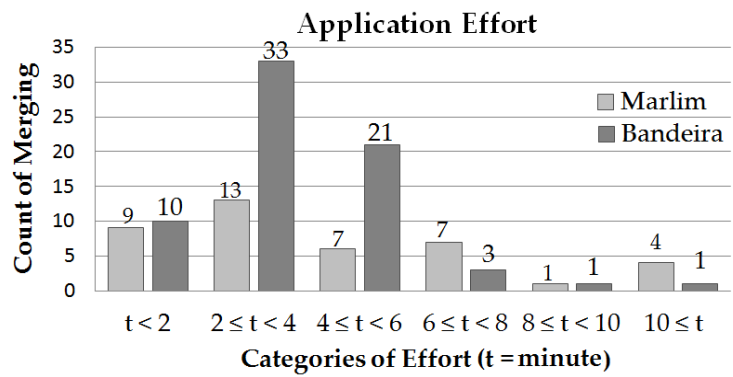

Figure 15: Histogram of the application effort measures

With these two previous features in mind, an important finding was observed: the application effort tends to reduce as developers become more familiar with technical issues rather than application domain issues. This finding is supported by the fact that developers invested more effort in Marlim project than in Bandeira project. After a careful analysis, the main reason was that the developers were more familiar with composition issues. That is, 30 percent of the cases had effort higher than 6 minutes, rather than the 7.24 percent ones in the Bandeira project. It is important to point out that: (1) both projects had a similar level of complexity; (2) the members of the development team had a similar level of knowledge about the meaning of application domain elements; and (3) the teamwork was the same throughout the both projects. Therefore, the application effort tends to decrease as the developers gained experience with the activities considering key steps to apply the composition technique, i.e., match the input models, resolve the conflicting changes, and then combine the input model elements.

**Detection Effort (diff)**

This section investigates the variable concerning the effort to detect the inconsistencies of the output composed model. Table 22 shows a descriptive statistic about the effort spent to detect inconsistencies. A careful analysis indicated that some interesting features were happing. First, the more experienced developers in both modeling and IBM RSA spend 23.2 percent less effort to detect inconsistencies than less experienced developers. This observation was derived from the comparison of the medians in the Marlim and Bandeira cases. This finding was possible to reach because the same development team firstly worked in the Marlim project and after this in the Bandeira. Observing the values of the mean computed this affirmation is still reasserted. In this case, the more experienced developers invested 38.57 percent less effort to detect inconsistency than less experienced developers, compared 7.57 and 4.65.

Second, the higher the number of teamwork members, the higher the effort to localize inconsistencies. This outstanding finding is supported by the comparison of the medians of the projects with high versus low number of developers. Comparing the number of teamwork members of the projects, we could observe that the developers of the Marlim and Bandeira project, both with 7 developers, invested a higher amount of effort to detect inconsistencies than the developers of the GeoRisc and PlanRef (with 2 and 4 developers, respectively). For example, the developers spent 49.46 percent more effort to detect inconsistencies in the Marlim project than in GeoRisc project, compared the medians 6.55 and 3.31, respectively. This striking observation was also reinforced when we compared the Marlim and PlanRef. That is, Marlim's developers spent 64.27 percent more effort to localize the inconsistencies, compared the medians 6.55 and 2.34, respectively. Therefore, the projects with a higher number of developers had to invest the double of effort to localize the inconsistencies.

Third, a remarkable finding is that the higher the number of inconsistencies in behavioral models, the higher the effort to detect inconsistencies. Even though, the Alope project had a low number of developers, a considerable number of inconsistencies were concentrated in behavioral models like sequence diagrams. The chief problem highlighted by developers was that the *behavioral models* require an additional effort to go through the flows of execution. For example, an association in a structural model (e.g., class diagram) represents essentially one relationship between two classes. On the other hand, in a behavioral model (e.g., sequence diagram) that represents the interaction between the instances of these classes; this simple association may be represented by $n$ interactions (i.e., messages

| Cases | N | Mean | SD | Min | 25th | Med | 75th | Max |
|---|---|---|---|---|---|---|---|---|
| Marlim | 63 | 7.57 | 5.1 | 0.54 | 2.45 | 6.55 | 12.49 | 16.54 |
| Bandeira | 86 | 4.65 | 2.39 | 0.36 | 2.37 | 5.03 | 6.38 | 9.21 |
| GeoRisc | 24 | 3.66 | 1.52 | 1.32 | 2.67 | 3.31 | 4.16 | 7.39 |
| PlanRef | 44 | 2.91 | 1.75 | 1.04 | 1.39 | 2.34 | 4.12 | 7.15 |
| Alope | 6 | 12.37 | 4.2 | 5.26 | 8.25 | 13.15 | 16.36 | 17.37 |

N = number of compositions, SD = standard deviation, Min = minimum, 25th = first quartile; Med = median, 75th: third quartile, Max: maximum.

Table 22: Descriptive statistics for detection effort

exchanged between the objects). The problem is that developers must check each interaction. This problem is enlarged with the need to check the consistencies between the class diagram and the sequence diagram. For example, there is a message from an object A to an object B in the sequence diagram, but there is no relationship between the class A and B in the class diagram. Even worse, sometimes the method corresponding to such message does not even exist in the class B. Another typical inconsistency is that a concrete class A becomes abstract, however, its instance remains represented in the sequence diagram. Thus, developers had an additional effort to examine the consistency between the structural and behavioral model.

Another observation is that the higher the distribution of inconsistencies in different modules, the higher the effort to identify them. In the case studies, the systems were strongly decomposed in conceptual areas. This unit of modularization brings together application domain concepts in a same space. The problem arises when the inconsistencies in a conceptual area give rise to an abundance of inconsistencies, and hence affecting many other model elements located in other conceptual areas as a ripple effect. This propagation is inevitable as there are usually some relationships between these units of modularization. Hence, developers must be able to identify inconsistencies in model elements of conceptual areas that they do not know. Note that during the case studies the developers created diagrams related to a specific functionality of the system (specified in case uses), and these diagrams were grouped in a conceptual are (something like a package). Thus, the lack of knowledge about the model elements in unknown conceptual area led developers to invest an extra effort to pinpoint the inconsistencies.

**Resolution Effort (g)**

This section investigates the variable concerning the effort to resolve the inconsistencies in the output composed model. Table 23 shows a descriptive statistic of the inconsistency resolution effort. The main outstanding feature is that the developers invest more effort to resolve inconsistencies rather than to both apply the model composition technique and detect the inconsistencies. This can be explained based on some evidences.

First, in Marlim project, for example, the teamwork members spent 64.91 percent more effort resolving inconsistencies than applying the model composition technique. This difference comprises the comparison between the medians 3.2 (application) and 9.12 (resolution). This difference becomes more explicit when we consider the values of the mean. This evidence is reinforced in Bandeira project. The resolution of inconsistencies consumes 80.31 percent more effort than the application of the composition technique, compared the medians 3.2 (application) and 9.12 (resolution). The difference between the application and resolution effort becomes stronger when we consider the value of the mean i.e., jumping significantly their values from 64.91 percent to 88.40 percent (in Marlim) and from 80.31 percent to 88.35 percent (in Bandeira).

Second, in Marlim project, the inconsistency resolution consumed 28.17 percent more effort than the inconsistency detection. This comprises the difference between the medians 6.55 and 9.12. The results in Bandeira project followed the same trend. Developers spent 66.99 percent more effort with inconsistency resolution than with inconsistency detection, compared the medians 5.03 and 15.24. Considering the mean, this difference of effort becomes more evident, leaping abruptly from 28.17 percent to 81.44 percent (in Marlim) and from 66.99 percent to 83.42 percent (in Bandeira). Analyzing the collected data from the GeoRisc and Alope project, this observation is confirmed. For example, the resolution effort is 82.98 percent and 54.96 percent higher than the detection effort in GeoRisc and Alope, respectively. On the other hand, in Alope project, the resolution and detection effort were practically equal. Therefore, the collected data suggest that teamwork members tend to spend more effort resolving inconsistency rather than applying the model composition technique and detecting inconsistencies.

Another striking feature is that the experience acquired by the developers did not help to minimize the inconsistency resolution effort. Although more experienced developers have invested less effort to compose the input models and detect inconsistencies, their additional experience did not help significantly to minimize the inconsistency resolution effort. For example, in Bandeira project, more experienced developers spent 40.15 percent more effort to resolve inconsistency than less experienced developers from Marlim project, compared the medians 9.12 and 15.24. The main reason is that more experienced developers tend to be more cautious than less experienced ones, and hence they tend to invest more time analyzing the impact of the resolution of each inconsistency.

| Cases | N | Mean | SD | Min | 25th | Med | 75th | Max |
|--------|-----|-------|-------|-------|-------|-------|-------|--------|
| Marlim | 31 | 40.79 | 74.79 | 3.09 | 4.13 | 9.12 | 11.33 | 246.25 |
| Bandeira | 8 | 28.06 | 28.04 | 5.55 | 8.17 | 15.24 | 41.44 | 95.44 |
| GeoRisc | 16 | 25.86 | 13.75 | 5.12 | 17.70 | 19.45 | 42.5 | 53.33 |
| PlanRef | 44 | 2.86 | 1.92 | 1.2 | 2.03 | 2.33 | 2.52 | 10.41 |
| Alope | 5 | 31.04 | 12.75 | 16.21 | 16.21 | 29.20 | 46.8 | 55.4 |

*N = number of compositions, SD = standard deviation, Min = minimum, 25th = first quartile; Med = median, 75th: third quartile, Max: maximum.*

Table 23: Descriptive statistics for resolution effort

## 4.2.4.2.
## RQ2.4: Influential Factors on Composition Effort

Some factors influence the effort of composing large-scale design models in real-world settings. This section analyzes the side effects of these factors on the composition effort variables.

## 4.2.4.2.1.
## The Effects of Conflicting Changes

A careful analysis of the results pointed out that the production of the intended model is affected by the presence of different types of change categories in the delta model. These changes would be the addition, removal, modification, and derivation of model elements. The current composition algorithms are not able to effectively accommodate these into a base model; mainly, when these changes occur simultaneously. We described the most common categories of changes identified throughout the study and after analyzing their effects:

• *Addition:* model elements are inserted into base model; for example, a stereotype <<instance>> was added to the directed relationship between the *ProductionSystem* and *EditPSDialogOptionsAction*.

• *Removal*: a model element in the base model is removed; for example, the class *PSElementGroup* is removed;

• *Modification*: a model element has some properties modified; for instance, the class *PSElementGroup* becomes abstract. For this, the property *isAbstract* has its value modified from *false* to *true*.

*Derivation*: model elements are refined to accommodate new changes and/or moved to other ones. For example, the class *ProductionSystem* is refined into two new classes: *ProductionAction* and *ProductionPanel*. The method *ProductionSystem.runProduction()* is inserted into *ProductionAction.* The attribute *ProductionSystem.productionTime* is inserted into *ProductionPanel*. This type of modification can be seen as a 1:N modification.

Developers and researchers recognize that evaluable software should adhere to the Open-Closed principle (Meyer, 1997) as evolutions become easier. This principle states "software should be open for extensions, but closed for modifications." However, this observation did not occur in all the cases as modifications and derivations of model elements happened as well. In our study, the open-closed principle was more closely adhered by the evolutions dominated by additions rather than any other one. In this case, developers invested low effort compared to other cases. This suggests that the closer to the Open-Closed principle the change is, the lower the composition effort.

On the other hand, evolution scenarios that do not follow the Open-Closed principle required more effort to produce the intended model, $M_{AB}$. This finding was identified when the change categories simultaneously occur in the delta model; hence, compromising the composition for some extent. This extra effort was due to the incapability of the matching algorithm to identify the similarities between the input model elements given the presence of widely scoped changes. In the Marlim project, for example, the composition techniques were not able to execute the compositions by about 17 percent (11/64) of the evolution scenarios. This required developers to recreate the models manually. In the Bandeira project, by about 10 percent (10/95) of the composition cases did not produce an output model as well, or the composed model produced had to be thrown away due to the high amount of inconsistencies.

In particular, we also observed that *the refinement (1:N) of model elements in the delta model caused severe problems*. A practical example of this refinement encompassed the direct relationship between *PSDiagramOptionsDialog* and *MarlimInputData*, named as *input*. This relationship was decomposed into (1) a direct relationship between *PSDiagramOptionsDialog* and *StatusPanel*, (2) the class *StatusPanel*; and (3) the aggregation between *StatusPanel* and *MarlimInputData*. In this case, the relationship (1:3) was not identified. This problematic scenario was also noticed during the refinement of some classes belonging to the MVC (Model-View-Controller) architecture style into a set of more specialized ones. In both cases, the name-based, structural model comparison was unable to recognize the 1:N composition relations between the input model elements. However, we have observed these conflicts do not only happen when developers perform modifications, removals, or refinements in parallel, but also when developers insert new model elements. This finding was noted from the fact that although evolutions following the Open-Closed principle had reduced the developers' effort, they still caused too frequent undetected inconsistencies.

Developers were often unable to localize inconsistencies that did not affect the model elements created by them. Even worse, the composition algorithms were unable to identify that overlapping changes might cause "cross-semantic inconsistency." That is, the semantic attributed to a model element conflict with another one assigned to the same (other) element. A very concrete example of semantic inconsistencies in our case studies was when UML stereotypes used to attribute new semantic to the model elements conflict with each other. The illustrative example shows two typical inconsistencies in our studies. For example, Steve attaches the stereotype <<MainClass>> to the class *EditPSDiagOptionAction*, while Bill attaches this attribute to *MarlimCore*. Hence, the algorithm does not detect that only one class can be defined as the main class.

We have noted that *these problems are more challenging to be detected when they occur in multi-valued properties defined in the UML metamodel* such as *Class.ownedOperation: Operation [*]*, which defines the methods of a class, or *Class.extension: Extension [*]*, which specifies the stereotypes applied to a class. For example, Bill attaches the stereotype <<instance>> to the directed relationship (B2.step 2) from *MarlimCore* to *EditPSDiagOptionsAction*, while Peter attaches the stereotype <<use>> to this relationship (P2.step 3). As these stereotypes are not present in ancestor version (*V1*), the algorithm incorrectly brings both to the new version (*V4*). One of the reasons for this is that the meaning of the stereotypes are often not taken into account during compositions—either because the semantics of these stereotypes are rarely represented or either because the composition algorithms are unable to infer that the stereotypes <<instance>> and <<use>> are semantically contradicting. However, developers must tame this problem.

Still considering the conflicting changes between Bill and Peter, whatever the change accepted — if the class *PSElemenGroup* is transformed into an abstract class, or if it is removed — inconsistencies will emerge when the Steve's changes are applied to *PSElemenGroup*. For example, Steve creates an inheritance relationship between the classes *PSElemenGroup* and *Production* (a concrete class). If the class *PSElemenGroup* is abstract, then a semantic inconsistency emerges because *PSElemenGroup* has an inheritance relationship with a concrete class *Production*. Note that this inconsistency is not related to the modeling language as the UML metamodel hinder inheritance relationship from the abstract class to concrete one. This inconsistency is because object-oriented programming like Java does not permit this type of relationships. On the other hand, if the class *PSElemenGroup* is removed, then a static semantic inconsistency arises because the inheritance relationship refers to a class that no longer exists.

Thus, we have observed that the current state-of-the practice composition techniques superficially support the evolution categories. For accuracy reasons, this implies that developers need innovative techniques supporting restructuring changes and identifying the ripple effects of the semantic added to the model elements. Moreover, developers know that these problems (from structural to semantic inconsistencies) may happen in practice. However, they neither know their side effects nor grasp the meaning of the changes. To demonstrate this distinct side effect more clearly, let us take a closer look at the illustrative example in Figure 11, Figure 12, and Figure 13. As a prerequisite to produce the composed model, it is necessary to match the input model elements, which are suffering the effects of the changes performed by Peter, Steve, and Bill. For this, the composition technique identifies the similarities between the model elements. With addition based evolutions, the conflicting changes are identified because of the superimposition of changes: the composition algorithm detects that two contradicting values were attributed to a particular property defined in the language metamodel (e.g., *isAbstract* or *isDerived*). For example, Bill modifies the value of the property *return type* of the method *MarlimCore.handleInvalidOutput()* from *void* to *Status* (B3.step 4), while Steve modifies it to *String*. Similarly, Bill transforms the concrete class *PSElemenGroup* into an abstract class (B3.step 3), while Peter removes this class (P2.step 4).

Therefore, although the composition algorithm is effective to detect the changes, it is unable to identify whether the differences are caused by a simple (or multiple) modification, removal, or even refinement of model elements. Having more semantically richer information about the type of the changes, developers might detect and earlier resolve the conflicts. This would increase the number of correctly composed models as this semantic information aided those developers in making better-informed decisions.

With this in mind, to alleviate these problems would be necessary to grasp the actual meaning of the model elements (in the base model and delta model) and the impact of the change categories on their quality issues (e.g., comprehensibility and correctness). However, the current name-based, structural model comparison strategy has demonstrated to be ineffective to recognize intricate equivalence relationships between the model elements. The meaning of the model elements is rarely represented in a formal way. Hence, the definition of the correspondence between the input model elements is essentially based on a signature-based approach (Reddy et al., 2005). In doing so, the developers have to address some false positives and false-negative definitions of correspondence between the input model elements. However, the problem is rarely resolved without causing any negative effects on the developers' effort and expected characteristics of the design models e.g., correctness (Table 4).

Consequently, it was particularly challenging for developers to perform the compositions, or even for modeling managers, authorize the execution of the compositions. The developers are reluctant to compose the input models, and hence all potential benefits (e.g., gains in productivity) of the use composition in collaborative software modeling are compromised. In these cases, the current composition techniques are not effective to compose design models in collaborative model evolution.

## 4.2.4.2.2.
## Conflict Management

The detection of all possible semantic conflicts between two versions of a model is an undecidable problem (Mens, 2002), as many false positive conflicts can appear. To reduce this problem, some previous works have recommended reducing the size of the delta model in order to reduce the number conflicts (Perry et al., 2001). However, this approach does not ameliorate in fact the complexity of the changes. That is, the problem is not essentially the number of conflicts that the size of the delta can cause, but the complexity of the conflicts. To alleviate the effort to resolve the conflicts, we narrowed down the scope of the conflicts. For this, the delta model became to represent one or two functionalities of a use case in particular. Hence, the conflicts became more manageable and reasonable. Following this strategy, we were able to reduce the number and complexity of the conflicts. In practical terms, this complexity was minimized by reducing the number of functionalities implemented in the delta model. That is, the compositions had a smaller scope.

On the other hand, sometimes the changes with broader scope were inevitable in the delta model. This was, for example, the case when the models (e.g., class and sequence diagrams) were reviewed and meliorated for reasons of quality assurance. Unfortunately, this results in a decreased precision of the compositions due to the presence of non-trivial compositions. It is known that the domain independent composition algorithms cannot rely on the detailed semantics of the models being composed or on the meaning of changes. Instead of being able to identify all possible conflicts, the algorithms detect as many conflicts as possible, assuming an approximate approach. Consequently, developers need to deal with many false positive conflicts.

In practice, we noted that if the composition generates many conflicts, developers prefer throwing the models away (and investing more effort to recreate it after) to resolving all conflicts. Although the composition algorithm detects the conflicting changes created by developers in parallel, developers are unable to understand and proactively resolve these conflicts generated from non-trivial compositions. This can be explained by two reasons. First, the complexity of the conflicts affected the model elements. Second, the difficulty of understanding the meaning of the changes performed by other developers. More importantly, developers were unable to foresee the ripple effects of their actions.

This is linked to two very interesting findings. First, developers have a tacit assumption that the models to-be-composed will *not* conflict with each other, and a common expectation is that little effort must be spent to integrate models. Hence, developers tend to invest low effort to check whether the composition produced inconsistencies or not. Therefore, we can conclude that the need to throw the model away in order to recreate it after demonstrates the complexity of the problem.

We have observed that the developers spend more effort when inconsistency propagation occurs. Although it is well known that the spread of the inconsistencies lead developers to spend some additional time to detect and resolve them, we have observed that this extra effort is due to, in part, the developers produce the inconsistencies are not the same to detect and resolve them. Note that in general inconsistencies are produced from the conflict resolution process performed incorrectly. This can be explained based on some reasons.

First, it is not always clear for developers that any inconsistency was produced. This perception is only realized along the project when the inconsistencies have already been resolved. Second, the inconsistencies tend to "keep alive" during the project because developers do not always detect and resolve the inconsistencies when they appear—either because they do not know which models are affected by the inconsistencies or either because the inconsistencies do not affect the use purpose of the models created by them.

In the first case, developers are concerned with the models under their responsibility i.e., models that they must produce. However, they feel comfortable to resolve inconsistencies localized in models that they are not under their responsibility. The main reason is that developers need to understand use cases (or scenarios) describing the functionalities represented in the diagrams. For a perfect understanding, developers should often grasp *business rules* and *design rules*, which define the domain elements and their constraints. That is, developers should know about the company business before resolving the inconsistencies. This represents one of the impairments to resolve the inconsistencies when they are detected. Another finding is that to resolve the inconsistencies, developers need sometimes to grasp the reasons why a composition was realized in one way and not in an expected manner.

In the second case, developers obligatorily spend effort to resolve inconsistencies that compromise the main purpose of use of the design models e.g., communication, but rarely to solve the inconsistencies that damage secondary purpose e.g., prediction. Developers do not solve all inconsistencies due to time constraint. Consequently, they live with inconsistencies in practice. In our case studies, the models were used for improving the communication between the developers. Although other inconsistencies might be resolved, only the inconsistencies that jeopardize the comprehensibility of the models were necessarily solved. For example, the layout of the models was an ever-present concern during the modeling. This means that developers invested time to arrange the elements in the model to ensure a good understanding of the features. Therefore, all inconsistencies that affect this layout must be resolved; otherwise, the purpose of use of the model is compromised. We can conclude that, although it is desired to keep models without inconsistencies only the inconsistencies that affect the purpose of use of the models are resolved.

### 4.2.4.2.3.
### Social Factors

The reputation of the developers influences the resolution of conflicting changes. We observed this finding during the observational study, interviews, and analyzing the change history in the repository. Recall that a developer can accept and reject a change of a second developer. This situation can be illustrated in turns of our motivating example. The developers Peter and Bill have distinct levels of experience. Peter is less experienced than Steve. Thus, if Peter performs a change that conflicts with another carried out by Steve (and he is not sure about how to resolve them) then he accepts the changes performed by Steve. That is, given that Peter is indecisive, he relies on the Steve's reputation.

Reputation can be seen as the opinion (or a social evaluation) of a member of the development team toward other developer. We have identified two types of reputation: *technical* and *social*.

Technical reputation refers to the level of knowledge considering issues related to the technology and tools used in the company such as the composition tool, IDEs, CASE tools, and version control systems. This type of reputation is mainly acquired solving daily problems. Social reputation refers to the position held by the members of the development team (e.g., senior developer). More experienced teamwork members (e.g., senior ones) influence less experienced members (e.g., novice ones). This happens mainly because the experienced ones are the human face of the development projects, making important project decisions, and coordinating teams.

Knowing that the reputation of the developers might affect the conflict resolution, we investigated which reputation would cause more influence. For this, eight developers were interviewed. The data collected suggests that technical knowledge causes more influence on decision making than social reputation. More specifically, 75 percent of the developers (6/8) reported that the technical reputation would influence more developers' decisions than social one.

### 4.2.5.
### Limitations of Related Work

We contrast this work with previous studies considering empirical studies, development effort, composition techniques, and modeling language as follows.

*Empirical Studies.* It is well known that empirical studies in model composition are severely lacking (Uhl, 2008; France & Rumpe, 2007). Some authors have contributed toward clarifying how conflicts emerge and how they are tamed in artificial scenarios. For the most part, these works have considered limited composition scenarios compared to the scenarios evaluated in this work. Still, the most of them do not consider effort as the investigation variable.

The observational study in (Perry et al., 2001), for example, investigates the change history of a legacy system to delineate the boundaries of (and to understand the nature of) the problems considering the software development in parallel. The authors considered only one observational study and all work was concentrated in level of code. Another example would be the experimental report in (Altmanninger et al., 2009). That study analyzes the challenges in merging different versions of one model, proposes an initial categorization of typical changes, and identifies resulting conflicts from the compositions. Although interesting, the current empirical studies do not evaluate composition effort. Still, the findings are normally collected from artificial and limited case tests rather from realistic composition scenarios. Finally, some previous works (Mens, 2002; Whittle & Jayaraman, 2010; Dingel et al., 2008) reinforce the need for empirical studies in model composition.

Considering two empirical studies in model-driven development (Hutchinson et al., 2011a; Hutchinson et al., 2011b), Hutchinson and colleagues presents some initial results from a twelve-month empirical research study of model driven engineering (MDE). More specifically, they document a set of technical, organizational, and social factors that apparently influence organizational responses to MDE (Hutchinson et al., 2011a). In (Hutchinson et al., 2011b), they describe the practices of three commercial organizations concerning MDE approach to their software development. The main contribution is a range of lessons learned, reporting the importance of social factors instead of technical factors on the relative success, or failure, of the adoption of MDE in practice. The authors do not mention any problem concerning model composition during these qualitative studies. This does not mean it is not a problem in practice since they take a much broader view and ask questions that are more general about the role and effectiveness of MDE.

On the other hand, in (Uhl, 2008), Uhl points out that composition of enterprise artefacts is not a trivial issue. Most because it requires the composition of graphical views, forms, dialogs, and depends on "friendly" views to tame all conflicts between the multiple models. Hence, developers end up avoiding model composition and adopting pessimistic locking of design models. Therefore, our results can be seen as the first to empirically investigate RQ2.3 and RQ2.4 using the state-of-the-practice composition technique in industry.

*Development Effort.* A major contribution of our work is the investigation of composition effort as a critical factor for the acceptance of the composition techniques in practice. Some previous works have also demonstrated that the effort is a critical factor during the software development (Jorgensen, 2005). Usually the effort is based on *ad hoc* estimation (Farias et al., 2011; Jorgensen, 2005). Jorgensen (Jorgensen, 2005) highlights that effort estimation is still a real, open problem due to the lack of empirical evidences about the effort required to perform development tasks. In fact, estimating effort based on the expert judgment is the most common approach today. Even worse, these feedbacks are often diverging or overoptimistic. When we consider this problematic in the context of composition, the problem is aggravated. However, little has been done to investigate this problem.

*Composition Techniques*. Model composition is a very active research field in many research areas such as synthesis of state charts (Ellis & Gibbs, 1989), weaving of aspect-oriented models (Whittle et al., 2009; Klein et al., 2006; Whittle & Jayaraman, 2010), governance and management of enterprise design models (Norris & Letkeman, 2011), software configuration management (Whitehead, 2007), composition of software product lines (Jayaraman et al., 2007), and composition of design models (Nejati et al., 2007; Epsilon, 2011). For this reason, several academic and industrial composition techniques have been proposed such as MATA (Whittle et al., 2009), Kompose (Kompose, 2011), Epsilon (Epsilon, 2011), IBM RSA (IBM, 2011), and so on. With this in mind, some observations can be done.

First, these initiatives focus only on proposing the techniques instead of also demonstrate their effectiveness. Consequently, qualitative and quantitative indicators considering these techniques are still incipient. In addition, the situation is accentuated considering effort indicators. This lack hinders mainly the understanding of their side effects. Second, their chief motivation is to provide a systematic algorithm. Unfortunately, these approaches do not offer any insights or empirical evidences whether developers might reach the potential benefits claimed by using composition techniques in practice. Although some techniques are interesting approaches, they are fundamentally flawed because of the large number of false positives that will be produced for large-scale systems. Nevertheless, the effort required for the user to understand and correct composition inconsistencies will ultimately prove to be too great. The current study takes a different approach. It aims to provide a precise assessment of composition effort in real life context, quantifying effort and identifying the influential effort.

Next, current works tend to investigate on the proactive detection and earlier resolution of conflicts. Most recently, Brun (Brun et al., 2011a) proposes an approach, namely Crystal, to help developers identify and resolve conflicts early. The key contributions are that conflicts are more common than would be expected, appearing overlapping textual edits but also as subsequent build, and test failures. In a similar way, Sarma (Sarma et al., 2011) proposes a new approach, named Palantír, based on the perception of workspace awareness, on the detection and earlier resolution of a larger number of conflicts. Based on two laboratory experiments, the authors confirmed that the use of the Palantír reduced of the number of unresolved conflicts. Although these two approaches are interesting studies, the earlier detection does alleviate the problem of model composition. The problem is the same, but is only reported more quickly. In addition, they appear to be overly restrictive to the code, not leading to broader generalizations at modeling level. Lastly, they neither make consideration about the effort to compose the artefacts used nor investigate the research questions in vivo case studies.

*Modeling Language.* There has been more research on evaluating the use of UML models (and its extensions) rather than the effort of composing them. These studies notably aimed at evaluating modeling languages in terms of some quality attributes such as comprehensibility (Lange & Chaudron, 2006), interpretation (Nugroho et al., 2008), and maintainability (Dzidek et al., 2008) rather than the composition effort. Additionally, most existing works have focused attention on exploring different quality issues considering UML models and understanding its appropriateness in mainly artificial scenarios. However, none of them attempt to understand how these quality issues may be affected during compositions and to examine a set of wider issues about the effort on composing these models in real-life scenarios. Some these issues include: are these quality issues of the UML models affected during the composition? In which composition tasks should the developer invest more effort? What is the trade-off between the composition tasks in practice? What are the characteristics of the UML models that help developers to compose them?

To sum up, there has been very limited empirical research evaluating the effort of composing large-scale design models in literature. Even worse, nothing has been done to both understand and describe the influential factors that can jeopardize the potential benefits of using composition techniques in industry. In particular, there are four critical gaps in current understanding. Firstly, the lack of practical knowledge on the effort of applying composition techniques, detecting and resolving inconsistencies in practice. More importantly, the lack of a trade-off analysis about three effort variables (Section 4.2.3.4.3). Secondly, a precise understanding about the influential factors of composition effort is lacking. Next, the lack of understanding of how technical and social factors can affect composition effort. Last, the absence of evaluation of important aspects in model composition beyond modeling languages and composition techniques. Some of these aspects would be such as the potential benefits of good practice of software modeling, merging in pair (two or more developers work together to compose the input models), inconsistency management, and strategies to allocate tasks to minimize the composition effort.

### 4.2.6.
### Concluding Remarks of the Second Study

Model composition is a key mechanism to support the evolution of design models in large-scale software projects. In particular, this mechanism is essential to promote collaborative work of separate development teams whereas increasing their productivity. Thus, developers naturally become concerned about the quality of the software evolutions produced (i.e., the composed models) and the effort invested by the teamwork members. However, there is a lack of empirical studies evaluating model composition effort in practice. This means that little empirical findings can be converted into practical knowledge to the industry. Developers have no guidance on how to reduce model composition effort and the number of emerging model inconsistencies.

This study represents the first in vivo exploratory study to evaluate the *effort* that developers invest to compose design models (RQ2.3) and to identify and analyze the *factors* that affect developers' effort (RQ2.4). In our study, a best-of-breed model composition technique was applied to evolve industrial design models along 297 evolution scenarios. Developers conducted the work during 56 weeks, which resulted in more than 2 million compositions of model elements. We investigated the composition effort in this sample, and analyzed the side effects of key factors that affected the effort of applying the composition technique as well as detecting and resolving inconsistencies. All conclusions from RQ2.3 and RQ2.4 were drawn from quantitative and qualitative analyses based on the use of metrics, interviews, and observational studies.

We summarize the findings related to RQ2.3 as follows: (1) the application effort measures do not follow an *ad hoc* distribution and, rather, it assumed a distribution pattern; (2) the application effort tends to reduce as developers become more familiar with technical issues rather than application domain issues;

(3) the more experienced developers spend 23.2 percent less effort to detect inconsistencies than less experienced developers; and (4) the higher the number of inconsistencies in behavioral models, the higher the effort to detect inconsistencies. Additionally, we also present four findings with respect to RQ2.4 as follows: (1) the production of the intended model is strictly affected by the presence of different types of change categories in the delta model; (2) the closer to the Open-Closed principle the change is, the lower is the composition effort; (3) evolution scenarios that do not follow the Open-Closed principle required more effort to produce the intended model; and (4) the refinement (1:N) of model elements in the delta model caused severe composition problems and hence increased the composition effort.

Although there is a significant amount of quantitative and qualitative evidence supporting our findings previously mentioned, further empirical studies are still required to check whether they are observed in other contexts with different subjects. For example, we need to better understand if the composition effort is alleviated when developers compose well-modularized input models. There is some expectation that design models with an improved modularization can aid the composition techniques to accommodate the changes in the base model. Another two interesting investigation points would be: (1) Do developers invest more effort to compose behavioral models (e.g., sequence diagrams) than structural models (e.g., component diagrams)? (2) Do developers invest more effort to resolve semantic inconsistencies than syntactic ones? It is by no means obvious that, for example, developers invest less effort to resolve inconsistencies related to the well-formedness rules of the language metamodel than to resolve inconsistencies considering the meaning of the model elements.

Finally, we hope that the issues outlined throughout the thesis encourage other researchers to replicate our study in the future under different circumstances. Moreover, we also hope that this work represents a first step in a more ambitious agenda on better supporting the model composition tasks.

# 5
# Effort on the Detection of Inconsistency

Modeling languages, such as UML (OMG, 2011) and its extensions, provide different types of models (e.g., class and sequence diagrams) to represent complementary views of a software system. These models define the system structure and behavior so that design decisions can be properly understood. Developers will implement these complementary models later. Examples of these complementary models would be UML sequence diagrams and UML class diagrams. It is well known that, in practice, these models are created and used by different developers in parallel and often suffer from the inconsistency problems (Lange, 2007a; Apel et al., 2011; Mens, 2002;). These inconsistencies are mainly caused by the mismatch between the overlapping parts of complementary models and by the lack of formal semantics to prevent these contradictions (Lange et al., 2006a; Lange et al., 2004). Consequently, developers must invest some effort to detect and properly deal with these inconsistencies (Farias et al., 2011); otherwise, misinterpretation caused by inconsistencies could be transformed into defects in code.

Different modeling languages support different forms of modular decomposition and may influence how developers detect or neglect inconsistencies (Farias et al., 2010a). This might be particularly the case with aspect-oriented modeling (AOM) (Clarke & Banaissad, 2005; Clarke, 2001) as it intends to improve design modularity of otherwise crosscutting concerns. Current research in AOM varies from UML extensions (Losavio et al., 2009; Chavez et al., 2002; Clarke & Banaissad, 2005) to alternative strategies for model weaving. Unfortunately, nothing has been done to investigate whether aspect-oriented models can alleviate the burden of dealing with model inconsistencies. Someone might hypothesize that they might help developers to understand the design before implementing it. Others could also postulate that the improved modularization would reduce the effort to detect inconsistencies and minimize misinterpretations arising between multiple design models.

Unfortunately, it is by no means obvious whether these assumptions hold or not. First, it may be the case that additional constructs in AO models to support a superior modularization lead to detrimental effects on design understanding. Second, it is still not clear if an aspect affecting multiple join points can increase the inconsistency detection and improve the model interpretation. Third, developers might get "distracted" by the global reasoning motivated by the presence of crosscutting relations (Filman & Friedman, 2000; Clarke & Walker, 2001) between classes and aspects. At last, developers might even invest more effort using AO models while examining all points that are crosscut by the aspects (Farias et al., 2010a).

In this context, the goal of this chapter is to investigate the effects of the design modeling languages on the following quality notions: detection, social, syntactic, and semantic ones. This Chapter, therefore, reports a controlled experiment aimed at investigating the impact of aspect-oriented (AO) modeling on: (1) the rate of inconsistency detection; (2) the developers' effort to detect these inconsistencies; and (3) developers' misinterpretation rate. The use of AO models was contrasted with the use of OO models in a particular context: the use and understanding of design models by developers needed to produce the corresponding implementation. The results supported by statistical tests and qualitative analysis, show that AO models alleviated the effort to detect inconsistencies. Nevertheless, it reduced neither inconsistency detection rate nor misinterpretation rate.

Other findings were also reported. For instance, we observed that the downsides of AO modeling were largely caused by the degree of aspect quantification (Filman & Friedman, 2000). That is, the higher the number of modules affected by an aspect, the lower the inconsistency detection rate and the higher the misinterpretation rate. Moreover, we observed that developers tended to detect inconsistencies more quickly in AO models when the scope of aspect pointcuts was narrow. Equally relevant was the finding that the number of crosscut relationships influences the creation of the "intended model." To the best of our knowledge, our results are the first to pinpoint the potential (dis)advantages of AO modeling in imprecise multi-view modeling.

The remainder of this chapter is organized as follows. Section 5.1 presents background. Section 5.2 describes the study methodology. Section 5.3 and Section 5.4 are the main contributions — the experimental results and their discussion itself. Section 5.5 compares the study with the related work and, Section 5.6 discusses the threats to validity. Finally, Section 5.7 gives some conclusions.

## 5.1.
## Background

This background is complementary to the explanations described in Chapter 2. Inconsistency detection has been studied for many years in software engineering (Lange et al., 2006a; Lange et al., 2004) and in other related disciplines. In fact, developers often need to detect conflicting information between artifacts during the software development process. In the context of our study, we investigate if developers are more able to detect inconsistencies in AO models rather than OO models used to communicate design decisions.

### 5.1.1.
### Aspect-Oriented Modeling

As previously mentioned in Chapter 2, aspect-oriented modeling (AOM) languages aim at improving the modularity of design models by supporting the modular representation of concerns that cut across multiple software modules.

The modularization of such crosscutting concerns is achieved by the definition of a new model element, called aspect. In general, the notation enables to explicitly distinguish between aspects and classes. An aspect can crosscut several classes in a system. These relations between aspects and other modules are called crosscutting relationships.

This aim is achieved in different ways in the AOM techniques. The current proposed approaches e.g., (Klein et al., 2006) are mainly aimed at supporting innovative weaving process for base and aspect models. That is, they aim at expressing and simulating the weaving relations between the base model and aspectual model elements. Approaches that are more conservative propose UML profiles (Losavio et al., 2009; Chavez & Lucena, 2002; Stein et al., 2002) for supporting the modeling aspect-oriented design. These techniques are more aligned to AOP models, such as those realized by AspectJ (AspectJ, 2011) and dialects.

Given the goal of our work (Section 5.2.1), we opt for evaluating the impact of aspect-oriented UML profiles on inconsistency detection processes. This choice can be explained by some reasons. First, real developers use UML profiles for AO modeling instead of any other AO modeling technique. Second, these profiles have the advantage of supporting classical AOP concepts at a more abstract level (Losavio et al., 2009; Aldawud et al., 2003; Chavez & Lucena, 2002). This means that AO key concepts are usually represented via conventional extension mechanisms of the Unified Modeling Language (UML), such as stereotypes. This alternative avoids classical side effects related to the learning curve in a controlled experiment like ours. Otherwise, it would not be possible to investigate the causal relationships between the dependent and independent variables (Section 5.2.6) without any high overhead to the subjects involved.

Another reason is that UML is the standard for designing software systems. The use of stereotypes reduces the gap between subjects with low experience and ones with more experience (Ricca et al., 2010). The other consequence of using UML profiles for AOM is that the model reading technique used by the subjects would not be more influenced by new notation issues. As UML profiles are supported by academic and commercial modeling tools, such as IBM Rational Software Architect and Borland Together, developers are familiar with stereotype notations. Moreover, the learning curve of the current state-of-the-art of AOM is not a trivial task for developers in early adoption of aspect-oriented programming.

Finally, UML profiles for aspect-oriented design is the approach more common for structural and behavioral diagrams. Therefore, the interpretation of the models is exclusively influenced by the use of the concepts in object-oriented and aspect-oriented modeling. Based on these reasons, the AOM language used in our study is a UML profile (Losavio et al., 2009; Aldawud et al., 2003; Chavez & Lucena, 2002). **Erro! Fonte de referência não encontrada.**Figure 16 presents an illustrative example of the models used in our study: a class and a sequence diagram of the AOM language used in our study: (A) and (B) represent the conflicting structural diagrams, while (C) and (D) represent the structural and sequence diagrams without inconsistencies. The notation supports the visual representation of aspects, crosscutting relationships and other AOM concepts. The stereotype <<aspect>> represents an aspect, while the dashed arrow decorated with the stereotype <<crosscut>> represents a crosscutting relationship. Inner elements of an aspect are also represented such as pointcut (<<pointcut>>) and advice. An advice adds behavior before, after, or around the selected join points (Clarke & Walker, 2005; Clarke & Walker, 2001). The stereotype associated with an advice indicates when (<<before>>, <<after>> or <<around>>) a join point is affected by the aspect. The join point is a point in the base element where the advice specified in a particular pointcut is applied.

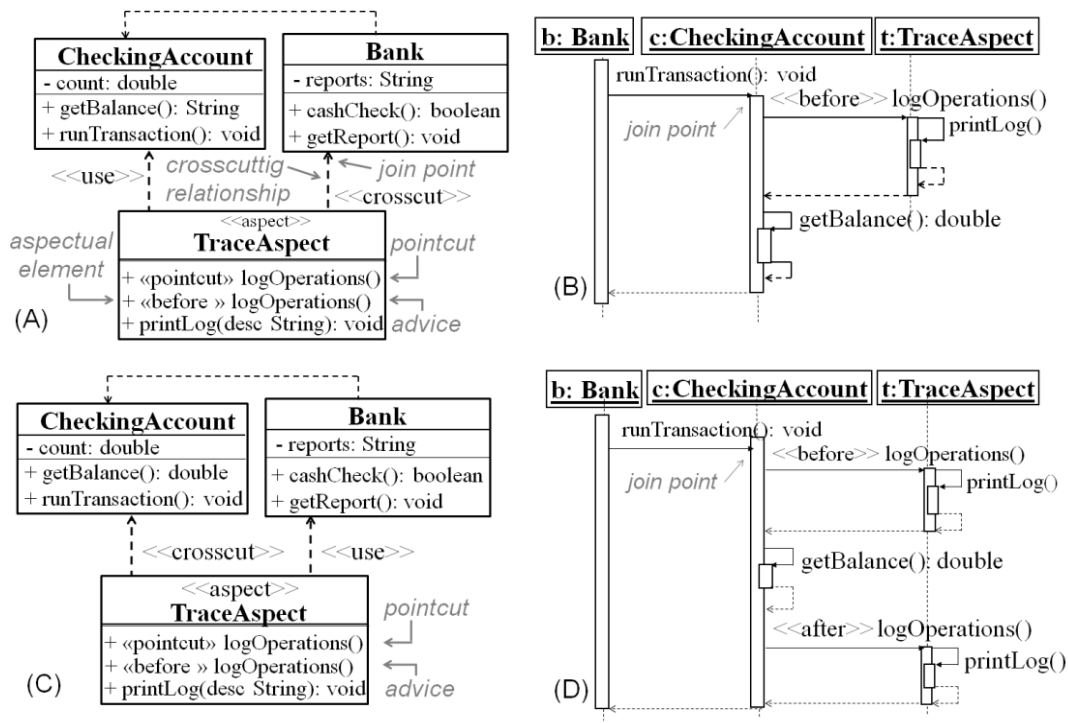

Figure 16: An illustrative example of aspect-oriented models used

## 5.1.2.
## Model Inconsistency

Model inconsistency was previously discussed in Chapter 2. However, it is discussed again due to the need for further details to investigate the research questions addressed in this Chapter. Additionally, it is only discussed here due to readability issues.

Model inconsistency is often the case that complementary diagrams of a software system, such as class and sequence diagrams, inevitably have conflicting information (Langes & Chaudron, 2004). If software developers do not detect and properly deal with these inconsistencies the potential benefits of using design models can be compromised. This means that, for instance, gains in productivity and design understandability will be hindered. Consequently, developers must invest some considerable effort to detect these inconsistencies. Two broad categories of the most common inconsistencies are: (1) syntactic inconsistencies, which arise when the models do not conform to the modeling language's metamodel; and (2) semantic inconsistencies, in which the meaning of one or more model elements does not match with that of the actual design model. Our study focused on semantic inconsistencies because they cannot be automatically identified with tool support (Lange & Chaudron, 2006a). Moreover, they are usually the main cause of design misinterpretation (Wohlin et al., 2000).

Occurrences of semantic inconsistencies are particularly very common when class and sequence diagrams are used in conjunction with a system (Lange & Chaudron, 2006a; Lange & Chaudron, 2004). This is probably due to the fact they are the most used UML models in practice (Doring & Parsons, 2006) and represent the same concepts under different perspectives. These are the key reasons governing the selection of these diagrams in our experimental study. Moreover, we have particularly selected semantic inconsistencies that are: (i) detectable by developers (Lange & Chaudron, 2004), and (ii) difficult or impossible to detect automatically. The reason for the latter is that the semantics of model elements are rarely expressed in a formal manner. Semantic inconsistencies are even more difficult to detect in multi-view modeling (Kitchenham et al., 2008). Semantic inconsistencies arise in multi-view models when they have overlapping parts. For instance, objects exchange messages in sequence diagrams, while these messages represent methods in the class diagram. In addition, a message from one object to another means that the first object calls a method that is provided by the second object. Other forms of overlapping elements occur in aspect-oriented models. There are several forms of multi-view inconsistencies and we discuss below how they can manifest in both OO and AO models. This thesis aims at inconsistencies that have been documented elsewhere (Lange et al. 2004) and used in a previous empirical study (Lange et al. 2006). The inconsistencies used in this study are described as follows:

1) *Conflicting relationships*: this inconsistency occurs when the presence or the nature of a relationship diverge in structural and behavioral models. For instance, according to the sequence diagram, the advice of an *aspect A* crosscuts the behavior of *class B*; however, the semantics of the advice in *A* dictates when the class diagram should have either a <<*crosscut*>> or a <<*use*>> relationship between *A* and *B*. For example, Figure 16 presents this kind of inconsistency. The aspect *t:TraceAspect* crosscuts the *c:CheckingAccount* objects (Figure 16.B). In this case, the relationship between *TraceAspect* and *CheckingAccount* should be <<*crosscut*>> instead of <<*use*>> (Figure 16.C) given the logging semantics of the advice *logOperations()*. In the structural diagram (Figure 16.A), the aspect *TraceAspect* has a <<*use*>> relationship with the class *CheckingAccount* instead of <<*crosscut*>> relationship.

2) *Messages with different return types*: the return type of a message m from an object *A* to an object *B* does not match with the return type of the method *M* in the corresponding class *B* in the class diagram. For instance, the method *CheckingAccount.getBalance* has conflicting return types: string in the class diagram and double in the sequence diagram. A similar conflict can occur with the return type of an around advice (Losavio et al., 2009; Aldawud et al., 2003; Chavez & Lucena, 2002) and the return type from a method execution being advised by the latter.

3) *Object without class/aspect*: an object in a sequence diagram does not have a corresponding class or aspect in the class diagram.

4) *Weaving in a wrong element*: an aspect *A* weaves advice into model element *B* in the sequence diagram, but in the class diagram does not exist any crosscutting relationship from *A* to *B*.

5) *Message without name*: a message between objects in the sequence diagram does not have a name.

6) *Message without method*: a message from an object of class *A* to an object of class *B* does not correspond to any method of the class *B* in the class diagram.

7) *Message with wrong return type*: the return type of a message *X* from an object of class *A* to an object of class *B* does not match with the return type of the method *X* of the class *B* in the class diagram.

8) *Message in the wrong direction*: there is a message from an object of class *A* to an object of class *B*, but the method corresponding to the message is a member of class *A* instead of class *B*.

9) *Class without meaning*: a class does not have any semantic value in the class diagram.

10) *Instance of abstract class*: an abstract class is used in the sequence diagram as object.

Although the behavioral and structural diagrams are syntactically correct, the contradicting information makes the models semantically incorrect. Note that if developers do not detect these inconsistencies, they will likely transform them into defects in code due to the misinterpretation. For example, a developer might take in consideration the specification of the method *CheckingAccount.getBalance* in the structural diagram (i.e., string as return type), whereas other developer might consider the specification in the sequence diagram (double as return type). Consequently, this can give rise to unexpected behavior in the code as a method can expect a string as return type instead of double (Mens, 2002). This contradicting information between the models may lead to static and behavioral inconsistencies in code.

## 5.1.3.
## Inconsistency Detection Effort

Developers detect inconsistencies when they identify conflicting information in the models and, then, possibly report that the models cannot be implemented. This decision often relies on "guessing" the semantics of model elements. To reach this conclusion, developers need to invest some effort: the time to go through the model and infer that the models suffer from inconsistencies. There is currently very limited knowledge regarding the amount of effort required to detect inconsistencies. Anecdotal evidence from companies suggests that the effort is significant (Farias et al., 2011), but nothing can be conjectured considering AO models in comparison to OO models.

There are some tools to support the visualization of crosscutting relation effects in class diagrams (Clarke & Walker, 2005). There are also tools to generate a woven sequence diagram (Klein et al., 2006) or even integrating or simulating the effects of composing state machines. The use of these tools was not included in our study for several reasons. First, the nature of the investigated conflicts would require that developers undergo model inspection anyway. In fact, the focus of our study is to investigate whether developers can pinpoint inconsistencies and understand the design decisions when producing the corresponding implementation. Second, even though the use of these tools might reduce or exacerbate the generation of specific categories of inconsistencies in AO models, it was not our goal to evaluate particular tools. More importantly, these tools are not used in practice yet; either because they are not robust enough to be applied in real-world settings, or because they are not intuitive to be used in practice. Hence, their use would impose severe threats the validity of our experimental results.

## 5.2.
## Study Methodology

This section presents the main decisions underlying the experimental design of the controlled experiment, which adheres to guidelines of empirical studies (Kitchenham et al. 2008; Wohlin et al. 2000). We chose controlled experiment due to the same reasons discussed in Section 4.1.1.

### 5.2.1.
### Experiment Definition

We formulate the goal of this study using the GQM template (Wohlin et al. 2000) as follows:

*Analyze* AO and OO modeling techniques

*for the purpose of* investigating the impact

*with respect to* detection effort and misinterpretation

*from the perspective of* developers

*in the context of* multi-view design models.

Therefore, this is related to research question RQ3, as stated in Chapter 1:

- **RQ3:** What is the effect of design decomposition techniques in particular with respect to misinterpretation, inconsistency rate, inconsistency detection effort, and inconsistency resolution effort?

Regarding the quality notions defined in Chapter 3, we study how design modeling languages affect six quality notions, namely: syntactic, semantic, pragmatic, social, effort, and detection ones. Based on this, we refine the research question into three more specific research questions. Thus, we focus on the following research questions:

**RQ3.1:** Does AO model affect the efficiency of developers to detect multi-view model inconsistencies?

**RQ3.2:** Does AO model influence the effort invested by developers to detect model inconsistencies?

**RQ3.3:** Do AO models lead to a different misinterpretation rate as compared to OO models?

The context selection is representative of situations where developers implement classes (or aspects) based on design models. The experiment was conducted within two postgraduate courses at the Pontifical Catholic University of Rio de Janeiro (PUC-Rio) and Federal University of Bahia (UFBA). In both courses, AO modeling and OO modeling were taught in the first year of Master and Doctoral programs in Computer Science. Therefore, all the subjects (18) hold a Master's or Bachelor's degree, or equivalent. In addition, eight (8) professionals from three companies also participated in the experiment. Most of the professionals held a Master's or Bachelor's degree.

## 5.2.2.
## Hypothesis Formulation

*First Hypothesis.* The first research question investigates whether developers by using AO models produce a lower (or higher) inconsistency detection rate than by using OO models. Usually developers do not indicate the presence of existing inconsistencies in multi-view models (Lange et. al., 2006). The main reason is that they can make implicit assumptions about the correct design decisions based on previous experience. Moreover, they might feel forced to produce an implementation even in the presence of inconsistency. Thus, our intuition is that developers identify fewer inconsistencies in AO models than OO models because they might get distracted by the global reasoning motivated by the presence of additional crosscutting relations in the models. Consequently, they may have a higher number of implicit assumptions to assemble the "big picture" of a system. However, it is by no means obvious that this hypothesis hold. Perhaps, the increased modularity of AOM models may help developers to switch more quickly between the behavioral and structural views while implementing their aspects. Consequently, developer may localize more inconsistencies than in OO models. Theses hypotheses are summarized as follows:

**Null Hypothesis 1, $H_{1-0}$**: The inconsistency detection rate in AO models is equal or higher than in OO models.

$H_{1-0}$: DetectionRate (AO) $\geq$ DetectionRate (OO)

**Alternative Hypothesis 1, $H_{1-1}$**: The inconsistency detection rate in AO models is lower than in OO models.

$H_{1-1}$: DetectionRate (AO) $<$ DetectionRate (OO)

*Second hypothesis.* The second research question investigates whether developers invest less (or more) effort to detect inconsistencies in AO models than in OO models. The superior modularity of AO models may help developers to better match and contrast the structural and behavioral information about the crosscutting relations. In this case, developers may switch more quickly between the behavioral and structural views while systematically implementing their aspects. Thus, our expectation is that the higher the number of crosscutting relationships (an aspect crosscutting a wider scope) in the model, the lower the effort to detect inconsistencies. This assumption is based on the superior ripple effects of inconsistencies observed in AO models when model composition techniques are applied (Farias et al., 2010a). This propagation can directly affect the effort in detecting inconsistencies, since developers, facing the complexity of the propagations, avoid doing any implementation. That is, by using AOM developers tend to get more quickly convinced about the severity of multi-view inconsistencies. This means that they are more likely to report them and not going forward on the design implementation. However, it is not clear whether this intuition holds because, at first, developers may examine all model elements affected (or not) by the inconsistencies, or even the inconsistencies to some extent may even be confined in the aspectual elements. This leads to the second null and alternative hypothesis as follows:

**Null Hypothesis 2, $H_{2-0}$**: The effort to detect inconsistencies in AO models is equal or higher than in OO models.

$H_{2-0}$: EffortToDetect (AO) $\geq$ EffortToDetect (OO)

**Alternative Hypothesis 2, H$_{2-1}$:** The effort to detect inconsistencies in AO models is lower than in OO models.

**H$_{2-1}$:** EffortToDetect (AO) < EffortToDetect (OO)

*Third hypothesis.* The third research question investigates whether developers' misinterpretation rate (MisR) is higher (or lower) in AO models than in OO models. The chief reason of the disagreement between developers' interpretations is the contradicting understanding of the design models. They are often caused by inconsistencies emerging from the mismatches between the diagrams specifying the multiple, complementary views of the software system (Lange & Chaudron, 2006a; Farias et al., 2010a). Contradicting design models make it difficult for developers to think alike and, hence, producing code with the same semantics. The key reason is that software implementation widely depends on cognitive factors. Someone can consider that additional AOM concepts, such as crosscutting relationships or aspects, may negatively interfere in a common understanding of design models by different developers. For instance, developers need to precisely grasp the actual meaning of the crosscutting relations (in addition to all other relations), and when they are actually established during the system execution. Then, as developers have to examine all join points affected by the aspects, their extra analyses can increase the opportunities of diverging interpretations. However, this expectation might not hold because the crosscutting modularity may improve the overall understanding of the design when compared to pure OO models. This would lead to the following null and alternative hypotheses:

**Null Hypothesis 3, H$_{3-0}$:** The misinterpretation rate (MisR) in AO models is equal or higher in AO models than in OO models.

**H$_{3-0}$:** MisR(AO) $\geq$ MisR(OO)

**Alternative Hypothesis 3, H$_{3-1}$:** The misinterpretation rate in AO models is lower than in OO models.

**H$_{3-1}$:** MisR(AO) < MisR(OO)

### 5.2.3.
### Selection of Subjects

Subjects (18 students and 8 professionals) were selected based on two key criteria: the level of theoretical knowledge and practical experience related to software modeling and programming. The subjects studied in educational systems that place a high value on key principles of software modeling and programming. In addition, the subjects were exposed to more than 120 hours of courses (lectures and laboratory) exclusively dedicated to software design, software modeling, OO programming, and AO software development. It can be considered they underwent an intensive modeling-specific and programming training. As far as practical knowledge is concerned, the main selection criterion was that subjects had, at least, 2 years of experience with software modeling and programming acquired from real-world project settings.

### 5.2.4.
### Experiment Design

The design of this study was a *paired comparison design*. All subjects were submitted to two treatments (AO and OO modeling) to allow us to compare the matched pairs of experimental material. The subjects were randomly assigned and equally distributed to the treatments. The distribution followed a within-subjects design in which all subjects served in the two treatments. Each treatment had a printed questionnaire with five multiple-choice questions. That is, the subjects did not make use of modeling tools to understand and answer the questions. Although it was generally accepted nowadays that the current state-of-the-art of AOM (such as (Klein et al., 2006)) should be always used with a tool, the use of any kind of tool would certainly add some bias to the collected data: the subjects would be influenced by the different maturity and usability degrees of AO and OO modeling tools. Hence, we would end up comparing the tools instead of modeling languages. Moreover, we emphasize that the focus of this work is on the current state-of-the-practice of AOM instead of the state of the art of AOM, as briefly justified in Section 5.1.1. By doing so, the first treatment had only questions with AO models while the second one had only questions with OO models. The subjects were assigned randomly and equally distributed to these treatments so that the effects of the order could be discarded. Therefore, the experimental design of this study was by definition a balanced design.

To minimize the "gain in information" from one treatment to another one, the models used in the study were fragments of real class and sequence diagrams. Hence, the subjects had no prior information and no accumulated knowledge about the semantics of the model elements. In addition, each pair of structural and behavioral models had different kinds of inconsistencies, and the meanings of their elements were completely different. Therefore, we can assume that the performance of subjects was not influenced by the treatments of previous questions.

### 5.2.5.
### Operation and Material

*Operation.* In both treatments, the subjects received a pair of corresponding class (structural) and sequence (behavioral) diagrams similar to the models presented in Figure 16. They were asked how they would implement particular classes (or aspects) based on these diagrams. That is, rather than stimulated to review or inspect the diagrams, the subjects were encouraged to implement particular model elements (classes or aspects). Our goal is to identify how developers deal with contradicting information between complementary models in the context of concrete software engineering tasks. The subjects should choose, then, the most appropriated implementations between the five possible answer options. In each question, although the subjects were responsible for registering the time invested in each question ("start time" and "end time"), they were properly managed to avoid bias in the collected data. They were also stimulated to justify their answers on the answer sheet, but this part of the time was not counted. In total, ten questions were answered. After the experiment, the subjects were also interviewed to clarify the answers and results.

*Material.* Table 24 describes some design characteristics for the OO and AO models used in the study. For example, in the first task, the AO model had seven classes and one aspect, seven relationships between the classes and aspect, and six crosscutting relationships. Additionally, it is important to highlight three points: (1) every pair of OO or AO class and sequence diagrams had two kinds of inconsistencies, (2) research questions were investigated in all tasks of the experiment, and (3) the AO models vary with respect to the number of crosscutting relationships. The reason for the latter decision is that we suspect that these relationships might affect the variables (i.e., inconsistency detection rate) and detection effort) of this study (Section 5.2.6). The inconsistencies were always related to contradictions between the class and sequence diagrams. That is, there was conflicting information between those diagrams, as the examples given in Section 5.1.1.

Considering the answer options in each question, they were planned according to the following schema. The first answer option is according to the class diagram while the second one is just in concordance with the sequence diagram. The third answer option is based on the combination of the information presented in both diagrams. The fourth one is incorrect considering all two diagrams. All questions had a fifth answer option where the subjects could indicate that an inconsistency was detected in the models. The subjects were encouraged to carefully explain their answers, but those careful explanations are not part of the time required to solve the task.

| Task | Treatment | Class Diagram | | | | Sequence Diagram | |
|------|-----------|------|------|------|------|------|------|
| | | #CA | #RC | #AT | #OP | #O | #M |
| 1 | OO | 7 | 6 | 18 | 27 | 6 | 7 |
| | AO | 8 | 11(6) | 5 | 16 | 7 | 13 |
| 2 | OO | 8 | 6 | 16 | 23 | 6 | 6 |
| | AO | 6 | 5(1) | 9 | 19 | 5 | 10 |
| 3 | OO | 4 | 4 | 4 | 16 | 4 | 7 |
| | AO | 5 | 4(1) | 6 | 14 | 5 | 10 |
| 4 | OO | 4 | 4 | 6 | 12 | 5 | 10 |
| | AO | 6 | 7(2) | 7 | 20 | 6 | 11 |
| 5 | OO | 4 | 4 | 11 | 13 | 5 | 7 |
| | AO | 5 | 5(2) | 7 | 14 | 5 | 8 |

#CA: the number of classes or /and aspects;
#RC: the number of UML relationships or crosscutting relationships
#AT: the number of attributes. #OP: number of operations.
#O: the number of objects or instance of aspects. (n): number of aspects.
#M: the number of messages between the classes and aspects.

Table 24: Measures of the diagram used in the study

## 5.2.6.
## Variables and Quantification Method

The independent variable of this study is the choice of the modeling language. It is nominal and can assume two values: AO modeling and OO modeling. We investigate the effects of this independent variable on following dependent variables.

*Inconsistency detection rate (Rate).* This variable is intended to measure the overall rate of inconsistencies detected by all subjects (RQ4.1). It represents the ratio of the number of subjects that detect inconsistencies in a question divided by the number of subjects that answer the question without notifying the presence of inconsistency. Note that subjects detect inconsistencies when they explicitly indicate that they are unable to achieve a suitable implementation from the conflicting diagrams.

*Inconsistency detection effort (Effort).* It represents the mean of time (minutes) spent by the subjects to detect inconsistencies in a question (RQ4.2).

*Misinterpretation rate (MisR).* This variable represents the degree of variation of the answers (RQ4.3). That is, it measures the concentration of the answers over the four possible alternatives (the fifth alternative represents the detection of inconsistency). Our concern is if the differences in (un)detected inconsistency affects the design interpretation of the subjects. An undetected inconsistency is not necessarily problematic (Lange & Chaudron, 2006a) if all subjects have the same interpretation. For example, if the 26 subjects have the same answer (e.g., the alternative "A") for a question, then the inconsistencies in the diagrams did not lead to misinterpretations (MisR = 1). On the other hand, if the developers' answers spread equally over the four alternatives, then the

$$MisR(k_0, \ldots, k_{K-1}) = 1 - 2 \frac{\sum_{0 \leq i < K} k_i i}{N(K-1)} \qquad (1)$$

Where:

$K$: the number of alternatives for a question

$k_i$: the number of times alternative i was selected,
    where $0 \leq i < K$ and   (for all i : $0 \leq i < K - 1 : k_i \geq k_{i+1}$)

$N$: the sum of answers over all alternatives: $N = \sum_{0 \leq i < K} k_i$

inconsistencies cause serious misinterpretations (MisR = 0). That is, the misinterpretation rate is 0 (zero) if the answers are distributed equally over all options, and 1 (one) if the answers are concentrated only one answer option. This variable can be measured as follows (Lange et al., 2004).

## 5.2.7.
## Operation

*Preparation phase.* The subjects (students and professionals) were not aware about the research questions (and hypotheses) of our study in order to avoid biased results. The motivation of the students was to gain extra points for their grades. The results obtained in the questionnaire had no effect on their grades. The professionals received the same questions as a printable questionnaire. All subjects received a refresher training to be sure of their familiarity with the modeling concepts used in the study.

*Execution phase.* The experiment tasks were run within two courses at two different Brazilian universities (PUC-Rio and UFBA). Both runs were carried out in a classroom following typical exam-like settings. However, because of time constraints and location, the professionals run the experiment in their work environment. However, the experiment was carefully controlled. All subjects received 10 questions and the answer sheets. It is important to point out that there was no time pressure for the subjects, but they were rigorously supervised to correctly register the time. Therefore, we are confident that the time was recorded properly. For clarification reasons, the subjects were encouraged to justify their answers. After finishing the experiment, the subjects filled out a questionnaire to collect their background i.e., their academic background and work experience.

## 5.2.8.
## Analysis Procedures

*Quantitative Analysis.* The normal distribution of the collected data was checked using the Shapiro-Wilk and Kolmogorov-Smirnov test (Devore et al., 1999; Wohlin et al., 2000). The three hypotheses were tested using the parametric paired t-test and the non-parametric Wilcoxon test. Both methods compare two related samples or repeated measurements on a single sample to assess whether their population means differ (Devore et al., 1999). All hypotheses were tested considering a significance level of 0.05 (p-value < 0.05). The null hypotheses were rejected when the p-value was lower than 0.05.

*Qualitative Analysis.* Qualitative data were collected from two sources: think aloud answer sheet comments and interviews. The comments were expressed in a free-text field in which the subjects could report anything to explain their answer. In addition, some questions were prepared and asked to developers in interview sessions. Interview guidance with relatively open questions was prepared and all sessions were audio recorded with the permission of the subjects

### 5.2.9.
### Qualitative Data

*Interviews.* A semi-structured interview approach (Wohlin et al., 2000) was chosen following a funnel model, in which one initial open question is told and then directed towards to more specific one. It was organized in topics with open and closed questions. They were organized in such a way that research questions could be exploited. An interview guide was created based on the authors' experience and the study design. The interviews were recorded and transcribed into text. All subjects were selected for the interviews. It was assured that only anonymous data would be presented externally. Each interview lasted from 30 to 55 minutes, depending on how talkative the subjects were.

*Observational Study.* In order to investigate how the tasks in the experiment were performed, extensive observations were conducted through two different approaches. First, the authors run the experiment. This allowed a more effective observation and monitoring of the tasks of the subjects. Second, to obtain an additional feedback from the subjects, they were encouraged to write down the rationale used to answer the questions.

### 5.3.
### Experimental Results

This section discusses the experimental results related to the research questions RQ4.1, RQ4.2, and RQ4.3 (Section 5.2.1). All hypotheses were tested at the significance level of α = 0.05 and the findings were derived from both descriptive statistics and statistical inference.

### 5.3.1.
### RQ4.1: Detection Rate in AO and OO models

*Descriptive Statistics.* The first research question investigates if developers detect more (or less) inconsistencies in AO models or OO models. Contradicting the expected AOM superiority, the collected data indicate that developers tend to detect more inconsistencies in OO models than in their AO counterparts. Table 25 provides evidence for this observation through descriptive statistics of the collected data. The superior detection rate in OO models manifests in terms of both means and medians. As far as the latter in concerned, the median of the detection rate is 0.35 in AO models and 0.5 in OO models. This difference represents a superiority of 42.85 percent in favor of OO models. This observation is reinforced by analyzing the means of the detection rate. Developers detected, on average, 43.24 percent more inconsistencies in OO models (0.53) than AO models (0.37). These results suggest that OO models enable developers to identify more inconsistencies than AO models. As a consequence, classical UML-based modeling for crosscutting modularity (Section 5.1.1) do not necessarily imply on more effective inconsistency detection according our observations. This contradicts somehow the intuition that the improved modularity of AOM helps developers to localize inconsistencies (Section 5.1.2).

| Variable | Treatment | Mean | St Dev | Min. | 25th | Med. | 75th | Max | %diff |
|---|---|---|---|---|---|---|---|---|---|
| Detection | AO | 0.37 | 0.09 | 0.23 | 0.29 | 0.35 | 0.46 | 0.54 | 43.24 |
| | OO | 0.53 | 0.11 | 0.38 | 0.42 | 0.5 | 0.67 | 0.69 | |
| Effort | AO | 5.28 | 1.67 | 4 | 4.08 | 4.22 | 7 | 7.8 | 19.69 |
| | OO | 6.32 | 1.57 | 4.33 | 5.06 | 6.08 | 7.71 | 8.65 | |
| MisR | AO | 0.51 | 0.07 | 0.38 | 0.45 | 0.52 | 0.57 | 0.58 | 37.25 |
| | OO | 0.7 | 0.07 | 0.62 | 0.64 | 0.69 | 0.77 | 0.81 | |

St Dev: standard deviation, diff: difference

Table 25: Descriptive statistics

*Hypothesis Testing.* We check whether this result is statistically significant by trying to reject the first null hypothesis $H_{1-0}$ in the five experimental tasks (Table 26). Since the Shapiro-Wilk and Kolmogorov-Smirnov normality tests (Devore et al., 1999) suggest that the data are normally distributed, the paired t-test was applied to test $H_1$. This strategy allowed us to realize a pairwise comparison of the distributions and check if there exists a significant difference between AO and OO models with respect to detection rate. Pairwise p-values and mean differences across pairs for each measure are reported in (Table 26). The mean differences between pairs of AO and OO models indicate the direction in which the result is significant. For example, considering the varying detection rate for AO and OO models, the mean difference is negative (-0.16); in addition, the p-value (0.015) is less than 0.05, our selected level of significance. This implies that the detection rate in AO models was statistically lower than in OO models. Given this unexpected result, we were encouraged to apply the non-parametric Wilcoxon test to eliminate any threats to statistical conclusion validity. The low value of the p-value collected (0.031) also confirmed the aforementioned conclusion. Hence, there is sufficient evidence to reject the null hypothesis, and conclude that there is a difference between the detection rates in AO and OO models at the 0.05 level of significance.

| Variables | Treatment | Paired t-test | | | Wilcoxon |
| | | $t$ | $p$-value | Mean Difference | $p$-value |
|---|---|---|---|---|---|
| Detection | AO | 4.03 | 0.015 | - 0.16 | 0.031 |
| | OO | | | | |
| Effort | AO | 3.1 | 0.036 | - 1.48 | 0.033 |
| | OO | | | | |
| MisR | AO | 2.94 | 0.042 | - 0.192 | 0.029 |
| | OO | | | | |

*with 4 degree of freedom, a significance level of $\alpha = 0.05$

Table 26: Hypotheses testing

## 5.3.2.
## RQ4.2: Detection Effort in AO and OO models

*Descriptive Statistics.* The second research question investigates the effort that developers must invest to detect inconsistencies in AO and OO models. The gathered data in Table 25 indicate that developers spend more effort to detect inconsistencies in OO models than AO models. The mean of detection effort is 5.28 (minutes) in AO models and 6.32 in OO models. This comprises a representative increase of 19.69 percent against plain UML models. This lower effort on the use of AOM is also observed comparing the medians. The detection effort ranges from 4.22 (minutes) in AO models to 6.08 in OO models, which represents an increase of 44.07 percent in the latter case. This difference suggests that users of AOM tend to realize faster that: (i) a particular multi-view conflict exists, and (ii) such a conflict will compromise the implementation of the intended design. This phenomenon would confirm our initial intuition that the superior modularity of AO models accelerates inconsistency detection. In fact, during the interviews, the subjects (18) reported that the manifestation of inconsistencies in crosscutting relations is an influential factor on the conflict detection. According to them, such inconsistencies are perceived more quickly than other non-AOM inconsistencies. They noticed they were keener to match and contrast the structural and behavioral information governing the crosscut relations. Therefore, developers often report conflicting crosscutting relations as the reason for not progressing towards the implementation. This implies that although developers detect fewer inconsistencies in AO models, they spend less effort to localize them.

*Hypothesis Testing.* We also check if the finding above is statistically significant as follows. The Shapiro-Wilk and Kolmogorov-Smirnov certified the normal distribution of the measure (Devore et al., 1999). Therefore, the paired t-test was also applied to test H2 and evaluate RQ4.2. Table 26 shows the pairwise p-values and mean differences across pairs for each measure. Recall that the mean differences between pairs of AO and OO models indicate the direction in which the result is significant. The detection effort in AO and OO groups presented a negative value for the mean difference (-1.48), while p-value (0.036) is less than 0.05. The non-parametric Wilcoxon was also applied, which confirmed the above results given the p-value equal to 0.033. This enables us to infer that the average difference for detection effort between AO and OO models is not zero and that there is significant evidence that AO models required lower detection effort than in the OO counterparts.

### 5.3.3.
### RQ4.3: Misinterpretation Rate in AO and OO models

*Descriptive Statistics.* The third research question investigates whether AO models lead to a higher or lower misinterpretation rate than OO models. Table 25 shows the descriptive statistics to the misinterpretation measures of AO and OO models. Recall that MisR varies between zero and one and that MisR = 1 indicates that developers did not have misinterpretation. On the other hand, MisR = 0 indicates that the developers' answers spread equally over the four different alternatives, which represent the most serious misinterpretations. The data revealed that the use of in OO models led to less misinterpretation (higher MisR value) than AO models. The misinterpretation rate was 37.25 percent lower in OO models; the mean was 0.51 in AO groups against 0.7 in OO groups. This upward trend was also observed in the medians: 0.52 in AO models against 0.68 in OO models, comprising an increase of 32.69 percent. The results suggest that the presence of inconsistencies in AO models entails a higher detrimental impact on model interpretation by developers than in OO models. Our initial expectation that by using contradicting AO design models would lead the number of diverging interpretations of the participants was confirmed. During the interviews and examining the answer sheets, the subjects (22) reported that the need to scan all join points affected by the aspects increased the likelihood of different interpretations by developers.

*Hypothesis Testing.* We analyze the strength of the result testing H3 as follows. As in the previous analysis, the paired t-test was applied to test H3 as the measures assumed a normal distribution. Table 26 shows the pairwise p-values and mean differences across pairs for each measure. As the mean difference is negative (-0.192) and p-value (0.042) is less than 0.05, we can conjecture that there is significant evidence that the number of diverging interpretations in AO models is statistically higher than in OO models. We also applied the non-parametric Wilcoxon test (Devore et al., 1999) to check this conclusion. The p-value (0.029) also assumed a low value ($p < 0.05$). Therefore, as the p-value is less than 0.05 and the mean difference is negative, we can conclude that: there is evidence that the MisR in AO models is significantly lower than in OO models. Therefore, we reject the null hypothesis $H_{3-0}$.

## 5.4.
## Discussion

This section highlights particular characteristics of the design modeling languages that more influenced the dependent variables. The answer sheets, interviews, and observational study were instrumental in this investigation. We have identified four main outstanding findings, which are described as follows.

*Higher Aspect Quantification and Lower Inconsistency Detection.* First, aspects with higher quantification (Filman & Friedman, 2000) harmed inconsistency detection (RQ4.1) and the model interpretation (RQ4.3) by developers. We observed that when an aspect had six crosscutting relationships (see Table 24) and, therefore, affected multiple join points (11, in this case), the subjects spend more time performing global reasoning. The analysis of several aspect effects in the structural diagrams made developers often to neglect the analysis of behavioral interactions at each specific join point in the behavioral diagrams. According to the interviewees, this effect distracts away developers from observing possible inconsistencies between the structural and behavioral views. This finding is also confirmed by complementary data analyses. We observed, for example, that the inconsistency detection rate in OO models was 71 percent higher than in AO models when the latter were composed of aspects with high quantification; in these circumstances, the mean in OO models was 0.65 compared to 0.38 in AO models. An explanation for this phenomenon can be derived from the interviews and the observational study. We noticed that 20 subjects explicitly reported that they felt distracted by the presence of high density of crosscutting relationships in the models.

*Overlapping Information about Crosscutting Relationships.* Conversely, we observed that the subjects tended to detect more quickly inconsistencies in AO models when the scope of aspect pointcuts was narrow. In these cases, developers invested effort in only confronting structural and behavioral information about the crosscutting relations. According to the subjects, they could observe inconsistencies more quickly in AO models because structural diagrams often express the type of an advice (i.e., before, after or around), which is also a behavioral information that is present in the sequence diagram. Then, they could easily identify inconsistencies between: (i) the types of advices in the class diagram, and (ii) when a particular message was being advised by the aspect in the sequence diagram.

*Crosscutting Relationships and Diverging Mental Models of the "Big Picture."* Data analysis seems to suggest that uniform interpretation of AO models by different developers is harder to achieve than in OO models. According to the comments from the subjects, they often faced difficulties to create a "big picture" view from the conflicting class and sequence diagrams. This view represents a mental model reflecting how software developers perceive the problem, think about it, and solve it by producing the expected code from the diagrams. This understanding shapes the actions of the developers and defines the approach to guide the design realization in the code. In particular, the developers apparently had diverging mental models when the model inconsistencies were sourced in the crosscutting relationships. In these cases, developers came up with very different solutions for realizing crosscutting relationships in the code. They provided different answers on which and when the advice should affect the base model elements. Consequently, the communication from designers to programmers seems to be more sensitive to inconsistencies in aspect-oriented models.

*The Level of Model Detail Matters.* Given the presence of inconsistencies in the diagrams, developers usually consider the sequence diagrams as the basis for the design implementation. Note that in this case the developers do not report the presence of inconsistency. This phenomenon can be explained based on some reasons observed during the interviews and the observational study. First, sequence diagrams often present a higher number of details than the class diagrams. Thus, the lower level of abstraction leads the software developers to be more confident to the behavioral diagrams than structural diagrams. Next, sequence diagrams are closer to the final implementation; hence, developers become confident that the information present in the sequence diagram is the correct one compared to the class diagram. As a result, it means that when models are used to guide the implementation of design decisions, inconsistencies in behavioral diagrams have a superior detrimental effect than those in class diagrams.

This finding is useful for improving quality assurance procedures in some activities in model-driven software development as, for example, model review.

Model review is a well-known, effective way to minimize defects in code. Nevertheless, it is not clear for developers what diagram should be reviewed at first. By using this finding, developers can put the focus on the behavioral diagrams rather than the structural diagrams. In practice, this information is important because the preference of the behavioral diagrams can result in action that is more effective. Since model review requires some considerable effort to examine and define the focus of the analysis, it usually receives some criticism. By using this finding, developers can also tame or improve this problem.

*Identifying Fewer Inconsistencies in Less Time.* Developers identify fewer inconsistencies in AOM than in OOM. However, they spend less effort to detect it in AOM. Note that when developers identify an inconsistency, they have two options: they report that they detected an inconsistency or try to overcome the problem based on their experience, but will give a wrong answer at the end. Based on this, we have observed that developers report more often the presence of inconsistency in AO models (compared to OO models) than try finding any other solution. On the other hand, by using OO models developers try answering the question even observing the presence of inconsistency.

During the interviews, it was possible to observe the main reason why developers stop in AOM and go ahead in OOM: inconsistencies in AOM cause more severe doubts to developers than in OOM. Hence, developers do not feel comfortable using their experience to overcome the inconsistency problems given the problem at hand. It is important to point out that the subjects identify fewer inconsistencies in AOM not because they spent less time but because it is seen as a "wicked problem." In doing so, we observed that the subjects are more afraid of dealing with problems in AO models rather than OO models. Finally, given that multi-view design models usually have inconsistencies (Lange et al., 2004), this can mean that classical UML extensions for AOM (Section 5.1.1) need to be carefully employed. The observed results of our study suggest that developers might insert more defects into code. This can be motivated for two reasons: (1) low inconsistency detection (Section 5.3.1), and (2) high disagreement on design interpretations (Section 5.3.3).

## 5.5.
## Limitations of Related Work

Aspect-oriented modeling supports early separation of otherwise crosscutting concerns in software design. Concerns are separated to improve, for example, the interpretation of design decisions governing crosscutting concerns by developers before the implementation is accomplished. In practice, AOM will be considered useful compared to traditional modeling techniques if the claimed improved modularity actually leads to practical benefits, such as reduction of inconsistency detection effort and misinterpretations. Unfortunately, it is well known, as previously mentioned, that empirical studies of AOM are rare in the current literature, which confirms that it is still in the craftsmanship era (France & Rumpe, 2007).

Research has been mainly carried out in two areas: (1) defining new AOM techniques, and (2) proposing new weaving mechanisms. First, several authors have proposed new modeling languages, focusing on the definition of constructs, such as <<aspect>> and <<crosscut>>. These constructs represent concepts of aspect-orientation as UML-based extensions (Clarke & Walker, 2005; Chavez & Lucena, 2002; Aldawud et al., 2003; Stein & Hanenberg, 2002). In addition, (Clarke and Baniassad, 2005) make use of UML templates to specify aspect models. The chief motivation of some works is to provide a systematic method for weaving aspect and base models e.g., (Whittle et al., 2010; Klein et al., 2006; Jézéquel, 2008). Klein (Klein et, al, 2006) presents a semantic-based aspect weaving algorithm for hierarchical message sequence charts (HMSC). They use a set of transformations to weave an initial HMSC and a behavioral aspect expressed with scenarios. Moreover, the algorithm takes into account the compositional semantics of HMSCs.

Most of empirical studies on aspect-orientation are performed at the code level. For example, Hanenberg (Hanenberg et al., 2009) compares the time invested by developers to implement crosscutting concerns using object-oriented and aspect-oriented programming techniques. Other studies focus on the assessment of aspect-oriented programming under different perspectives, such as stability (Ferrari et al., 2010; Greenwood et al., 2007) and fault-proneness (Lasavio et al., 2009; Burrows et al., 2010). However, empirical studies of AOM (such as (Farias et al., 2010a)) have not been conducted, in particular in the context of modeling inconsistencies (or defects). Only the literature on OO modeling does highlight that empirical studies have been done on identifying defects in design models (Lange & Chaudron, 2004). Lange (Lange & Chaudron, 2006a) investigates the effects of defects in UML models. The two central contributions were: (1) the description of the effects of undetected defects in the interpretation of UML models, and (2) the finding that developers usually detect more certain kinds of defects than others do.

In conclusion, there are two critical gaps in the current understanding about AOM: (1) the lack of practical knowledge about the developers' effort to localize inconsistencies, and (2) the lack of empirical evidence about the detection rate and misinterpretations when understanding AO models.

## 5.6.
## Threats to Validity

*Internal validity.* Inferences between our independent variable and the dependent variables are internally valid if a causal relation involving these two variables is demonstrated (Wohlin et al., 2000). Our study met the internal validity because: (1) the temporal precedence criterion was met; (2) the covariation was observed, i.e., the dependent variables varied accordingly, when the independent changed; and (3) there is no clear extra cause for the detected covariation. Our study satisfied all these three requirements for internal validity.

*External validity.* It refers to the validity of the obtained results in other broader contexts (Wohlin et al., 2000). Thus, we analyzed whether the causal relationships investigated in this study could be held over variations in people, treatments, and other settings. Some characteristics were identified that strongly contributed for this purpose. First, the subjects used: (1) a practical AOM technique to perform the tasks; and (2) the design models were fragments of real-world models. Second, the reported controlled experiment was rigorously performed, in particular, when compared to previously reported controlled experiments (Lange et al., 2006; Ricca et al., 2010).

*Construct Validity.* It concerns the degree to which inferences are warranted from the observed cause and effect operations included in our study to the constructs that these instances might represent. All variables of this study were quantified using a suite of effort metrics or indicators that were previously defined and independently validated in experiments of inconsistency detection (Lange, 2007). Moreover, the concept of effort used in our study is well known in the literature (Jorgensen, 2005; Menzies et al., 2006; Grimstad & Jorgensen, 2007; Jorgensen et al., 2008) and its quantification method was reused from previous work (Lange & Chaudron, 2006a). Therefore, we are confident that the quantification method used is correct, and the quantification was accurately performed.

*Statistical Conclusion Validity.* We evaluated the statistical conclusion validity checking if the independent and dependent variables were submitted to suitable statistical methods. Experimental guidelines were followed to eliminate this threat (Wohlin et al., 2000): (1) the assumptions of the statistical tests (paired t-test and Wilcoxon) were not violated; (2) collected datasets were normally distributed; (3) the homogeneity of the subjects' background was assured; (4) the quantification method was properly applied; and (5) statistical methods were used. The Kolmogorov-Smirnov and Shapiro-Wilk tests (Devore et al., 1999) were used to check how likely the collected sample was normally distributed.

## 5.7.
## Concluding Remarks

This study reports an empirical investigation about the impact of alternative design decompositions on the inconsistency detection rate, the effort to detect inconsistencies, and the misinterpretation rate. We observed that developers detected fewer inconsistencies in AO decompositions than OO decompositions. The reason is that they got more distracted by the global reasoning motivated by the presence of crosscutting relations and overlooked the negative effects of existing model inconsistencies. According to the subjects, complex-crosscutting collaborations between modules led developers to unconsciously make assumptions that are more implicit about the correct design decisions. Consequently, aspects with higher quantification were the cause of the low detection rate of inconsistencies. Second, developers spent less effort using AO models to detect each inconsistency than in OO models. This was mainly due to the higher degree of overlapping information in structural and behavioral views of AOM. Third, the developers presented a superior rate of misinterpretation in AO models mostly thanks to the additional number of modeling concepts (e.g., crosscut relationships and aspects). They also had to examine all join points affected by the aspects. This extra analysis increased the degree of disagreement by developers while interpreting AO models and producing the code. It is important to highlight that all the aforementioned findings were independent of inconsistencies being assessed.

We should point out that empirical studies in AOM are in its initial stage and there is very little practical knowledge that can be used to determine the effectiveness of the current AOM approaches on improving design understanding. This study represents the first controlled experiment that addresses this issue. Although we are confident that the collected results are very concrete, significant results, further empirical studies are still required to test the hypotheses in other contexts. This is essential to better understand whether the results of this study hold (or not) in a broader context. In further studies, some questions should be investigated: what will it be the impact of quantification on the misinterpretation rate? Which will inconsistencies cause a higher misinterpretation rate? What is the effort to repair AO models with elevated quantification rate? Will we collect the same results by using larger design models? Finally, we hope that the issues outlined throughout the Chapter encourage researchers to replicate our study in the future under different circumstances.

# 6
# Effort on the Resolution of Inconsistency

The goal of this Chapter is to evaluate the effects of model stability and design modeling language on the inconsistency resolution effort. For this, two studies are realized. The first study (Section 6.1) is an exploratory study that analyzes and reports the effects of model stability on the effort required to resolve inconsistencies, and its impact on the inconsistency rate. These inconsistencies emerged when three well-known composition algorithms (such as *override*, *merge,* and *union*) were applied in evolution scenarios of three software product lines. The results, supported by statistical tests, show that model stability was an effective indicator of severe inconsistencies and high resolution effort of inconsistency.

The second exploratory study (Section 6.2) reports the impact of modeling language on the inconsistency rate and the resolution effort. More specifically, it investigates whether aspect-orientation reduces the resolution effort as improved modularization may help developers to better restructure the model. Similar to the previous study, it uses model composition to express the evolution of design models along six releases of a software product line. The composition algorithms (i.e., override, merge, and union algorithms) were also applied. The AO and non-AO composed models produced were compared in terms of their inconsistency rate and effort to solve the identified inconsistencies. The findings reveal specific scenarios where aspect-orientation properties, such as obliviousness and quantification, result in a lower (or higher) resolution effort.

## 6.1.
## Effect of Model Stability on Inconsistency Resolution

As previously mentioned, the composition of design models can be defined as a set of activities that should be performed over two input models, $M_A$ and $M_B$, in order to produce an output intended model, $M_{AB}$. To put the model composition in practice, software developers usually make use of composition heuristics (Clarke, 2001) to produce $M_{AB}$. These heuristics match the model elements of $M_A$ and $M_B$ by automatically "guessing" their semantics and then bring the similar elements together to create a "big picture" view of the overall design model.

The problem is that, in practice, the output composed model ($M_{CM}$) and the intended model ($M_{AB}$) often do not match (i.e., $M_{CM} \neq M_{AB}$). Since, $M_A$ and MB conflict with each other in some way, producing some syntactic and semantics inconsistencies in $M_{CM}$. Consequently, software developers should be able to anticipate composed models that are likely to exhibit inconsistencies and transform them into $M_{AB}$. In fact, it is well known that the derivation of $M_{AB}$ from $M_{CM}$ is considered an error-prone task (France & Rumpe, 2007). The developers do not even have practical information or guidance to plan this task. Their inability is due to two main problems.

First, developers do not have any indicator pointing which $M_{CM}$ should be reviewed (or not), given a sequence of output composed models produced by the software development team. Hence, they have no means to identify or prioritize parts of design models that are likely to have a higher density of inconsistencies. They are often forced to go through all output models produced or assume an overoptimistic position i.e., all output composed models produced is a $M_{AB}$. In both cases, the inadequate identification of an inconsistent $M_{CM}$ can even compromise the evolution of the existing design model ($M_A$) as some composition inconsistencies can affect further model compositions.

Second, model managers are unable to grasp how much effort the derivation of $M_{AB}$ from $M_{CM}$ can demand, given the problem at hand (Norris & Letkman, 2011). Hence, they end up not designating the most qualified developers for resolving the most critical effort-consuming cases where severe semantic inconsistencies are commonly found. Instead, unqualified developers end up being allocated to deal with these cases. In short, model managers have no idea about which $M_{CM}$ will demand more effort to be transformed into a $M_{AB}$. If the effort to resolve these inconsistencies is high, then the potential benefits of using composition heuristics (e.g., gains in productivity) may be compromised.

The literature in software evolution highlights that software remaining stable over time tends to have a lower number of flaws and require less effort to be fixed than its counterpart (Kelly, 2006; Molesini et al., 2009). However, little is known whether the benefits of stability are also found in the context of the evolution of design models supported by composition heuristics. This is by no means obvious for us because the software artifacts (code and models) have different level of abstraction and are characterized by alternative features. In fact, design model has a set of characteristics (defined in language metamodel expressing it) that are manipulated by composition heuristics and can assume values close to what it is expected (or not) i.e., $M_{CM} \approx M_{AB}$. If the assigned value to a characteristic is close to one found in the intended model, then the composed model is considered stable concerning that characteristic. For example, if the difference between the coupling of the composed model and the intended model is small, then they can be considered stable considering coupling.

Although researchers recognize software stability as a good indicator to address the two problems described above in the context of software evolution, most of the current research on model composition is focused on building new model composition heuristics (e.g., (Clarke & Walker, 2001; Kompose, 2010; Nejati et al., 2007). That is, little has been done to evaluate stability as an indicator of the presence of semantic inconsistencies and of the effort that, on average, developers should spend to derive $M_{AB}$ from $M_{CM}$. Today, the identification of critical $M_{CM}$ and the effort estimation to produce $M_{AB}$ are based on the evangelists' feedback that often diverge (Mens, 2002).

This section, therefore, presents an initial exploratory study analyzing stability as an indicator of composition inconsistencies and resolution effort. More specifically, we are concerned with understanding the effects of the model stability on the inconsistency rate and inconsistency resolution effort. We study a particular facet of model composition: the use of model composition when adding new features to a set of models for three realistic software product lines. Software product lines (SPLs) commonly involve model composition activities (Jayaraman et al., 2007; Thaker et al., 2007; Apel et al., 2009) and, while we believe the kinds of model composition in SPLs are representative of the broader issues, we make no claims about the generality of our results beyond SPL model composition. Three well-established composition heuristics (Clarke & Walker, 2001), namely override, merge and union, were employed to evolve the SPL design models along eighteen releases. SPLs are chosen because designers need to maximize the modularization of features allowing the specification of the compositions. The use of composition is required to accommodate new variabilities and variants (mandatory and optional features) that may be required when SPLs evolve. That is, in each new release, models for the new feature are composed with the models for the existing features. We analyze if stability is a good indicator of high inconsistency rate and resolution effort.

Our findings are derived from 180 compositions performed to evolve design models of three software product lines. Our results, supported by statistical tests, show that stable models tend to manifest a lower inconsistency rate and require a lower resolution effort than their counterparts. In other words, this means that there is significant evidence that the higher the model stability, the lower the model composition effort.

In addition, we discuss scenarios where the use of the composition heuristics became either costly or prohibitive. In these scenarios, developers need to invest some extra effort to derive $M_{AB}$ from $M_{CM}$. Additionally, we discuss the main factors that contributed to the stable models outnumber the unstable one in terms of inconsistency rate and inconsistency resolution effort. For example, our findings show that the highest inconsistency rates are observed when severe evolution scenarios are implemented, and when inconsistency propagation happens from model elements implementing optional features to ones implementing mandatory features. We also notice that the higher instability in the model elements of the SPL design models realizing optional features, the higher the resolution effort. To the best of our knowledge, our results are the first to investigate the potential advantages of model stability in realistic scenarios of model composition. We therefore see this study as a first step in a more ambitious agenda to empirically assess model stability.

The remainder of the chapter is organized as follows. Section 6.1.1 describes the main concepts and knowledge that are going to be used and discussed throughout the Chapter. Section 6.1.2 presents the study methodology. Section 6.1.3 discusses the study results. Section 6.1.4 compares this work with others, presenting the main differences and commonalities. Section 6.1.5 highlights some threats to validity. Finally, Section 6.1.6 presents some concluding remarks and future work.

### 6.1.1.
### Background

This Section presents the fundamental concepts to a correct understanding of the contributions presented in this Chapter. To this end, the concepts of model stability, composition heuristics, and model inconsistency will be discussed.

### 6.1.1.1.
### Model Stability

According to (Kelly, 2006), a design characteristic of software is stable if, when compared to other, the differences in the metric associated with that characteristic are regarded small. In a similar way in the context of model composition, $M_{CM}$ can be considered stable if its design characteristics have a low variation concerning the characteristics of $M_{AB}$. In (Kelly, 2006), Kelly studies stability from a retrospective view i.e., comparing the current version to previous ones. In our study, we compare the current model and the intended model.

We define low variation as being equal to (or less than) 20 percent. This choice is based on previous empirical studies (Kelly, 2006 on software stability that has demonstrated the usefulness of this threshold. For example, if the measure of a particular characteristic (e.g., coupling and cohesion) of the $M_{CM}$ is equal to 9, and the measure of the $M_{AB}$ is equal to 11. So $M_{CM}$ is considered stable concerning $M_{AB}$ (because 9 is 18% lower than 11) with respect to the measure under analysis. Following this stability threshold, we can systematically identify weather (or not) $M_{CM}$ keeps stable considering $M_{AB}$, given an evolution scenario. Note that threshold is used more as a reference value rather than a final decision maker. The results of this study can regulate it, for example. The differences between the models are computed from the comparison of measures of each model characteristic calculated with a suite of metrics described in Chapter 3 and Table 27.

We adopt the definition of stability from (Kelly, 2006) (and its threshold) due to some reasons. First, it defines and validates the quantification method of stability in practice. This method is used to examine software systems that have been actively maintained and used over a long term. Second, the quantification method of stability has demonstrated to be effective to flag evolutions that have jeopardized the system design.

Third, many releases of the system under study were considered. This is a fundamental requirement to test the usefulness of the method. As such, all these factors provided a solid foundation for our study. These metrics were used because previous works (Farias et al., 2008a; Medeiros et al., 2010; Guimarães et al., 2010; Kelly, 2006; Farias, 2011) have already observed the effectiveness of these indicators for the quantification of software stability. Knowing the stability in relation to the intended model it is possible to identify evolution scenarios, where composition heuristics are able to accommodate upcoming changes effectively and the effort spent to obtain the intended model. The stability quantification method is presented later in Section 6.1.2.4.

| Type | Metric | Description |
|---|---|---|
| Size | NClass | The number of classes |
| | NAttr | The number of attributes |
| | NOps | The number of operations |
| | NInter | The number of interfaces |
| | NOI | The number of operations in each interface |
| Inheritance | DIT | The depth of the class in the inheritance hierarchy. |
| | InhOps | The number of operations inherited. |
| | InhAttr | The number of attributes  inherited. |
| Coupling | DepOut | The number of elements on which a class depends. |
| | DepIn | The number of elements that depend on this class. |

Table 27: Metrics used

## 6.1.1.2.
## Composition Heuristics

As previously mentioned in Section 2.4, composition heuristics rely on two key activities: *matching* and combining the input model elements (Farias et al., 2010a; Farias et al., 2010b; Clarke, 2001, Reddy et al., 2006). Usually they are used to *modify*, *remove*, and *add* features to an existing design model. This work focuses on three state-of-practice composition heuristics: override, merge, and union (Clarke & Walker, 2001; Clarke & Walker, 2005). These heuristics were chosen because they have been applied to a wide range of model composition scenarios such as model evolution, ontology merge, and conceptual model composition. In addition, they have been recognized as effective heuristics in evolving product-line architectures e.g., (Farias et al., 2010a). In the following, we briefly define these three heuristics, and assume $M_A$ and $M_B$ as the input two models. The input model elements are corresponding if they can be identified as equivalent in a matching process. Matching can be achieved using any kind of standard heuristics, such as *match-by-name* (Oliveira et al., 2009a; Oliveira et al., 2009b; Reddy et al., 2005).

The design models used are typical UML class and component diagrams, which have been widely used to represent software architecture in mainstream software development (Ambler, 2005; Fowler, 2003; Dennis et al., 2007; Lüders et al., 2000). In Figure 17, for example, *R2* diagram plays the role of the base model ($M_A$) and *Delta(R2,R3)* diagram plays the role of the delta model ($M_B$). The components *R2.BaseController* and *Delta(R2,R3).BaseController* are considered as equivalent. We defer further considerations about the design models used in our study in Section 6.1.2.3. The composition heuristics considered in our study were override, merge, and union. These heuristics were previously discussed in Section 2.4.1. Figure 17 shows two input models and two composed models produced following the override and merge heuristics, respectively. Figure 18 shows the intended model and the composed model produced following the union heuristic.

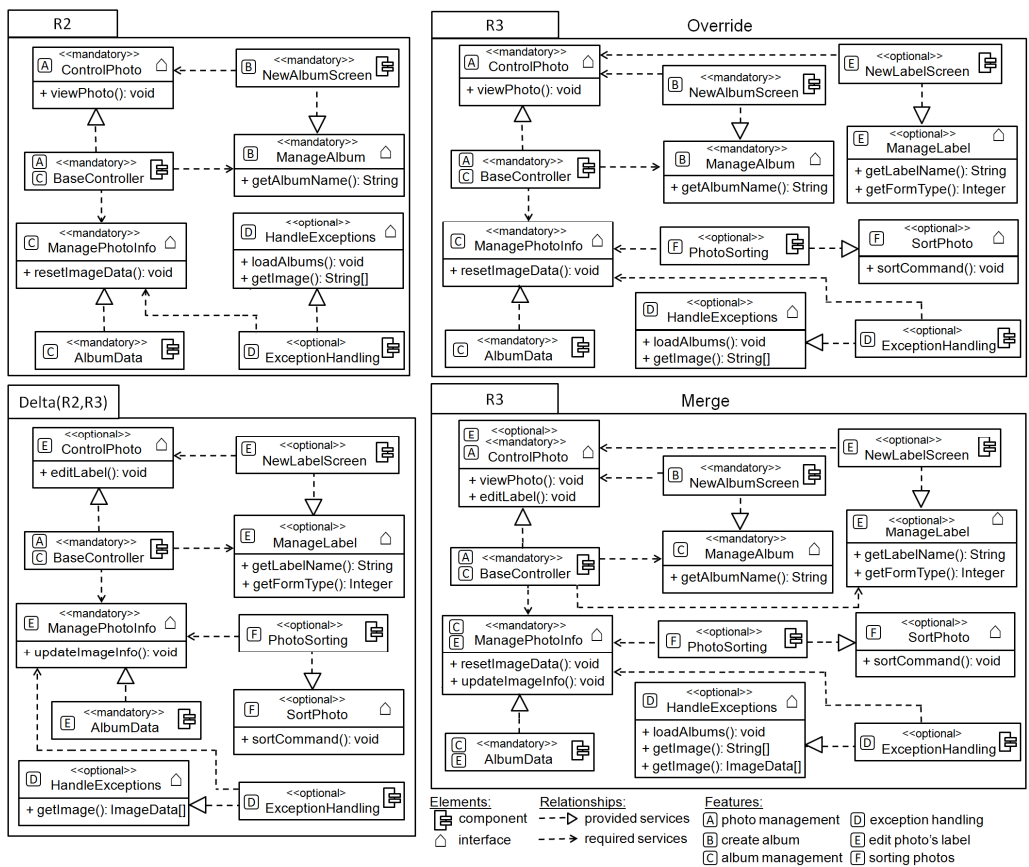

Figure 17: Example of composition of the Mobile Media product line

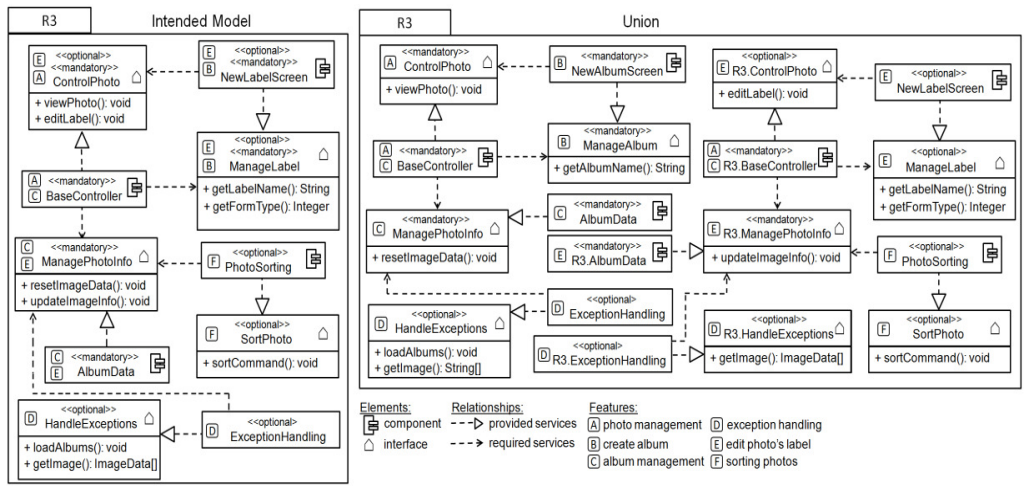

Figure 18: The intended and composed model produced following the union heuristic

### 6.1.1.3.
### Model Inconsistency

Inconsistencies emerge in the composed model when its properties assume values other than those expected, as previously defined in Section 3. These values can affect the syntactic and semantic properties of the model elements. Usually the undesired values come from conflicting changes that were incorrectly realized (Samar et al., 2011). We can identify two broad categories of inconsistencies: (i) *syntactic inconsistencies*, which arise when the composed model elements do not conform to the modeling language's metamodel; and (ii) *semantic inconsistencies*, which mean that static and behavioral semantics of the composed model elements do not match those of the intended model elements.

In our study, we take into account syntactic inconsistencies that were identified by the IBM Rational Software Architecture's model validation mechanism (IBM RSA, 2011). For example, this robust tool is able to detect the violation of well-formedness rules defined in the UML metamodel specification (OMG, 2011). In order to improve our inconsistency analysis, we also considered the types of inconsistencies shown in Table 28, which were checked by using the SDMetrics tool (Wust, 2011). In particular, these inconsistencies were used because their effectiveness has been demonstrated in previous works (Farias et al., 2008a; Farias et al., 2010a; Farias et al., 2012d). In addition, both syntactic and semantic inconsistencies were manually identified as well. All these procedures were followed in order to improve our confidence that a representative set of inconsistencies were tackled by our study.

Many instances of these inconsistency types (Table 28) were found in our study. For example, the static property of a model element, *isAbstract*, assumes the value *true* rather than *false*. The result is an abstract class where a concrete class was being expected. Another typical inconsistency considered in our study was when a model element provides (or requires) an unexpected functionality or even requires a functionality that does not exist.

The absence of this functionality can affect other design model elements responsible for implementing other functionalities, thereby propagating an undesirable ripple effect in the resulting composed model. For example, the *AlbumData* does not provide the service "Update Image Information" because the method *updateImageInfo():void* is not present in the *ManagePhotoInfoInterface*. Hence, the *PhotoSorting* component is unable to provide the service "*SortingPhotos.*" This means that the feature "*SortingPhoto*" (feature 'F' in Figure 17) – a critical feature of the software product line – is not correctly realized. On the other hand, this problem is not present in Figure 17 (merge), in which the *AlbumData* implement two features (C, model management, and E, edit photo's label). We defer further discussion about the examples and the quantification of these types of inconsistencies to Section 6.1.2.4.

| Metric | Description |
|---|---|
| NFCon | The number of functionality inconsistencies. |
| NCCon | The number of model elements that are not compliance with the intended model. |
| NDRCOn | The number of dangling reference inconsistencies. |
| NASCon | The number of abstract syntax inconsistencies. |
| NUMECon | The number of non-meaningful model elements |
| NBFCon | The number of behavioral feature inconsistencies. |

Table 28: The inconsistencies used in our case study

## 6.1.2.
## Study Methodology

This section presents the main decisions underlying the experimental design of our exploratory study. To begin with, the objective and research questions are presented (Section 6.1.2.1). Next, the study hypotheses are systematically stated from these research questions (Section 6.1.2.2). The product lines used in our studies are also discussed in detail as well as their evolutionary changes (Section 6.1.2.3). Then, the variables and quantification methods considered are precisely described (Section 6.1.2.4). Finally, the method used to produce the releases of the target architectures is carefully discussed (Section 6.1.2.5). All these methodological steps were based on practical guidelines of empirical studies (Wohlin et al., 2000; Basili, 2007; Kitchenham et al., 2008; Kitchenham, 2006; Shadish et al., 2006).

### 6.1.2.1.
### Objective and Research Questions

This study essentially attempts to evaluate the effects of model stability on two variables: the *inconsistency rate* and *inconsistency resolution effort*. These effects are investigated from concrete scenarios involving design model compositions so that practical knowledge can be generated. In addition, some influential factors are also considered into precisely revealing how they can affect these variables. With this in mind, the objective of this study is stated based on the GQM template (Basili, 1994) as follows:

*analyze* the stability of design models

*for the purpose of* investigating its effect

*with respect to* inconsistency rate and resolution effort

*from the perspective of* developers

*in the context of* evolving design models with composition heuristics

In particular, this study aims at revealing the stability effects while evolving composed design models on inconsistency rate and the inconsistency resolution effort. Therefore, we address research question RQ4, as stated in Section 1.3:

- **RQ4:** What is the impact of design characteristics on the inconsistency rate and inconsistency resolution effort?

Considering the quality notions defined in Chapter 3, we study whether the syntactic and semantic quality notions of a model affects the effort and resolution quality notions. We refine the research question into two research questions. Thus, we focus on the following two research questions:

- **RQ4.1:** What is the effect of stability on the *inconsistency rate*?

- **RQ4.2:** What is the effect of stability on the *developers' effort*?

### 6.1.2.2.
### Hypothesis Formulation

*First Hypotheses: Effect of Stability on Inconsistency Rate (RQ5.1).* In the first hypothesis, we speculate that a high variation of the design characteristics of the design models may lead to a higher incidence of inconsistencies; since, it increases the chance for an incorrect manipulation of the design characteristic by the composition heuristics. In fact, modifications from severe evolutions may lead the composition heuristics to be ineffective or even prohibitive. In addition, these inconsistencies may also propagate. As a higher incidence of changes is found in unstable models, we hypothesize that unstable models tend to have a higher (or equal to) inconsistency rate than stable models. The first hypothesis evaluates whether the inconsistency rate in unstable models is significantly higher (or equal to) than in stable models. Thus, our hypotheses are summarized as follows:

**Null Hypothesis 1, $H_{1-0}$:**

Stable design models have similar or higher inconsistency rate than unstable design models.

$H_{1-0}$: Rate(stable design models) $\geq$ Rate(unstable design models).

**Alternative Hypothesis 1, $H_{1-1}$:**

Stable design models have a lower inconsistency rate than unstable design models.

$H_{1-1}$: Rate(stable design models) < Rate(unstable design models)

By testing the first hypothesis, we evaluate if stability is a good indicator to identify the most critical $M_{CM}$ in term of inconsistency rate from a sequence of $M_{CM}$ produced from multiple software development teams. Hence, developers can then review the design models having a higher density of composition inconsistencies. We believe that this strategy is a more effective one than going through all $M_{CM}$ produced or assuming an overoptimistic position where all $M_{CM}$ produced is a $M_{AB}$.

*Second Hypothesis: Effect of Stability on Developer Effort (RQ5.2).* As previously mentioned, developers tend to invest different quantity of effort to derive $M_{AB}$ from $M_{CM}$. Today, model managers are unable to grasp how much effort this transformation can demand. This variation is because developers need to resolve different types of problems in a composed model, from a simple renaming of elements to complex modifications in the structure of the composed model. In fact, the structure of the composed models may be affected in different ways during the composition e.g., creating unexpected interdependences between the model elements. Even worse, these modifications in the structure of the model may cause ripple effects i.e., inconsistency propagation between the model elements. The introduction of one inconsistency can often lead to multiple other inconsistencies because of a "knock-on" effect. An example would be the inconsistency whereby a client component is missing an important operation in the interface of a server component. This semantic inconsistency leads to a "knock-on" syntactic inconsistency if another component requires the operation. In the worst case, there may be long chains of inconsistencies all derived from a single inconsistency. Given a composed model at hand, developers need to know if they will invest little or too much effort to transform $M_{CM}$ into $M_{AB}$, given the problem at hand. Based on this knowledge, they will be able to prioritize the review of the output composed models and to better comprehend the effort to be invested e.g., reviewing the models that require higher effort first and those requiring less effort after. With this in mind, we are interested in understanding the possible difference of effort to resolve inconsistencies in stable and unstable design models. The expectation is that stable models require a lower developers' effort to produce the output intended model. This expectation is based on the speculation that unstable models may demand more restructuring modifications than stable models; hence, requiring more effort. This leads to the second null and alternative hypotheses as follows:

**Null Hypothesis 2, $H_{2-0}$:**

Stable models require similar or higher effort to resolve inconsistencies than unstable models.

**$H_{2-0}$:** Effort(stable models) $\geq$ Effort(unstable models).

**Alternative Hypothesis 2, $H_{2-1}$:**

Stable models tend to require a lower inconsistency resolution effort than unstable ones.

**$H_{2-1}$:** Effort(stable models) $<$ Effort(unstable models).

By testing the first hypothesis, we evaluate if stability is a useful indicator to identify the most critical effort-consuming cases in which severe semantic inconsistencies in architectural components are more often. This knowledge helps model mangers to allocate qualified developers to overcome the composition inconsistencies in $M_{CM}$.

### 6.1.2.3.
### Target Cases: Evolving Product-Line Design Models

*Model Composition for Expressing SPL Evolution.* We have applied the composition heuristics to evolve design models of three realistic SPLs for a set of evolution scenarios (Table 29). That is, the compositions are defined to generate the new releases of the SPL design models. These three SPLs are described below and soon after the evolution scenarios are presented. The first target case is a product-line called MobileMedia, whose purpose is to support the manipulation of photos, music, and videos on mobile devices. The last release of its design model consists of a UML component diagram with more than 50 component elements. Figure 17 and Figure 18 show a practical example of the use of composition to evolve this SPL.

The second SPL, called Shogi Game, is a board game, whose purpose is to allow users to move, customize pieces, save, and load game. All the movements of the pieces are governed by a set of well-defined rules. The last SPL, called Checkers Game, is a board game played on an eight by eight-squared board with twelve pieces on each side. The purpose of Checkers is to essentially move and capture diagonally forwards.

The reason for selecting these SPLs in our evaluation is manifold. Firstly, the models are well designed. Next, 12 releases of Mobile Media's architectural models were produced by independent developers using the model composition heuristics. These releases are produced from five evolution scenarios. Note that an evolution is the production of a release from another one e.g., from R1 to R2 (Table 28). In addition, 12 releases of Shogi's and Checkers' architectural models were available as well. In both cases, six releases were produced from five evolution scenarios. Together the 36 releases provide a wide range of SPL evolution scenarios to enable us to investigate our hypotheses properly. These 36 releases were produced from the evolution scenarios described in Table 29. Secondly, these releases were available for our investigation and had a considerable quantity of structural changes in the evolution scenarios.

Another reason to choose these SPLs is that the original developers are available to help us to validate the identified list of syntactic and semantic inconsistencies. In total, eight developers worked during the development of the SPLs used in our study being three developers from the Lancaster University (UK), two from the Pontifical Catholic University of Rio de Janeiro (Brazil), two from University of São Paulo (Brazil), one from Federal University of Pernambuco (Brazil). These are fundamental requirements to test our hypotheses in a reliable fashion. Moreover, each SPL has more than one hundred modules and their architecture models are the main artifact to reason about change requests and derive new products. The SPL designs were produced by the original developers without any of the model composition heuristics under assessment in mind. It helped to avoid any bias and entailed natural software development scenarios. . In total, eight developers worked during the development of the SPLs used in our study being three developers from the Lancaster University (UK), two from the Pontifical Catholic University of Rio de Janeiro (Brazil), two from University of São Paulo (Brazil), and one from Federal University of Pernambuco (Brazil).

|  | Release | Description |
|---|---|---|
| **Mobile Media** | R1 | MobilePhoto core (Figueiredo et al, 2008) |
|  | R2 | Exception handling included |
|  | R3 | New feature added to count the number of times a photo has been viewed and sorting photos by highest viewing frequency.<br>New feature added to edit the photo's label |
|  | R4 | New feature added to allow users to specify and view their favorite photos |
|  | R5 | New feature to keep multiple copies of photos |
|  | R6 | New feature to send photo to other users by SMS |
| **Checkers Game** | R1 | Checkers Game core |
|  | R2 | New feature to indicate the movable pieces |
|  | R3 | New feature to indicate possible movements |
|  | R4 | New feature to save and load the game |
|  | R5 | New feature added to customize the pieces |
|  | R6 | New feature added to log the game |
| **Shogi Game** | R1 | Shogi Game core |
|  | R2 | New feature to customize pictures |
|  | R3 | New feature to customize pieces |
|  | R4 | New feature to indicate the piece movement |
|  | R5 | New feature to indicate the movable pieces |
|  | R6 | New feature to allow the users to save and load the game |

Table 29: Descriptions of the evolution scenarios

Finally, these SPLs have a number of other relevant characteristics for our study, such as: (i) proper documentation of the driving requirements; and (ii) different types of changes were realized in each release, including refinements over time of the architecture style employed. After describing the SPLs employed in our empirical studies, the evolution scenarios suffered by them are explained in Table 29.

### 6.1.2.4.
### Measured Variables and Quantification Method

*First Dependent Variable.* The dependent variable of hypothesis 1 is the inconsistency rate. It quantifies the amount of composition inconsistencies divided by the total number of elements in the composed model. That is, it allows computing the density of composition inconsistencies in the output composed models. This metric makes it possible to assess the difference between the inconsistency rate of stable models and unstable models (H1). It is important to point out that inconsistency rate is defined from multiple inconsistency metrics (Oliveira, 2008a).

*Second Dependent Variable.* The dependent variable of the hypothesis 2 is the inconsistency resolution effort, $g(M_{CM})$—that is, the number of operations (creations, removals, and updates) required to transform the composed model into the intended model. We compute these operations because they represent the main operations performed by developer to evolve software in real-world settings (Mens, 2002). Thus, this computation represents an estimation of the inconsistency resolution effort. The collected measures of inconsistency rate are used to assess if the composed model has inconsistencies after the composition heuristic is applied ($diff(M_{CM},M_{AB}) > 0$). Then, a set of removals, updates, and creations were performed to resolve the inconsistencies. As a result, the intended model is produced and the inconsistency resolution effort is computed.

*Independent Variable.* The independent variable of the hypotheses 1 and 2 is the Stability (S) of the output composed model ($M_{CM}$) with respect to the output intended model ($M_{AB}$). The Stability is defined in terms of the Distance (D) between the measures of the design characteristics of $M_{CM}$ and $M_{AB}$.

$$\text{Distance}(x, y) = \frac{|Metric(x) - Metric(y)|}{Metric(y)} \tag{1}$$

Where:

*Metric* are the indicators defined in Table 1

*X* is the output composed model, $M_{CM}$

Y is the output intended model, $M_{AB}$

Table 27 defines the metrics used to quantify the design characteristics of the models, while Formula 1 shows how the Distance is computed. The Stability can assume two possible values: 1, indicating that $M_{CM}$ and $M_{AB}$ are *stable*, and 0, indicating that $M_{CM}$ and $M_{AB}$ are *unstable*. $M_{CM}$ is stable concerning $M_{AB}$ if the distance between $M_{CM}$ and $M_{AB}$ (considering a particular design characteristic) assumes a value equal (or lower than) to 0.2. That is, if $0 \leq \text{Distance}(M_{CM}, M_{AB}) \leq 0.2$), then Stability($M_{CM}, M_{AB}$) = 0. On the other hand, $M_{CM}$ is *unstable* if the distance between $M_{CM}$ and $M_{AB}$ (regarding a specific design characteristic) assumes a value higher than 0.2. That is, if Distance($M_{CM}, M_{AB}$) > 0.2), then Stability($M_{CM}, M_{AB}$) = 0. We use this threshold to point out the most severe unstable models. For example, we check if architectural problems happen even in cases where the output composed models are considered stable. In addition, we also analyze the models that are closer to the threshold. Formula 2 shows how the measure Stability is computed.

$$Stability(x, y) = \begin{cases} 1, if\ 0 \leq Distance(x, y) \leq 0.2 \\ 0, if\ Distance(x, y) > 0.2 \end{cases} \tag{2}$$

For example, $M_{CM}$ and $M_{AB}$ have the number of classes equals to 8 and 10, respectively (i.e., NClass = 8 and NClass = 10). To check the stability of $M_{CM}$ regarding this metric, we calculate the distance between $M_{CM}$ and $M_{AB}$ considering the metric NClass as described below.

$$\text{Distance}(M_{CM}, M_{AB}) = \frac{|NClass(M_{CM}) - NClass(M_{AB})|}{NClass(M_{AB})} = \frac{|8 - 10|}{10} = 0.2$$

As the Distance($M_{CM}$,$M_{AB}$) is equal to 0.2, then we can consider that $M_{CM}$ is equal to 1. Therefore, $M_{CM}$ is stable considering $M_{AB}$ in terms of the number of classes. Elaborating on the previous example, we can now consider two design characteristics: the number of classes (NClass), the afferent coupling (DepOut), and the number of attributes (NAttr). Assuming DepOut($M_{CM}$) = 12, DepOut($M_{AB}$) = 14, NAttr($M_{CM}$) = 6, and NAttr($M_{AB}$) = 7, the Distance is calculated as follows.

$$Distance(M_{CM}, M_{AB}) = \frac{|DepOut(M_{CM}) - DepOut(M_{AB})|}{DepOut(M_{AB})} = \frac{|12 - 14|}{14} = 0.14$$

$$Distance(M_{CM}, M_{AB}) = \frac{|NAttr(M_{CM}) - NAttr(M_{AB})|}{NAttr(M_{AB})} = \frac{|7 - 9|}{9} = 0.22$$

Therefore, $M_{CM}$ is stable concerning $M_{AB}$ in terms of NClass and DepOut. However, $M_{CM}$ is unstable in terms of NAttr. In this example, we evaluate the stability of $M_{CM}$ considering three design characteristics, which was stable in two cases. As developers can consider various design characteristics to determine the stability of the $M_{CM}$, we define the Formula 3 that calculates the overall stability of $M_{CM}$ with respect to $M_{AB}$. Refining the previous example, we evaluate the stability of $M_{CM}$ considering two additional design characteristics: the number of interfaces (NInter) and the depth of the class in the inheritance hierarchy (DIT). Supposing that NInter($M_{CM}$) = 15, NInter($M_{AB}$) = 17, DIT($M_{CM}$) = 11, and DIT($M_{AB}$) = 13, the Distance is calculated as follows.

$$Distance(M_{CM}, M_{AB}) = \frac{|NInter(M_{CM}) - NInter(M_{AB})|}{NInter(M_{AB})} = \frac{|15 - 17|}{17} = 0.11$$

$$Distance(M_{CM}, M_{AB}) = \frac{|DIT(M_{CM}) - DIT(M_{AB})|}{DIT(M_{AB})} = \frac{|11 - 13|}{13} = 0.15$$

In both cases, $M_{CM}$ is stable as 0.11 and 0.15 are $\geq 0$ and $\leq 0.2$. Investigating this overall stability, we are able to understand how far the measures of the design characteristics of $M_{CM}$ in relation to $M_{AB}$ are. The overall stability of $M_{CM}$ in terms of NClass, DepOut, NAttr, NInter, and DIT is calculated as follows. As the overall stability is equal to 0.2, we can consider that $M_{CM}$ is stable considering $M_{AB}$.

$$Stability(x,y)_{overall} = 1 - \frac{\sum_{k=0}^{j-1}(Stability_k)}{j}$$

(3)

*Legend:*

$j$: number of metrics used (e.g., 10 metrics in case of Table 1)

$$Stability(x,y)_{overall} = 1 - \frac{\sum_{k=0}^{4}(Stability(x,y))}{5}$$

$$\sum_{k=0}^{4}\big(Stability(x,y)\big) = \frac{|NClass(M_{CM}) - NClass(M_{AB})|}{NClass(M_{AB})}$$

$$+ \frac{|DepOut(M_{CM}) - DepOut(M_{AB})|}{DepOut(M_{AB})} + \frac{|NAttr(M_{CM}) - NAttr(M_{AB})|}{NAttr(M_{AB})}$$

$$+ \frac{|NInter(M_{CM}) - NInter(M_{AB})|}{NInter(M_{AB})} + \frac{|DIT(M_{CM}) - DIT(M_{AB})|}{DIT(M_{AB})}$$

$$= 0.2 + 0.14 + 0.22 + 0.11 + 0.11 \qquad \text{(applying the Formula 2)}$$

$$= \quad 1 \quad + \quad 1 \quad + \quad 0 \quad + \quad 1 \quad + \quad 1 \quad = \quad 4$$

Then,

$$Stability(x,y)_{overall} = 1 - \frac{4}{5} = 1 - 0.8 = 0.2$$

## 6.1.2.5.
## Evaluation Procedures

### a. *Target Model Versions and Releases*

To test the study hypotheses, we have used the releases described in Table 29. Our key concern is to investigate these hypotheses considering a larger number of realistic SPL releases as possible in order to avoid bias of specific evolution scenarios.

*Deriving SPL Model Releases.* For each release of the three product-line architectures, we have applied each of the composition heuristics (override, merge, and union) to compose two input models in order to produce a new release model. That is, each release was produced using the three algorithms. Similar compositions were performed using the override, merge, and union heuristics to help us to identify scenarios where the SPL design models succumb (or not). For example, to produce the release 3 (R3) of the Mobile Media, the developers combine R3 with a delta model that represents the model elements that should be inserted into R3 in order to transform it into R4. For this, the developers use the composition heuristics described previously. A practical example about how these models are produced can be seen in Figure 17 and Figure 18.

*Model Releases and Composition Specification.* The releases in Table 29 were in particular selected because visible and structural modifications in the architectural design were carried out to add new features. For each new release, the previous release was changed in order to accommodate the new features. To implement a new evolution scenario, a composition heuristic can remove, add, or update the entities present in the previous model release. During the design of all releases, a main concern was to maximize good modeling practices in addition to the design-for-change principles. For example, assume that the mean of the coupling measure of $M_{CM}$ and $M_{AB}$ is equal to 9 and 11, respectively. So $M_{CM}$ is stable regarding $M_{AB}$ (because 9 is 18% lower than 11). Following this stability threshold, we can systematically identify if the $M_{CM}$ keeps stable over the evolution scenarios.

## b. Execution and Analysis Phases

*Model Definition Stage.* This step is a pivotal activity to define the input models and to express the model evolution as a model composition. The evolution has two models: the base model, $M_A$, the current release, and the delta model, $M_B$, which represents the changes that should be inserted into $M_A$ to transform it into $M_{CM}$, as previously discussed. Considering the product-line design models used in the case studies, $M_B$ represents the new design elements realizing the new feature. Then, a composition relationship is specified between $M_A$ and $M_B$ so that the composed model can be produced, $M_{CM}$.

*Composition and Measurement Stage.* In total, 180 compositions were performed, being 60 in the Mobile Media, 60 in the Shogi Game and 60 in the Checkers Game. The compositions were performed manually using the IBM RSA (IBM RSA, 2011; Norris & Letkeman, 2011). The result of this phase was a document of composition descriptions, including the gathered data from the application of our metrics suite and all design models created. We used a well-validated suite of inconsistency metrics applied in previous work (Oliveira et al., 2008; Farias et al., 2010a; Farias et al., 2010b; Medeiros et al., 2010; Guimaraes et al., 2010; Farias, 2011a, Farias et al., 2011b) focused on quantifying syntactic and semantic inconsistencies. The syntactic inconsistencies were quantified using the IBM RSA's model validation mechanism. The semantic inconsistencies were quantified using the SDMetrics tool (Wust, 2011). In addition, we also check both syntactic and semantic inconsistencies manually because some metrics e.g., "the number of non-meaningful model elements" depend on the meaning of the model elements and the current modeling tools are unable to compute this metric.

The identification of the inconsistencies was performed in three review cycles in order to avoid false positives and false negatives. We also consulted the developers as needed, such as checking and confirming specific cases of semantic inconsistencies. On the other hand, the well-formedness (syntactic and semantic) rules defined in the UML metamodel were automatically checked by the IBM RAS's model validation mechanism.

*Effort Assessment Stage.* The goal of the third phase was to assess the effort to resolve the inconsistencies using the quantification method described in Section 6.1.2.4. The composition heuristics were used to generate the evolved models, so that we could evaluate the effect of stability on the model composition effort. In order to support a detailed data analysis, the assessment phase was further decomposed in two main stages. The first stage is concerned with pinpointing the inconsistency rates produced by the compositions (H1). The second stage aims at assessing the effort to resolve a set of previously identified inconsistencies (H2). All measurement results and the raw data are available in Appendix A.

### 6.1.3.
### Results

This section reports and analyzes the data set obtained from the experimental procedures described in the previous section. The findings of this work are derived from both the numerical processing of this data set and the graphical representation of interesting aspects of the gathered results. Then, Section 7.1.3.1 elaborates on the gathered data in order to test the first hypothesis (H1). Lastly, Section 7.1.3.2 discusses the collected data related to the second hypothesis (H2).

### 6.1.3.1.
### H1: Stability and Inconsistency Rate

#### c. *Descriptive Statistics*

This section describes aspects of the collected data with respect to the impact of stability on the inconsistency rate. For this, descriptive statistics are carefully computed and discussed. The understanding of these statistics is a key step to know the data distribution and grasp the main trends. To go about this direction, not only the main trend was calculated using the two most used statistics to discover trends (mean and median); the dispersion of the data around them was also computed mainly making use of the standard deviation. Note that these statistics are calculated from 180 composition scenarios i.e., with 60 compositions applied to the evolution of MobileMedia SPL, 60 compositions applied to the Shogi SPL, and 60 compositions applied to the Checkers SPL. From this bunch of evolution scenarios, we are confident that the collected data are representative to be analyzed using descriptive statistics.

Table 30 shows descriptive statistics about the collected data regarding inconsistency rate. Figure 19 depicts the box-plot of the collected data. By having carried out a thorough analysis of this statistic, we can observe the positive effects of high level of stability on the inconsistency rate. In fact, we observed only harmful effects in the absence of stability. The main outstanding finding is that inconsistency rate in stable design model is lower than in unstable design model. This result is supported by some observations described as follows

| Variables | Groups | N | Min | 25th | Median | 75th | Max | St. Dev. |
|-----------|--------|---|-----|------|--------|------|-----|----------|
| Inconsistency Rate | Stable | 78 | 0 | 0.11 | 0.31 | 0.78 | 3.86 | 0.84 |
| | Unstable | 102 | 0.17 | 1.64 | 3.86 | 6.88 | 9.21 | 2.63 |

N: number of composed models, St. Dev.: Standard Deviation

Table 30: Descriptive statistics of the inconsistency rate

First, the median of inconsistency rate in stable models is considerably lower than in unstable models. That is, a mean of 0.31 in relation to the intended model instead of 3.86 presented by unstable models. This means, for example, that stable SPL models present no inconsistencies in some cases. On the other hand, unstable models probably hold a higher inconsistency rate than that presented by stable models. This comprises normally 3.86 inconsistencies in relation to the intended model. This implies, for example, that if the output composed model is unstable, then there is a high probability of having inconsistencies in these models.

Stable models have a favorable impact on the inconsistency rate. More importantly, its absence has harmful consequences for the number of inconsistencies. These negative effects are evidenced by the significant difference between the number of inconsistencies in stable and unstable models. If, for example, one SPL developer has to work with an unstable model, then he or she will certainly have to handle 91.9 percent more inconsistencies, compared the medians 0.31 (stable) and 3.86 (unstable). In fact, stable models tend to have just 8.1 percent of the inconsistencies that are found in unstable models, compared the medians 0.31 (stable) and 3.86 (unstable). One of the main reasons is because *inconsistency propagations* are found in unstable models more frequently. This means that developers must check all model elements so that they can identify and manipulate the composed model so that the intended model can be obtained.

Another interesting finding is that the inconsistencies tend to be quite close to the central tendency in stable models, with a standard deviation equals to 0.84. On the other hand, in unstable models these inconsistencies tend to spread out over a large range of values. This is represented by a high value of the standard deviation that is equal to 2.63. It is important to point out that to draw out valid conclusions from the collected data it is necessary to analyze and possibly remove outliers from the data.

Outliers are extreme values assumed by the inconsistency measures that may influence the study's conclusions. To analyze the threat of these outliers to the collected data, we made use of box-plots. According to (Wohlin et al., 2000; Basili, 2007), it is necessary to verify whether the outliers are caused by an extraordinary exception (unlikely to happen again), or whether the cause of the outlier can be expected to happen again. Considering the first case, the outliers must be removed, and in the latter, they should not be removed. In our study, some outliers were identified; however, they were not extraordinary exceptions since they could happen again. Consequently, they were left in the collected data set as they do not affect the results.

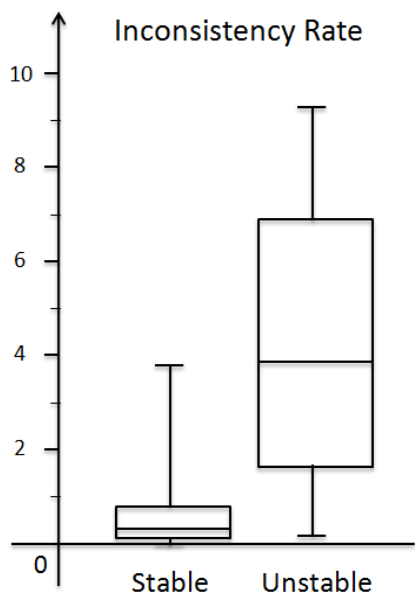

Figure 19: Box-plot of inconsistencies

### d. Hypothesis Testing

We performed a statistical test to evaluate whether in fact the difference between the inconsistency rates of stable and unstable models are *statistically significant*. As we hypothesize that stable models tend to exert a lower inconsistency rate than unstable models, the test of the mean difference between stable and unstable groups will be performed as one-tailed test. In the analyses, we considered significance level at 0.05 level (p ≤ 0.05) to indicate a true significance.

*Mann-whitney test.* As the collected data violated the assumption of normality, the non-parametric Mann-Whitney test was used as the main statistical test. The results produced are U' = 7.21, U = 744, z = 9.33 and p < 0.001. The *p-value* is lower than z and 0.05. Therefore, the null hypothesis of no difference between the rates of inconsistency in stable and unstable models ($H_{1-0}$) can be rejected. That is, there is sufficient evidence to say that the difference between the inconsistency rates of stable and unstable models are statically significant. Table 31 depicts that the mean rank of inconsistency rate for unstable models  are higher than that of stable models. As Mann-Whitney test (Wohlin, 2000) relies on ranking scores from lowest to highest, the group with the lowest mean rank is the one that contains the largest amount of lower inconsistency rate. Likewise, the group with the highest mean rank is the group that contains the largest amount of higher inconsistency rate. Hence, the collected data confirm that unstable models tend to have a higher inconsistency rate than the stable design models.

| Variable | Groups | N | Mean Rank | Rank Sum | SC | t-value* | *p* |
|----------|--------|---|-----------|----------|-----|----------|-----|
| Resolution effort | Stable | 78 | 46,99 | 3665 | - 0,698 | - 13 | < 0.001 |
| | Unstable | 102 | 123,77 | 12625 | | | |

*with 178 degree of freedom, SC: Spearman's Correlation

Table 31: Mann-whitney test and Spearman's correlation analysis

*Correlation.* To examine the strength of the relationship (the correlation coefficient) between stability and inconsistency rate, the Spearman's correlation (SC) test was applied (see Table 31). Pearson's correlation is not used because the data sets are not normally distributed. Note that this statistic test assumes that both variables are independent; i.e., is neither dependent on, causes nor influences the other. The correlation coefficient takes on values between -1 and 1. Values close to 1 or -1 indicate a strong relationship between the stability and inconsistency rate. A value close to zero indicates a weak or non-existent relationship.

As can be seen in Table 31, the t-test of significance of the relationship has a low *p-value*, indicating that the correlation is significantly different from zero. Spearman's correlation analysis resulted in a negative and significant correlation (SC = - 0.71). The negative value indicates an inverse relationship. That is, as one variable increases, the other decreases. Hence, composition inconsistencies tend to manifest more often in unstable models than stable models. The above correlation suggests that whereas the stability of product-line architectures decreases the inconsistency rate in their models increases.

Therefore, the results suggest that, on average, stable models have *significantly* lower inconsistency rate than unstable design models. Therefore, we are confident that the results confirm a strong indication of correlation between stability and inconsistency rate. Consequently, the null hypothesis ($H_{1-0}$) can be rejected and the alternative hypothesis ($H_{1-1}$) confirmed.

## e. Discussion

*The Effect of Severe Evolution Categories.* After discussing how the dataset is grouped, grasping the main trends, and studying the relevance of the outliers, the main conclusion is that stable models tend to present a lower inconsistency rate than unstable models. This finding can be seen as the first step to overcome the lack of practical knowledge about the effects of the model stability on the inconsistency rate in realistic scenarios of model evolution supported by composition heuristics. Some previous studies e.g., (Kelly, 2006; Kemerer & Slaughter, 1999; Eman et al., 2002; Perry, 1998; Berzins, 1994, Yang et al., 1992) also check similar insights on the code level. These studies report a positive association between low variation of coupling and size with stability.

We have noticed that although the input design models ($M_A$ and $M_B$) are well structured, they are the target of widely scoped inconsistencies in certain model composition scenarios. These widely scoped inconsistencies are motivated by unexpected modifications in specific design characteristics of the design models such as coupling and cohesion. These scenarios occurred mainly when composition heuristics accommodate unanticipated, severe changes from $M_A$ to $M_B$. The most complicate changes observed are those related to the refinement of the MVC (Model-View-Controller) architecture design of the SPLs used in this study.

Another observation is that the composition heuristics (override, merge, and union) are not effective to accommodate these changes from $M_A$ to $M_B$. The main reason is that the heuristics are unable to "restructure" the design models in such way that these changes do not harm static or behavioral aspects of the design models. These harmful changes usually emerge from a set of ever-present evolving change categories, such as *modification* of the model properties and *derivation* of new model elements (e.g., components or classes) from other existing ones.

In the first category, *modification*, model elements have some properties affected. This is typically the case when a new operation conflicts with an operation previously defined. In Figure 17 and Figure 18, for example, the operation *getImage()* in the interface *R2.HandleException* had its return type, *String[]*, conflicting with the return type, *ImageData[]* of the interface *Delta(R2,R3). HandleException*. Another example is the component *ManageAlbum* that had its name modified to *ManageLabel* to express semantic alterations in the concepts used to realize the error handling feature. Only one of the names and return types can be accepted, but the two modifications cannot be combined. Both cases are scenarios in which the heuristics are unable to correctly pick out what element must be renamed and what return type must be considered. The problem is that detection and decision of these inconsistencies demand a thorough understanding of: (i) what the design model elements actually mean as well as the domain terms "Album" and "Label"; and (ii) the expected semantics of the modified method. In addition, semantic information is typically not included in any formal way so that the heuristics can infer the most appropriated choice. Consequently, the new model elements responsible for implementing the added features are presented with overlapping semantic values and unexpected behaviors. Interestingly, this has been the case where existing optional as well as alternative features are involved in the change.

In the second category, *derivation*, the changes are a little more severe. Architectural elements are refined and/or moved in the model to accommodate the new changes. Differently from the previous category, the affected architectural elements are usually mandatory features because this kind of evolution in software product lines is mainly required to facilitate the additions of new variabilities or variants later in the project. Unfortunately, in this context of more widely scoped changes, the heuristic-based composition heuristics have demonstrated to be ineffective.

A concrete example of this inability in our target cases was the refinement of the MVC architecture style of the MobileMedia SPL in the third evolution scenario. In practical terms, the central architectural component, *BaseController*, was broken into other controllers such as *PhotoListController, AudioController, VideoController* and *LabelController* to support a better manipulation of the upcoming media like photo, audio, video and the label attached to them. This design rigidness to accommodate four new specific controllers (by refining the previous general one) contributed significantly to the instability of the output composed model. This is partially due to the name-based model comparison policy in the heuristics, which are unable to recognize more intricate equivalence relationships between the model elements. Indeed, this comparison strategy is very restrictive whenever there is a correspondence relationship 1:N between elements in the two input models. That is, it is unable to match the upcoming four controllers with the previous one, *BaseController*.

A practical example of this category of relationship (1:N) encompassed the required interface *ControlPhoto* (release 3) of the *AlbumListScreen* component. This interface was decomposed into two new required interfaces *ControlAlbum* and *ControlPhotoList* (release 4), thereby characterizing a relationship 1:2. For this particular case, the name-based model comparison should be able to "recognize" that *ControlAlbum* and *ControlPhotoList* are equivalent to *ControlPhoto*. However, in the output model (release 4), the *AlbumListScreen* component provides duplicate services to the environment giving rise to a severe inconsistency.

*Inconsistency Propagation.* After addressing the hypotheses and knowing that instabilities have a detrimental effect on the density of inconsistencies, we analyze whether the local where they arise (i.e., architectural elements realizing mandatory, alternative or optional features) can cause some unknown side effects. Some interesting findings were found, which is properly discussed as follows.

To begin with, instability problems are more harmful when they take place in design model elements realizing mandatory features. This can be explained by some reasons. First, the *inconsistency propagation* is often higher in the model elements implementing mandatory features than in alternative or optional features. When inconsistencies arise in elements realizing optional and alternative features they also tend to naturally cascade to elements realizing mandatory features. Consequently, the mandatory features end up being the target of inconsistency propagation. Based on the knowledge that mandatory features tend to be more vulnerable to ripple effects of inconsistencies, developers must structure product-line architectures in such a way that inconsistencies can keep precisely "confined" in the model elements where they appear. Otherwise, the quality of the products extracted from the SPL can be compromised as the core elements of the SPL can suffer from problems caused by incorrect feature compositions. The higher the number of inconsistencies, the higher the chance of them to continue in the same output model, even after an inspection process performed by a designer. Consequently, the extraction of certain products can become error-prone or even prohibitive.

The second interesting insight is that the higher the instability in alternative and optional features, the higher the inconsistency propagation to mandatory features. However, the propagation in the inverse order (i.e., from alternative and optional to mandatory features) seems to be less common. In Figure 17 (override), a practical example can be seen. The instability in mandatory features, *Album and Photo Management*, compromises the optional feature, *Edit Photo's Label*. The *NewLabelScreen* component (optional feature) has its two services i.e., *getLabelName()* and *getFormType()* (specified in the interface *ManageLabel*) compromised. The reason is that the required service *editLabel()* cannot be provided by the *BaseController* (mandatory feature). Thus, the "edit photo' label" feature can no longer be provided due to problems in the mandatory feature "album and photo management."

For example, in the fourth evolution scenario of the Checkers Game, the optional feature, *Customize Pieces*, is correctly glued to the R4 using the *override* heuristic so that the new release, R5, can be generated. The problem is that the inconsistencies emerging in the architectural component, *Command,* are propagated to the architectural elements *CustomizePieces* and *GameManager*. Thus, the mandatory feature "piece management" implemented by the Command is affecting the optional feature "customize pieces" implemented by the components *CustomizePieces* and *GameManager.* Although the optional feature, *Customize Pieces*, has been correctly attached to the base architecture, the composed models will not have the expected functionality related to the customization of pieces.

### 6.1.3.2.
### H2: Stability and Resolution Effort

#### a. Descriptive Statistics

This section discusses interesting aspects of the collected data concerning the impact of stability on the developers' effort. The knowledge derived from them helps to understand the effects of model stability on the inconsistency resolution effort. In a similar way to the previous section, we calculate the main trend and the data dispersion. Table 32 provides the descriptive statistics of sampled inconsistency resolution effort in stable and unstable model groups. Figure 20 graphically depicts the collected data by using box-plot. To begin with our discussion, we first compare the median values of the inconsistency resolution effort of the both stable and unstable groups. We can observe that the median of the stable models (equals to 6) is much lower than that one of unstable models (equals to 111).

| Variables | Groups | N | Min | 25th | Median | 75th | Max | St. Dev. |
|---|---|---|---|---|---|---|---|---|
| Resolution effort | Stable | 78 | 0 | 3,50 | 6 | 13 | 46 | 10.29 |
| | Unstable | 102 | 4 | 27 | 111 | 229.25 | 368 | 106.7 |

N: number of composed models, St. Dev.: Standard Deviation

Table 32: Descriptive statistics of the resolution effort

This superiority of the unstable models is also observed in the mean and standard deviation, which represent the main trend and dispersion measures, respectively. The gathered results, therefore, indicate that stable models claim less resolution effort than unstable models. This means that developers tend to perform a lower amount of tasks (creations, removals, and modifications) to transform the composed model into the intended model. Although we have observed some outliers e.g., the maximum value (368) registered in unstable models, they are not an extraordinary exception as they could happen again. Consequently, they were left in the collected data set, as they do not tamper the results.

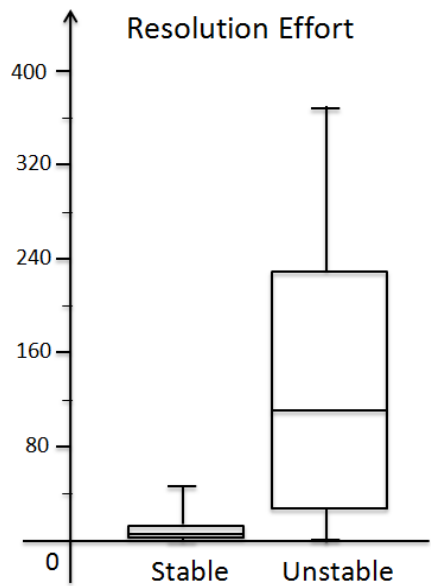

Figure 20: Box-plot of resolution effort in relation to the intended model

## b. Hypothesis Testing

Given the difference between the mean and median described in the descriptive statistical analysis, statistical tests are applied to assess whether in fact the difference in effort to fix unstable model and stable model is statistically significant. We conjecture that stable models tend to require a lower inconsistency resolution effort than unstable models. Hence, a one-tailed test is performed to test the significance of the mean difference between stable and unstable groups. Again, in the analyses we considered significance level at 0.05 level ($p \leq 0.05$) to indicate a true significance.

*Mann-Whitney test.* As the dataset does not respect the assumption of normality, we use the non-parametric Mann-Whitney test was used as the main statistical test as well as it was done in the first hypothesis. However, the Mann-Whitney test was only applied to the effort measures needed to transform the composed model into the intended model. The results of the Mann-Whitney test produced are U' = 7.372, U = 584, z = 9.79 and $p < 0.001$. The p-value is lower than *z* and 0.05, therefore, the null hypothesis can be rejected. In other words, there exists a difference between the efforts required to resolve inconsistencies in stable and unstable model groups. In fact, there is substantial evidence pointing out the difference between the median measures of the two groups.

Table 33 shows that the difference between the mean ranks is significant. The mean of rank in stable models consists of about 38 of the mean rank in unstable models. As the Mann-Whitney test relies on ranking scores from lowest to highest, the group with the lowest mean rank is the one that requires the highest incidence of lowest effort. Likewise, the group with the highest mean rank is the group that contains the largest occurrence of higher effort needed. Hence, the collected data show that unstable models that are not stable tend to have higher effort than the stable models.

| Variable | Groups | N | Mean Rank | Rank Sum | SC | t-value* | $p$ |
|---|---|---|---|---|---|---|---|
| Resolution effort | Stable | 78 | 46,99 | 3665 | - 0,698 | - 13 | < 0.001 |
| | Unstable | 102 | 123,77 | 12625 | | | |

*with 178 degree of freedom

Table 33: Mann-whitney test and Spearman's correlation analysis

*Correlation Analysis.* As the gathered data do not follow a normal distribution, we cannot apply the Pearson's correlation analysis. An alternative way was to apply the Spearman's correlation (SC) test to measure the strength of the linear relationship (the correlation coefficient) between stability and inconsistency resolution effort. Table 33 provides the results of the Spearman's correlation test. The low p-value $< 0.001$ indicates that the correlation significantly departs from zero. Remember that Spearman's correlation value close to 1 or -1 indicates a strong relationship between the stability and effort. On the other hand, a value close to 0 indicates a weak or non-existent relationship. The results (SC = - 0.698) suggest that there is a negative and significant correlation between the two variables. This implies that whereas the stability increases the effort to resolve inconsistency decreases.

Hence, stable models required much lesser effort to be transformed into the intended model than unstable models. Based on such results, we can reject the null hypothesis ($H_{2-0}$), and accept the alternative hypothesis ($H_{2-1}$): stable models tend to require lower effort to resolve composition inconsistency than unstable models.

## c. Discussion

*The Effect of Instability on Resolution Effort.* We have observed that the higher instability in optional and alternative features, the higher the resolution effort. This increased effort is due to instabilities in optional features cause inconsistencies in model elements implementing mandatory features. In practice, inconsistencies in architectural elements realizing optional features tend to affect the structure of model elements realizing mandatory features. The reason is that some relationships are (or not) introduced between architectural elements realizing mandatory and optional features during the composition. These undesired dependences favor the inconsistency propagation. Consequently, developers must invest some additional effort to resolve the inconsistencies. The effort is to restructure the composed model. That is, instability in optional features tends to jeopardize some properties of the architectural elements realizing the mandatory features, which requires some unexpected effort. That is, it is required to resolve a cascading chain of inconsistencies, and usually this process should be applied recursively until all inconsistencies have been resolved. This is typically the case scenario when inconsistencies of operations with earlier operation, the heuristic can therefore remove the earlier operation and add the new one, or vice-versa.

We have identified that this higher effort to resolve inconsistencies is due to the syntax-based composition heuristics being unable to deal with occurring semantic conflicts between the model elements of mandatory and optional features. As a result, inconsistencies are formed. In Figure 17, for example, the component *BaseController* requires services from a component *NewALbumScreen* that provides just one mandatory feature "create album" rather than from a component that provides two features: "create album" and "edit photo's label." This is because releases R2 and R3 use different component names (*R2.NewAlbumScreen* and *R3.NewLabelScreen*) for the same purpose. That is, they implement the mandatory feature *Create Album* in components with contracting names.

A syntax-based composition is unable to foresee these kinds of semantic inconsistencies, or even indicate any problem in *BaseController* as the component remains syntactically correct. From R2 to R3, the domain term *Album* was replaced by *Label*. However, the purely syntactical, match-by-name mechanism is unable to catch and incorporate this simple semantic change into the composition heuristic. To overcome this, a semantic-based approach would be required to allow, for example, a systematic semantic alignment between these two domain terms. Consequently, the heuristics would be able to properly match *R2.NewAlbumScreen* and *R3.NewLabelScreen.*

Still in Figure 17, the architectural model R3, which was produced following *merge heuristic*, contains a second facet of semantic problem: *behavioral inconsistency*. The component *ExceptionHandling* provides two services with the same purpose, *getImage():String[]* and *getImage():ImageData[]*. However, they have different semantic values. This contrasting characteristic is emphasized by the different return types, *String[]* and *ImageData[]*. However, in this case, the inconsistency got confined in the optional feature rather than propagating to model elements implementing mandatory features. To resolve the problem, the method *getImage():String[]* should be removed. In total, only one operation is performed. Thus, these inconsistencies can be only pinpointed by resorting to sophisticated semantics-based composition, which relies on the action semantics of the model elements. According to (Mens, 2002), the current detection of behavioral inconsistency is just based on complex mathematical, program slicing, and program dependence graphs. Unfortunately, none of them is able to systematically compare behavioral aspects of components neither realizing two features nor even composing them properly. Even worse, the composition techniques would be unable to match, for example, *ManageAlbum* and *ManageLabel* interface

*The Effect of Multiple Concerns on Resolution Effort.* Another finding is that the higher the number of features implemented by a model element, the higher the resolution effort. We have observed that model elements realizing multiple features tend to require more inconsistency resolution effort than those realizing just one feature. The reason is that the models elements realizing multiple features tend to receive a higher number of upcoming changes to-be accommodated by the composition heuristics than ones realizing a single feature. These model elements become more vulnerable to the unpredictable effects of the severe evolution categories. This means that developers tend to invest more effort to resolve all possible inconsistencies.

In fact, a higher number of inconsistencies has been observed in 'multiple-featured' components rather than in 'single-featured' components. As developers cannot foresee or even precisely identify all ripple effects of these inconsistencies through other model elements, the absence of stability can be used as a good indicator of inconsistency. Let us consider the *BaseController*, the central controller in MobileMedia architecture that implements two features (see Figure 17). The collected data show that the *BaseController* was modified in almost all evolution scenarios because it is a pivotal architectural component in the *model-view-control* architectural style of the SPL MobileMedia. Unfortunately, the changes cannot be properly realized in all cases. In addition, we observe that *BaseController*'s inconsistencies affect other four components, namely *NewLabelScreen*, *AlbumListScreen*, *PhotoListScreen*, *PhotoViewScreen*, and *AddPhotoToAlbumScreen*. All these affected components require the provided services by the BaseController.

Moreover, we notice that the *BaseController* had a higher likelihood to receive inconsistencies from other model elements than any other components. The reason is that it also depends on many other components to provide the services of the multiple features. For example, *BaseController* can be harmed by inconsistencies arising from the components *ManageAlbum, ManagePhotoInfo,* and *ControlPhoto*. This means that, at some point, *BaseController* can no longer provide its services because it was probably affected by inconsistencies located in these components.

It is interesting to note that *NewAlbumScreen* is also affected by an inconsistency that emerged from *AlbumData*, as it requires the service (*viewPhoto*) provided by the *BaseController* in the interface, *ControlPhoto* that cannot be accessed. The main reason is that the service, *resetImageData*(), specified in the interface *ManagePhotoInfo* can no longer be provided by the component *AlbumData*, compromising the serviced offered in the interface *ControlPhoto*. Since *BaseController* is not able to correctly provide all services defined in the provided interface *ControlPhoto*, it is also re-affected by an inconsistency that previously arose from it. This happens because *NewAlbumScreen* does not provide the services described in the interface *ManageAlbum*. This phenomenon represents cyclic *inconsistency propagation*. Understanding this type of phenomenon, designer can examine upfront and more precisely the design models in order to localize undetected cyclic dependence between the model elements.

Another striking observation is that optional features are also harmed by this propagation on the mandatory features. For example, the *PhotoSorting* component (realizing optional feature "sorting photos") is unable to provide the service, *sortCommand()*, specified in the interface *SoftPhoto*. This is due to the absence of the required service, *resetImageData()* from the *ManagePhotoInfo* interface, which the mandatory feature "album management." In practical terms, it indicates that undesired effects in features can be due to some unexpected instabilities in the mandatory features. In collaborative software development, for example, this is a typical problem because the model elements implementing different features are developed in parallel, but they rarely prepared upfront to-be composed. Hence, developers should invest some considerable effort to properly promote the composition.

### d. Some Additional Considerations

*Quantification Method.* We are aware that there are pros and cons in studying either an overall indicator or a single metric of design stability. In (Kelly, 2006), she defines a single metric of design stability and then uses this method as an indicator of good practices of design. This study is performed in retrospective i.e., analyzing software artifacts that evolved over a long term. On the other hand, this thesis has a different goal that is to evaluate whether the "most severe instabilities" may be related to model composition effort. We conjecture that the most severe instability can be identified considering a greater number of design characteristics. This will be also analyzed during the empirical studies.

If we consider only one single design characteristic, we will have at least two problems: (i) first, we will potentially ignore severe instabilities that affected other design characteristics, and (ii) second, we will end up artificially concluding those variations of a single characteristic (e.g., high number of methods or high number of attributes) always represents severe design instabilities. Then, we opted for following a strategy, commonly adopted nowadays e.g., (Marinescu, 2004; Lanza & Marinescu, 2006), to detect significant design problems through a combination of multiple measures rather than a single metric.

*Effectiveness of the Threshold.* As previously mentioned in Chapter 2, we have also analyzed whether the threshold defined in (Kelly, 2006) is also valid in the context of this study. To this end, we analyze whether the threshold (0.2) jeopardizes the results (or not). More specifically, we study whether small differences around a threshold of 0.2 can produce different results. After a careful analysis of the collected data, we conclude that our conjecture stated in Section 2.6.1 is confirmed. That is, the threshold of 0.2 was effective for the purpose of this study. The main reason is that the threshold did not harm the identification of severe cases of inconsistency rate and resolution effort. This can be confirmed by analyzing, for instance, the data in Table 30: the inconsistency rates of the stable group and instable group are significantly different considering the median (0.31 against 3.86); the same pattern of significant difference applies to the other cases (25th and 75th columns). Again, the same pattern is observed in Table 32 for resolution effort. This means that the threshold considered (0.2) can clearly separate the composed models into groups of stable and unstable models; since, their measures concentrate in the opposite extremes. This confirms that we are able to consistently implement our strategy of studying the impact of models with the most severe instabilities (i.e., ones where more than 20% of the design characteristics varied considerably) rather than analyzing the different degrees of instabilities.

### 6.1.4.
### Limitations of Related Work

To the best of our knowledge, our results are the first to empirically investigate the relation between quality notions and model composition effort in a broader context. In (Farias et al., 2011b), we initially investigated the research questions addressed in this Chapter, but they were evaluated in a smaller scope. This work, therefore, represents an extension of the results obtained previously. The main extensions can be described as follows: (1) two more case studies were performed i.e., the evolution studies with the Shogi and Checkers SPLs. This implies that the number of composition jumped from 60 to 180; (2) new lessons learned were obtained from a broader study; and (3) the size of the sample data was higher than the previously found; hence, the hypotheses might be better tested.

We have observed not only a wide variety of model composition techniques Nejati et al., 2007; Clarke, 2001; Reddy, et al., 2005; Lange & Chaudron, 2006a; OMG, 2011; Kompose, 2011; Norris & Letkeman, 2011; Whittle & Jayaraman, 2010; France et al., 2007; Fleury et al., 2007) have been created, but also some previous works (Farias et al., 2011b; Nagappan et al., 2010) have demonstrated that stability is a good predictor of defects (Nagappan et al., 2010) and the presence of good designs (Kelly, 2006). However, none of them has directly investigated the impact of stability on model composition effort.

The lack of empirical evidence hinders the understanding of the side effects peculiar to stability on developers' effort. Consequently, developers in industrial projects have to rely solely on feedback from experts to determine "the goodness" of the input models and their compositions. In fact, according to several recent observations the state of the practice in model quality assessment indicates that modeling is still in the craftsmanship era and this problem is even more accentuated in the context of model composition (France & Rumpe, 2007; Dingel et al., 2008; Farias et al., 2008; Molesini et al., 2009; Mens, 2002; Berzins, 1994; France et al., 2006; Dzidek et al., 2008).

The current model composition literature does not provide any support to perform empirical studies in model composition effort (France & Rumpe, 2007; Farias et al. 2010a), or even to evaluate the effects of model stability on composition effort. In (France & Rumpe, 2007), the authors highlight the need empirical studies in model composition to provide insights about how deal with ever-present problems such as conflicts and inconsistencies in real world settings. In (Mens, 2002), Mens also reveals the need of more "experimental researches on the validation and scalability of syntactic and semantic merge approaches, not only regarding conflict detection, but also regarding the amount of time and effort required to resolve the conflicts." Without empirical studies, researchers and developers are left without any insight about how to evaluate model composition in practice. For example, there is no metric, indicator, or criterion available to assess the UML models that are merged through, for instance, the UML built-in composition mechanism (i.e., package merge) (Dingel et al., 2008; OMG, 2011).

There are some specific metrics available in the literature for supporting the evaluation of model composition specifications. For instance, Chitchyan and colleagues (Chitchyan et al., 2009) have defined some metrics, such as scaffolding and mobility, to quantify quality attributes of compositions between two or more requirements artifacts. However, their metrics are targeted at evaluating the reusability and stability of explicit descriptions of model composition specifications. In other words, their work is not targeted at evaluating model composition heuristics. Boucke and colleagues (Boucke et al., 2006) also propose a number of metrics for evaluating the complexity and reuse of explicitly defined compositions of architectural models. Their work is not focused on heuristic-based model composition as well. Instead, we have focused on analyzing the impact of stability on the effort to resolve emerging inconsistencies in output models. Therefore, existing metrics (such as those described in (Lange & Chaudron, 2006a; Lange & Chaudron, 2006b; Nugroho et al., 2008)) cannot be directly applied to our context.

Although we have proposed a metric suite for quantifying inconsistencies in UML class diagrams (Farias et al., 2008a) and then applied these metrics to evaluate the composition of aspect-oriented models and UML class diagrams (Farias et al., 2010a), nothing has been done to understand the effects of model stability on the developers' effort. We therefore see this study as a first step in a more ambitious agenda to support empirically the assessment of model composition techniques in general.

Finally, some previous works investigate the effect of using UML diagrams and its profiles with different purposes. In (Briand et al., 2005), Briand looked into the formality of UML models and its relation with model quality and comprehensibility. In particular, Briand and colleagues investigated the impact of using OCL (Object Constraint Language (OMG, 2011)) on defect detection, comprehension, and impact analysis of changes in UML models. In (Ricca et al., 2010), Ricca carried out a series of four experiments to assess how developer´s experience and ability influence Web application comprehension tasks supported by UML stereotypes. Although they have found that the use of UML models provide real benefits for typical software engineering activities, none has investigated the peculiarities of UML models in the context of model composition.

## 6.1.5.
## Threats to Validity

Our exploratory study has obviously a number of threats to validity that range from internal, construct, statistical conclusion validity threats to external threats. This section discusses how these threats were minimized and offers suggestions for improvements in future study.

### 6.1.5.1.
### Internal Validity

Inferences between our independent variable (stability) and the dependent variables (inconsistency rate and composition effort) are internally valid if a causal relation involving these two variables is demonstrated (Brewer, 2000; Shadish et al., 2002). Our study met the internal validity because: (1) the *temporal precedence* criterion was met, i.e., the instability of design models preceded the inconsistencies and composition effort; (2) the covariation was observed, i.e., instability of design models varied accordingly to both inconsistencies and composition effort; and (3) there is no clear extra cause for the detected covariation. Our study satisfied all these three requirements for internal validity.

The internal validity can be also supported by other means. First, the detailed analysis of concrete examples demonstrating how the instabilities were constantly the main drivers of inconsistencies presented in this study. Second, our concerns throughout the study to make sure that the observed values in the inconsistency rates and composition effort were confidently caused by the stability of the design models. However, some threats were also identified, which are explicitly discussed below.

First, due to the exploratory nature of our study, we cannot state that the internal validity of our findings is comparable to the more explicit manipulation of independent variables in controlled experiments. This exceeding control employed to deal with some factors (i.e., with random selection, experimental groups, and safeguards against confounding factors) was not used because it would significantly jeopardize the external validity of the findings.

Second, another threat to the internal validity is related to the imperfections governing the measurements of inconsistency rate and resolution effort. As the measures were partially calculated in a manual fashion, there was the risk that collected data would not be always reliable. Hence, this could lead to inconsistent results. However, we have mitigated this risk by establishing measurement guidelines, two-round data reviews with the actual developers of the SPL design models, and by engaging them in discussions in cases of doubts related to, for instance, the semantic inconsistencies.

Next, usually the *confounding* variable is seen as the major threat to the internal validity (Shadish et al., 2002). That is, rather than just the independent variable, an unknown third variable unexpectedly affects the dependent variable. To avoid c*onfounding* variables in our study, a pilot study was carried out to make sure that the inconsistency rate and composition effort were not affected by any existing variable other than stability. During this pilot study, we tried to identify which other variables could affect the inconsistency rate and resolution effort such as the size of the models.

Another concern was to deal with the *experimenter bias.* That is, the experimenters inadvertently affect the results by unconsciously realizing experimental tasks differently that would be expected. To minimize the possibility of experimenter bias, the evaluation tasks were performed by developers, which that know neither the purpose of the study nor the variables involved. For example, developers created the input design models of the SPLs without being aware of the experimental purpose of the study. In addition, the composition heuristics are automatically applied and are algorithms explicitly and independently defined by others. Consequently, the study results can be more confidently applied to realistic development settings without suffering influences from experimenters.

Finally, the randomization of the subjects was not performed because it would require simple task simple software engineering task. Hence, this would undermine the objective of this study.

### 6.1.5.2.
### Statistical Conclusion Validity

We evaluated the statistical conclusion validity checking if the independent and dependent variables (Section 6.1.2.4) were submitted to suitable statistical methods. These methods are useful to analyze whether (or not) the research variables covary (Cook et al., 1979; Shadish et al., 2006). The evaluation is concerned on two related statistical inferences: (1) whether the presumed cause and effect covary, and (2) how strongly they covary (Cook et al., 1979; Shadish et al., 2006). Considering the first inferences, we may improperly conclude that there is a causal relation between the variables when, in fact, they do not. We may also incorrectly state that the causal relation does not exist when, in fact, it exists. With respect to the second inference, we may incorrectly define the magnitude of covariation and the degree of confidence that the estimate warrants (Shadish et al., 2006).

*Covariance of cause and effect.* We eliminated the threats to the causal relation between the research variables studying the normal distribution of the collected sample. Thus, it was possible to verify if parametric or non-parametric statistical methods could be used (or not). For this purpose, we used the Kolmogorov-Smirnov test to determine how likely the collected sample was normally distributed. As the dataset did not assume a normal distribution, nonparametric statistics were used (Section 6.1.2.1 and Section 6.1.2.2.). Hence, we are confident that the test statistics were applied correctly; as the assumptions of the test statistics were not violated.

*Statistical significance.* Based on the significance level at 0.05 level ($p \leq 0.05$), Mann-Whitney test was used to evaluate our formulated hypotheses. The results collected from this test indicated $p < 0.001$. This shows sufficient evidence to say that the difference between the inconsistency rates (and composition effort) of stable and unstable models are statically significant. The correlation between the independent and dependent variables is also evaluated. For this, Spearman's correlation test was used. The low collected p-value ($< 0.001$) indicated that there is a significant correlation between the inconsistency rate and stability as well as composition effort and stability. In addition, we followed some general guidelines to improve conclusion validity (Wohlin et al., 2000). First, a high number of compositions were performed to increase the sample size, hence improving the statistical power. Second, experienced developers used more realistic design models of SPLs, state-of-practice composition heuristics, and robust software modeling tool. These improvements reduced "errors" that could obscure the causal relationship between the variable under study. Consequently, it brought a better reliability for our results.

### 6.1.5.3.
### Constructs Validity

Construct validity concerns the degree to which inferences are warranted from the observed cause and effect operations included in our study to the constructs that these instances might represent. That is, it answers the question: "Are we actually measuring what we think we are measuring?" With this in mind, we evaluated (1) whether the quantification method is correct, (2) whether the quantification was accurately done, and (3) whether the manual composition threats the validity.

*Quantification method.* All variables of this study were quantified using a suite of metrics, which was previously defined and independently validated (Farias et al. 2010a; Kelly, 2006; Medeiros et al., 2010; Guimaraes et al.; 2010). Moreover, the concept of stability used in our study is well known in the literature (Kelly, 2006) and its quantification method was reused from previous work. The inconsistencies were quantified automatically using the IBM RSA's model validation mechanisms and manually by the developers through several cycles of measurements and reviews. In practice, the developers' effort is computed by "time spent." However, the "time spent" is a reliable metric when used in controlled experiments. Unfortunately, controlled experiments require that the software engineering tasks are simple; hence, it harms the objective of our investigation (Section 6.1.2.1) and hypotheses (Section 6.1.2.2). Moreover, we have observed in the examples of recovering models that, in fact, the "time spent" is actually greater for unstable models than stable models, independently of the type of inconsistencies. In addition, the number of syntactic and semantic inconsistencies was always higher in unstable models than stable models.

*Correctness of the Quantification.* Developers worked together to assure that the study does not suffer from construct validity problems with respect to the correctness of the compositions and application of the suite of metrics. We checked if the collected data were in line with the objective and hypotheses of our study. It is important to emphasize that just one facet of composition effort was studied: the effort to evolve well-structured design models using composition heuristics. The quantification procedures were carefully planned and followed well-known quantification guidelines (Wohlin et al., 2000; Basili et al., 1999; Kitchenham et al., 2008; Kitchenham et al., 2006).

*Execution of the Compositions.* Another threat that we have controlled is if by using manual composition threats validity since we might unintentionally avoids conflicts. We have observed that the manual composition helps to minimize problems that are directly related to model composition tools. There are some tools to compose design models, such as IBM Rational Software Architect. However, the use of these tools to compose the models was not included in our study for several reasons. First, the nature of the compositions would require that developers understood the resources/details of the tools. Second, even though the use of these tools might intentionally reduce (or exacerbate) the generation of specific categories of inconsistencies in the output composed models, it was not our goal to evaluate particular tools. Therefore, we believe that by using a model composition tool would impose more severe threats to the validity of our experimental results. Finally, and more importantly, we don't think the manual composition would be a noticeable problem to the study for many reasons, including: (i) even if the conflicts were unconsciously avoided, we deeply believe that the heuristics should be used as "rules of thumb" (guidelines) even if tool support is somehow available, and (ii) we have reviewed the produced models, at least, three times in order to ensure that conflicts were injected accordingly; in the case they still made their way to the models used in our analysis, they should be minimal.

### 6.1.5.4.
### External Validity

External validity refers to the validity of the obtained results in other broader contexts (Mitchell & Jolley, 2001). That is, to what extent the results of this study can be generalized to other realities, for instance, with different UML design models, with different developers and using different composition heuristics. Thus, we analyzed whether the causal relationships investigated in this study could be held over variations in people, treatments, and other settings.

As this study was not replicated it in a large variety of places, with different people, and at different times, we made use of the theory of proximal similarity (proposed by Donald T. Campbell (Campbell & Russo, 1998)) to identify the degree of generalization of the results. The goal is to define criteria that can be used to identify similar contexts where the results of this study can be applied. Two criteria are shown as follows. First, developers should be able to make use of composition heuristics (Section 7.1.1.2) to evolve UML design models such as UML class and component diagrams. Second, developers should also be able to apply the inconsistency metrics described previously and use some robust software modeling tool e.g., IBM RSA (Norris & Letkeman, 2011; IBM RSA, 2011).

Given that these criteria can be seen as ever-present characteristics in mainstream software development, we conclude that the results of our study can be generalized to other people, places, or times that are more similar to these requirements. Some characteristics of this study contributed strongly to its external validity as follows. First, the reported exploratory study is realistic and, in particular, when compared to previously reported case studies and controlled experiments on composing design models (Dingle et al., 2008; Chitchyan et al., 2009; Farias et al., 2010a; Whittle & Jayaraman, 2010; Briand et al., 2005; Clarke & Walker, 2001; Norris & Letkeman, 2011). Second, experienced developers used: (1) state-of-practice composition heuristics to evolve three realistic design models of software product lines; (2) industrial software modeling tool (i.e., IBM RSA) to create and validate the design models; and (3) metrics that were validated in previous works (Farias et al., 2010b). Finally, this work investigates only one facet of model composition: the use of model composition heuristics in adding new features to a set of design models for three realistic software product lines.

## 6.1.6.
## Concluding Remarks

Model composition plays a pivotal role in many software engineering activities e.g., evolving SPL design models to add new features. Hence, software designers are naturally concerned with the quality of the composed models. Our study, therefore, represents a first exploratory study to empirically evaluate the impact of stability on model composition effort. More specifically, the focus was on investigating whether the presence of stable models reduces (or not) the inconsistency rate and composition effort. In our study, model composition was exclusively used to express the evolution of design models along eighteen releases of three SPL design models. Three state-of-practice composition heuristics have been applied, and all were discussed in detail throughout this chapter.

The main finding was that the model stability is a good indicator of composition inconsistencies and resolution effort. More specifically, we found that stable models tend to minimize the inconsistency rate and alleviate the model composition effort. This observation was derived from statistical analysis of the collected empirical data that have shown a significant correlation between the independent variable (stability) and the dependent variables (inconsistency rate and effort). Moreover, our results also revealed that instability in design models would be caused by a set of factors as follows. First, SPL design models are not able to support all upcoming changes, mainly unanticipated incremental changes. Next, the state-of-practice composition heuristics are unable to semantically match simple changes in the input model elements, mainly when changes take place in crosscutting requirements. Finally, design models implementing crosscutting requirements tend to cause a higher number of inconsistencies than the ones modularizing their requirements more effectively. The main consequence is that the evolution of the design models using composition heuristics can even become prohibitive given the effort required to produce the intended model.

As future work, we will replicate the study in other contexts (e.g., evolution of statecharts) to check whether (or not) our findings can be extended to different evolution scenarios of design models supported by composition heuristics. We also consider exploring varieties of our stability metrics. We also wish to improve understanding if design models with superior stability have some gain (or not): (i) when produced from other composition heuristics, and (ii) on the effort localizing the inconsistencies. It would be useful if, for example, intelligent recommendation systems could help the developers to indicate the best heuristic to-be applied to a given evolution scenario or even recommending how the input model should be restructured to prevent inconsistencies. Finally, we hope that the issues outlined throughout the evaluation encourage other researchers to replicate our study in the future under different circumstances and that this work represents a first step in a more ambitious agenda on better supporting model composition tasks.

## 6.2.
## Impact of Design Language on Inconsistency Resolution Effort

This section aims at evaluating the impact of design modeling languages such as AO and non-AO modeling on the inconsistency resolution effort. The hypothesis investigated is that aspect-orientation may alleviate the effort of inconsistency resolution to some extent. Aspect-orientation provides an improved modularity and that more effective modularization may help developers to deal with the inconsistencies, thus minimizing the resolution effort. However, it is by no means obvious that this hypothesis holds. It may be, for instance, that inconsistencies in aspect-oriented models have a detrimental effect on the resolution effort because inconsistencies aspectual elements may require the developers to examine all points in the model crosscut by the aspects.

With this in mind, the goal of this section is to report on an exploratory empirical study that aimed at providing evidence to support or refute this hypothesis. To this end, we again make use of model composition to add new features to a set of models in a software product line, called Mobile Media.

We investigate this hypothesis in the context of SPLs evolution because they commonly involve model composition activities (Jayaraman et al., 2007; Thaker et al., 2007) and, while we believe the kinds of model composition in SPLs are representative of the broader issues, we make no claims about the generality of our results beyond SPL model composition. We show the results for model compositions of six releases of an SPL. In each release, models for the new feature are composed with the models for existing features. For each release, we analyze both the quantity and nature of the composition inconsistencies. Furthermore, we compare two versions of the SPL models — one which uses aspect-oriented modeling and one which does not.

The results show that higher inconsistency rates were observed in the presence of aspects when they had a higher degree of quantification. On the other hand, this problem did not entail more effort on inconsistency resolution. We also found that higher degree of obliviousness tended to yield compositions of AO composed models that are closer to the intended compositions. To the best of our knowledge, our results are the first to empirically investigate the potential advantages of aspects during modeling phase. Despite a wide variety of technical approaches to AOM e.g., MATA (Whittle & Jayaraman, 2010) and Kompose (Kompose, 2011), to-date there been almost no empirical evaluation of AOM. We therefore see this study as a first step in a more ambitious agenda to empirically assess aspect-oriented modeling.

The remainder of the study is organized as follows. Section 6.2.1 introduces the main concepts and knowledge that are going to be used and discussed throughout this section. Section 6.2.2 we present the methodology. Section 6.2.3 discusses the composition analysis effort. Section 6.2.4 contrasts this work with others, highlighting the commonalities and differences. Section 6.2.5 analyzes the threats to validity. Finally, Section 6.2.6 presents some concluding remarks and future work.

### 6.2.1.
### Aspect-Oriented Modeling for Architectural Models

Model composition applies both to development with and without aspect-oriented modeling (Clarke & Walker, 2005). This study compares the inconsistency resolution effort in both cases. AOM languages aim at improving separation of concerns by supporting the modular representation of concerns that cut across multiple software modules. Crosscutting concerns are represented by a new model element, called *aspect*. The goal of AOM is, therefore, to provide software developers with the means to express aspects and crosscutting relationships in their models. There are AOM languages for modeling aspects at many levels of abstraction, ranging from use cases and architectural design to detailed designs. As far as the solution space is concerned, aspects are usually first expressed in architectural models.

Figure 21 is an illustrative example of the architectural AOM language (Garcia et al., 2009) used in this study (Section 6.2.3). We chose this AOM language because: (i) we selected architectural models as our focus due to the availability of existing industrial models; (ii) the AOM language has been widely used in other contexts (such as modularization of crosscutting concerns (Sant'Anna, 2008)) and is therefore mature (Garcia et al., 2009).

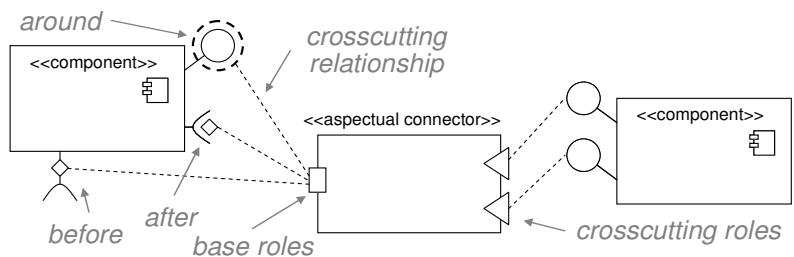

Figure 21: AOM language for architectural models

The notation supports the visual symmetric representation of aspect-oriented software architectures. The target modeling approach consists of an extension of the UML's component diagram (OMG, 2011). In order to put the composition in practice, we should consider the properties of model elements defined in the UML metamodel specification in this diagram. Thus, the properties of the model elements considered were component (name, provided interface, and required interface), interface (name, operation, and attribute), operation (name, return type, and parameters), attribute (name and type), relationship (source and target), crosscutting relationship, and join-points. Therefore, the composition algorithms are fine-grained due to take into account these properties in each composition.

The notation provides explicit elements for expressing different forms of component-aspect collaborations, which are represented by aspectual connectors. Aspectual connectors are illustrated by rectangles in Figure 21. They define which components, interfaces or specific operations are affected by a component modularizing a crosscutting concern. Aspectual connectors are associated with crosscutting relationships represented by dashed arrows. The notation also supports the visual modeling of specific pointcut designators (e.g., advising all the provided interfaces) and sequencing operators (after, before, and around). For the sake of simplicity in this study, only aspectual connectors and crosscutting relationships will be represented in the models of our case study; all the other visual details have been omitted from here on.

### 6.2.2.
### Study Methodology

This section describes the study definition, the target application, the evaluation method used for computing model composition effort, and the other study procedures in our exploratory study.

### 6.2.2.1.
### Objective and Research Questions

This study attempts to evaluate the impacts of aspect-oriented modeling on two variables: the *inconsistency rate* and *inconsistency propagation*. These effects are evaluated from evolution scenarios considering compositions of architectural models. Additionally, some scenarios are described in which the influence of AO models on effort is precisely described. With this in mind, the objective of this study is stated based on the GQM template (Basili et al., 1994) as follows:

*Analyze* design modeling techniques

*for the purpose of* investigating their effects

*with respect to* inconsistency rate and inconsistency propagation

*from the perspective of* developers

*in the context of* evolution of architectural models

Specially, this study aims at discovering the inconsistency rate, resolution effort, and revealing scenarios where these inconsistencies propagate, affecting multiple model elements. Therefore, we address research question RQ3, as stated in Chapter 1:

- **RQ3:** What is the effect of design decomposition techniques in particular with respect to misinterpretation, inconsistency rate, inconsistency detection effort, and inconsistency resolution effort?

Regarding the quality notions defined in Chapter 3, we study whether the syntactic and semantic quality of a design model affects the effort and resolution quality notions. We refine RQ4 into two more research questions. Thus, we focus on the following research questions:

- **RQ3.4:** Does the composition of AO models produce a higher inconsistency rate than non-AO models?

- **RQ3.5:** What is the impact of AO modeling on the way inconsistencies propagate in the output model?

These research questions were investigated considering the inconsistencies described in Section 5.1.2 and Section 6.1.1.3.

### 6.2.2.2.
### Hypotheses Formulation

Aspect-oriented modeling has been a topic of research for at least ten years (Clarke & Walker, 2005; Clarke & Banaissad, 2005). However, there is currently very limited knowledge as to how aspects, when incorporated in input models, affect the model composition effort. In particular, there is no understanding if the composition of aspect-oriented models affects the emergence of inconsistencies in the output composed models.

*First Hypothesis: Impact of Aspect on Inconsistency Rate.* Our first null hypothesis assumes that the inconsistency rate in output AO composed models is equal or higher than in output non-AO composed models. As aspect orientation tends to improve the modularization of design models, the alternative hypothesis states that the inconsistency rate in AO models is lower than in non-AO models. This would lead to the following null and alternative hypotheses:

**Null Hypothesis 1, $H_{1-0}$:** The inconsistency rate (Rate) in AO models is equal or higher than in non-AO models.

$H_{1-0}$: Rate(AO) $\geq$ Rate(non-AO).

**Alternative Hypothesis 1, $H_{1-1}$:** The inconsistency rate (Rate) in AO models is lower than in non-AO models.

$H_{1-1}$: Rate(AO) $<$ Rate(non-AO).

Given that inconsistency tends to propagate in a composed model (Farias et al., 2010a). That is, the introduction of one inconsistency can often lead to multiple other inconsistencies because of a "knock-on" effect. An example would be the inconsistency whereby a composed component is missing an important operation. This semantic inconsistency leads to a "knock-on" syntactic inconsistency if another component requires the operation. In the worst case, there may be long chains of inconsistencies all derived from a single inconsistency. Studying such propagation effects is important because propagation directly affects the effort in resolving inconsistencies e.g., a propagation chain of length $n$ may be actually fixed by resolving a single inconsistency rather than the expected $n$ inconsistencies. Thus, we are interested in understanding the possible inconsistency propagation patterns in AO and non-AO models (RQ4.5). Similar to the previous hypothesis, it is assumed that inconsistency equally spread through output (non-)AO models. This leads to the second null and alternative hypotheses as follows:

> **Null Hypothesis 2, $H_{2-0}$:** The inconsistency propagation in AO models is equal or higher than in non-AO models.
>
> **$H_{2-0}$:** Prop(AO) $\geq$ Prop(non-AO).
>
> **Alternative Hypothesis 2, $H_{2-1}$:** The inconsistency propagation in AO models is lower than in non-AO models.
>
> **$H_{2-1}$:** Prop(AO) < Prop(non-AO).

To test the hypotheses, metrics were used to quantify inconsistency rate, the propagation, and the effort to resolve the inconsistencies when they spread through model elements. Aforementioned, these metrics are presented in Chapter 3. The metrics were applied to both non-AO and AO models of an evolving software product line described in the next section.

### 6.2.2.3.
### Case Study: Evolving an SPL

Model composition can be applied in different contexts and with different purposes. We have selected a particular scenario to test our study hypotheses: the use of model composition to express the evolution of software product line (SPL) architecture.

*Model Composition for Expressing SPL Evolution.* Model compositions were defined to generate the new releases of the SPL architecture model. That is, the composition algorithms (override, merge, and union) were used to define how each architecture model ($M_A$) of an SPL release and the new model increments ($M_B$) were going to be combined to generate the new architecture SPL release ($M_{AB}$). The first input model ($M_A$) represents the current architecture of an SPL release, while the second input model ($M_B$) represents the delta capturing the modifications to the base model ($M_A$). The output model ($M_{AB}$) generated by the application of the composition algorithm represents the next SPL release.

*MobileMedia: the Target SPL.* A product line, called Mobile Media (Figueiredo et al., 2008), of 6 kLOC was selected to be the target case of the evaluation. The purpose of the MobileMedia SPL is to manipulate photos, music, and videos on mobile devices. In (Figueiredo et al., 2008), it is possible to find a fine-grained description about its characteristics and how its evolution happened. The reasons for selecting this system in the evaluation are described as follows. First, the developers of the MobileMedia SPL are the responsible for creating its architecture design models. Second, two versions of the same product line and the respective architectural models were available for our investigation: an AO version and a non-AO version. This is a fundamental requirement to test the hypotheses (Section 6.2.2.2). Third, the last release of the architectural design has more than one hundred modules, and its architectural models are the main artifact to reason about change requests and derive new products. Fourth, the architectural models were produced by the original developers, which do not have any of the model composition algorithms under assessment in mind, thereby avoiding any bias and entailing a more natural software development scenario. Fifth, the architectural models ($M_A$) and the increment models ($M_B$) were conceived with the modularity and changeability as key drivers. Sixth, we had available seven fully documented evolution scenarios, which could be expressed with model compositions (examples are given later).

Finally, Mobile Media met a number of other equally-important requirements, such as: (1) proper documentation of the driving requirements; (2) the system evolved for more than three years, and the more recent releases have more than 100 modules; (3) different types of change were realized in each release, including refinements of the architecture style employed, (4) the system has been successfully used in other studies involving empirical evaluation of OO and AO implementations (Figueiredo et al., 2008), and (5) the original developers were available to help us with the production and analysis of the composed models and the intended models. As such, all these factors provided a solid foundation for our study.

### 6.2.2.4.
### Quantifying Inconsistency Rate and Resolution Effort

The goal is to quantify: (i) the number of inconsistencies, and (ii) the activities required to transform the output composed model into an output intended model. The analysis of the results relies on an inconsistency measure, called inconsistency rate (Rate), to quantify the amount of composition inconsistencies divided by the total number of elements in the output model. That is, inconsistency rate allows computing the density of composition inconsistencies in the output composed models. Using this metric, we may quantify the inconsistency rate in AO and non-AO models, and analyze the differences between them (H1). Note that the inconsistency rate is defined from multiple inconsistencies, which can be found in Section 6.1.1.3.

The resolution effort consists of the number of operations that should be performed to transform an output composed model into an output intended model. We compute the number of *creations*, *removals*, and *modifications* needed to realize this transformation. That is, this computation represents an estimation of the resolution effort ($g(M_{CM})$). After we collect the $g(M_{CM})$ measure, we performed an inspection of the output model to check if there was any occurrence of inconsistency propagation. This enabled us to check if the presence of aspects in the input models had any impact on the way composition inconsistencies were propagated (H2). In order to come up with a suitable characterization of the measures of the compositions and the MobileMedia SPL releases, we defined a basic formalism for the metric space of composition effort as follows.

A metric space is a set M equipped with a real-valued function CE(w,s) defined for all w, s $\epsilon$ M. Let M = {$R_{i,x,y}$, i = 1,…,n; x = override, merge; y = left, right}, where:

- *n* is a finite natural number representing the model release;
- *left* and *right* represent the direction of the composition relationship in the override algorithm.

For example, $R_{3,merge,right}$ represents the Release 3 that was produced by merging: Release 2 $+_{merge}$ Delta(Release 2, Release 3) $\rightarrow$ Release 3. Delta(Release 2, Release 3) represents the model elements that should be merged with Release 2 to transform it into Release 3, as previously discussed. In practical terms, the Delta represents the evolution to be inserted into the previous release. On the other hand, $R_{3,merge,left}$ would be Delta(Release 2,Release3) $+_{merge}$ Release 2 $\rightarrow$ Release 3 (the inverse order can also be represented with an asterisk). Therefore, the reader should note that the order of override-based composition might produce different output composite models (Dingel et al., 2008). Each model of a $R_{i,x,y}$ can be characterized by observing its syntactic and semantic properties. If we have a high inconsistency rate in an evolution scenario, then this implies a higher effort to resolve inconsistencies.

## 6.2.2.5.
## Evaluation Procedures

Once the case study was selected (Section 6.2.2.3) and the inconsistency resolution metrics were defined (Section 6.2.2.4), we needed to undergo a number of specific evaluation procedures. They are discussed in the following.

### a. Target Model Versions and Releases

We have used both non-AO and AO versions of the Mobile Media models in order to test the study hypotheses (Section 6.2.2.2). These two model versions of the same system enabled us to identify if the presence of aspects in the input models had positive or negative effects on the quality of the output model.

*Deriving AO and non-AO Model Releases.* For each release of Mobile Media, we have applied each of the composition algorithms described in Section 2.3. That is, we have used the merge algorithm to compose two input AO models in order to produce a new AO release model; similarly, we applied the merge strategy to compose two input non-AO models in order to produce the next non-AO release model. We performed similar compositions with override and union algorithms. The goal was to identify if the outcomes, in terms of inconsistency rate and propagation (hypotheses), were the same or different. All the releases of the non-AO and AO versions realized exactly the same SPL features and variability points. They also underwent the same evolution scenarios, ranging from changes in heterogeneous mobile platforms and additions of many alternative and optional features (Figueiredo et al., 2008). Non-AO models were represented by conventional UML component models, while AO models were represented using the AOM language described in Section 6.2.1.

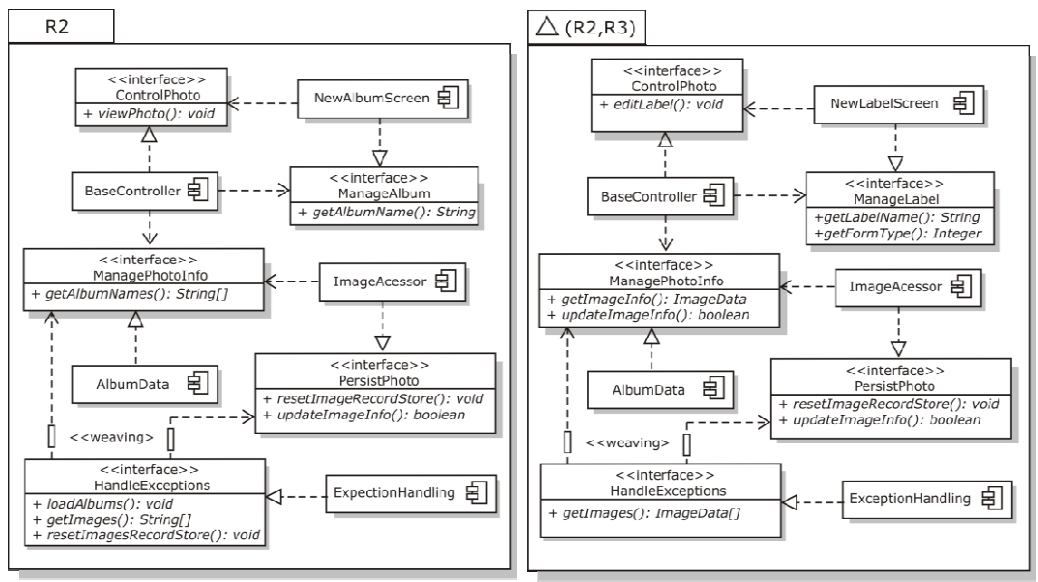

Figure 22: The input models: the AO base and AO delta model

In fact, AOM is used in this work to represent the aspect-oriented model releases of the SPL under study. For example, in Figure 22, in addition to have interfaces (e.g., *PersistPhoto*), components (e.g., *ImageAcessor* and *AlbumData*), we also have aspectual components, such as the *ExceptionHandling* aspect.

Moreover, we can also have some relationships: realization (e.g., between the components *BaseController* and *ControlPhoto*), dependency (e.g., between the component *NewAlbumScreen* and the interface *ControlPhoto*), and crosscutting (e.g., between the aspectual component *ExceptionHandling* and the component *PersistPhoto*, in which the service *loadAlbums*(): void is woven into the component). The notation used in this work to express the architectural models has been used in other works (Figueiredo et al., 2008; Garcia et al., 2009) and has shown to be effective for its purpose.

*Model Releases and Composition Specification.* We considered six releases of MobileMedia (Figueiredo et al., 2008) in this study. They were selected because they were the ones where the changes implied visible modifications in the architectural design. For each new release, the previous release was modified in order to accommodate the features to be modified, inserted, or removed. To implement a new evolution scenario, a model composition specification can remove, add, derive, or modify the entities present in the previous release. During the design of all releases, a main concern was to follow best practices of modeling.

### b. Execution and Assessment Phases

The execution and assessment of the study were structured in three main steps, which are described in the following.

*Model Refactoring Phase.* The model refactoring is a pivotal activity to define the input models and, hence, to express the model evolution as an explicit model composition relationship. To this end, MobileMedia's architectural models were initially refactored to specify the delta itself and to represent the change scenarios as composition relationships. To create the delta model it is necessary to identify the differences between the releases models and then gather them into the input model. To go about this, we took into account an evolution description created by the original modelers involved in a previous study (Figueiredo et al., 2008). These descriptions specify in-depth the modifications needed to realize each evolution scenario (from one release to another). They allowed us to identify how the model elements were changed. For example, in the second evolution description, the Delta(R2,R3) were based on the description such as: the interface *ControlPhoto* — realized by *BaseController* — had the method *edilLabel*(): void added (see Figure 22). Another example would be the change concerning the name of the interface *ManageLabel* to *ManageAlbum*. Thus, all model elements of the Delta(R2,R3) are derived from one evolution description, which ensures that the input model specification is free of bias.

*Composition and Measurement Phase.* From one release to another, 6 compositions were produced: 3 compositions following override, merge, and union from the current release to delta, and 3 compositions in the inverse direction. We considered 5 evolution scenarios for the non-AO version as well as the AO version of the Mobile Media, totaling 60 compositions. The result of this phase was a document of composition descriptions, including the gathered data from the application of our metrics suite. Figure 22 presents partial input models being used in one of the releases, while Figure 23 and Figure 24 represent examples of composition based on merge, override, and union, respectively.

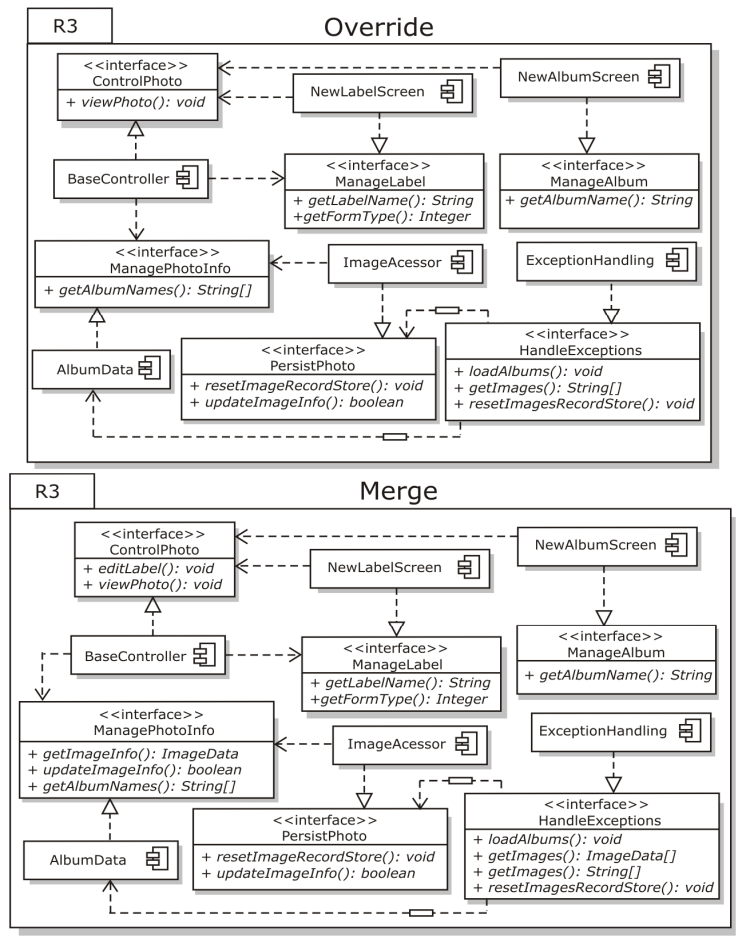

Figure 23: Output AO models produced by override and merge algorithms

Figure 24 is the intended result of the composition (or intended model). As well-validated metrics for model composition are not available yet, we used a set of inconsistency metrics defined in our previous work (Farias et al., 2008a). The inconsistencies (and their effects) were identified manually using such inconsistency metrics. The identification of the inconsistencies was performed in 5 review cycles in order to avoid false positives/negatives. We also consulted the Mobile Media developers when needed, such as checking and confirming specific cases of semantic inconsistencies.

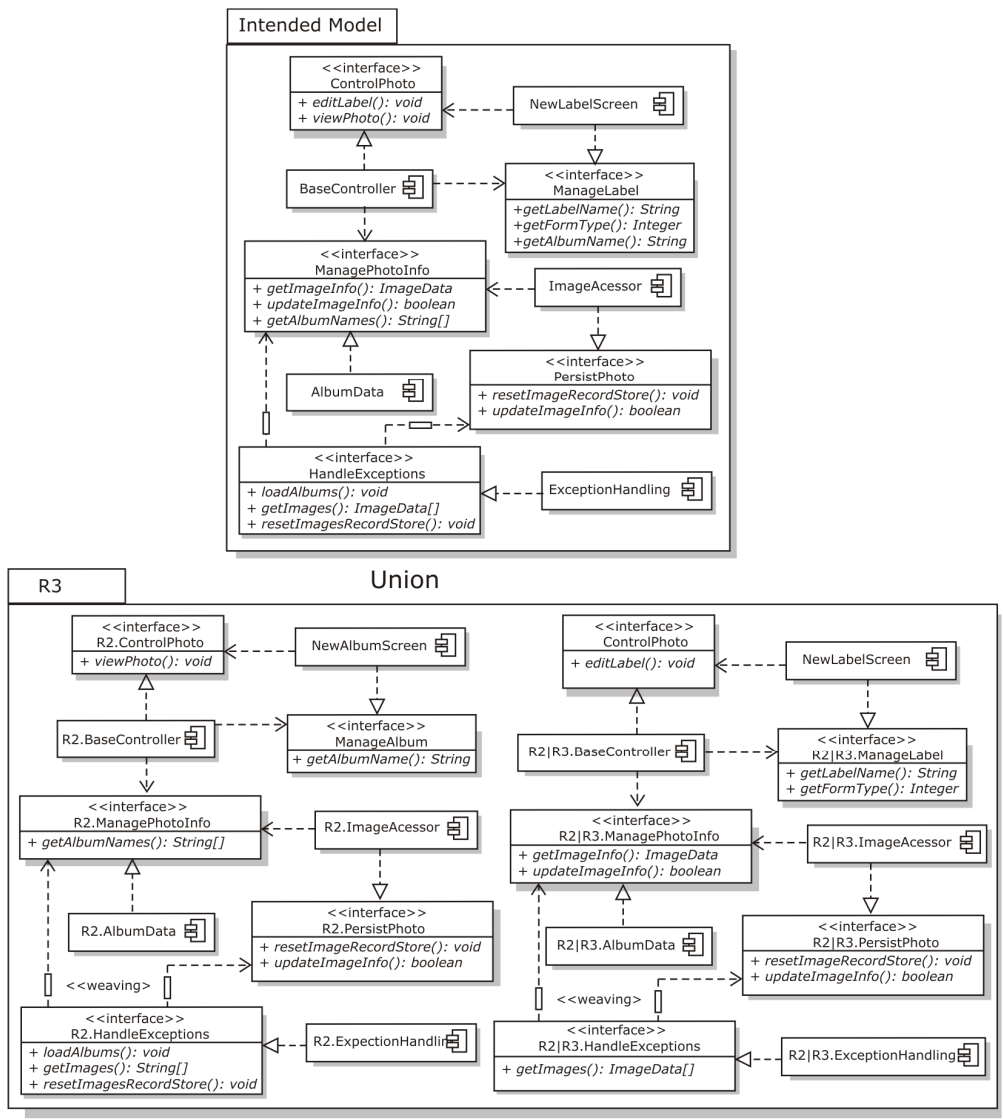

Figure 22: AO intended model (from Figure 22) and AO output model produced following the union heuristic

*Effort Assessment Phase.* The goal of the third phase was to assess the effort to resolve the inconsistencies using the metrics described previously. The composition algorithms were used to generate the evolved models, so that we could assess the impact of aspects on the model composition effort. In order to support a detailed data analysis, the assessment phase was further decomposed in two main stages. The first stage (Section 6.2.3.1) is concerned with pinpointing the inconsistency rates produced by composition of either non-AO or AO (H1). The second stage (Section 6.2.3.2) aims at assessing the effort to resolve a set of previously identified inconsistencies and whether (or not) the use of aspect has a higher impact on the way composition inconsistencies are propagated (H2). We analyzed how inconsistency rate differs across the releases in order to detect potential benefits and drawbacks of using AOM in the input models. We have decided to focus the discussions on the merge and override algorithms, because the union algorithm did not present any additional interesting insight. However, all measurement results were considered during the study analysis.

### 6.2.3.
### Composition Effort Analysis

This section presents the results collected during the investigation of the RQ3.4 and RQ3.5 to both the AO and non-AO output models realizing each SPL release. Histograms are used to provide an overview of the data gathered in the measurement process. These histograms allow us to analyze the impact of aspects on study variables: inconsistency rate, inconsistency propagation, and inconsistency resolution effort. Each histogram focuses on the application of a particular composition algorithm. The Y-axis presents the values gathered for a particular metric. The X-axis specifies the evolution scenarios.

Note that each pair of bars is attached to a pair of values, with the first capturing the performance of the AO version and the second capturing the non-AO one. The lower the value, the better is the performance of the modeling approach used. It is important to highlight that the results shown in the histograms were gathered with respect to the entire model. Based on the inconsistencies identified by the inconsistency rate metric, Section 6.2.3.1 discusses the findings related to the first hypothesis (H1). Section 6.2.3.2 relies on the metric for quantifying model recovery effort in order to support the analysis of the second hypothesis (H2).

### 6.2.3.1.
### H1: Aspects and Inconsistency Rate

Figure 25 illustrates the results for the inconsistency rate obtained following the override algorithm. Figure 26 shows the results of the same metric for the merge algorithm. The first observation allows us to conclude that the inconsistency rate measures have favored aspect-orientation in both merge and override cases and for most of the evolution scenarios. This implies that the tally of inconsistencies to some extent is decreased whenever aspects are present in the models to-be-composed. The presence of aspects in the input models produced lower inconsistency rate than aspect-free models when the override algorithm is applied in both directions (right and left (represented by the *-columns)). For example, the inconsistency rate decreases from 1.72 (non-AO version) to 1.33 (AO version) in Scenario 2, which represents a reduction of 22.6% in favor of aspect-orientation. Similarly, the inconsistency rate decreases from 0.59 to 0.41 when the composition is performed in the left direction, which represents a reduction of 30%.

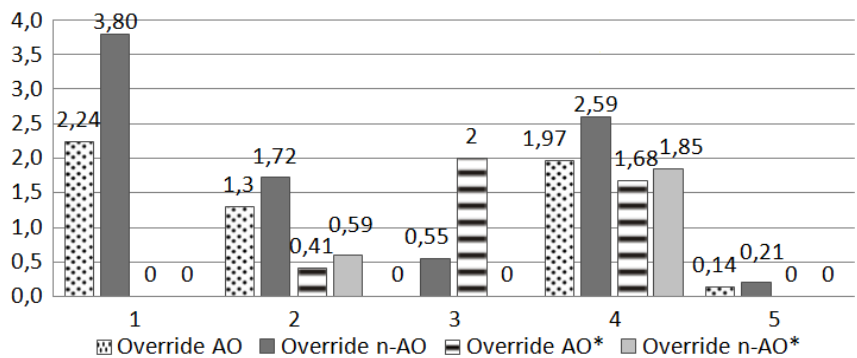

Figure 23: Inconsistency rate produced by the override algorithm

Moreover, it is well known that the higher the number of model elements that take part in compositions, the higher the likelihood of inconsistencies being generated. Nevertheless, the AO versions still had lower absolute measures of inconsistencies. For example, the absolute measure decrease from 38 (non-AO version) to 36 (AO version) in Scenario 2, which represents a reduction of 5.2% in favor of aspect-orientation. Similarly, the inconsistency rate decreases from 13 to 11 in the inverse order, which represents a reduction of 15.3%. The only case where aspect-free models led to a close inconsistency was the application of the merge algorithm in the second release; this special case is discussed in the following section.

The main reason for the superiority of the AO models is that changes, reified by the delta model, tend to be confined in fewer modules due to the superior modularization of crosscutting features in AO models. The confinement of modifications to aspects, in turn, leads to a better localization of both syntactic and semantic inconsistencies, thereby making them easier to detect and address in the output models. Therefore, we refute the null hypothesis $H_{1-0}$ and confirm the alternative hypothesis $H_{1-1}$.

We have noticed that the decrease of inconsistencies observed in the AO models is potentially influenced by two factors: (i) *quantification,* the higher the quantification of aspects in input models, the higher the inconsistency rate measures, and (ii) *obliviousness*, the higher the degree of obliviousness, the lower the inconsistency rate measures in the output models. Another predominant factor in the emergence of high inconsistency rates was the nature of the change. Independently of the degree of obliviousness and quantification in AO models, the nature of the change directly affected the inconsistency rate observed in the output models. In the following, we elaborate these issues further and discuss examples that support each of these findings.

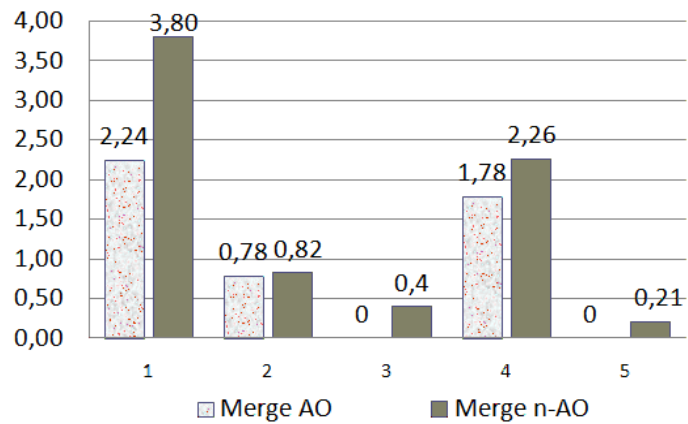

Figure 24: Inconsistency rate produced by the merge algorithm

### a. *Obliviousness and Quantification*

We have observed that quantification (Filman & Friedman, 2000) influenced the inconsistency rate measures. The presence of aspects with lower quantification (in the input models) led to fewer syntactic and semantic inconsistencies in the output models. When aspects were being used, for example, to encapsulate domain–specific features, a lower inconsistency rate manifested in the output models. On the other hand, we also observed that when a conflict arises in aspects with higher quantification (in the input models), higher rates of syntactic and semantic inconsistencies occurred in the output models. Therefore, the quantification mechanism may (or may not) improve inconsistency rate results.

This category of aspects is the case where the aspects work as glue between a few elements in the base model and the changes realized by the delta model. Aspects with a higher degree of quantification, such as exception handling (Figure 22, Figure 23, and Figure 24), affect the input base model in many places (join points). This was exactly the case in Scenario 2, where the non-AO version (Rate = 0.82) has a measure close to the AO version (Rate = 0.78) (Figure 22). Higher quantification increases the aspect scope and, therefore, the likelihood of aspects interfering with each other. When the merge algorithm was applied, the exception handling aspect (Figure 23) led to undesired superimpositions with other aspectual behaviors advising the same join points.

The overall inconsistency rate (Rate measure) has been usually lower in the AO version because most of the aspects were not affecting more than three elements. By overall rate, we mean the average of inconsistencies considering all the model elements. However, a careful analysis of the number of inconsistencies in individual model elements (e.g., a particular component) reveals some interesting information. The composition output of AO models consistently caused an increase on the number of inconsistencies for some specific model elements. For example, this can be observed in Scenario 4, when the highest number of inconsistencies emerged in both non-AO and AO versions. Despite the significant Rate difference favoring the AO version, the component *BaseController* presented an increase (Rate = 38) in relation to *BaseController* of the non-AO version (Rate = 24). We noted that this problem occurred in situations where the components were affected by two aspects or more in the delta model. In other words, when a base component had a high density of join points shared by multiple aspects; it generated a higher number of inconsistencies.

An additional interesting finding was that the composition of AO models tended to manifest fewer inconsistencies when the obliviousness degree of the base elements was higher. We have noted that the creation of new aspects (via the delta model) for encapsulating new features implies that the modules in the input base model are more oblivious to the modification being implemented in the release. This observation holds for both mandatory and varying(optional or alternative) features. Consequently, the combination of the AO modules tended to ripple fewer inconsistencies in the output models.

This finding implies that the presence of obliviousness is a good indicator that the model composition at hand will better adhere to the Open-Closed principle (Meyer, 1988). This principle states "software should be open for extensions, but closed for modification." AO modeling conformed more closely to this principle in scenarios where the behavior in the new aspect (part of the Delta model) is more independent of the affected elements in the base model. Release 3 illustrates this finding. For instance, the *AlbumData* component demanded modifications in the non-AO version of Release 3 in order to include the feature of sorting photos by highest viewing frequency. On the other hand, the AO counterpart required no modification in this component. The reason was that new components and the *PhotoSorting* aspect in the delta model modularly implemented the feature.

The open-closed principle was more closely adhered by the composition of AO models than non-AO models. However, this observation did not occur in all the cases. In general, this principle was fully achieved only when the delta model was adding new elements to the base models. The other types of changes realized by the delta model exerted more specific implications in the rate of inconsistencies detected in the output models. This issue is discussed in the following section.

### b. The Effect of the Change Category

A careful analysis of the results has pointed out that the inconsistency rate is strictly affected by the category of changes to be applied to the base model. We identified four types of changes throughout our target SPL study:

- *Addition:* new model elements are inserted into base model; for instance, the new method *getFormType()* is inserted into the provided interface, named *ManageLabel*, of the component *NewLabelScreen* (Figure 23).
- *Removal*: a model element in the base model is removed; for example, the required interface *ControlPhoto* of the component *AlbumListScreen* is removed in the fourth Mobile Media release;

- *Modification*: a model element has some properties modified; for instance, the component *NewAlbumScreen* (Release 1) has its name modified to *NewLabelScreen* in Release 2.

- *Derivation*: model elements are refined and/or move to accommodate the changes; for example, the provided interface *ControlPhoto* (with 14 methods) of the component *BaseController* (Release 3) has some methods moved to the provided interface *ControlPhoto* of the component *PhotoController* (Release 4).

*Additions.* As previously discussed in the previous section, the use of aspects has contributed to produce an output model with much lower inconsistency rate when the evolution scenarios were dominated by *additions*. This finding is supported by the low inconsistency rate in Scenarios 3 and 5. The main reason is that the created aspects (in the delta model) modularize the changes and insert them into the target model elements, without requiring their modifications. In these cases, we also observed that lower Rate measures were observed in the AO models when the override algorithm is used and performed in the left direction. For all the other compositions, the inconsistency rate of the AO releases was equal or lower than the non-AO releases.

A concrete example of the superiority of the AO version was the decrease of the inconsistency rate from 3.8 to 2.24 in Scenario 1. This was due to the aspectual component, included in this release (via the delta model), which advises 9 methods: (i) three of them in the interface *ManagePhotoInfo* of the component *AlbumData*; and (ii) 6 of them in the interface *PersistPhoto* of the *ImageAcessor*. This led to a Rate decrease in the interface *PersistPhoto* from 11 (non-AO version) to 4 (AO version). In the same way, the *ManagePhotoInfo* had its inconsistency rate decreased from 9 to 6.

*Modifications, Removals and Derivations.* We could not find a recurring Rate pattern (in favor of AO or non-AO versions) when modification was being realized. The AO version performed better in certain cases, while the non-AO version was better in others. On the other hand, the inconsistency rate was slightly higher in non-AO models when removals and derivations were applied. We also observed that a very high inconsistency rate occurred simultaneously in both AO and non-AO models when the change scenario was complex. This was the case when the change scenario involved a blend of modifications, removals, and derivations. More specifically, this occurred in Scenario 4, when there is a significant architectural change: a single controller was restructured as a set of specialized controllers, for example.

Therefore, the heuristic composition algorithms were inefficient in widely scoped architecture evolution, such as the refinement of the MVC (Model-View-Controller) architecture style of Mobile Media. This is also due in part to the name-based model comparison, which is not able to recognize more intricate equivalence relationships between the model elements. This comparison strategy is very restrictive whenever there is a 1:N correspondence relationship between elements in the two input models. An example of the 1:N relationship category encompassed the required interface *ControlPhoto* (Release 3) of the *AlbumListScreen* component. This interface was decomposed into two new required interfaces *ControlAlbum* and *ControlPhotoList* (Release 4), thereby characterizing a 1:2 relationship. In this particular case, the name-based model comparison should be able to "recognize" that *ControlAlbum* and *ControlPhotoList* are equivalent to *ControlPhoto*. However, in the output model (Release 4), the *AlbumListScreen* component provides duplicated services to the environment giving rise to an inconsistency. However, even in these cases the aspect orientation presented a lower inconsistency rate (e.g., see Scenario 4 in Figure 27 and Figure 28).

It is known that a higher number of model elements may lead to a higher inconsistency rate when the composition is put in practice. However, this was not the case with aspect-orientation. For instance, let us consider the fourth scenario. Although fewer composed elements (25) were observed in the non-AO version, the latter presents a higher Rate measure (2.59). On the other hand, the AO version has a higher number of compositions (27), but the inconsistency rate is lower (Rate = 1.97). A real example would be the *PhotoViewScreen* component, which decreased the number of inconsistencies from 3 (non-AO version) to 1 (AO version).

## 6.2.3.2.
## H2: Aspects and Inconsistency Propagation

We focus our discussion about inconsistency propagation on the analysis of model recovery effort, the resolution effort ($g(M_{CM})$) measure (Section 6.2.2.4). This $g(M_{CM})$ measure is a useful indicator to support the analysis of the presence (or absence) of inconsistency propagation ($H_2$) in both AO and non-AO models. The higher the effort of recovering the output model (towards the intended composed model), the higher the chance of inconsistency propagation being observed in the output model. Figure 27 depicts the recovery effort measures to transform the output model produced by the override algorithm in the intended model. Similarly, Figure 28 shows the results of the same metric for the merge algorithm. The structure of the histograms follows those in the previous section.

We have concluded that aspects indeed affect the manner of the inconsistencies spread over the output models. We identified a number of recurring inconsistencies in the AO models, which did not occur in the non-AO models. In general, some inconsistencies specific to aspect orientation were caused by a conflict (or several) arising at a single aspect and spreading through all the affected elements in the base model. Therefore, we have found that there is a sensible difference on the way composition inconsistencies are propagated in non-AO and AO models. Therefore, we refute the null hypothesis $H_{2-0}$ and confirm the alternative hypothesis $H_{2-1}$.

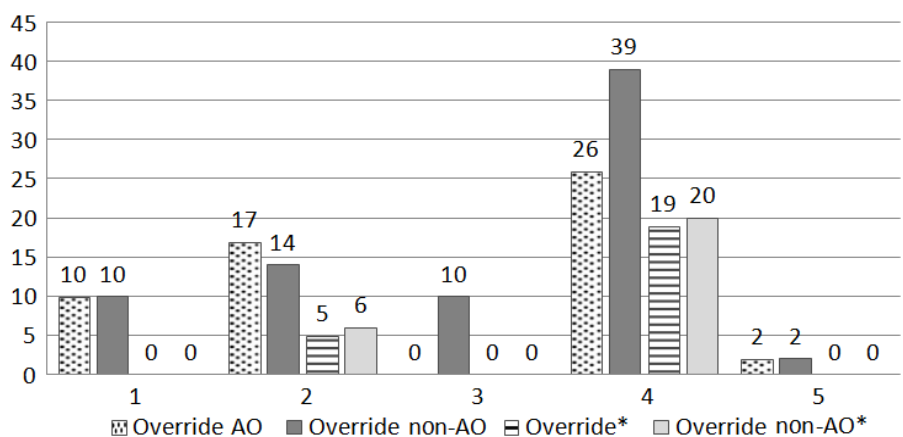

Figure 25: Inconsistency resolution effort to recover the output model produced by override algorithm

### a. Quantification and Model Recovery Effort

According to previous discussion, aspects with higher quantification contribute to higher inconsistency rates in AO models. An inspection of the output models, however, pointed out that this problem occurred because these aspects led to higher inconsistency propagation manifesting during the model composition process. Surprisingly, increase the inconsistency rates in AO models does not imply in more effort to transform the output composed into the intended composed model. In other words, the finding is that a high degree of quantification does not lead to more effort to recover the output model. The $g(M_{CM})$ measure often tends to be similar in AO and non-AO models.

This phenomenon can be illustrated, for example, in Scenario 2 (Figure 28), where the AO version presents an inconsistency rate closer to (Rate = 0.78) than the non-AO version (Rate = 0.82). However, the model resolution effort is equal to 9 for both AO and non-AO versions (Figure 28). This was the case of inconsistencies arising in a reusable exception handling aspect (modified by the delta model). When inconsistencies arose in such an aspect, they spread over all the model elements directly advised by the aspect. During the model recovery process, there was a need to fix only the inconsistency in the specification of the exception handling aspect.

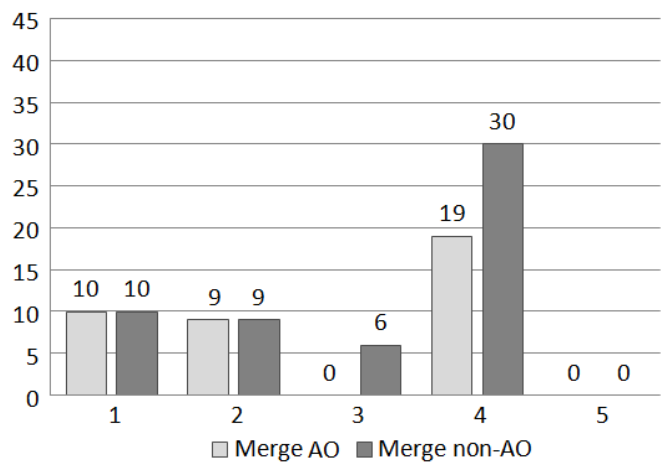

Figure 26: Effort to recover the output model produced by merge algorithm

Therefore, although AO and non-AO versions present different inconsistency rates in certain evolution scenarios (e.g., Scenario 1 in Figure 28), the effort to recover the output model from the inconsistencies in both versions is similar. The effort directly depends on how instances of inconsistencies are interrelated. Propagation channels of inconsistencies were more common in AO models as discussed above. For example, despite aspect orientation exhibiting an inconsistency rate close to the non-AO inconsistency rate in Scenario 2 (Figure 27 and Figure 28), the inconsistency resolution effort is similar to non-AO models. Thus, when the inconsistency that is responsible for propagation is identified and resolved, all inconsistencies are indirectly resolved as well.

### b. Shared Join Points and Cyclic Propagation

We have noticed that when an inconsistency emerged in a highly coupled base module (e.g., a controller in Mobile Media), it led to a higher degree of inconsistency propagation in the AO versions than the non-AO versions. This problem was particularly observed when the highly coupled base module was the source of join point shadows shared by multiple aspects. For instance, we have analyzed the inconsistency channels triggered by an inconsistency arising in the *BaseController,* a central model element in the Mobile Media architecture. We observed that the inconsistency propagation affected four components in the non-AO version, namely *AlbumListScreen, PhotoListScreen, PhotoView Screen,* and *AddPhotoToAlbumScreen.* However, the propagation affected three additional modules (aspects) in the AO version.

The *HandleExceptions* interface had a method signature modified from *String[] getImages(String record-Name)* to *ImageData[] getImages(String record-Name).* However, the *R1.HandleExceptions* incorrectly overrides *Delta(R1,R2).HandleExceptions.* As a result, this method was incorrectly present into the output model, which gives rise to some functionality inconsistencies. This propagation was spread through the component *AlbumData*, because the aspect is no longer able to introduce the expected method *ImageData[] getImages(String record Name)* into the provided interface *ManagePhotoInfo* of *AlbumData*. Consequently, *AlbumData* does not provide any expected service to the environment. Hence, inconsistencies are propagated through the component *BaseController* and *ImageAcessor.*

It is interesting to note that *ImageAcessor* is also affected by an inconsistency that emerged from *AlbumData*. As *ImageAcessor* requires the service (*ImageData[] getImages(…)*) provided by the interface *ManagePhotoInterface*, it is not able to correctly provide the all services defined in the provided interface *PersistPhoto*. Hence, the *AlbumData* is also re-affected by an inconsistency that previously arose from it. This phenomenon represents the cyclic conflict propagation. On the other hand, this propagation is solved in the composition $R_{2,overide,left}$ due to the Delta*(R1,R2).HandleExceptions* override the `R1.HandleExceptions`, decreasing the inconsistency rate from 1.3 in R2,overide,right to 0.41 in R2,overide,left.

### 6.2.4.
### Limitations of Related Work

Model composition is a very active research field in many domains, including database integration (Bernstein & Melnik, 2007), composition of web services (Milanovic & Malek, 2004), merging of statecharts (Nejati et al., 2007) , model composition in product lines (Jayaraman et al., 2007), composition of UML models (Dingel et al., 2008; Clarke & Walker, 2005; Farias et al., 2010), aspect-oriented modeling (Whittle et al., 2009; Klein et al., 2006), and AO composition of models (Reddy et al., 2006; Cottenier et al., 2007). However, there is little related work focusing on the quantitative and qualitative assessment of AOM. In general, most of the research on the interplay of AOM and model composition rest on subjective assessment criteria. Even worse, they lead to dependence on experts who have built up an arsenal of mentally held indicators to evaluate the growing complexity of models in general (France & Rumpe, 2007; Lange et al., 2006a, Lange et al., 2006b). Consequently, the truth is that modelers ultimately rely on feedback from experts to determine "how well" the input models and their compositions can be. According to (Figueiredo et al., 2008), the state of the practice in assessing model quality provides evidence that modeling is still in the craftsmanship era and when we assess model composition this problem is accentuated.

More specifically, to the best of our knowledge, researchers have neglected the assessment of how aspects affect model composition effort. The need for assessing models during a model composition process has neither been pointed out nor proposed by current model composition techniques (Cottenier et al., 2008;

Nejati et al., 2007; Reddy et al., 2006; Apel et al., 2011; IBM RSA, 2011). For example, the UML built-in composition mechanism, namely package merge (OMG, 2011; Dingel et al., 2008), does not define metrics or criteria to assess the merged UML models. Moreover, it has been found to be incomplete, ambiguous, and inconsistent (OMG, 2011).

The lack of quantitative and qualitative indicators for model compositions hinder the understanding of side effects peculiar to certain model composition strategies (in the presence of aspects or not). Many different types of metrics have been developed during the past few decades for different UML models. These metrics have certainly helped designers analyze their UML models to an extent. However, as researchers' focus has shifted to the activities related to model management (such as model composition, evolution, and transformation), the shortcomings, and limitation of UML model metrics have become more apparent. Some authors (Fenton & Pfleeger, 1996; Lorenz & Kidd, 1994; Chidamber & Kemerer, 1994) have proposed a set of metrics that can be applied to measure UML models' properties. These works have shown that their measures satisfy some properties expected for good measures of design models. However, these metrics cannot be employed to assess problems that may arise in a model composition process such as semantic inconsistencies.

There are some specific metrics available in the literature for supporting the evaluation of model composition specifications. For instance, Chitchyan and colleagues (Chitchyan et al., 2009) have defined some metrics to quantify the effort to specific compositions between two or more requirements models, such as scaffolding and mobility. However, their metrics are targeted at evaluating the reusability and stability of explicit model composition specifications. Boucké and colleagues (Bouke et al., 2006) propose a number of metrics for evaluating the complexity and reuse of architectural model compositions. However, in this study, we have focused on the evaluation of heuristic composition algorithms, such as merge and override, where explicit model compositions are not provided up front. In addition, we have focused on analyzing the impact of aspects on the effort to resolve emerging inconsistencies in output models. Therefore, existing metrics (such as those described in (Chitchyan et al., 2009; Bouke et al., 2006)) cannot be directly applied to our context.

### 6.2.5.
### Threats to Validity

The exploratory study obviously has a number of threats to validity that range from (Wohlin et al., 2000): (i) the use of single target application and a single AOM language, to (ii) the use of specific metrics to compute the conflict resolution effort. Obviously, more investigations involving other case studies with compositions of larger UML models are required. We observed that the number of properties and details (i.e., granularity) of the model elements taken into consideration throughout the compositions affect directly the composition results. Consequently, it is necessary to observe that, to generalize our findings, other types of model with different levels of abstraction are needed to make further investigation.

Further empirical evaluations are indeed fundamental to confirm or refute our findings in other real-world design settings involving UML model compositions. However, it was never our goal to conduct a controlled study. Our investigation represents a first stepping-stone, where a number of initial findings can be used to drive the experimental designs of more controlled studies in the future.

### 6.2.6.
### Conclusions and Future Work

Model composition is one of the pillars of AOM, and it is an operation intended to be used in many software development activities. Hence, software designers naturally become concerned about the quality of the composed models. This study represents a first exploratory study to assess the potential advantage of aspect-orientation in reducing conflict resolution effort. In our study, model composition was used to express the evolution of architectural models along six releases of a software product line. Three canonical algorithms for heuristic model composition have been applied, and two of them were discussed in detail in this study. As expected, we found that the presence of aspects in input models improved modularization and, therefore, tended to better localize inconsistencies.

We have also observed: (i) a higher degree of obliviousness between base models and aspects led to a significant decrease of inconsistencies when compared to the non-AO model counterparts, and (ii) aspects with higher quantification were the cause of higher inconsistency rates in AO models. Another interesting finding was that, even in scenarios where the inconsistency rate of AO models was close to (or higher than) the inconsistency rate of non-AO models, conflict resolution effort was similar in AO and non-AO models. This means that the time spent in recovering output AO models from emerging inconsistencies is, at least, similar to non-AO models. All these findings were independent of the specific composition algorithms being assessed. These results provide some initial indication that aspect-orientation may alleviate conflict resolution effort.

We should point out that assessing the benefit of AOM in model composition is in its initial stage and there is little experience that can be used to determine the feasibility of current approaches. This study represents a first exploratory study that investigates the impact of aspects on conflict resolution effort. However, further empirical studies are still required to evaluate the impact of AO modeling on model composition in real-world settings. We also need to better understand if aspect orientation provides some gain or not: (i) when applied to other composition algorithms, and (ii) with respect to the time spent to identify the inconsistencies rather than the effort to resolving them. We hope that the issues outlined throughout the study encourage researchers to replicate our study in the future under different circumstances.

# 7
# Conclusions

This thesis addresses several limitations of the current literature with respect to empirical evaluation of model composition effort. An overall research question has been formulated to specify the scope of this thesis: How can the composition of design models be evaluated with respect to developers' effort? This overall question was further decomposed into four specific research questions (Section 1.3); the goal was to explicitly investigate specific dimensions of model composition effort. Even though many contributions have been presented in the previous chapters, overall conclusions need to be drawn and much work remains to be done. Therefore, this chapter: (i) summarizes the main topics studied (Section 7.1) to address our research questions, (ii) refines the contributions previously discussed (Section 7.2), and (iii) gives directions for future work (Section 7.3).

## 7.1.
## Summary

Model composition plays a pivotal role in many software engineering activities. Moreover, software modeling is increasingly becoming a collaborative work. However, a clear understanding of the effort required for composing design models is still a challenging task. Developers need to know how to quantify this effort and grasp the possible factors that influence it. To address these issues, a systematic evaluation approach for model composition effort and a range of empirical studies are crucial.

Most existing work on model composition proposes new composition techniques (Sarma et al., 2011; Epsilon, 2011; Whittle et al., 2009). In addition, as far as the assessment of such techniques is concerned, nothing has been done so that an evaluation framework for model composition can be proposed. Even worse, there is no empirical study aimed at understanding how certain software modeling factors affect model composition effort in practice. As a result, developers are left without any evaluation framework and practical knowledge about how to identify model composition problems and alleviate the developers' effort.

We believe that without practical knowledge derived from empirical investigations (rather than conflicting advice of evangelists (Norris & Letkeman, 2011)), it is not possible to realize well-informed improvements on techniques and strategies for model composition. It would be not possible, for example, to tame the side effects of the influential factors - such as the composition technique, the design decomposition, and model stability - more effectively. With this in mind, we investigate four research questions (Section 1.3) and confront the results collected from them. Thus, developers can be aware of the overall cost of composing design models and identify means to ameliorate this cost.

In this context, this thesis proposes a quality model (RQ1) derived from our experience of conducting a series of empirical studies. This quality model identifies three relevant factors: the model composition techniques, the design decomposition technique, and model stability. More importantly, the quality model identifies a series of quality notions, including semantic, syntactic, social, and so on. This framework for evaluating model composition has guided all empirical investigations performed in this thesis. We believe that this quality model also serves as a guideline for other researchers to select procedures and metrics while evaluating how the same or different influential factors affect the model composition. Given the unifying terminology of our quality model, it also enables to map, contrast, and bring together findings from different empirical studies on model composition effort.

After defining the quality model (RQ1), we started investigating the effects of specific model composition techniques on the developers' effort (RQ2). More specifically, we evaluate the effects of some specification-based and heuristic-based composition techniques on the developers' effort and the correctness of the output composed models. This evaluation is performed based on a set of empirical studies including one controlled experiment, five industrial case studies, observational studies, and interviews. The combination of these studies allows to build a body of knowledge about the effort that developers invest to compose design models. The results, supported by statistical analyses, contradict the intuition by disclosing that specification-based techniques neither reduce the developers' effort nor assure the correctness of the compositions when compared to the heuristic-based techniques.

Following the studies of the four research questions, we investigate the effects of alternative design decompositions (e.g., OOM and AOM) on the effort to detect inconsistencies (RQ3). We performed one controlled experiment, five industrial case studies, observational studies, and interviews to understand these effects. This allowed us to study RQ3 from different perspectives. The results, also supported by a complete statistical analysis, show that aspect-oriented modeling neither increased the inconsistency detection rate nor improve the interpretation of the models. However, developers invested less effort to detect inconsistencies in AO models than in OO models.

Lastly, we investigate the effort that developers spend to resolve inconsistencies (RQ4). For this, we study the influence of modeling languages and model stability on the inconsistency rate and on the effort to resolve these inconsistencies. From two quasi-experiments in the context of the evolution of design models, the results revealed that aspect-oriented design models had a higher inconsistency rate than non-AO ones. However, the inconsistency resolution effort required by AO models was lower than the OO models. The model stability has shown to be a good indicator of high density of inconsistency and resolution effort. That is, unstable models tended to present a higher inconsistency rate and require a higher effort to transform the output composed model into an output intended model. All results were supported by statistical tests.

## 7.2.
## Contributions

We claim that evaluation of model composition must not only be based on conventional design attributes. Model composition evaluation must be oriented by the effort that developers should invest to produce an output intended model. This research work defined an evaluation approach that promotes effort as an explicit measurement unit, thereby filling the gap between experimental investigations and the influential factors that affect the composition effort. Additionally, we applied this new evaluation approach in a series of empirical studies in order to evaluate the effects of the influential factors on: (i) the effort to apply composition techniques, (ii) the effort to detect inconsistencies, and (iii) the effort to resolve inconsistencies.

After investigating the four research questions in the previous Chapters, we refine the contributions of this work stated in Chapter 1.

1. *A quality model for model composition effort* (RQ1). As previously mentioned in Chapter 1, the central topic of this thesis is the empirical evaluation of effort on composing design models. Therefore, we first define quality notions for model composition effort to be applied in this thesis (Section 3.5.2). We selected and extended existing quality models for software modeling in the context of model composition. In total, seven quality notions were introduced in the proposed quality model, namely syntactic, semantic, social, effort, application, detection, and resolution. The syntactic, semantic, and social quality notions were tailored from the previous studies, while the effort, application, detection, and resolution quality notions were proposed in this thesis. We believe that these quality notions together are effective to comprise a basic quality model for model composition effort. The quality model was defined in four levels following a metamodeling approach. Its main practical contribution is to guide researchers and developers in two main contexts: (i) the adoption of a unifying terminology related to the evaluation of model composition effort – this adoption enables the comparison of different studies and their findings, and (ii) the selection of metrics for structuring empirical studies on model composition (Section 3.5.3). In fact, this model has driven all studies in Chapters 4, 5, and 6; we observed that this model was effective to support our evaluation of different facets of model composition effort through the empirical studies. For instance, the quality model was instantiated to select metrics as well as structuring the procedures required to evaluate how the influential factors affect model composition effort.

2. *Practical knowledge on model composition effort (RQ2,3,4).* To address RQ2, RQ3, and RQ4, we apply the quality model to assess the effects of the composition factors on the model composition effort. Empirical knowledge was reported from a series of experimental studies including: two controlled experiments, five industrial case studies, three quasi-experiments, more than fifty interviews, and observational studies. The chief contributions were practical knowledge about the impact of the influential factors on: (i) the effort to apply model composition techniques (Chapter 4), (ii) the effort to detect inconsistencies (Chapter 4 and 5), and (iii) the effort to resolve inconsistencies (Chapters 4, 5, and 6). Moreover, practical knowledge about how to: (i) evaluate the developers' effort, (ii) reduce the likelihood of emerging inconsistencies, and (iii) tame the side effects of the influential factors are defined in the previous Chapters 4, 5, and 6. An overview of the generated knowledge is emphasized as follows:

**Model Composition Techniques**

a) Developers tend to spend less effort by using the heuristic-based techniques rather than the specification-based techniques. In fact, the heuristic-based techniques required less effort to apply them, detect inconsistencies, and resolve inconsistencies. Consequently, the general composition effort invested by developers was lower. The traditional algorithms required less effort than the IBM RSA, which in turn required less than the Epsilon.

b) The specification-based technique did not reduce the inconsistence rate whereas also got higher measures than the heuristic-based techniques. Developers were not more effective to produce the output intended model by using the specification-based composition techniques. This finding did not confirm the claims reported in the current literature that such techniques significantly reduce the number of inconsistencies compared to the heuristic-based composition techniques (Epsilon, 2011; Kolovos et al., 2011; Kompose, 2011; Whittle et al., 2009). This finding indicates that developers should more carefully use specification-based techniques.

c) The specification-based techniques added undesired difficulties to specify the similarity between the input model elements. In particular, it was challenging for developers to proactively write down match and merge rules, which were able to produce an output intended model. Severe compositions dominated by relations of the type many-to-many (N:N) between the input model elements characterized the most effort-consuming scenarios. In short, the specification-based technique demonstrated to be a highly intensive manual task and more prone to errors. This leads to the insight that developers should be equipped with heuristics that, for instance, automatically recommend relations between elements of the input models.

d) The aforementioned results also lead to three lessons: (1) the model composition techniques should be more flexible to express different categories of changes; (2) the techniques should represent the conflicts between the input models in more innovative views and report them as soon as they arise; and (3) new composition techniques could be a mixture of specification-based and heuristic-based techniques.

a) **Design Decomposition Techniques**The technique used for design decomposition, such as object-orientation and aspect-orientation, definitely has a profound impact on model composition effort. For instance, developers tend to detect more inconsistencies in OO models than in their AO counterparts. Therefore, AO models explicitly representing crosscutting modularity do not necessarily imply on more effective inconsistency detection. This contradicts somehow the intuition that the improved modularity of AO models would help developers to localize inconsistencies. Therefore, developers of AO designs should be more conscious that the increased number of abstractions in AO models requires more attention from them while revising the output composed models.

b) Developers tend to invest more effort to detect inconsistencies in OO models than in AO models. In fact, developers tend to report more often the presence of inconsistency in AO models (compared to OO models) instead of trying to find any other solution. On the other hand, by using OO models, developers try to provide more often the corresponding implementation even observing the presence of inconsistencies. That is, the superior modularity of AO models accelerates inconsistency detection. Therefore, this implies that although developers detect fewer inconsistencies in aspect-oriented models, they spend less effort to localize them.

c) Developers localized more quickly inconsistencies in AO models when the scope of aspect pointcuts is narrow, thereby confronting structural and behavioral information about the crosscutting relations. This faster localization happened because the similarity between advices represented in structural and behavioral diagram allowed an "easy transition" between the two diagrams. This leads to the insight that developers should, whenever it is possible, avoid wildcards in their pointcuts and break them down in more explicit pointcut expressions. This strategy seems to improve the readability and consistency detection in AO models.

d) AO models with inconsistencies tend to cause a higher number of misinterpretations compared to the OO counterparts. The presence of the inconsistencies cause a detrimental effect due to the nature of the AO constructs. In fact, the need to scan all join points affected by the aspects increased the likelihood of different interpretations by developers. Therefore, we confirmed our initial expectation that by using contradicting AO design models would lead to a higher number of diverging interpretations of the participants. Therefore, developers working on parallel on aspect-oriented design should be more conscious about the increased likelihood of different design interpretations by the team members.

e) Developers tend to consider the sequence diagrams as the basis for the design implementation, as it is closer to the final implementation of the method (or advice) bodies; hence, developers become confident that the information present in the sequence diagram is the correct one compared to the class diagram. That is, the lower level of abstraction of this diagram leads the software developers to be more confident into the behavioral diagrams than the structural ones. Therefore, inconsistencies in behavioral diagrams tend to have a superior detrimental effect than those in class diagrams.

**Design Characteristics**

a) A number of design characteristics, such as coupling and size, play a role in the stability characteristic of an evolving design. We have observed that the inconsistency rate and the inconsistency resolution effort in stable design models are significantly lower than in unstable design models. The model stability has demonstrated to be a good indicator of inconsistency rate and inconsistency resolution effort. This also leads to the insight that developers should also invest upfront on applying well-known design principles to improve the stability of each new delta model to be composed. This is going to save cost involved in resolving critical inconsistencies later.

b) The location where the inconsistencies emerge is important. For instance, inconsistencies are more harmful when they take place in design model elements realizing mandatory features of software product lines. Because inconsistency propagation is often higher in model elements implementing mandatory features than in alternative or optional features. When inconsistencies emerge in elements realizing optional and alternative features they also tend to naturally propagate to elements realizing mandatory features. Consequently, the mandatory features end up being the target of inconsistency propagation. This observation further confirms the importance of structuring well key modules of a system in order to avoid instability and critical inconsistencies later.

c) Developers must structure product-line architectures in such a way that inconsistencies can keep precisely "confined" in the model elements where they appear. Otherwise, the quality of the products extracted from the SPL can be compromised; as the core elements of the SPL can suffer from problems caused by incorrect feature compositions. The higher the number of inconsistencies, the higher the chance of them to continue in the same output model, even after an inspection process performed by a designer. Consequently, the extraction of certain products can become error-prone or even prohibitive.

## 7.3.
## Future Works

This section categorizes the areas where future work is still required such as composition technologies, additional quality notions and heuristics, formal foundations, and additional empirical investigations. These areas are discussed below.

**Improvement of Model Composition Technologies**

We can highlight two main areas in which supporting tools would be pivotal to improve model composition in the context of real-world projects: support for improved awareness in collaborative model composition activities; and automated detection and resolution of inconsistencies.

First, it would be useful to investigate and develop model composition tools that support developers with awareness about model composition activities being performed in parallel. These tools should be able to make developers conscious about relevant changes in the design model elements. This improvement is important because developers should be able to identify conflicting changes earlier than the model composition time. Therefore, future work in this area will be focused on including support for "awareness" in model composition tools, such as IBM RSA and Kompose (Kompose, 2011).

Second, the current software modeling tools should support the anticipation, detection, and resolution of the most critical inconsistencies. Since, it is particularly challenging for developers to detect and resolve severe inconsistencies without any guidance (or recommendations) supported by tools. Therefore, as a future work in this direction, the model composition tools might incorporate, for instance, the use of model stability as an indicator of severe inconsistencies emerging in the output composed models. After the detection of inconsistencies, a recommendation system should assist the developers to resolve the inconsistencies.

**Additional Quality Notions**

The proposed quality model for model composition effort was defined based on the limitations of existing quality models and from empirical studies. A possible direction for future research related to the quality model is to go further in its application in different contexts. By doing so, new empirical studies might be planned and carried out to evaluate the quality model considering the different purposes of using model composition. In this thesis, the quality model was mainly evaluated in the context of changing and reconciling of deign models (Section 3.5.3), but the model may be applied to support the analysis of overlapping design models. In this context, quality notions such as social and effort quality should be investigated.

**Formal Foundations**

The specification of the metrics and the quality model in this thesis is informal. Therefore, we cannot state that their definitions are, for instance, mathematically sound and fully free of ambiguities. We believe that a formal foundation for the metrics and the quality model is a useful additional step in the future. For example, the metrics could be formalized using set theory and theoretically evaluated using systematically criteria from the measurement theory.

**Additional Empirical Investigations**

We can highlight at least two requirements for replications of the studies performed in this thesis.

First, even though the results of the studies ($RQ_{2,3,4}$) were statistically significant, the studies were limited with respect to the types of design models and inconsistencies analyzed. More types of inconsistencies and models should be analyzed in replications of our studies. This would allow us to confront the collected data with the new data. Another proper way to go is to investigate the effects of inconsistency propagation on the inconsistency detection rate, detection effort, and the degree of misinterpretation of the design model. In this study, we have observed that inconsistencies in AO models led to a superior misinterpretation compared to OO models. However, further studies should be performed to evaluate, for example, whether the inconsistencies are in fact converted into a higher number of implementation defects in AO programming rather than OO programming. That is, we are going to investigate if inconsistencies in design level are converted into defects in code. Moreover, it would be great to investigate the effects of key properties in AO modeling such as obliviousness and quantification on the inconsistency detection rate, detection effort, and misinterpretation.

Second, although the results (RQ2) were also statistically significant, the study considered small design models and a low number of subjects. Thus, the results may have been threatened by the size of the design models or by level of experience of the subjects. Therefore, future works might replicate the study by considering more experienced subjects and more complex design models.

# 8